# Imperial College London

# Mechanical Properties of La0.6Sr0.4Co0.2Fe0.8O3 Fuel Cell Electrodes

## by Zhangwei Chen

April 2014

Supervised by: Prof. Alan Atkinson, Dr. Finn Giuliani

## *to my beloved family*

*"Love is patient, love is kind. It does not envy, it does not boast, it is not proud. It does not dishonour others, it is not self-seeking, it is not easily angered, it keeps no record of wrongs. Love does not delight in evil but rejoices with the truth. It always protects, always trusts, always hopes, always perseveres. Love never fails."*

*1 Corinthians 13:4-8*

# Abstract

Mixed ionic-electronic conductive (MIEC) perovskite material $La_{0.6}Sr_{0.4}Co_{0.2}Fe_{0.8}O_{3-\delta}$ (LSCF6428) is a promising candidate for the cathode in intermediate temperature solid oxide fuel cells (IT-SOFCs). Understanding the three dimensional (3D) microstructural characteristics of such a material is crucial to its application because they predominately determine the performance and durability of the porous cathodes and hence of the SOFCs. They affect the overall cathode kinetics and thus the electrochemical reaction efficiency, as well as the mechanical properties, which dominate the lifetime of SOFCs. It is necessary to balance the trade-off between the electrochemical performance, which is improved by high porosity and minimal sintering, and the ability to withstand mechanical constraints, which is improved by the opposite.

To date LSCF6428 has been widely investigated on subjects of microstructure-related electrochemical performance, while little work has been reported on the mechanical properties and their correlation with the 3D microstructures. The main purpose of this research was to study the mechanical properties (i.e. elastic modulus, hardness and fracture toughness) of LSCF6428 cathode films and bulk samples fabricated by high temperature sintering, and to evaluate the effect of 3D microstructural parameters on elastic modulus, and the Poisson's ratio where applicable, by means of both experimental and numerical methods.

Room-temperature mechanical properties were investigated by nanoindentation of porous bulk samples and porous films sintered at temperatures from 900 to 1200 °C. A spherical indenter was used so that the contact area was much greater than the scale of the porous microstructure. The elastic modulus of the bulk samples was found to increase from 33.8 to 174.3 GPa and hardness from 0.64 to 5.32 GPa as the porosity decreased from 45 to 5 vol% after sintering at 900 to 1200 °C. Densification under the indenter was found to have little influence on the measured elastic modulus. The residual porosity in the nominally dense sample was found to account for the discrepancy between the elastic moduli measured by nanoindentation and by impulse excitation. Based on the optimisation of a commercial LSCF6428 ink formulation, crack-free films of acceptable surface roughness for indentation were also prepared by sintering at 900 to 1200 °C. It was shown that reliable measurements of the true properties of the films were obtained by data extrapolation provided that the effects from both surface roughness and substrate were minimised to neglected levels within a certain range of indentation depth to film thickness ratio (which was 0.1 to 0.2 in this study).

The elastic moduli of the films and bulk materials were approximately equal for a given porosity.

Based on the crack length measurements for Berkovich-indented samples, the fracture toughnesses of bulk LSCF6428 were determined to increase from 0.51 to 0.99 MPa·m$^{1/2}$, after sintering at 900 to 1200 °C.

The microstructures of films before and after indentation were characterised using FIB/SEM slice and view technique and the actual 3D microstructure models of the porous films were reconstructed based on the tomographic data obtained. Finite element modelling of the elastic modulus of the resulting microstructures showed excellent agreement with the nanoindentation results. The 3D microstructures were numerically modified at constant porosity by applying a cellular automaton algorithm based method, so that the influence on elastic modulus of factors other than porosity could be evaluated. It was found that the heterogeneity of the pore structure has a significant influence on the elastic properties computed using mechanical simulation.

# Acknowledgements

First of all, I would like to express my sincere appreciation to my two supervisors, Prof. Alan Atkinson and Dr. Finn Giuliani, who took me on as their student about three years ago, for their valuable guidance, encouragement and excellent advices throughout my PhD studies in Imperial College London. I would also particularly thank Prof. Alan Atkinson for his essential help during my period of applying PhD study at Imperial before I came, and for the wonderful opportunities he has later made possible for me to present my work in a number of national and international workshops and conferences. My special gratitude also goes to Dr. Xin Wang, for his kind support and discussions throughout my PhD research.

The colleagues and staff of the Department of Materials (particularly in the Fuel Cell Group and the Centre of Advanced Structural Ceramics) at Imperial College London are also acknowledged for their technical assistance: Miss Soo-na Lee and Mr Samuel Taub for help and discussions with lab work and material issues, Dr. Luc Vandeperre for discussion on finite element modelling and Dr. Vineet Bhakhri for extensive help in nanoindentation. Special thanks are due to Prof. Nigel Brandon and Dr. Farid Tariq from the Department of Earth Science and Engineering of Imperial, for providing computing software and facilities and valuable discussions on simulation work. The help and training received from staff in materials characterisation laboratory for SEM, XRD, FIB and sample preparations are also appreciated. My gratitude also goes to some of my friends who are currently or were previously studying or working at Imperial, for their constant spiritual encouragement.

The Chinese Scholarship Council, Imperial College London and Department of Materials are gratefully acknowledged for financial support of my life and studies here. Supergen Consortium is also thanked for funding the project.

Last but not least, my deepest gratitude goes to my dearest wife, parents and brother, and to my brothers and sisters in Chinese Church in London and Lubiao Fellowship, without whose love, encouragement, support, sharing and prayers this work would not have been possible.

# Declaration of Originality

I herewith declare that the work in this thesis is my own, and that all else has been appropriately referenced.

Zhangwei Chen

# Declaration of Copyright

# Table of Contents

# List of Figures

# List of Tables

# List of Abbreviations

| | |
|---|---|
| 2D | two dimensional |
| 3D | three dimensional |
| AFC | alkaline fuel cell |
| BEI | backscattered electron imaging |
| CA | cellular automaton |
| CGO | gadolinium-doped ceria |
| DA | degree of anisotropy |
| ECD | equivalent circular diameter |
| EDX | energy dispersive X-ray analysis |
| ESD | equivalent spherical diameter |
| FE | finite element |
| FIB | focused ion beam |
| GIS | gas injection system |
| HFW | horizontal field of view |
| IET | impulse excitation test |
| IM | indentation microfracture |
| IT-SOFC | intermediate temperature solid oxide fuel cell |
| LSC | $La_{1-x}Sr_xCoO_{3-\delta}$ |
| LSCF | lanthanum strontium cobalt iron oxide ($La_{1-x}Sr_xCo_{1-y}Fe_yO_{3-\delta}$) |
| LSCF6428 | $La_{0.6}Sr_{0.4}Co_{0.2}Fe_{0.8}O_{3-\delta}$ |
| LSCF-$i$ ($i$=9~12) | LSCF6428 film sintered at $i\times100$ ˚C |
| LSCF-$i$M ($i$=9~12) | LSCF-$i$ with computationally modified microstructure |
| LSM | lanthanum strontium manganite |
| MCFC | molten carbonate fuel cell |
| MIEC | mixed ionic-electronic conductivity |
| MSA | minimum solid area |
| OSG | organo-silicate glass |
| PAFC | phosphoric acid fuel cell |
| PEMFC | polymer electrolyte membrane fuel cell |
| PSD | particle size distribution |
| RA | Ramakrishnan and Arunachalam model |
| RVE | representative volume element |
| RF | reaction force |
| RT | room temperature |
| SEI | secondary electron imaging |
| SEM | scanning electron microscopy |
| SOFC | solid oxide fuel cell |
| TEC | thermal expansion coefficient |
| TEM | transmission electron microscopy |
| TPB | triple phase boundary |
| TPBL | TPB length |
| TGA | thermogravimetric analysis |
| VOI | volume of interest |
| XRD | X-ray diffraction. |
| YSZ | yttria-stabilised zirconia |

# 1  Introduction

This Chapter is intended to introduce the background of the project, i.e. the fundamentals of fuel cell and solid oxide fuel cell (SOFC) technology, the actual issues this study will address, and the aims of the study.

## 1.1  Fuel Cell Technology

Fuel cells are energy conversion devices which produce electricity directly from the electrochemical reactions of fuels and air (oxygen) without intermediate mechanical or combustion processes, which makes them a promising power generation technology with high efficiency and low environmental impact. They are quite different from conventional batteries since they do not store energy or require recharging [2]. Fuel cells can have various architectures, such as planar, tubular and monolithic. However they are generally made up of three segments which are sandwiched together: anode, electrolyte and cathode. Two chemical reactions (usually reduction of oxygen and oxidation of fuel) occur at the interfaces of the three different segments. The net result of the two reactions is that fuel is consumed, water or carbon dioxide is created, and an electric current is created, which can be used to power electrical devices. Fig. 1-1 shows the working principle of an individual fuel cell [1]. Advantages of fuel cell systems compared with other power generation systems include: higher energy conversion efficiency, much lower pollutant production, minimal sitting restriction, no moving parts, quiet and safe operation, long-term stability, and relatively low cost [3].

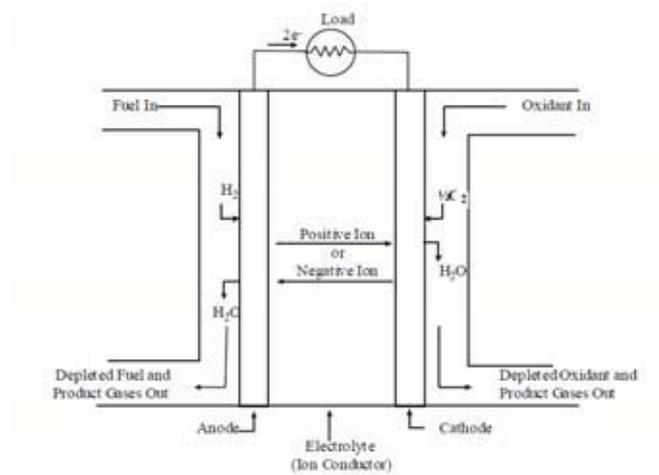

*Fig. 1-1 Schematic of an individual fuel cell, from [1].*

Fuel cells are classified according to the choice of electrolyte and fuel, which in turn determine the electrode reactions and the type of ions that carry the current across the electrolyte. A variety of fuel cell systems have been developed since they were first invented. Table 1-1 lists their electrolytes, oxidant, fuel and operating temperature [2, 4].

*Table 1-1 Fuel cell types and operating conditions*

| Fuel Cell Type | Oxidant | Flow Direction | Fuel | Operation Temperature (°C) |
|---|---|---|---|---|
| Solid Oxide (SOFC) | $O_2$ (air) | $O^{2-}$ cathode→anode | Any fuel | 500-1000 |
| Molten Carbonate (MCFC) | $O_2$ (air) | $CO_3^{2-}$ cathode→anode | Hydrocarbons | 650 |
| Phosphoric Acid (PAFC) | $O_2$ (air) | $H^+$ anode→cathode | Pure $H_2$ (hydrocarbons) | 200 |
| Polymer Electrolyte Membrane (PEMFC) | Pure $O_2$ | $H^+ + H_2O$ anode→cathode | Pure $H_2$ | 60-100 |
| Alkaline (AFC) | Pure $O_2$+$H_2O$ | $OH^-$ cathode→anode | Pure $H_2$ | 60-120 |

## *1.2  Solid Oxide Fuel Cells*

The fuel cell type of interest in this report is the SOFC, where the fuel used is typically hydrogen ($H_2$). A schematic of an SOFC is given in Fig. 1-2.

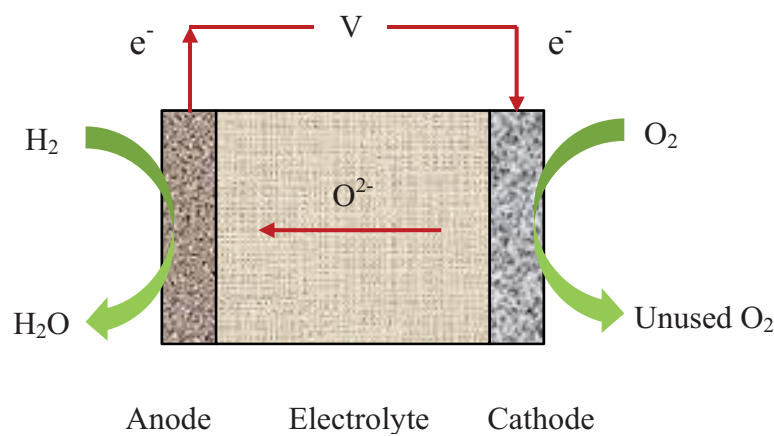

Fig. 1-2 Schematic representation of a SOFC operation

The electrolyte is a dense ceramic solid that exhibits high oxygen ionic conductivity and low electronic conductivity. Either side of the electrolyte is a porous electrode exhibiting mixed ionic-electronic conductivity. Oxygen reduction occurs at the cathode and the ionic species pass through the electrolyte to the anode where the fuel is oxidised [3]. The chemical energy of the fuel, $H_2$, is converted into electrical power resulting in water being produced as waste. The reactions are given in Equations 1.1-1.3 [5].

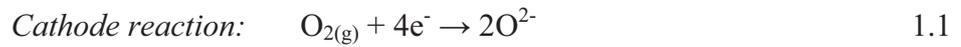

*Cathode reaction:* $\qquad O_{2(g)} + 4e^- \rightarrow 2O^{2-}$ $\qquad\qquad$ 1.1

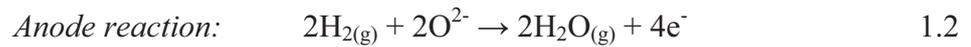

*Anode reaction:* $\qquad 2H_{2(g)} + 2O^{2-} \rightarrow 2H_2O_{(g)} + 4e^-$ $\qquad\qquad$ 1.2

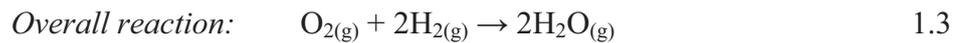

*Overall reaction:* $\qquad O_{2(g)} + 2H_{2(g)} \rightarrow 2H_2O_{(g)}$ $\qquad\qquad$ 1.3

## 1.3 Importance of Mechanical Properties of Cathodes and Conflict with Electrochemical Performance

As the central interest of this study, an SOFC cathode material must achieve and sustain desirable mechanical properties (elastic modulus and fracture toughness for example) after fabrication and during application. Cathodes produced with lower strength are likely to fail under mechanical stresses [6], which can arise from: manufacturing (residual stresses); thermal expansion coefficients difference; temperature gradients; oxygen activity gradients; and external mechanical loading [7]. An operating cathode is exposed to a gradient of oxygen pressure, where damage including cracks and fractures may occur if the mechanical strength is not high enough, which also results in degeneration of electrochemical properties. Moreover, as the cathode layer is joined together with the electrolyte layer or interlayer, poor interfacial adhesion can also induce delamination.

Both the mechanical properties and electrochemical performance of a cathode are basically controlled by its microstructural properties, as well as being directly related to the bulk properties of the mixed conductor, such as vacancy concentration, vacancy diffusivity, surface area, oxygen surface exchange kinetics and connectivity of the mixed conductor. The microstructure affects the overall electrode kinetics and thus the electrochemical reaction efficiency [8], as well as the mechanical properties, which dominate the durability and thus the lifetime of SOFCs, but in different ways. It is therefore necessary to balance the trade-off between the electrochemical performance, which is improved by high porosity resulting from minimal sintering, and the ability to withstand mechanical constraints, which is improved by the opposite, as depicted in Fig. 1-3.

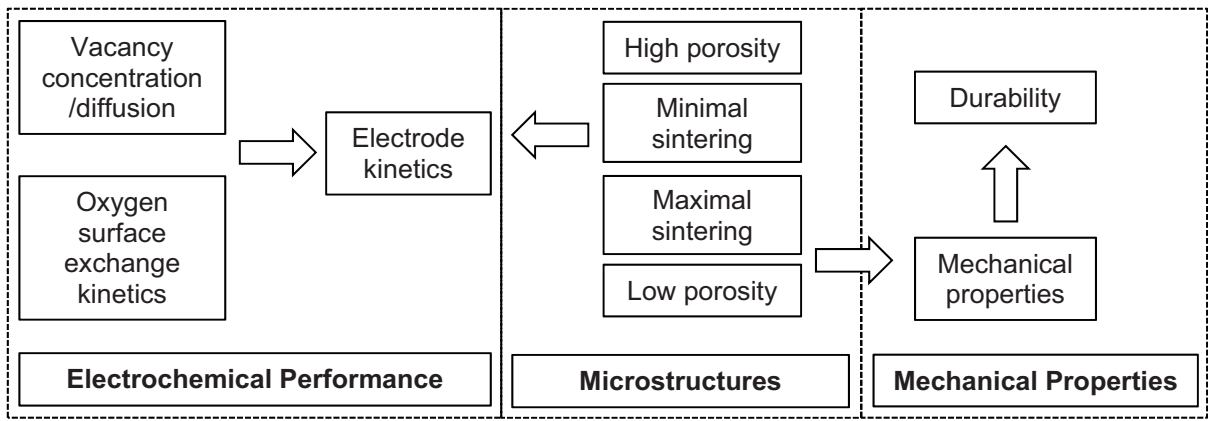

*Fig. 1-3 Inter-relationships between microstructures, mechanical properties and electrochemical performance of SOFC electrodes.*

## 1.4  Research Objectives and Thesis Outline

Mixed ionic-electronic conductive perovskite material $La_{0.6}Sr_{0.4}Co_{0.2}Fe_{0.8}O_{3-\delta}$ (LSCF6428) is a promising candidate for the cathode in intermediate temperature solid oxide fuel cells (IT-SOFCs). To date LSCF6428 has been widely investigated on subjects of microstructure-related electrochemical performance, while little work has been reported on the mechanical properties and their correlation with the 3D microstructures. The main purpose of this research was to study the mechanical properties (i.e. elastic modulus, hardness and fracture toughness) of porous thin LSCF6428 cathode films and bulk samples fabricated by high temperature sintering, and to evaluate the effect of 3D microstructural parameters on elastic modulus and the Poisson's ratio by means of both experimental testing such as nanoindentation and numerical approaches using FIB/SEM tomography-based 3D reconstruction and finite element modelling.

The thesis outline is detailed as follows: the first chapter is the introduction which includes the research background and objectives. A comprehensive literature review is made in the second chapter about the current perovskite materials to be used as cathodes in SOFCs. Recent development of the mechanical and microstructural characterisations of partially sintered ceramic bulk and films is also summarised. Particular attention is paid to LSCF6428 perovskite material and the porous cathodes made thereby. Chapter 3 describes the methodologies and techniques employed in this study, among which nanoindentation and FIB/SEM tomography-based 3D reconstruction are emphasised.

Five chapters are devoted to the results and discussion in this thesis. Chapter 4 investigates important physical properties such as density, porosity, and 2D microstructures of the LSCF6428 films and bulk specimens as well as CGO bulk specimens. Efforts are made to reduce the defects in the porous LSCF6428 films fabricated, so that the films are morphologically suitable for nanoindentation tests. In Chapter 5 and Chapter 6, room-temperature mechanical properties including elastic modulus, hardness and fracture toughness were investigated respectively mainly by indentation of porous bulk samples and porous films sintered at temperatures from 900 to 1200 °C. It shows how reliable measurements on porous ceramic films can be made by appropriate nanoindentation experiments and analysis. Possible influence of microstructures, particularly porosities, on the mechanical characterisations was evaluated. FIB/SEM slice and view technique was applied to examine the subsurface microstructural changes, such as plastic deformation, crack morphologies, and film/substrate interface condition. Chapter 7 deals with the reconstruction of actual 3D microstructures of the films based on FIB/SEM tomography and FEM of these actual models. Problems encountered during image acquisition, processing and analysis are addressed. Quantification and analysis of the important microstructural parameters for the solid and pore phases in each type of film microstructure is conducted using advanced 3D quantification packages. The elastic moduli computed by mechanical simulation using FEM of the actual 3D microstructures agreed well with the nanoindentation-determined data. Finally in Chapter 8, in order to evaluate the factors other than porosity influencing the elastic properties of the microstructures, mechanical simulation and the quantification of relevant microstructural parameters were carried out both on the digital film microstructures and the artificially created microstructures which were numerically modified at constant porosity by applying a cellular automaton algorithm based method.

The last chapter summarises the main results and findings and the major conclusions of the study. Future work on further exploration of the microstructure modification evaluation is also suggested.

## *Chapter 1 References*

1.      EG&G Technical Services Inc.: **Fuel Cell Handbook (Seventh Edition)**; 2004.
2.      Steele BCH: **Materials for IT-SOFC stacks: 35 years R&D: the inevitability of gradualness?** *Solid State Ionics* 2000, **134**(1-2):3-20.
3.      Dusastre V, Kilner JA: **Optimisation of composite cathodes for intermediate temperature SOFC applications**. *Solid State Ionics* 1999, **126**(1-2):163-174.
4.      Larminie J, Dicks A: **Fuel Cell Systems Explained**, 2nd edn: Wiley; 2003.
5.      Lashtabeg A, Skinner SJ: **Solid oxide fuel cells - a challenge for materials chemists?** *Journal of Materials Chemistry* 2006, **16**(31):3161-3170.
6.      Bellon O, Sammes NM, Staniforth J: **Mechanical properties and electrochemical characterisation of extruded doped cerium oxide for use as an electrolyte for solid oxide fuel cells**. *Journal of Power Sources* 1998, **75**(1):116-121.
7.      Atkinson A, Selçuk A: **Mechanical behaviour of ceramic oxygen ion-conducting membranes**. *Solid State Ionics* 2000, **134**(1-2):59-66.
8.      Adler SB: **Electrode Kinetics of Porous Mixed-Conducting Oxygen Electrodes**. *J Electrochem Soc* 1996, **143**(11):3554-3564.

# 2 Literature Review

## 2.1 Perovskite-structured Materials

Some of the perovskite-structured materials have been widely used as SOFC cathodes. It is important to first understand the fundamentals of the perovskite crystal structure for a better design and optimisation of the properties and performance of the cathode components.

### 2.1.1 Perovskite Crystal Structures

Perovskite-structured materials, which were named after the mineral $CaTiO_3$, are a family of oxides with a general composition $ABO_3$, in which there is one formula unit per unit cell and A and B are cations with a total valence of +6 [1]. An A-site cation located at the corners of the unit cell, is an alkali, an alkaline earth or a rare earth cation (such as La, Sr, Ca and Ba, etc.) with lower valence possessing larger volume and coordinating with twelve oxygen anions while B is a transition metal cation (such as Co, Ni, Fe, and Mn, etc.) located in the centre which is smaller and has six-coordinated oxygen anions. Fig. 2-1 shows a typical and idealised structure of a close packed cubic perovskite $ABO_3$.

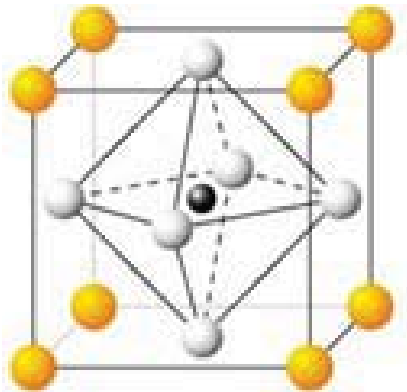

*Fig. 2-1 Schematic of idealised perovskite unit cell [2]. Yellow spheres represent A-site cations, dark sphere represents B-site cations and gray spheres represent oxygen anions*

Most perovskite structures are distorted and do not possess the above typical cubic unit cell form. Common distortions such as cation displacements within the octahedra and tilting of the octahedra are related to the properties of the A and B substituted atoms. This cubic form can only appear in the case when the Goldschmidt tolerance factor $t$ related to the ratio of their atomic radii as shown in Equation 2.1 satisfies the condition $1.0 < t < 0.75$ [3], the degree of distortion in $ABO_3$ perovskites can be determined according to $t$ as follows:

$$t = \frac{R_A + R_O}{\sqrt{2} \cdot (R_B + R_O)} \hspace{3cm} 2.1$$

where $R_A$, $R_B$, and $R_O$ are the effective ionic radii of A-site, B-site, and oxygen ions, respectively.

## 2.1.2 Mixed Ionic-Electronic Conductivity

Perovskite materials exhibit many interesting properties such as ferroelectricity, magnetoresistance, mixed ionic-electronic conductivity and superconductivity. The most significant feature employed for the application of SOFC cathodes is the mixed ionic-electronic conductivity.

The partial substitution of A-site cations by other metal cations with lower valences often brings about the formation of oxygen vacancies, which results in ionic conductivity. The concentration of oxygen vacancies can also be adjusted by substituting ions of similar sizes but lower valences at B sites. As a result, the materials have not only high ionic conductivity because of the high concentration of oxygen vacancies but also good electronic conductivity because of the mixed-valence states [4].

## 2.2 Lanthanum-based Cathode Materials

Over the last two decades, lanthanum-based perovskite materials with general formula $La_{1-x}Sr_xCo_yFe_{1-y}O_{3-\delta}$, denoted as LSCF, have been extensively studied due to their promising mixed ionic-electronic conductivity [5] and high oxygen surface exchange rate [6] for applications in cathodes for intermediate-temperature solid oxide fuel cells (IT-SOFCs) [7] and oxygen separation membranes [8].

## 2.2.1 LSC and LSM

Early cathode materials including lanthanum cobaltite ($LaCoO_3$) [9-11], and lanthanum manganite ($LaMnO_3$) [9, 12-14] based perovskites have been found to have poor electrochemical performance. However, this can be significantly improved by A-site doping with an alkaline earth like Sr in the perovskite structures. The ionic conductivity of $La_{1-x}Sr_xCoO_{3-\delta}$ (LSC) is very high (0.004-0.1 $S.cm^{-1}$), whereas the $La_{1-x}Sr_xMnO_{3-\delta}$ (LSM) has electronic conductivity of 300 $S.cm^{-1}$ but poor ionic conductivity (0.8-5×10$^{-7}$ $S.cm^{-1}$) [15, 16].

An attractive SOFC cathode candidate material requires not only high mixed ionic-electronic conductivity for excellent electrochemical performance purpose, but also appropriate compatibility with the adjacent electrolyte within the SOFC. LSC has very high thermal expansion coefficient (TEC $\geq 20 \times 10^{-6}$ K$^{-1}$), especially at high temperatures, due to oxygen loss and resulting cation repulsion, as well as increases in B-site ionic radii due to temperature- and charge compensation- induced reduction of $Co^{4+}$, $Co^{3+}$ [17]. This can cause serious thermomechanical issues such as cracking when the material is used alongside existing electrolytes with very different thermal expansion coefficients. Substitution of Fe on B-site in this material has been proven to significantly reduce the TEC, which made the perovskite $La_{1-x}Sr_xCo_{1-y}Fe_yO_{3-\delta}$ (LSCF) one of the most attractive cathode materials for SOFC [7, 17, 18].

### 2.2.2 Introduction to LSCF and LSCF6428

The discovery of significant oxygen ionic conductivity of LSCF by Teraoka *et al.* in the late 1980s [8, 19], has made this perovskite family potential candidates as oxygen separation membrane materials [8] and mixed-conducting cathodes for SOFCs [7].

### 2.2.2.1 Crystal Structures

Most LSCF compositions do not take the idealised cubic perovskite structure at room temperature. Extensive studies on the important properties of LSCF have been carried out by Tai and co-workers in 1995 [7, 17]. They characterised the crystal structures of the LSCF perovskite family and revealed that the degree of rhombohedral distortion (from an ideal cubic perovskite) decreased and the crystal structure changed toward cubic as x > 0.6. At high Fe and/or low Sr contents, the orthorhombic structure is preferable due to the Goldschmidt factor. However, addition of Sr increases the stability region of the rhombohedral phase to higher Fe contents[17]. At high Sr concentrations, cubic perovskite phases dominate [7]. Temperature also influences the stability of different structures; for example, $La_{0.6}Sr_{0.4}Co_{0.2}Fe_{0.8}O_3$ (LSCF6428) has a rhombohedral perovskite structure at room temperature [7, 20], with a transition to cubic perovskite occurring at approximately 773K [20].

### 2.2.2.2 Electrical Conductivities

Tai *et al.* [7, 17] examined LSCF electronic conductivity versus temperature and showed that the electronic conductivity increased to a maximum then decreased as the temperature increased, as shown in Fig. 2-2. The magnitude of the observed conductivity maximum increased and it shifted to lower temperatures with increasing Sr concentration. The maximum electronic conductivities ranging from 200-330 S.cm$^{-1}$ for La$_{1-x}$Sr$_x$Co$_{0.2}$Fe$_{0.8}$O$_3$ compositions with $x$ = 0.2-0.4 were found in the temperature range of 600-800°C in air. Furthermore, LSCF6428 had the highest electronic conductivity with the peak value of 330 S.cm$^{-1}$ at 550°C [7], suggesting it as a promising cathode material.

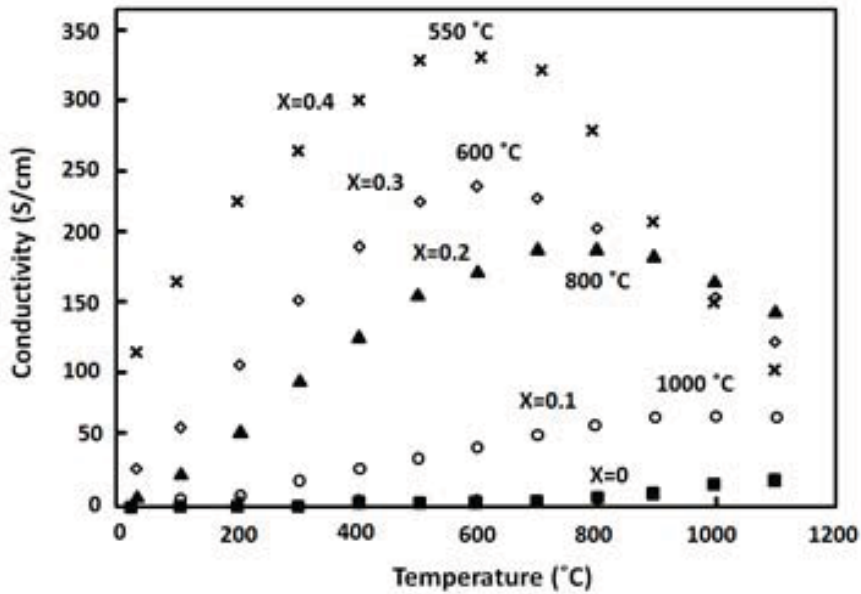

*Fig. 2-2 Electronic conductivities of La$_{1-x}$Sr$_x$Co$_{0.2}$Fe$_{0.8}$O$_3$ as functions of temperature and Sr content (mol) in air, after [7].*

### 2.2.2.3 Thermo-mechanical Properties

As mentioned earlier, B-site doping of Fe with Co can largely reduce LSC's TEC, by as much as 40% [17]. This allows the resultant LSCF compositions to be used with existing electrolytes, such as yttria-stabilised zirconia (YSZ) and gadolinium-doped ceria (CGO) [6]. Tai *et al.* [7] also found that the rhombohedral (high temperature) phase always exhibited a higher TEC than the orthorhombic phase. It appeared that the TEC had little dependence on Sr content but increased with Co content. The TEC of LSCF6428 in the temperature range of 373-973K was found to be 15.3×10$^{-6}$ K$^{-1}$, which was below the average value for LSCF materials with different amounts of Sr-doping.

The thermal cycle stability and cell lifetime of SOFCs can be significantly degraded if there is a large thermal expansion mismatch between LSCF cathodes and the conventional YSZ electrolytes [21]. Moreover, most cobalt-containing cathode materials are known to react with zirconia-based electrolytes. Therefore, the use of LSCF as cathode materials in conjunction with YSZ electrolyte produces highly resistive zirconates, such as SrZrO$_3$ [22, 23]. The introduction of CGO as an additional interlayer between the LSCF and YSZ interface has been proved to effectively mitigate these problems [24, 25]. In our current study, CGO pellets were made to be substrates for the ink deposition of LSCF6428.

### 2.2.2.4 Ionic Conductivity

To initiate the electrochemical reaction with the fuel flux on the anode side, oxygen molecules absorbed by cathode must be first catalysed to be anions and continue to permeate through cathode and electrolyte materials. Therefore a porous ionic conductive cathode is desired. Oxygen anion vacancies can be created by doping with aliovalent cations (such as Sr-doping for LSCF), which results in extrinsic oxygen anion vacancies and oxygen ion conductivity of materials. LSCF materials have excellent potential for oxygen permeability [8, 26], which makes LSCF possible cathode candidates for SOFC. Furthermore, the oxygen permeation rate through LSCF increased with the Sr or Co doping content, suggesting that the oxygen vacancy concentration within LSCF was the main factor controlling the oxygen permeability [19].

In another work by Teraoka $et$ $al.$ [5], ionic and electronic conductivities of LSCF were measured. They reported that LSCF materials were good mixed conductors with ionic transport number $10^{-2}$ - $10^{-4}$. At 1073 K, for instance, ionic conductivity ranged in the order of 1-$10^{-2}$ S.cm$^{-1}$ while electronic conductivity was around $10^2$ S.cm$^{-1}$. On the other hand, Anderson $et$ $al.$ [27] found that upon reduction LSCF ($0 < x < 0.4$) at 800 °C achieved a maximum value of ionic conductivity of about 0.3 S.cm$^{-1}$.

## 2.3 Elastic Properties of Porous Ceramics Partially Sintered from Powders

The classical ceramic processing route generally involves the consolidation of raw powders by sintering at elevated temperatures which leads to desired porosity for obtaining either key application parameters which increase with increasing porosity such as surface area, or better mechanical performance which can be improved by decreasing porosity (i.e. increased sintering degree) such as strength and elastic modulus. Fabrication methods other than partially sintering of powder, e.g. melt casting, sintering with pore former additives, colloidal processing and lamination etc., would often result in very different microstructure architectures (such as tubular pores resulting from injection moulding and laminar character by rapid prototyping [28]) so that the microstructure dependence of elastic properties would vary significantly, and hence they are out of the scope of the current study and are not discussed here.

The elastic properties of partially sintered two-phase (i.e. pore and solid phases) porous ceramics depend on the microstructural parameters of the specimens. Relevant aspects include porosity and geometrical characteristics of the pores, solid phase and the inter-particle necks. Their influences have been widely studied [29-34], despite the fact that the property-microstructure correlation is very complex and there may not have precise definition of some of the features due to the complexity of real microstructures where pores or particles are highly interconnected and vary within a given body so that challenges arise when distinguishing the beginning and end of individual pore/particle [35].

In order to correlate porosity to properties of porous materials, a variety of modelling approaches have been developed, including mainly the following three categories, and most of which treated the porous solids as two-phase microstructures with the second phase being pores in terms of estimating the effective elastic modulus [30, 31]:

(i) *Empirical/semi-empirical methods*: A number of equations, either empirical or semi-empirical, have been developed based on fitting of experimentally measured data, generally with the porosity being the main variable accompanied by one or two material-dependent fit constants, such as zero-porosity theoretical elastic modulus. More recently a thorough overview was given by Willi et al. [36] on most modulus-porosity relations ever proposed, of which the theoretical background, structure and merit were systematically discussed and compared. Examples of these commonly used expressions include the linear relationship first developed by Fryxell and Chandler [37] for polycrystalline BeO modulus-porosity analysis; a simple exponential relationship proposed by Spriggs [38] for fitting experimentally measured modulus of ceramics; and some other non-linear expressions, such as the Coble-Kingery non-linear relation [30] for solids containing isolated pores, the Phani-Niyogi power-law relation [39, 40] to better describe the modulus-porosity data of porous solids over a wide range of porosity and a non-linear relation proposed by Hasselman [41] to overcome the drawback of Spriggs form's resulting modulus not being 0 when porosity reaches 100 %. However, most of the relations have been rigorously derived only for porous materials with dilute distributions of spherical pores and few of them were related to non-porosity parameters of the microstructures. Moreover, although the models above could be useful in the way of predicting properties within the range of reasonable extrapolation, the understanding of the mechanisms affecting properties was limited.

(ii) *Mechanistic/micromechanics based methods* [42]: This category includes the differential [43] and self-consistent methods [44, 45], which focus more on rigorous evaluation of the specific mechanisms, e.g. mechanics, for the porosity effects on the specific property. They generally extend experimentally measured data for microstructures containing a small porosity to higher porosities. Although more insight into mechanisms can be gained by these models, their predictive capabilities are quite limited due to the rigorously idealised pore characters of spheroidal shapes (e.g. spherical or ellipsoidal) with unspecified stacking. They typically can only treat a very limited number of cases in a narrow way which is quite different to that for real microstructures.

(iii) *Geometry-based methods*: particularly those based on the minimum solid area (MSA) [46, 47] by purely geometrical reasoning to approximate the porosity dependence of elastic moduli using load-bearing concepts, of the form $E/E_0 = e^{bP}$, where $E_0$ is the dense solid elastic modulus, $b$ is a geometrical constant and $P$ the porosity. However, the expression applies strictly to microstructures with stacking of spherical pores or particles although it has general prediction ability over a range of porosities. In addition, the common drawback of approaches described in categories (ii) and (iii) is that the microstructure corresponding to a particular formula/prediction is not definitely known. Therefore, agreement or disagreement with data can neither confirm nor reject a particular model [48].

Besides these methods mentioned above, the elastic properties could also be precisely estimated, in theory, by computationally solving the equations of elasticity for numerical microstructure models [49]. However, in practice the accuracy of results is greatly limited by large statistical variations and insufficient resolution, which also need further investigation and will be a major concern of our current study.

Moreover, as mentioned above, there exists a large amount of literature dealing with porosity dependence of elastic modulus of porous solids and results reported show that the ultimate significance of the dependence of modulus on porosity is self-evident. Particularly the elastic modulus of a two-phase (solid/pore) porous microstructure relies primarily on the pore volume fraction of the microstructure. It is therefore not the aim of this study to use these formulas to estimate/predict elastic modulus of porous solids. But rather, its concern was more specifically that how factors other than porosity would affect the elastic modulus and the Poisson's ratio. In other words, how the elastic modulus would behave upon the modification of porous microstructures at constant porosity, because the scatter of mechanical property measurements are also related to the effects of heterogeneity and anisotropy of the porous specimens, which have been however almost universally neglected. This will be further discussed in subsequent chapters.

## *2.4 Measurement of Mode I Fracture Toughness for Partially Sintered Ceramic Films Using Indentation Microfracture (IM) Method*

Mode I fracture toughness ($K_{Ic}$) is a material property which describes the ability of a material containing a pre-existing crack to resist the crack propagation for fracture in mode I. It is therefore an important parameter to know for a material for the prevention of fracture failure since the occurrence of cracks and defects is not completely avoidable during processing, fabrication or application [50].

The indentation microfracture (IM) method [51] (which will be detailed in Chapter 3), using either nanoindentation or microindentation where applicable, is based on the estimation of radial cracking taking place on the sample surface due to the material deformation caused by the loading of a sharp indenter. It is easy and straightforward to use but sometimes is unreliable due to the difficulty of crack identification and crack-length measurement. In spite of serious criticisms to the IM method for the determination of fracture toughness [52], it is still widely used as a simple, low-time consuming and inexpensive method.

Fig. 2-3 (a) and (b) show respectively the isometric schematics of the common three-dimensional crack systems (marked in blue colour) and deformed zones (in brown) beneath the residual imprints following indentation on tested samples with a Berkovich tip (radial/Palmqvist cracks) and a Vickers tip  (median or half-penny cracks) (after [53]).

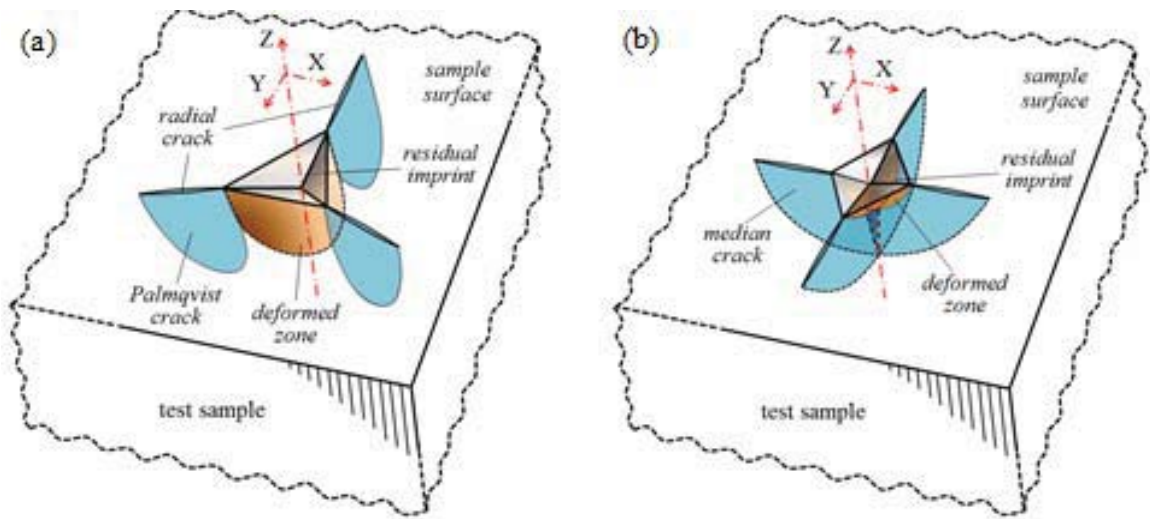

*Fig. 2-3 Isometric schematics showing three-dimensional crack systems and deformed zone beneath the residual imprint resulted from loading on the test sample with (a) a Berkovich tip - radial/Palmqvist cracks) and (b) a Vickers tip - median or half-penny cracks (after [53]).*

For accurate fracture toughness estimations, well-defined indentation-induced cracks are a prerequisite. In addition, for better cracking behaviour, the indents and cracks should not have sizes comparable to grains, but either much smaller or larger than the grain sizes, as found by Anstis *et al.* [54]. Larger grains can act as bridges behind the crack tip and contribute to the toughness, or even give rise to *R*-curve behaviour, as observed on coarse-grained alumina [55] and silicon nitride [56].

Recently increasing attention has been given to the determination of film fracture toughness using nanoindentation following the IM method using sharp indenters. Most of the previous work in the literature was based on dense thin coatings [51, 57-60] in the absence of porosity, for which the morphologies of indentation-induced cracks varied a lot, ranging from annular cracks to interfacial delamination. In addition, particularly extensive work has been reported regarding fracture in the nano-porous low-*k* thin films [61-64], but little has been reported for micro-porous films. In comparison to the longer crack lengths generated in thicker films at the same load, Morris *et al.* [61] have ascribed the shortened crack lengths for thinner films to the strain energy loss to the substrate during Berkovich indentation.

It is important to emphasise that the indentation-based fracture toughness calculations (as detailed in Chapter 3) were all established using stress analysis. Despite the possible existence of the similar crack patterns in both porous bulk specimens and films, it is worth noting that the toughness equation was developed for bulk materials and therefore its applicability for assessing film fracture toughness remains disputable [63]. Nevertheless, it has been directly used without necessary verification to assess the thin layer fracture toughness by several workers, as reported in [58, 61, 62, 65].

Some researchers have also developed energy-based approaches to determine fracture toughness by indentation of thin coatings on substrates. One energy-based model was initially proposed by Li *et al.* [57, 66] based on extrapolating the loading curve when there is a fracture-induced pop-in discontinuity, as shown in Fig. 2-4. However, this approach was not possible for the current nanoindentation load-displacement curves as they showed no pop-in events.

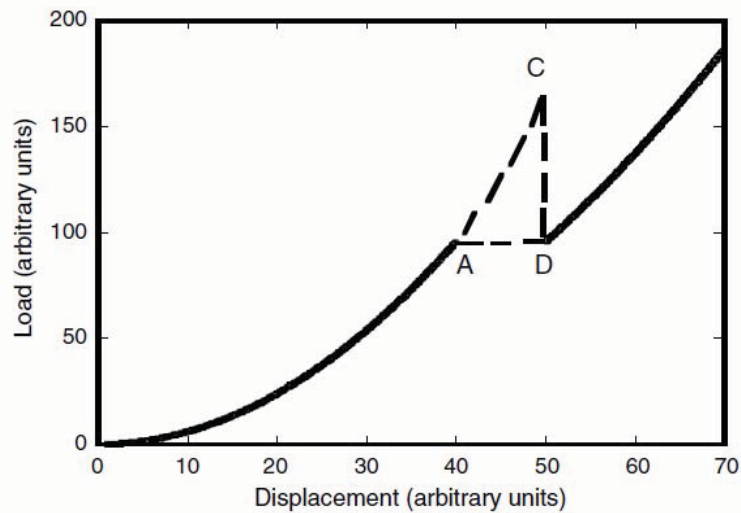

*Fig. 2-4 Schematic of the load vs. displacement method to determine the fracture dissipated energy. The area ACD is assumed to be fracture dissipated energy [57]*

Recently Chen *et al.* [67, 68] have developed another energy-based method for load-displacement curves absent of discontinuity Fig. 2-5 (a). However, this approach is claimed to be limited to dense brittle coatings displaying "picture-frame" cracks confined to the residual imprint contact area (arrowed in Fig. 2-5 (b)) and is not applicable to films with through-thickness cracks as observed in the current study.

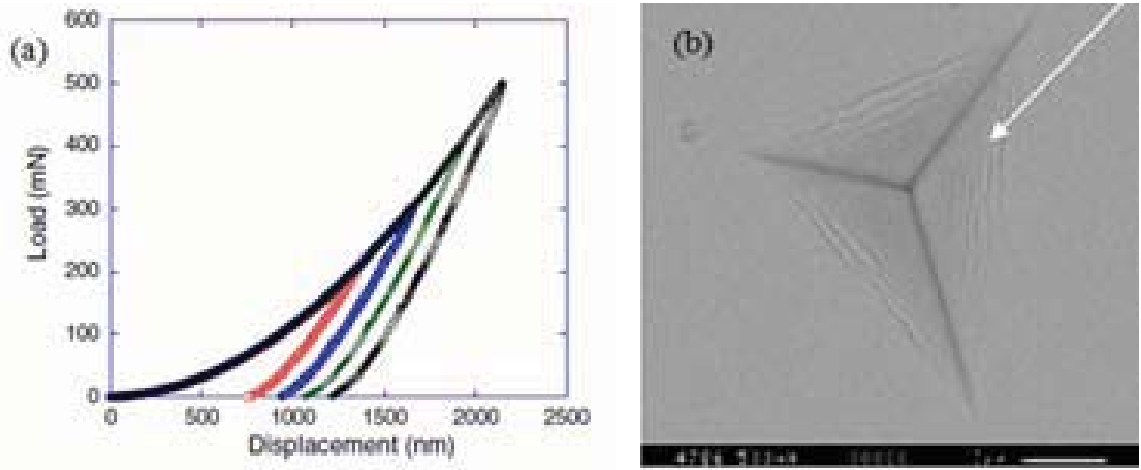

*Fig. 2-5 (a) Indentation load-displacement curves without obvious excursions, (b) SEM micrograph showing the resulted "picture-frame" cracks for monolayer coated glass with a 240 nm indium tin oxide cap layer at 500 mN load [67, 68]*

## 2.5 Nanoindentation of Porous/Thin Materials

The current study is intended to characterise the mechanical properties, particularly the elastic modulus, hardness and fracture toughness of LSCF6428 porous films which were sintered at varying temperatures in the range from 900 to 1200 °C. It is worth noting that mechanical properties of such thin films of several tens of microns thickness are not possible to be measured by conventional macroscopic characterisation methods, such as bending, tensile or compression methods. A more sophisticated technique was desired for the measurements to be achieved. The nanoindentation technique has been developed and used extensively to measure the mechanical properties of small volumes of material including thin films, from the measured load-displacement data, due to the high spatial and depth resolution of the measurement [69, 70]. Nanoindentation was first developed in mid-1970s to measure the hardness of small volumes of material [71]. Significant work has been done (primarily on monolithic materials) by Oliver and Pharr in 1992 [69] based on the use of the Berkovich indenter tip. The feasibility of applying the same data analysis method in spherical nanoindentation was proposed also by Oliver and Pharr in 2004 [70]. Although nanoindentation has been proved an established technique for measurement of bulk, dense samples, its use for porous bulk and thin film samples with roughness deposited on dense substrates is still on an empirical basis and lacks an agreed and well-defined standard methodology. As there exist a number of challenging issues when applying this technique to thin films, the awareness of which may help establish a better methodology to determine the film-only mechanical properties for our LSCF6428 thin porous films.

Nanoindentation has been used to determine the mechanical properties of a wide variety of porous solids, such as bones, thermal barrier coatings and low-k dielectric thin films. In this section, as one of the main focuses of the current study, the application of nanoindentation method for measurement of mechanical properties of porous materials and thin films is reviewed. The fundamentals related to nanoindentation experiments and calculations can be found in Chapter 3.

### 2.5.1 Bones and Porous Biomaterials

The elastic properties, hardness and isotropy of individual bone components with varying microstructures have been studied using nanoindentation [72-75], mostly with sharp Berkovich indenters. These proved the technical feasibility and showed that indentation properties may be quite different from one bone type to another and the values obtained from macroscopic tests, such as bending, are often lower. However, no further insight regarding the effect of surface conditions or porous microstructures was revealed in these studies. Recently Ferguson *et al.* [76] have reviewed the nanoindentation of bones but the influence of porous microstructures was not considered. The elastic modulus and hardness of highly porous bioceramic scaffolds for bone tissue engineering prepared by Chowdhury *et al.* [77] were measured using Berkovich nanoindentation. Surface morphology and porosity were thought to be factors influencing the test results while no further evaluation of the relationship was made.

### 2.5.2 Porous Bulk Ceramics

The nanoindentation of porous ceramic materials has been little studied. For example, Kaul *et al.* [78] prepared porous SiC with wood precursors and measured the solid phase elastic modulus and hardness using Berkovich nanoindentation. The porosity, however, was used to assess the elastic modulus by applying conventional *E* vs. porosity fitting formula with little attention being paid to the evaluation of the nanoindentation behaviour. Most indentation studies have focused on microindentation with larger indentation load and displacement. Latella *et al.* [79] studied the contact damage and fatigue conditions in porous and dense liquid-phase-sintered alumina using spherical indentation. Results showed that increasing porosity induced a transition from an essentially brittle to a quasi-plastic response in the damage mode upon spherical indentation. More recently, Clément *et al.* [80] reported an inverse analysis to validate the spherical microindentation data of highly porous plaster materials, but no further detailed correlation between the microstructures and mechanical data was obtained.

### 2.5.3 Low-k Dielectric Porous Polymers

In the integrated circuit industry, the so called low-k dielectric highly porous (porosity > 30%) materials have been developed with pores on the nanometre scale. Organo-silicate glasses (OSGs) are one of these materials for which mechanical stability was assessed using nanoindentation. Vella *et al.* [63, 64] measured various mechanical properties of OSG thin films using nanoindentation with sharp indenters. Although the importance of porosity was mentioned related to the mechanical properties, the relationships between them were not developed. However, the nature of porosity of OSG materials appears to be very different from that in conventionally sintered powders, with their pores less than 10 nm rather than micrometre-pores in other porous materials. Very different from micron-scale pores, open or closed, the pores in OSGs are created by methyl inclusions on a molecular scale, which cannot be easily discerned even using high resolution transmission electron microscopy (TEM). Xiang *et al.* [81] studied the mechanical properties of porous and fully dense low-k thin films using nanoindentation with a sharp indenter. Their results showed a significant effect from substrate on contact stiffness. The results further showed that densification of the porous material under indenters did not affect the elastic modulus significantly. By contrast, the hardness values of the porous films were affected by both substrate and densification effects.

### 2.5.4 Porous Coatings

Nanoindentation has been applied for ceramic coatings of the order of 100 microns thickness, such as thermal barrier coatings, which contain porous microstructures. Alcala *et al.* [82] have conducted spherical micro-indentation on plasma-sprayed Al-based cermet coatings. Although the substrate effect was reported to be eliminated due to shallow indentation, the effects of surface roughness, microstructural densification and porosity on the mechanical property measurements were not evaluated. Huang *et al.* [83] reported the influence of surface roughness and substrate on the elastic modulus and hardness measurements using a Berkovich tip of silica thin films (thickness = 880 nm) on silicon substrates. They concluded the indent depth should be kept less than one tenth of the film thickness to obtain film properties without the influence from substrate. However, no supporting model or methodology was produced. Furthermore, porosity was not taken into account to interpret the results. More recently, Dey *et al.* [84] have reported Berkovich nanoindentation on porous component layers of a SOFC, which consisted of a NiO-8YSZ anode, an 8YSZ electrolyte and a LSM cathode. They acknowledged the existence of large scatter in the nanoindentation mechanical data for both anode and cathode layers, which were considered to be related to their porous microstructures. However, no further quantitative microstructural evaluation was made to correlate porosity with the mechanical properties.

### 2.5.5 Main Issues in Nanoindentation of Films on Substrates

As mentioned earlier, nanoindentation has been widely used in the mechanical characterisation of bulk samples. While for thin films deposited on dense substrates, the methodology to accurately determine and thus interpret the intrinsic film-only properties has not been well established due to the uncertainties caused by experimental errors [81]. Research is still required to tackle a number of challenging sources of error, including the influence of the presence of substrates [85, 86], the densification effect of film upon indentation [87] and surface roughness of films [88], although extensive effort has been made to understand them [89-92].

A commonly employed rule of thumb to avoid effect from substrates is indenting to a maximum depth less than 1/10 of the film thickness [69]. However, this simple rule which mainly applies to thick films with flat surfaces is not necessarily valid for the highly porous thin films with prominent surface roughness, such as the LSCF6428 films investigated in the current study.

In particular, the effect of porosity on the elastic behaviour of partially sintered ceramic films has been rarely studied. Recent studies of porous films for microelectronic applications using nanoindentation with Berkovich indenters have been reported [81, 91]. However, their methodology is not suitable for typical partially sintered ceramic films with larger pore sizes. Moreover, the effect of porosity variation on elastic modulus and hardness was not considered in those studies. Their films were very smooth, flat and ductile polymers having low porosity (< 30%) and extremely small pores in the nm range. Therefore a sharp Berkovich indenter was able to deform a representative volume of the porous film. However, partially sintered ceramic films (such as the ones in the current study) often feature a relatively rough surface and have pores in the micron range and porosity as high as 50%. Furthermore they are not ductile and have a low elastic limit. The scale of their microstructures requires the use of a blunt indenter in order to deform a representative volume of material in the film. The brittleness also implies a densification mechanism by crushing, which is significantly different from the ductile deformation of polymer films.

## 2.6 Mechanical Characterisation of LSCF

### 2.6.1 Importance of Mechanical Properties

The functional components of SOFCs must withstand mechanical stresses during fabrication and operation in order to achieve long-term durability and reliability. Therefore, suitable mechanical properties are desired to prevent failures such as cracks, delamination and fractures due to mechanical stresses arising from thermal expansion coefficient differences, temperature gradients, oxygen pressure gradients and external mechanical loading [93, 94]. Such damage even if not mechanically catastrophic, often results in degradation of electrochemical performance. Therefore, it is imperative to understand the mechanical stability of these materials in the form in which they are deployed in applications which, in the case of a fuel cell or electro-catalyst, is as a porous film on a dense substrate. One of the key mechanical properties is the elastic modulus and this is one of the subjects of this study. However, most past relevant studies have concentrated on optimising the electrochemical performance. On the other hand, mechanical properties including Young's modulus, hardness, Poisson's ratio and fracture strength etc., as well as damage and failure mechanism in a variety of operating environments for conventional electrolyte and electrode materials, such as YSZ [93, 95-97] and CGO [98](electrolyte), LSM (cathode)[13, 99] and Ni/YSZ (anode)[100, 101], respectively, have attracted attention [102].

### 2.6.2 Studies of LSCF Mechanical Properties

To date, most of the studies on LSCF have concentrated on electrical properties [5, 103], oxygen permeability, diffusion and transport [104-106], degradation mechanisms [107, 108], thin film synthesis [109, 110] and applications [9, 111]. There are only a few reports of their mechanical properties and their relationship with microstructures. Room temperature mechanical properties including Young's and shear moduli, hardness, fracture toughness, biaxial flexure strength of nominally dense specimens of $La_{1-x}Sr_xCo_{0.2}Fe_{0.8}O_3$ (x = 0.2-0.8) were reported by Chou *et al.* [4]. Young's moduli of 151-188 GPa and shear moduli of 57-75 GPa were determined using resonance method, ball-on-ring mechanical testing and Vickers indentation, respectively. They found that increased Sr content could slightly increase the Young's and shear moduli, and there was no dependence of toughness on crack size. Young's modulus (measured using resonant ultrasound spectroscopy method), strain-stress behaviour (using four-point bending method), fracture strength and fracture toughness (using single edge notch beam or single edge V-notch beam method) of nominally dense $La_{0.5}Sr_{0.5}Co_yFe_{1-y}O_{3-\delta}$ ($0 \leqslant y \leqslant 1$) specimens in the temperature range of 20 to 1000°C have been studied by Lein *et al.* [112]. LSCF materials room temperature Young's moduli and fracture strengths measured were $130\pm1$ GPa and in the range of 107-128 MPa, respectively. The nonlinear stress-strain relationship observed by four-point bending at room temperature was inferred as a signature of ferroelastic behaviour of the materials. Above the ferroelastic to paraelastic transition temperature (~900 °C), the materials showed normal elastic behaviour, but due to high-temperature creep, a nonelastic response was also evident above ~800 °C. The ferroelasticity of perovskite materials had been reported and discussed earlier by Kleveland and co-workers [113, 114] and Faaland *et al.* [115]. Huang and colleagues [116, 117] have measured the mechanical properties of $La_{0.58}Sr_{0.4}Co_{0.2}Fe_{0.8}O_{3-\delta}$ using ring-on-ring bending and Berkovich microindentation. They found that the measured room temperature fracture load was nearly 40% higher than that at 800°C, due the ferroelasticity of LSCF at room temperature. More recently, Li *et al.* [118] utilised nanoindentation with a Vickers indenter to assess the mechanical properties of bulk $La_{0.58}Sr_{0.4}Co_{0.2}Fe_{0.8}O_{3-\delta}$, resulting in a much higher Young's modulus ($155\pm4$ GPa) and fracture toughness ($1.75\pm0.25$ MPa·m$^{0.5}$), compared to the results measured by Huang *et al.* [116, 117], with Young's modulus being $76\pm5$ GPa (measured using biaxial ring-on-ring bending test) and fracture toughness (using Berkovich microindentation) being $0.91\pm0.05$ MPa·m$^{0.5}$. The significant discrepancy between the two elastic moduli was possibly due to the fact that nanoindentation probed very locally the microscopic material which could be regarded as fully dense while the ring-on-ring test examined macroscopically the bulk volume specimen inclusive of residual porosity for measuring the apparent modulus. However, the reason for the large difference between the fracture toughnesses remains unclear.

Reported mechanical properties of LSCF materials are summarised in Table 2-1. The mechanical properties of individual layers in SOFC systems are significantly influenced by the microstructural features such as porosity arising from the fabrication process and the composition ratios in the materials.

*Table 2-1 Summary of mechanical characterisations of LSCF materials from the literature*

| Ref. | Composition | Sintering Condition | Relative Density | Grain size (μm) | Mechanical Properties (at Room Temperature) | Measurement Techniques |
|---|---|---|---|---|---|---|
| Chou *et al.* [4] | $La_{1-x}Sr_xCo_{0.2}Fe_{0.8}O_3$ ( $0.2 \leq x \leq 0.8$ ) | 1250°C in air for 4h; heating/cooling at 5°C/min | 95.4±0.2% (for x=0.4) | 2.9 (for x=0.4) | Young's modulus = 151–188 GPa | Ultrasonic/pulse-echo method |
| | | | | | Shear modulus = 57–75 GPa | |
| | | | | | Biaxial flexure strength = 40-160MPa | Mechanical testing machine |
| Lein *et al.* [112] | $La_{0.5}Sr_{0.5}Fe_{1-y}Co_yO_{3-\delta}$ ( $0 \leq y \leq 1$ ) | 1150 °C in air for 2-12h; heating at 200°C/h; cooling at 6,50,100°C/h | 97% (for y=0.75) | 1.4±0.3 (for y=0.75) | Young's modulus $\cong$ 130 GPa | Resonant ultrasound spectroscopy |
| | | | | | Fracture strength = 107-128 MPa | Four-point bending |
| | | | | | Fracture toughness = 1.16±0.12 MPa·m$^{0.5}$ | Single edge V-notch beam method |
| Li *et al.* [118] | $La_{0.58}Sr_{0.4}Co_{0.2}Fe_{0.8}O_{3-\delta}$ | 1200°C in air for 2h; heating/cooling at 5°C/min | 98.3% | 0.8 | Young's modulus = 155±4 GPa | Nanoindentation using a Vickers indenter |
| | | | | | Hardness = 8.6±0.3 GPa | |
| | | | | | Fracture toughness = 1.75±0.25 MPa·m$^{0.5}$ | |
| Huang *et al.* [116, 117] | $La_{0.58}Sr_{0.4}Co_{0.2}Fe_{0.8}O_{3-\delta}$ | 1200 °C in air for 3h; heating/cooling at 5°C/min | 96.6±0.2% | 0.6±0.2 | Young's modulus = 76±5 GPa | Biaxial ring-on-ring bending test |
| | | | | | Fracture strength = 109±9 GPa | |
| | | | | | Hardness = 6.3±0.4 GPa | Depth-sensitive microindentation |
| | | | | | Fracture toughness = 0.91±0.05 MPa·m$^{0.5}$ | |

However, the above literature data are all based on nominally dense bulk materials, which are quite different from the highly porous films used in most applications. Moreover, most studies have employed conventional macroscopic techniques to measure the mechanical properties, such as resonance methods, bending tests, compression and tensile tests, but these techniques are not applicable for thin films coated on substrates.

## 2.6.3 Mechanical Characterisation of Porous LSCF by Nanoindentation

There are no studies available in the literature in which nanoindentation has been used to characterise the micromechanical properties of porous LSCF samples, in bulk or thin film forms, apart from the indentation on highly dense bulk as reported in [116-118]. Porous films may behave mechanically very differently from bulk samples of the same materials/compositions, and therefore a method to obtain the film-only properties of actual films is desirable. However, the methodology for extracting "long range" mechanical properties from indentation of porous films has not been properly established. As mentioned earlier, previous studies on nanoindentation of porous ceramic films [82, 83], polymeric coatings [63, 64, 119] and highly porous bioceramic layers such as bones [72-74] have adopted simple rules-of-thumb, such as indenting to less than 10% of the film thickness. Furthermore, no significant work has been done on characterising elastic modulus and/or hardness as a function of porosity by indentation in partially sintered ceramics/films. As mentioned earlier, conventional methodologies for understanding and extracting mechanical properties might not necessarily applicable in the partially sintered ceramics/films due to their mechanically brittle characteristic and the existence of large volume fraction and size of pores as well as relatively rough surface features.

## 2.7 3D Microstructures of SOFC Electrodes

### 2.7.1 Importance of Electrode Microstructures

SOFC electrodes often possess complex 3D microstructural features which provide critical links between materials properties/processing routes and electrode performance and in turn affect the durability of the system. Important microstructural parameters include three-phase boundary length (TPBL, for composite electrodes), phase volume fraction (e.g. porosity), pore size, phase surface areas and interface areas, phase tortuosity, and other geometric factors. Excellent electrode candidates must balance a range of requirements.

In composite electrode systems containing both ionically and electronically conducting material phases in 3D space, the contact of a pore phase, an ionic phase and an electronic phase creates a three-phase boundary (TPB), where the electrochemical reaction takes place. However for MIEC materials, electrochemical reaction can be active at a broader gas/solid interface due to the presence of both electrons and ions in the single solid phase. Fig. 2-6 illustrates schematically the TPB in a composite electrode material (Ni-YSZ) and the two phase boundary in single phase MIEC electrode [120].

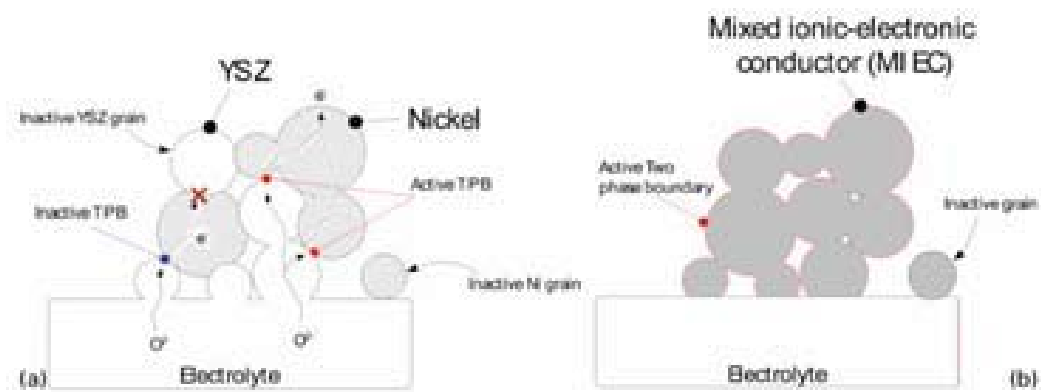

*Fig. 2-6 A schematic of transport of reacting species in (a) composite electrode material (Ni–YSZ) and (b) single phase MIEC electrode, from [120]*

The electrode microstructures have significant inter-relationships with the fabrication process. That is, the processing factors including particle size distribution [101, 121, 122], materials composition [123-125], deposition methods [126, 127] and sintering [128, 129], as well as the methods for microstructural control, such as pore former additive [130] and composite powders [131, 132], can affect the electrode microstructures. The mechanical properties [133], durability [131], and electrochemical performance [134, 135] including ionic and electronic conductivities, and polarisation [136-138] of electrodes during the full range of operation are in turn influenced by the microstructures. For example, high mechanical strength can be obtained by sufficient sintering which results in lower porosity; whereas electrochemical performance can be improved by the opposite. The polarisation characteristics of MIEC electrodes rely on the diffusion of ions, electrons and gas, as well as the electrochemical reaction rates on the gas/solid interface, which are strongly influenced by the microstructures.

It was shown that the polarisation resistance of the composite LSM-YSZ electrode was decreased by extending the TPBL which resulted in a much lower over-potential toward oxygen reduction in SOFC application than the pure LSM [139]. On the other hand, microstructures can also be affected during operation by redox cycling [140], load and thermal cycling [141], and long term sintering [142]. Changes in microstructure can result in numerous impacts, from reduced electrical or electrochemical performance [142] to catastrophic failure by cracking or delamination [143]. Therefore, the investigation and optimisation of microstructures has drawn many researchers' attention.

Microstructural features including porosity, pore surface area, TPBL (for composite electrodes), and tortuosity are considered to be the most critical factors influencing electrochemical efficiency. Both adsorption and surface diffusion of oxygen in the cathodes are dependent on the pore surface area, which can therefore control the oxygen reduction therein [144]. Electrochemical reaction is restricted to the TPB in composite electrodes, and the electrode resistance can be reduced by increasing TPBL [121]. Porosity is the volume fraction of the pores to the total volume in the microstructure. Oxygen flux must first diffuse via the pore network to the reaction sites. An ideal microstructure must have sufficient pore space for gas diffusion, while a partially dense microstructure may restrain the flow to the reaction sites, thus limiting the cathodic reaction [135]. Tortuosity is a property that quantifies the complexity of the path through which a diffusing particle must travel in order to reach a desired destination. For SOFCs, pore tortuosity is a dimensionless parameter defined as the fraction between typical diffusion path between two points through the pores and the straight distance between them when the oxygen flux travels through the porous cathode to reach the solid electrolyte [145]. A large tortuosity corresponds to a convoluted path for a given gas molecule to traverse in order to go from the gas stream to the reaction site [135].

### 2.7.2 Electrode Microstructure Study Based on SEM

Traditional methods of visualising the material microstructures rely on the simplification of 3D microstructures into 2D planar representation using optical microscopy and SEM. A number of studies have been carried out on the 2D microstructural characterisation of SOFC electrodes in order to correlate microstructures with electrochemical performance. Various structures of LSM cathodes were made by Weber et al. [146] through changing only the manufacturing procedure, and the microstructures were qualitatively analysed using SEM. Results showed that the microstructure change particularly at the cathode-electrolyte interface led to large variation in cathode polarisation resistance. Lee *et al.*[147] reported that higher inactive pore volume significantly reduced cell performance. Suzuki *et al.* [148] showed highly porous anode microstructure extensively improved the electrochemical performance. Processing methods were used to control the uniformity of porosity in the Ni-YSZ anode by Lee *et al.* [149] which was visualised by 2D imaging without quantitative analysis. Hagiwara *et al.* [150] described spray pyrolysis process for cathode fabrication to understand how a finely controlled cathode microstructure was realised. However, again the microstructure analysis was based on 2D imaging without further quantification. Without detailed morphological or topological information of 3D microstructures, many parameters are difficult to determine independently. For instance, from a 2D image, it is nearly impossible to measure the tortuosity (of either the pores or solids) which strictly depends on the details of 3D interconnected network. Also, limitations are imposed because stereological methods can only be accurate for isotropic microstructures [137]. In conventional multiphase electrodes, the connectivity of individual material phases cannot be determined in 2D images. More fundamentally, basic parameters such as the sizes and shapes of interconnected solid particles and pores are not possible to calculate in the absence of the topological data on the third dimension.

### 2.7.3 Electrode Microstructure Study by Modelling

On the other hand geometric modelling using mathematical methods taking into account various parameters such as thickness and porosity has been reported. Stereological analysis was typically used to establish the ideal and virtual 3D microstructures and parameters, by assuming the geometry to be that of regular shapes such as spheres, cylinders, plates and so on [151]. Joshi *et al.* [152] used the 2D lattice Boltzmann method to investigate multi-component gas transport in anode structures based on SEM micrograph images, but without understanding the 3D connectivity of the pore structures. A tree-like model of $La_{0.85}Sr_{0.15}MnO_3$ cathodes was presented by Tanner *et al.* [153], whose results suggested significant benefit, such as lower effective charge-transfer resistance, could be obtained with finer microstructures as long as the porosity is sufficient to ensure negligible concentration polarization. The effect of composite LSM–YSZ cathode geometric parameters on charge transfer was theoretically and experimentally analysed by Virkar *et al.* [136]. It has also been shown that a fine microstructure is effective in lowering the activation polarization, assuming that porosity is high enough to prevent concentration polarisation (related to the diffusion process of gases through the gas-filled pores of the electrode to reach reaction sites). However, a contradictory result was found by Maggio *et al.* [154] suggesting that maximum current density could only be achieved with an optimal range of pore sizes. Porous electrode theory was employed in those models assuming the homogeneous properties of the microstructures [155]. In most cases, all parameters predicted have been based on idealised geometric models without the actual 3D microstructural data [156-158].

While 3D microstructure models can be quickly generated by such simplification with much less computing power needed, their limited representation of actual complex 3D microstructures results in merely questionable approximations of the physical and geometrical properties of the porous electrodes, which are produced from real material processing routes. In addition, their narrow geometrical idealisation led to highly questionable ability to represent the real bulk transport/diffusion properties of interest. An accurate estimation of the material properties and behaviours can only be obtained by incorporating actual 3D microstructures as a basis of the model, and without the realisation of these 3D microstructures the assumptions and predictions of the models are hard to validate.

### 2.7.4 3D Microstructure Study Using FIB/SEM Tomography

The development and application of advanced 3D tomographic techniques have made it possible to analyse the real spatial microstructures of SOFC electrodes by getting access to a variety of valuable microstructural information based on 3D reconstruction, which can also be used to correlate with previous research for a better understanding of SOFC electrode performance so that better designs of electrodes can be developed [120, 159].

One of the most important and widely used tomographic techniques for 3D microstructure reconstructions of material microstructures is the dual-beam system, i.e. focused ion beam/scanning electron microscopy (i.e. FIB/SEM) (its detailed technical principles will be described in Chapter 3), which was first introduced to investigate SOFC components by Wilson *et al.* [160]. Based on the methodology of in-situ processing by FIB/SEM slice and view, they demonstrated the first 3D microstructure reconstruction of a Ni/YSZ anode and the potential of the technique for characterisation of triple phase contact in composite electrode materials. Geometric analysis was then applied to obtain TPBL, phase fractions and tortuosity. After this paper, coupled with the advancement of 3D data processing software, substantial efforts have been made with FIB/SEM tomography to acquire the detailed and integrated 3D data necessary to facilitate the full characterisation of the microstructure and performance of various porous electrodes, such as Ni/YSZ [128, 161, 162], Ni/CGO [163], LSM [135, 151], LSM/YSZ [164] and LSCF [165-167]. These investigations provide a great deal of quantitative information about the complexity of actual electrodes. 3D imaging and reconstruction provide high-resolution measurements of TPBLs, phase volumes, porosity, specific surface and interface areas, phase connectivity (percolation), and tortuosity. Such measurements can also be used to validate the simulated electrode microstructures.

### 2.7.5 3D Microstructure Study of LSCF Cathodes

As described above, with the aid of the advanced FIB/SEM tomography, some researchers have carried out in-depth investigations on 3D reconstruction to help fully understand the microstructures of LSCF cathode materials and their relationship with electrochemical performance. Gostovic and colleagues [135, 165] have reconstructed an actual 3D structure of a $La_{0.8}Sr_{0.2}Co_{0.2}Fe_{0.8}O_{3-\delta}$ cathode and its interface with a dense YSZ electrolyte. Various microstructural features were measured, including overall porosity, closed porosity, graded porosity, surface area, tortuosity, TPBL, and pore size.

Electrochemical impedance spectroscopy data were correlated to microstructures and thus the extent of dependence of charge transfer and adsorption on TPBL and of absorption related polarisation resistance on specific surface area were determined respectively. In Joos *et al.*'s work [166], based on the reconstructed 3D microstructure data of a $La_{0.58}Sr_{0.4}Co_{0.2}Fe_{0.8}O_{3-\delta}$ cathode and by the evaluation of qualified algorithms discriminating between porosity and electrode material, a sensitivity analysis of the greyscale threshold value of the essential parameters, i.e. the surface area, volume/porosity fraction and tortuosity was performed. The microstructural parameters were identified as specific surface area 6.3 $\mu m^{-1}$, porosity 47.3% and tortuosity of pore 1.97. The 3D microstructure of an $La_{0.6}Sr_{0.4}Co_{0.2}Fe_{0.8}O_{3-\delta}$ cathode was reconstructed by Matsuzaki *et al.* [167] and over-potential was predicted by the lattice Boltzmann method. Surface area estimated by the marching cube method showed fairly good agreement with those from stereology. Electron transfer to the adsorbed oxygen atom on the gas/solid interface was assumed to be the rate limiting step in the electrochemical modelling. The cathode over-potential prediction agreed well with the experimental data.

However, none of the above 3D microstructure characterisations of LSCF cathode materials was combined with CGO substrate and no correlation was made between their 3D microstructural features and mechanical properties. Possible influence of the modification of microstructures, from either actual fabrication or numerical simulation, on the changes of properties was not addressed in these studies, either.

## Chapter 2 References

1.  Wells AF: **Structural Inorganic Chemistry**. New York: Oxford University Press; 1984.

2.  [http://mrc.iisc.ernet.in/Research_Areas/01_Perovskite.htm](http://mrc.iisc.ernet.in/Research_Areas/01_Perovskite.htm)

3.  Bouwmeester HJMaB, A.J. : **Dense Ceramic Membranes for Oxygen Separation**. In: *Solid State Electrochemistry.* Edited by Gellings PJ, Bouwmeester HJM. New York: CRC Press; 1997: 481-553.

4.  Chou Y-S, Stevenson JW, TArmstrong TR, LPederson LR: **Mechanical Properties of La1-xSrxCo0.2Fe0.8O3-δ Mixed-Conducting Perovskites Made by the Combustion Synthesis Technique**. *Journal of American Ceramic Society* 2000, **83**(6):1457-1464.

5.  Teraoka Y, Zhang HM, Okamoto K, Yamazoe N: **Mixed ionic-electronic conductivity of La1-xSrxCo1-yFeyO3-δ perovskite-type oxides**. *Materials Research Bulletin* 1988, **23**(1):51-58.

6.  Katsuki M, Wang S, Dokiya M, Hashimoto T: **High temperature properties of La0.6Sr0.4Co0.8Fe0.2O3-δ oxygen nonstoichiometry and chemical diffusion constant**. *Solid State Ionics* 2003, **156**(3-4):453-461.

7.  Tai LW, et al.: **Structure and Electrical Properties of La1-xSrxCo1-yFeyO3. Part 2: The System La1-xSrxCo0.2Fe0.8O3**. *Solid State Ionics* 1995(76):273-283.

8.  Teraoka Y: **Oxygen-sorptive properties of defect perovskite-type La1- xSrxCo1-yFeyO3-δ**. *Chemistry Letters* 1985(9):1367.

9.  Minh NQ: **Ceramic Fuel Cells**. *Journal of the American Ceramic Society* 1993, **74**(3):563-588.

10. Adler SB: **Mechanism and kinetics of oxygen reduction on porous La1-xSrxCoO3-[delta] electrodes**. *Solid State Ionics* 1998, **111**(1-2):125-134.

11. van Doorn RHE, Burggraaf AJ: **Structural aspects of the ionic conductivity of La1-xSrxCoO3-[delta]**. *Solid State Ionics* 2000, **128**(1-4):65-78.

12. Brugnoni C: **SOFC cathode/electrolyte interface. Part I: Reactivity between La0.85Sr0.15MnO3 and ZrO2-Y2O3**. *Solid State Ionics* 1995, **76**(3-4):177.

13. Jiang Y, Wang S, Zhang Y, Yan J, Li W: **Electrochemical reduction of oxygen on a strontium doped lanthanum manganite electrode**. *Solid State Ionics* 1998, **110**(1-2):111-119.

14. Van Herle J: **A study on the La1-xSrxMnO3 oxygen cathode**. *Electrochimica Acta* 1996, **41**(9):1447.

15. De Souza RA: **Oxygen transport in La1-xSrxMn1-yCoyO3-[delta] perovskites: Part I. Oxygen tracer diffusion**. *Solid State Ionics* 1998, **106**(3-4):175.

16. De Souza RA: **A SIMS study of oxygen tracer diffusion and surface exchange in La0.8Sr0.2MnO3-[delta]**. *Materials letters* 2000, **43**(1-2):43.

17. Tai L-W, al. e: **Structure and electrical properties of La1-xSrxCo1-yFeyO3.Part 1. The system La0.8Sr0.2Co1-yFeyO3**. *Solid State Ionics* 1995, **76**:259-271.

18. Steele BCH: **Oxygen transport and exchange in oxide ceramics**. *Journal of Power Sources* 1994, **49**(1-3):1-14.

19. Teraoka Y: **Oxygen permeation through perovskite-type oxides**. *Chemistry Letters* 1985(11):1743.

20. Wang S, Katsuki M, Dokiya M, Hashimoto T: **High temperature properties of La0.6Sr0.4Co0.8Fe0.2O3-δ phase structure and electrical conductivity**. *Solid State Ionics* 2003, **159**(1-2):71-78.

21.     Ullmann H, Trofimenko N, Tietz F, Stöver D, Ahmad-Khanlou A: **Correlation between thermal expansion and oxide ion transport in mixed conducting perovskite-type oxides for SOFC cathodes**. *Solid State Ionics* 2000, **138**(1-2):79-90.

22.     Tu HY, Takeda Y, Imanishi N, Yamamoto O: **Ln0.4Sr0.6Co0.8Fe0.2O3-[delta] (Ln=La, Pr, Nd, Sm, Gd) for the electrode in solid oxide fuel cells**. *Solid State Ionics* 1999, **117**(3-4):277-281.

23.     Kindermann L, Das D, Nickel H, Hilpert K: **Chemical compatibility of the LaFeO3 base perovskites (La0.6Sr0.4)zFe0.8M0.2O3 - [delta] (z = 1, 0.9; M = Cr, Mn, Co, Ni) with yttria stabilized zirconia**. *Solid State Ionics* 1996, **89**(3-4):215-220.

24.     Mai A, Haanappel VAC, Tietz F, Stöver D: **Ferrite-based perovskites as cathode materials for anode-supported solid oxide fuel cells: Part II. Influence of the CGO interlayer**. *Solid State Ionics* 2006, **177**(19-25):2103-2107.

25.     Shiono M, Kobayashi K, Lan Nguyen T, Hosoda K, Kato T, Ota K, Dokiya M: **Effect of CeO2 interlayer on ZrO2 electrolyte/La(Sr)CoO3 cathode for low-temperature SOFCs**. *Solid State Ionics* 2004, **170**(1-2):1-7.

26.     Yamazoe N: **TPD and XPS study on thermal behavior of absorbed oxygen in La1-xSrxCoO3**. *Chemistry Letters* 1981, **10**(12):1767.

27.     Anderson HU, Tai L-W, Chen CC, Nasrallah MM: **Review of the structural and electrical properties of the LaSrCoFeO3 system**. *Proceedings of the Fourth International Symposium on Solid Oxide Fuel Cells* 1995:375-383.

28.     Reed JS: **Introduction to the Principles of Ceramic Processing**. Chichester: Wiley; 1988.

29.     Rice RW: **Porosity of Ceramics: Properties and Applications**: CRC Press; 1998.

30.     Coble R, Kingery W: **Effect of porosity on physical properties of sintered alumina**. *Journal of the American Ceramic Society* 1956, **39**(11):377-385.

31.     Dean EA, Lopez JA: **Empirical Dependence of Elastic Moduli on Porosity for Ceramic Materials**. *Journal of the American Ceramic Society* 1983, **66**(5):366-370.

32.     Herakovich C, Baxter S: **Influence of pore geometry on the effective response of porous media**. *Journal of Materials Science* 1999, **34**(7):1595-1609.

33.     Hardy D, Green DJ: **Mechanical properties of a partially sintered alumina**. *Journal of the European Ceramic Society* 1995, **15**(8):769-775.

34.     Green DJ, Nader C, Brezny R: **Elastic behaviour of partially-sintered alumina**. In: *Sintering of Advanced Ceramics Proc Symposium on Sintering of Advanced Ceramics Cincinnati, Ohio, 2-5 May 1988: 1988*; 1988: 345-356.

35.     Rice R: **Porosity of ceramics**. In.: Marcel Dekker, New York; 1998.

36.     Pabst W, Gregorová E, Tichá G: **Elasticity of porous ceramics—A critical study of modulus−porosity relations**. *Journal of the European Ceramic Society* 2006, **26**(7):1085-1097.

37.     Fryxell R, Chandler B: **Creep, strength, expansion, and elastic moduli of sintered BeO as a function of grain size, porosity, and grain orientation**. *Journal of the American Ceramic Society* 1964, **47**(6):283-291.

38.     Spriggs R: **Expression for effect of porosity on elastic modulus of polycrystalline refractory materials, particularly aluminum oxide**. *Journal of the American Ceramic Society* 1961, **44**(12):628-629.

39.     Phani KK, Niyogi SK: **Elastic Modulus-Porosity Relation in Polycrystalline Rare-Earth Oxides**. *Journal of the American Ceramic Society* 1987, **70**(12):C-362-C-366.

40.     Phani K, Niyogi S: **Young's modulus of porous brittle solids**. *Journal of Materials Science* 1987, **22**(1):257-263.

41.     Hasselman D: **On the porosity dependence of the elastic moduli of polycrystalline refractory materials**. *Journal of the American Ceramic Society* 1962, **45**(9):452-453.

42.  Hashin Z: **Analysis of composite materials—a survey**. *Journal of Applied Mechanics* 1983, **50**(3):481-505.

43.  McLaughlin R: **A study of the differential scheme for composite materials**. *International Journal of Engineering Science* 1977, **15**(4):237-244.

44.  Budiansky B: **On the elastic moduli of some heterogeneous materials**. *Journal of the Mechanics and Physics of Solids* 1965, **13**(4):223-227.

45.  Hill R: **A self-consistent mechanics of composite materials**. *Journal of the Mechanics and Physics of Solids* 1965, **13**(4):213-222.

46.  Rice RW: **Evaluation and extension of physical property-porosity models based on minimum solid area**. *Journal of Materials Science* 1996, **31**(1):102-118.

47.  Rice R: **Comparison of physical property-porosity behaviour with minimum solid area models**. *Journal of Materials Science* 1996, **31**(6):1509-1528.

48.  Roberts AP, Garboczi EJ: **Elastic Properties of Model Porous Ceramics**. *Journal of the American Ceramic Society* 2000, **83**(12):3041-3048.

49.  Poutet J, Manzoni D, Hage-Chehade F, Thovert J-F, Adler P: **The effective mechanical properties of random porous media**. *Journal of the Mechanics and Physics of Solids* 1996, **44**(10):1587-1620.

50.  Schiffmann KI: **Determination of fracture toughness of bulk materials and thin films by nanoindentation: comparison of different models**. *Philosophical Magazine* 2011, **91**(7-9):1163-1178.

51.  Harding D, Oliver W, Pharr G: **Cracking during nanoindentation and its use in the measurement of fracture toughness**. In: *Materials Research Society Symposium Proceedings: 1995*: Cambridge Univ Press; 1995: 663-663.

52.  Quinn GD: **On the Vickers indentation fracture toughness test**. *Journal of the American Ceramic Society* 2007, **90**(3):673.

53.  Hagan J, Swain MV: **The origin of median and lateral cracks around plastic indents in brittle materials**. *Journal of Physics D: Applied Physics* 1978, **11**(15):2091.

54.  Anstis G, Chantikul P, Lawn BR, Marshall D: **A critical evaluation of indentation techniques for measuring fracture toughness: I, direct crack measurements**. *Journal of the American Ceramic Society* 1981, **64**(9):533-538.

55.  Swanson PL, Fairbanks CJ, Lawn BR, MAI YW, HOCKEY BJ: **Crack-Interface Grain Bridging as a Fracture Resistance I, Mechanism in Ceramics: I, Experimental Study on Alumina**. *Journal of the American Ceramic Society* 1987, **70**(4):279-289.

56.  Lange F: **Fracture Toughness of Si3N4 as a Function of the Initial α-Phase Content**. *Journal of the American Ceramic Society* 1979, **62**(7-8):428-430.

57.  Li X, Diao D, Bhushan B: **Fracture mechanisms of thin amorphous carbon films in nanoindentation**. *Acta Materialia* 1997, **45**(11):4453-4461.

58.  Malzbender J, den Toonder JMJ, Balkenende AR, de With G: **Measuring mechanical properties of coatings: a methodology applied to nano-particle-filled sol–gel coatings on glass**. *Materials Science and Engineering: R: Reports* 2002, **36**(2–3):47-103.

59.  Hag AJ, Munroe P, Hoffman M, Martin P, Bendavid A: **Berkovich indentation of diamondlike carbon coatings on silicon substrates**. *Journal of Materials Research* 2008, **23**(7):1862-1869.

60.  Chicot D, Duarte G, Tricoteaux A, Jorgowski B, Leriche A, Lesage J: **Vickers Indentation Fracture (VIF) modeling to analyze multi-cracking toughness of titania, alumina and zirconia plasma sprayed coatings**. *Materials Science and Engineering: A* 2009, **527**(1–2):65-76.

61.  Morris DJ, Cook RF: **Indentation fracture of low-dielectric constant films: Part I. Experiments and observations**. *J Mater Res* 2008, **23**(9):2429.

62.  Morris DJ, Cook RF: **Indentation fracture of low-dielectric constant films: Part II. Indentation fracture mechanics model**. *J Mater Res* 2008, **23**(9):2443-2457.

63.  Volinsky AA, Vella JB, Gerberich WW: **Fracture toughness, adhesion and mechanical properties of low-k dielectric thin films measured by nanoindentation**. *Thin Solid Films* 2003, **429**(1):201-210.

64.  Vella J, Adhihetty I, Junker K, Volinsky A: **Mechanical properties and fracture toughness of organo-silicate glass (OSG) low-k dielectric thin films for microelectronic applications**. *International journal of fracture* 2003, **120**(1):487-499.

65.  Üçisik AH, Bindal C: **Fracture toughness of boride formed on low-alloy steels**. *Surface and Coatings Technology* 1997, **94–95**(0):561-565.

66.  Li X, Bhushan B: **Measurement of fracture toughness of ultra-thin amorphous carbon films**. *Thin Solid Films* 1998, **315**(1–2):214-221.

67.  Chen J, Bull SJ: **Indentation fracture and toughness assessment for thin optical coatings on glass**. *Journal of Physics D: Applied Physics* 2007, **40**(18):5401.

68.  Chen J: **Indentation-based methods to assess fracture toughness for thin coatings**. *Journal of Physics D: Applied Physics* 2012, **45**(20):203001.

69.  Oliver WC, Pharr GM: **An improved technique for determining hardness and elastic modulus using load and displacement sensing indentation experiments**. *Journal of Materials Research* 1992, **7**(6):1564-1583.

70.  Oliver WC: **Measurement of hardness and elastic modulus by instrumented indentation: Advances in understanding and refinements to methodology**. *Journal of Materials Research* 2004, **19**(1):3.

71.  Bulychev S, Alekhin V, Shorshorov M, Ternovskii A, Shnyrev G: **Determining Young's modulus from the indentor penetration diagram**. *Ind Lab* 1975, **41**(9):1409-1412.

72.  Rho J-Y, Tsui TY, Pharr GM: **Elastic properties of human cortical and trabecular lamellar bone measured by nanoindentation**. *Biomaterials* 1997, **18**(20):1325-1330.

73.  Turner CH: **The elastic properties of trabecular and cortical bone tissues aresimilar result from two microscopic measurement techniques.pdf**. *Journal of Biomechanics* 1999, **32**:437-441.

74.  Zysset PK, Edward Guo X, Edward Hoffler C, Moore KE, Goldstein SA: **Elastic modulus and hardness of cortical and trabecular bone lamellae measured by nanoindentation in the human femur**. *Journal of Biomechanics* 1999, **32**(10):1005-1012.

75.  Tricoteaux A, Rguiti E, Chicot D, Boilet L, Descamps M, Leriche A, Lesage J: **Influence of porosity on the mechanical properties of microporous β-TCP bioceramics by usual and instrumented Vickers microindentation**. *Journal of the European Ceramic Society* 2011, **31**(8):1361-1369.

76.  Ferguson VL, Olesiak SE: **Nanoindentation of bone**. *Handbook of Nanoindentation with Biological Applications, Pan Stanford Publishing, Singapore* 2010:185-238.

77.  Chowdhury S, Thomas V, Dean D, Catledge SA, Vohra YK: **Nanoindentation on Porous Bioceramic Scaffolds for Bone Tissue Engineering**. *Journal of Nanoscience and Nanotechnology* 2005, **5**(11):1816-1820.

78.  Kaul VS, Faber KT: **Nanoindentation analysis of the elastic properties of porous SiC derived from wood**. *Scripta Materialia* 2008, **58**(10):886-889.

79.  Latella BA, OConnor BH, Padture NP, Lawn BR: **Hertzian contact damage in porous alumina ceramics**. *Journal of the American Ceramic Society* 1997, **80**(4):1027-1031.

80. Clément P, Meille S, Chevalier J, Olagnon C: **Mechanical characterization of highly porous inorganic solids materials by instrumented micro-indentation**. *Acta Materialia* 2013, **61**(18):6649-6660.

81. Xiang Y, Chen X, Tsui TY, Jang J-I, Vlassak JJ: **Mechanical properties of porous and fully dense low-κ dielectric thin films measured by means of nanoindentation and the plane-strain bulge test technique**. *Journal of Materials Research* 2006, **21**(02):386-395.

82. Alcala J, Gaudette F, Suresh S, Sampath S: **Instrumented spherical micro-indentation of plasma-sprayed coatings**. *Materials Science and Engineering: A* 2001, **316**(1):1-10.

83. Huang X, Pelegri AA: **Nanoindentation Measurements on Low-k Porous Silica Thin Films Spin Coated on Silicon Substrates**. *Journal of Engineering Materials and Technology* 2003, **125**(4):361-367.

84. Dey T, Dey A, Ghosh PC, Bose M, Mukhopadhyay AK, Basu RN: **Influence of microstructure on nano-mechanical properties of single planar solid oxide fuel cell in pre-and post-reduced conditions**. *Materials & Design* 2014, **53**:182-191.

85. Chen X, Vlassak JJ: **Numerical study on the measurement of thin film mechanical properties by means of nanoindentation**. *Journal of Materials Research* 2001, **16**(10):2974-2982.

86. Bhattacharya AK, Nix WD: **Analysis of elastic and plastic deformation associated with indentation testing of thin films on substrates**. *International Journal of Solids and Structures* 1988, **24**(12):1287-1298.

87. Fleck NA, Otoyo H, Needleman A: **Indentation of porous solids**. *International Journal of Solids and Structures* 1992, **29**(13):1613-1636.

88. Jiang W-G, Su J-J, Feng X-Q: **Effect of surface roughness on nanoindentation test of thin films**. *Engineering Fracture Mechanics* 2008, **75**(17):4965-4972.

89. Tsui TY, Vlassak J, Nix WD: **Indentation plastic displacement field: Part I. The case of soft films on hard substrates**. *Journal of Materials Research* 1999, **14**(06):2196-2203.

90. Saha R, Nix WD: **Effects of the substrate on the determination of thin film mechanical properties by nanoindentation**. *Acta Materialia* 2002, **50**:23-38.

91. Chen X, Xiang Y, Vlassak JJ: **Novel technique for measuring the mechanical properties of porous materials by nanoindentation**. *Journal of Materials Research* 2006, **21**(03):715-724.

92. Bouzakis K-D, Michailidis N, Hadjiyiannis S, Skordaris G, Erkens G: **The effect of specimen roughness and indenter tip geometry on the determination accuracy of thin hard coatings stress–strain laws by nanoindentation**. *Materials Characterization* 2002, **49**(2):149-156.

93. Atkinson A, Selcuk A: **Mechanical behaviour of ceramic oxygen ion-conducting membranes**. *Solid State Ionics* 2000, **134**:59-66.

94. Bellon O, Sammes NM, Staniforth J: **Mechanical properties and electrochemical characterisation of extruded doped cerium oxide for use as an electrolyte for solid oxide fuel cells**. *Journal of Power Sources* 1998, **75**(1):116-121.

95. Selcus A: **Strength and Toughness of Tape-Cast Yttria-Stabilized Zirconia**. *Journal of American Ceramic Society* 2000, **83**(8):2029-2035.

96. Atkinson AaaS: **Crack Growth in Yttria-Stabilised Zirconia in Operating Environments**. *The 3rd European SOFC Forum* 1998:343-352.

97. Atkinson A, Kim J-S, Rudkin R, Taub S, Wang X: **Stress Induced by Constrained Sintering of 3YSZ Films Measured by Substrate Creep**. *Journal of the American Ceramic Society* 2011, **94**(3):717-724.

98.   Morales M, Roa JJ, Capdevila XG, Segarra M, Piñol S: **Mechanical properties at the nanometer scale of GDC and YSZ used as electrolytes for solid oxide fuel cells**. *Acta Materialia* 2010, **58**(7):2504-2509.

99.   Meixner DL, Cutler RA: **Sintering and mechanical characteristics of lanthanum strontium manganite**. *Solid State Ionics* 2002, **146**(3-4):273-284.

100.  Gutierrez-Mora F, Ralph JM, Routbort JL: **High-temperature mechanical properties of anode-supported bilayers**. *Solid State Ionics* 2002, **149**(3-4):177-184.

101.  Yu JH, Park GW, Lee S, Woo SK: **Microstructural effects on the electrical and mechanical properties of Ni-YSZ cermet for SOFC anode**. *Journal of Power Sources* 2007, **163**(2):926-932.

102.  Atkinson AS: **Mechanical properties of ceramic materials for solid oxide fuel cells**. *Electrochemical Society Transaction* 1997:671-680.

103.  Stevenson JW, Armstrong TR, Carneim RD, Pederson LR, Weber WJ: **Electrochemical Properties of Mixed Conducting Perovskites La1-xMxCo1-yFeyO3-δ (M = Sr, Ba, Ca)**. *Journal of The Electrochemical Society* 1996, **143**(9):2722-2729.

104.  Schlehuber D, Wessel E, Singheiser L, Markus T: **Determination of diffusion coefficients of oxygen vacancies in La0.58Sr0.4Co0.2Fe0.8O3-δ perovskite type oxides**. In: *Diffusion in Materials - Dimat2008.* Edited by Aguero A, Albella JM, Hierro MP, Philibert J, Trujillo FJP, vol. 289-292. Stafa-Zurich: Trans Tech Publications Ltd; 2009: 551-554.

105.  Carter S, Selcuk A, Chater RJ, Kajda J, Kilner JA, Steele BCH: **Oxygen transport in selected nonstoichiometric perovskite-structure oxides**. *Solid State Ionics* 1992, **53-56**(Part 1):597-605.

106.  Ishigaki T, Yamauchi S, Kishio K, Mizusaki J, Fueki K: **Diffusion of oxide ion vacancies in perovskite-type oxides**. *Journal of Solid State Chemistry* 1988, **73**(1):179-187.

107.  Mai A, Becker M, Assenmacher W, Tietz F, Hathiramani D, Ivers-Tiffée E, Stöver D, Mader W: **Time-dependent performance of mixed-conducting SOFC cathodes**. *Solid State Ionics* 2006, **177**(19-25):1965-1968.

108.  Yokokawa H, Tu H, Iwanschitz B, Mai A: **Fundamental mechanisms limiting solid oxide fuel cell durability**. *Journal of Power Sources* 2008, **182**(2):400-412.

109.  Tsai C-Y, Dixon AG, Ma YH, Moser WR, Pascucci MR: **Dense Perovskite, La1-xA′xFe1-yCoyO3- δ (A′ = Ba, Sr, Ca), Membrane Synthesis, Applications, and Characterization**. *Journal of the American Ceramic Society* 1998, **81**(6):1437-1444.

110.  Zydorczak B, Wu ZT, Li K: **Fabrication of ultrathin La0.6Sr0.4Co0.2Fe0.8O3-δ hollow fibre membranes for oxygen permeation**. *Chemical Engineering Science* 2009, **64**(21):4383-4388.

111.  Steele BCH: **Oxygen ion conductors and their technological applications**. *Materials Science and Engineering: B* 1992, **13**(2):79-87.

112.  Lein HL, Andersen ØS, Vullum PE, Lara-Curzio E, Holmestad R, Einarsrud M-A, Grande T: **Mechanical properties of mixed conducting La0.5Sr0.5Fe1−xCoxO3−δ (0⩽x⩽1) materials**. *Journal of Solid State Electrochemistry* 2006, **10**(8):635-642.

113.  Kleveland K, Orlovskaya N, Grande T, Moe AMM, Einarsrud M-A, Breder K, Gogotsi G: **Ferroelastic Behavior of LaCoO3-Based Ceramics**. *Journal of the American Ceramic Society* 2001, **84**(9):2029-2033.

114.  Orlovskaya N, Browning N, Nicholls A: **Ferroelasticity in mixed conducting LaCoO3 based perovskites: a ferroelastic phase transition**. *Acta Materialia* 2003, **51**(17):5063-5071.

115. Faaland S, Grande T, Einarsrud M-A, Vullum PE, Holmestad R: **Stress–Strain Behavior During Compression of Polycrystalline La1−xCaxCoO3 Ceramics**. *Journal of the American Ceramic Society* 2005, **88**(3):726-730.

116. Huang BX, Malzbender J, Steinbrech RW, Singheiser L: **Mechanical properties of La0.58Sr0.4Co0.2Fe0.8O3-δ membranes**. *Solid State Ionics* 2009, **180**(2-3):241-245.

117. Huang B, Chanda A, Steinbrech R, Malzbender J: **Indentation strength method to determine the fracture toughness of La0.58Sr0.4Co0.2Fe0.8O3-delta and Ba0.5Sr0.5Co0.8Fe0.2O3-delta**. *Journal of Materials Science* 2012, **47**(6):2695-2699.

118. Li N, Verma A, Singh P, Kim J-H: **Characterization of La0.58Sr0.4Co0.2Fe0.8O3−δ−Ce0.8Gd0.2O2 composite cathode for intermediate temperature solid oxide fuel cells**. *Ceramics International* 2013, **39**(1):529-538.

119. Shen L, Zeng K: **Comparison of mechanical properties of porous and non-porous low-k dielectric films**. *Microelectronic Engineering* 2004, **71**(2):221-228.

120. Shearing PRB, D J L; Brandon, N P: **Towards intelligent engineering of SOFC electrodes: a review of advanced microstructural characterisation techniques**. *International Materials Reviews* 2010, **55**:347-363.

121. Ostergard MJL: **Manganite-zirconia composite cathodes for SOFC influence of structure and composition**. *Electrochimica Acta* 1995, **40**(12):1971-1981.

122. Song H, Kim W, Hyun S, Moon J: **Influences of starting particulate materials on microstructural evolution and electrochemical activity of LSM-YSZ composite cathode for SOFC**. *Journal of Electroceramics* 2006, **17**(2):759-764.

123. Jiang SP: **Sintering behavior of Ni/Y2O3-ZrO2 cermet electrodes of solid oxide fuel cells**. *Journal of Materials Science* 2003, **38**:3775-3782.

124. Koide H, Someya Y, Yoshida T, Maruyama T: **Properties of Ni/YSZ cermet as anode for SOFC**. *Solid State Ionics* 2000, **132**(3-4):253-260.

125. Lee C-H, Lee C-H, Lee H-Y, Oh SM: **Microstructure and anodic properties of Ni/YSZ cermets in solid oxide fuel cells**. *Solid State Ionics* 1997, **98**(1-2):39-48.

126. Wang Y, Coyle T: **Solution Precursor Plasma Spray of Nickel-Yittia Stabilized Zirconia Anodes for Solid Oxide Fuel Cell Application**. *Journal of Thermal Spray Technology* 2007, **16**(5):898-904.

127. Yang YZ, Zhang HO, Wang GL, Xia WS: **Fabrication of Functionally Graded SOFC by APS**. *Journal of Thermal Spray Technology* 2007, **16**(5):768-775.

128. Jorgensen MJ: **Effect of sintering temperature on microstructure and performance of LSM–YSZ composite cathodes**. *Solid State Ionics* 2001, **139**:1-11.

129. Matsushima T, Ohrui H, Hirai T: **Effects of sinterability of YSZ powder and NiO content on characteristics of Ni-YSZ cermets**. *Solid State Ionics* 1998, **111**(3-4):315-321.

130. Suzuki T, Funahashi Y, Yamaguchi T, Fujishiro Y, Awano M: **Design and Fabrication of Lightweight, Submillimeter Tubular Solid Oxide Fuel Cells**. *Electrochemical and Solid-State Letters* 2007, **10**(8):A177-A179.

131. Kim S-D, Moon H, Hyun S-H, Moon J, Kim J, Lee H-W: **Performance and durability of Ni-coated YSZ anodes for intermediate temperature solid oxide fuel cells**. *Solid State Ionics* 2006, **177**(9-10):931-938.

132. Marinsek M, Zupan K, Macek J: **Preparation of Ni-YSZ composite materials for solid oxide fuel cell anodes by the gel-precipitation method**. *Journal of Power Sources* 2000, **86**(1-2):383-389.

133. Chou Y-S: **Microstructure and mechanical propertiesof Sm1−xSrxCo0.2Fe0.8O3**. *J Mater Res,* 2000, **7**:1505–1513.

134. Haanappel V, Mertens J, Rutenbeck D, Tropartz C, Herzhof W, Sebold D, Tietz F: **Optimisation of processing and microstructural parameters of LSM cathodes to improve the electrochemical performance of anode-supported SOFCs**. *Journal of Power Sources* 2005, **141**(2):216-226.

135. Smith JR, Chen A, Gostovic D, Hickey D, Kundinger D, Duncan KL, DeHoff RT, Jones KS, Wachsman ED: **Evaluation of the relationship between cathode microstructure and electrochemical behavior for SOFCs**. *Solid State Ionics* 2009, **180**(1):90-98.

136. Virkar AV, Chen J, Tanner CW, Kim J-W: **The role of electrode microstructure on activation and concentration polarizations in solid oxide fuel cells**. *Solid State Ionics* 2000, **131**(1-2):189-198.

137. Zhao F, Jiang Y, Lin GY, Virkar AV: **The effect of electrode microstructure on cathodic polarization**. In: *Seventh International Symposium of Solid Oxide Fuel Cell 2001; Pennington NJ*; 2001: 501-510.

138. Jiang SP: **Activation, microstructure, and polarization of solid oxide fuel cell cathodes**. *Journal of Solid State Electrochemistry* 2005, **11**(1):93-102.

139. Dusastre V, Kilner JA: **Optimisation of composite cathodes for intermediate temperature SOFC applications**. *Solid State Ionics* 1999, **126**(1-2):163-174.

140. Klemenso T, Appel CC, Mogensen M: **In Situ Observations of Microstructural Changes in SOFC Anodes during Redox Cycling**. *Electrochemical and Solid-State Letters* 2006, **9**(9):A403-A407.

141. Hsiao YC, Selman JR: **The degradation of SOFC electrodes**. *Solid State Ionics* 1997, **98**(1-2):33-38.

142. Simwonis D, Tietz F, Stöver D: **Nickel coarsening in annealed Ni/8YSZ anode substrates for solid oxide fuel cells**. *Solid State Ionics* 2000, **132**(3-4):241-251.

143. Primdahl S, Sørensen BF, Mogensen M: **Effect of Nickel Oxide/Yttria-Stabilized Zirconia Anode Precursor Sintering Temperature on the Properties of Solid Oxide Fuel Cells**. *Journal of the American Ceramic Society* 2000, **83**(3):489-494.

144. Robertson NL, Michaels JN: **Oxygen Exchange on Platinum Electrodes in Zirconia Cells: Location of Electrochemical Reaction Sites**. *Journal of The Electrochemical Society* 1990, **137**(1):129-135.

145. Tsai C-L, Schmidt VH: **Tortuosity in anode-supported proton conductive solid oxide fuel cell found from current flow rates and dusty-gas model**. *Journal of Power Sources* 2011, **196**(2):692-699.

146. A. Weber: **Interaction between microstructure and electrical properties of screen printed cathodes in SOFC single cells**. *J Electrochem Soc Jpn* 1996, **64**(6):582-589.

147. Lee JH, Heo JW, Lee DS, Kim J, Kim GH, Lee HW, Song HS, Moon JH: **The impact of anode microstructure on the power generating characteristics of SOFC**. *Solid State Ionics* 2003, **158**(3-4):225-232.

148. Suzuki T, Hasan Z, Funahashi Y, Yamaguchi T, Fujishiro Y, Awano M: **Impact of Anode Microstructure on Solid Oxide Fuel Cells**. *Science* 2009, **325**(5942):852-855.

149. Lee DS, Lee JH, Kim J, Lee HW, Song HS: **Tuning of the microstructure and electrical properties of SOFC anode via compaction pressure control during forming**. *Solid State Ionics* 2004, **166**(1-2):13-17.

150. Hagiwara A, Hobara N, Takizawa K, Sato K, Abe H, Naito M: **Microstructure control of SOFC cathodes using the self-organizing behavior of LSM/ScSZ composite powder material prepared by spray pyrolysis**. *Solid State Ionics* 2007, **178**(15-18):1123-1134.

151.  Gunda NSK, Choi H-W, Berson A, Kenney B, Karan K, Pharoah JG, Mitra SK: **Focused ion beam-scanning electron microscopy on solid-oxide fuel-cell electrode: Image analysis and computing effective transport properties**. *Journal of Power Sources* 2011, **196**(7):3592-3603.

152.  Joshi AS, Grew KN, Peracchio AA, Chiu WKS: **Lattice Boltzmann modeling of 2D gas transport in a solid oxide fuel cell anode**. *Journal of Power Sources* 2007, **164**(2):631-638.

153.  Tanner CW, Fung K-Z, Virkar AV: **The Effect of Porous Composite Electrode Structure on Solid Oxide Fuel Cell Performance**. *Journal of The Electrochemical Society* 1997, **144**(1):21-30.

154.  Maggio G, Ielo I, Antonucci V, Giordano N: **Morphological optimization of a SOFC anode based on theoretical considerations: a preliminary approach**. In: *the Second International Symposium on Solid Oxide Fuel Cells: 1991; Luxembourg*; 1991: 611-620.

155.  Tiedemann WH, Newman JS: **Porous electrode theory with battery applications**. *AIChE Journal* 1975, **21**:25-41.

156.  Deng X, Petric A: **Geometrical modeling of the triple-phase-boundary in solid oxide fuel cells**. *Journal of Power Sources* 2005, **140**(2):297-303.

157.  Wang C-T, Smith JM: **Tortuosity factors for diffusion in catalyst pellets**. *AIChE Journal* 1983, **29**(1):132-136.

158.  Fleig J, Maier J: **The Influence of Laterally Inhomogeneous Contacts on the Impedance of Solid Materials: A Three-Dimensional Finite-Element Study**. *Journal of Electroceramics* 1997, **1**(1):73-89.

159.  Golbert J, Adjiman CS, Brandon NP: **Microstructural modeling of solid oxide fuel cell anodes**. *Industrial & Engineering Chemistry Research* 2008, **47**(20):7693-7699.

160.  Wilson JR, Kobsiriphat W, Mendoza R, Chen H-Y, Hiller JM, Miller DJ, Thornton K, Voorhees PW, Adler SB, Barnett SA: **Three-dimensional reconstruction of a solid-oxide fuel-cell anode**. *Nature Materials* 2006, **5**(7):541-544.

161.  Shearing PR, Golbert J, Chater RJ, Brandon NP: **3D reconstruction of SOFC anodes using a focused ion beam lift-out technique**. *Chemical Engineering Science* 2009, **64**(17):3928-3933.

162.  Iwai H, Shikazono N, Matsui T, Teshima H, Kishimoto M, Kishida R, Hayashi D, Matsuzaki K, Kanno D, Saito M: **Quantification of SOFC anode microstructure based on dual beam FIB-SEM technique**. *Journal of Power Sources* 2010, **195**(4):955-961.

163.  H. Galinski JR, J. L. M. Rupp, A. Bieberle-Hutter,, L. J. Gauckler: **Ostwald ripening and oxidation kinetics of nicklel gadolinia doped ceria anodes**. *ECS Trans* 2009, **25**(2):2057-2060.

164.  Wilson J, Cronin JS, Rukes S, Duong A, Mumm D, Barnett SA: **Analysis of LSM-YSZ Composite Cathode Phase Connectivity Using Three-Dimensional Reconstructions**. *ECS Transactions* 2009, **25**(2):2283-2292.

165.  Gostovic D, Smith JR, Kundinger DP, Jones KS, Wachsman ED: **Three-Dimensional Reconstruction of Porous LSCF Cathodes**. *Electrochemical and Solid-State Letters* 2007, **10**(12):B214.

166.  Joos J, Carraro T, Weber A, Ivers-Tiffée E: **Reconstruction of porous electrodes by FIB/SEM for detailed microstructure modeling**. *Journal of Power Sources* 2010.

167.  Matsuzaki K, Shikazono N, Kasagi N: **Three-dimensional numerical analysis of mixed ionic and electronic conducting cathode reconstructed by focused ion beam scanning electron microscope**. *Journal of Power Sources* 2011, **196**(6):3073-3082.

# 3 Experimental Methods

This Chapter is intended to introduce the main techniques involved in sample preparation, property characterisation, measurements and analysis of the practical part in this work, including the principles of equipment and the analysis procedure of data. Particular attention is paid to nanoindentation, the FIB/SEM tomography and 3D reconstruction technology.

## 3.1 Sample Preparation

### 3.1.1 Bulk LSCF6428 and CGO Samples

Bulk samples of LSCF6428 and CGO in pellet form to be used for mechanical measurements and as substrates for ink deposition were fabricated using commercially available powders of $La_{0.6}Sr_{0.4}Co_{0.2}Fe_{0.8}O_{3-\delta}$ and $Ce_{0.9}Gd_{0.1}O_{2-\delta}$ (both provided by Fuelcellmaterials.com, Ohio, USA, the specification is shown in Table 3-1), which were pressed uniaxially in a circular stainless steel die of 25 mm diameter by applying approximately 7.3 tons of load (~150MPa of pressure) for 10 s. Care was taken to prevent fracture while the green samples were handled. Acetone was used to clean the die between each pressing process to eliminate cross-contamination. Moreover, in order to avoid cracking issues in the bulk resulting from heterogeneous packing densification during fabrication, isostatic pressing at a pressure of 300 MPa for 30 s was also applied after die-pressing. If cracks persisted in some cases, they might be further avoided by increasing the size of soft agglomerates and hence enhancing the densification when the powders were first isostatically pressed then re-ground with an agate mortar and pestle before die-pressing, then again isostatically pressed after die-pressing.

*Table 3-1 Manufacturer provided powder specification of LSCF6428 and CGO*

| Powder | Surface Area ($m^2/g$) | $D_{50}$ ($\mu m$) |
|---|---|---|
| LSCF6428 | 6.3 | 0.3-0.6 |
| CGO | 6.0 | 0.3-0.5 |

The green samples of LSCF6428 and CGO were then placed in an alumina crucible in the digital controlled furnace BRT-15/5 (ELITE, UK) and sintered in air at high temperature (900, 1000, 1100 and 1200 °C for LSCF6428 pellets and 1400 °C for CGO pellets) for 4 h with a heating and cooling rate of 300 °C/h to generate samples for different experimental purposes described in the sample characterisation section.

In this study, due to the high sensitivity of nanoindentation to surface roughness and for the sake of obtaining sufficiently reliable results close to the true properties, the bulk LSCF6428 specimen surface should be treated to be as smooth and flat as possible. Therefore a rigorous grinding and polishing process was applied to achieve the best possible surface quality. The polishing procedure that included several steps was performed as follows: first, the sintered LSCF6428 pellet was mounted on a copper stub using a low melting point resin. The sample was sequentially ground for ten minutes with SiC grinding papers of grit 240, 400, 800 and 1200. Polishing to a 0.25 μm finish was then followed using an automatic polisher with four fine polishing cloths (to prevent contamination) by separately adding a water-based diamond polishing suspension of successive particle size 6, 3, 1, and 0.25 μm. Each polishing step lasted 45 minutes and the sample was ultrasonically cleaned in distilled water for 10 minutes before changing polishing grade. The copper stub was heated to melt the resin and allow the sample to be removed after the final polishing step. The sample was then cleaned ultrasonically in acetone to remove any residue of either the mounting resin or the polishing compounds.

### 3.1.2 LSCF6428 Films

Three types of LSCF6428 inks were used in this study to produce the films.

(1) INK-A: an LSCF6428 ink (with particle surface area reported by the supplier to be 5.48 $m^2$/g, $D_{50}$ = 0.7 μm and $D_{95}$ = 6 μm) prepared and provided by ESL, UK.

(2) INK-B: an alternative LSCF6428 ink was also prepared, starting with the same LSCF6428 powders used for the fabrication of dense pellet samples. The powders were mixed with ink vehicle from Fuelcellmaterials.com (NexTech Materials Ltd., USA) by a volume fraction ratio of 74%:26% and then ball-milled using 2.5mm zirconia balls in a 250 ml container for 20 h. After milling, the ink was transferred to a large petri dish for subsequent deposition.

(3) INK-C1 and INK-C2: two other reformulated inks were prepared respectively by diluting INK-A 1:1 and 1:2 by volume with terpineol (Sigma, UK) and then ball-milled for 12 h to reach homogeneity. These two inks are referred to as INK-C1 and INK-C2, respectively, in this study.

The above inks were separately sealed in bottles and kept in cold storage such that their viscosities would not change significantly. They were then deposited on the polished dense CGO or LSCF6428 pellets using either screen printing with 250 mesh screen and 2.5 mm gap or tape casting with a perimeter mask of 40 μm height. The as-deposited films were then oven-dried at 100 °C for 12 h, followed by sintering in air at 900-1200 °C for 4 h with a heating and cooling rate of 5 ºC/min. An example of the final part of LSCF/CGO film/substrate is illustrated in Fig. 3-1.

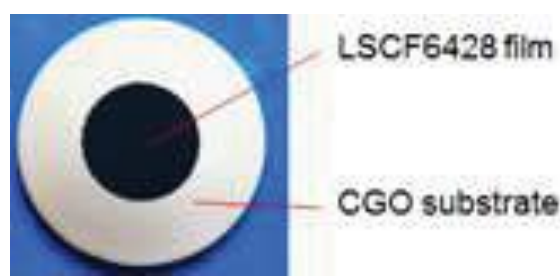

*Fig. 3-1 LSCF6428 film screen printed onto CGO substrate pellet, after sintering.*

## 3.2 Material Characterisation

### 3.2.1 Particle Size Distribution Measurement

The particle size and distribution of the powders and inks were measured using a Malvern particle analyser based on laser diffraction (i.e. low angle laser light scattering technique). The principle is that the diffraction angle and thus the amount of light scattered is inversely proportional to the particle size. The powders and inks were dispersed and stirred in deionised water and particles in the suspensions were measured by re-circulating the sample in front of a laser beam of optical wavelength. Using this equipment, the volume median diameter $D_{50}$ can be measured and the particle size distribution (i.e. frequency) obtained.

### 3.2.2 X-Ray Diffraction

In this study, a Philips PW 1729 diffractometer (operated at 40 kV and 40 mA) with a copper target producing Cu-K$_\alpha$ radiation was used to obtain XRD data and to confirm phase purity and for structural analysis. All scans were performed between 10°-100° ($2\theta$) with a step size of 0.02°/s and a dwell time of 10 seconds. The data analysis was performed using General Structure Analysis System and EXPGUI software [1, 2] based on Rietveld refinement which provided a method of analysing powder XRD patterns to obtain structural information (such as lattice parameters and space group) and compositional information (such as relative quantities of phases) based on a least squares refinement procedure [3, 4].

### 3.2.3 Density and Porosity Measurements for Bulk Samples

The bulk geometric densities, relative densities and open porosities of the samples sintered at different temperatures were measured using an Archimedes' Principle balance of 0.1 mg accuracy. Equation 3.1 was employed to calculate the densities.

$$\rho_b = \frac{m_1}{m_3 - m_2} \times \rho_l \qquad 3.1$$

where: $m_1$ = the mass of the dry sample; $m_2$ = apparent mass of the sample immersed in liquid; $m_3$ = mass of soaked sample; $\rho_l$ = density of liquid (in this case water). The samples were ultrasonically cleaned with ethanol for 10 min and dried at 90 °C for 10 h before the measurement. In the measurement, they were immersed in distilled water under vacuum for 1 hour to make sure all open pores were filled before recording $m_2$ and $m_3$. The sample's theoretical density $\rho_0$ (6.309 g/cm$^3$ for LSCF6428 and 7.225 g/cm$^3$ for CGO) was derived from the data in the online Inorganic Crystal Structure Database (ICSD) via Chemical Database Service [5] and the sample's relative density $\rho_r$ was recorded as a percentage of the bulk density relative to the theoretical density, from which the total porosity was also determined, i.e. $p = (1-\rho_r) \times 100\%$.

### 3.2.4 Porosity Measurement for As-sintered LSCF6428 Films

Due to the very small volume of thin LSCF6428 films after sintering at 900 - 1200 °C, it was not possible to apply the method explained above to measure their porosities. Instead, they were measured using the quantification module of Avizo 8.0 image processing software (VSG Co., USA) based on the actual 3D microstructural data of the films collected using the FIB/SEM tomography technique. Details will be discussed in the following chapters.

### 3.2.5 Surface Roughness Measurement for As-sintered LSCF6428 Films

The average surface roughness $R_a$, which is defined as the arithmetic average of the absolute vertical deviations of the roughness profile from its centerline through a prescribed sampling length, of the as-sintered LSCF6428 films was measured using an optical interference surface profiler OMP-0360G (Zygo, USA), which has the advantages of non-contact and high accuracy and is suitable to measure surface roughness of materials with highly porous fragile microstructures.

### 3.2.6 Ink Viscosity Measurement

The instantaneous viscosity of the LSCF6428 inks was measured by controlled-stress rheometer (CVO100D, Malvern Instruments Ltd, UK) at room temperature under shear rates in the range from 0 to 95 s$^{-1}$.

### 3.2.7 Mechanical Property Characterisation

### 3.2.7.1 Impulse Excitation Test (IET) for Elastic Modulus Measurement

The elastic modulus of the nominally dense LSCF6428 pellets was measured first using the impulse excitation technique (IET), which is a dynamic (high frequency) and macroscopic method. The experiment was carried out following British Standard EN 843-2:2006 [6], which involves measuring two natural vibrational frequencies of the sample.

According to the standard, a disc-shaped pellet, supported by a damping material at four equally spaced points on the nodal circle at approximately 0.7 of its radius, is excited by a light mechanical impulse at the nodal circle (between the support points) and centre, designated points P1 and P2. A microphone transducer placed at the vicinity of the sample transmitted the resulting sound vibrations of flexure mode and anti-flexure mode, to the signal processing unit, as illustrated in Fig. 3-2. The fundamental resonance frequencies were recorded and used to calculate Young's modulus and Poisson's ratio.

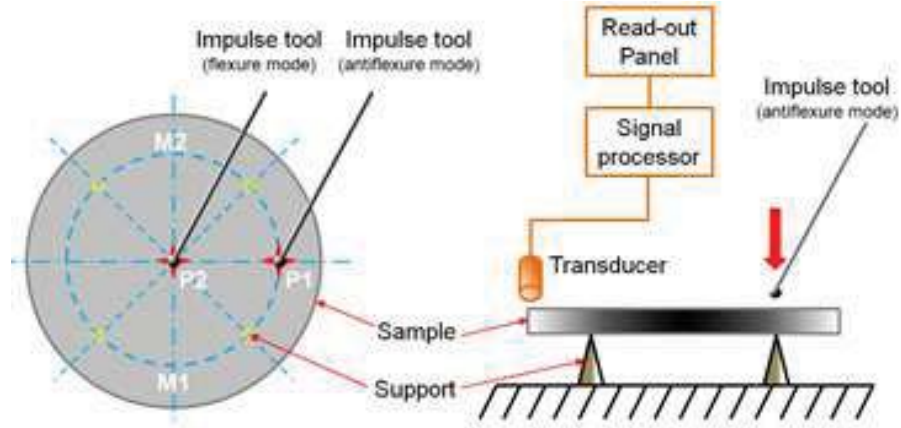

*Fig. 3-2 Schematic of impulse exciting test for disc shape sample with top view on the left and side view on the right (P1, P2: first and second mode impulse points; M1, M2: corresponding microphone transducer points).*

In this study, a commercial resonance system (GrindoSonic MK5, J.W. Lemmens, Belgium) was employed. The measurements were conducted on LSCF6428 dense samples 25 mm in diameter and approximately 2 mm thick. The samples were surface polished to meet the requirement of the standard, and four low-damping wooden rods, of identical length and with sharp tips, were vertically fixed on a foam substrate with their tips touching the four nodal points on the sample's under-surface to support the sample. The microphone transducer was held near points M1 and M2 when the impulse points P1 and P2 were lightly tapped respectively using a 0.5 mm diameter steel sphere attached to a plastic rod to excite the vibration.

After the two frequencies, $f_1$ and $f_2$, were measured, from which two dynamic Young's moduli, $E_1$, $E_2$, were calculated by the following equation [6]:

$$E_{i,i=1,2} = 12\pi f_i^2 d_i^2 m_i^2 \frac{1-\upsilon^2}{K_i^2 t_i^3} \qquad\qquad 3.2$$

where $d$ = the pellet diameter in mm; $t_p$ = the pellet thickness in mm; $m$ = the pellet mass in g; $f_1$ = the first mode frequency in Hz; $f_2$ = the second mode frequency in Hz; $K_1$ = the geometric factor for first vibration mode; $K_2$ = the geometric factor for second vibration mode; $v$ = the Poisson's ratio. The Poisson's ratio, along with values of $K_1$ and $K_2$ were determined according to the parameter tables from [6], as a function of the resultant ratios of $2t_p/d$ and $f_2/f_1$. The average value of dynamic Young's modulus, $E$, was determined by taking the average of $E_1$ and $E_2$.

### *3.2.7.2 Nanoindentation for Elastic Modulus and Hardness Measurement*

The nanoindentation technique was first developed in the mid-1970s to measure the hardness of small volumes of material [7]. Significant work has been done by Oliver and Pharr [8] with an improved analysis method (i.e. the Oliver-Pharr method) for determining elastic modulus and hardness from the indentation load-displacement data. The technique is also known as depth-sensing indentation and uses very small scale (submicron) tip indenters and control of load and displacement, with high spatial resolution to place the indenters, and provides real-time load-displacement data. These advantages allow mechanical property tests for small volume materials at under submicron scales, because the indentation area may only be a few square micrometres or even nanometres. For this reason, the method is used extensively to measure the mechanical properties of thin films with small structural feature size of a few microns [9]. Nanoindentation is an ideal technique to investigate the mechanical properties of LSCF6428 films that were produced previously.

The combined nanoindentation and microindentation instrument used in this project was the MML NanoTest platform from Micro Materials Limited, UK, as shown in Fig. 3-3. A polished sample was mounted on a cylindrical aluminium holder using a low melting resin. The holder was then mounted to the platform and the holder surface was perpendicular to the indenter tip.

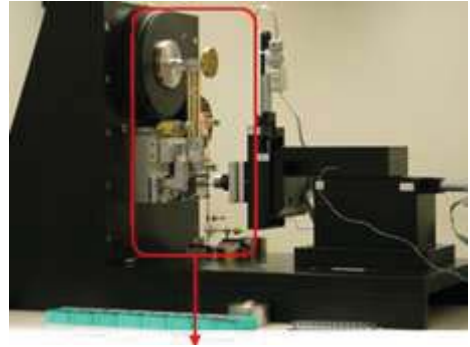

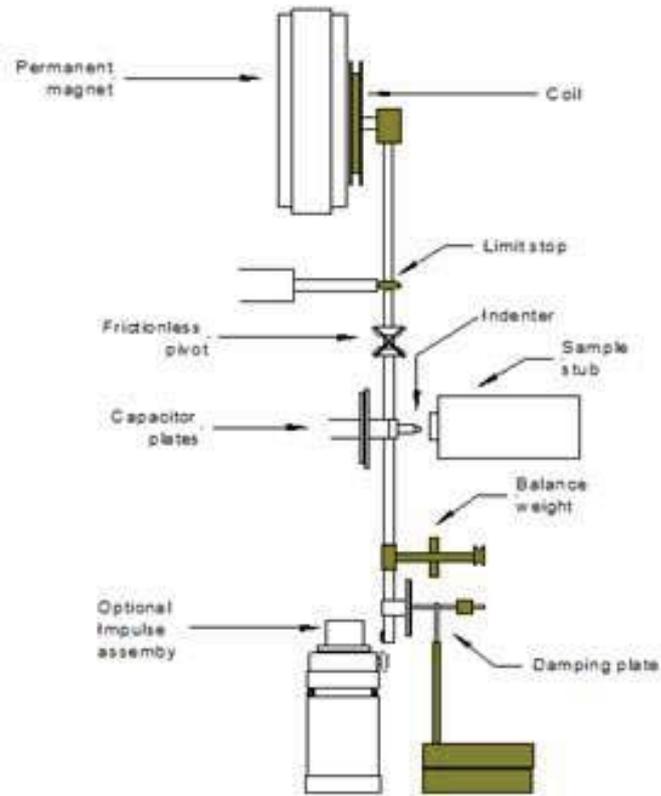

*Fig. 3-3 NanoTest machine and the schematic of the setup [10]*

The main functional components of a nanoindentation machine are the indenter tip, the sensors and actuators used to apply and measure the mechanical load and indenter displacement. The former is usually either a Berkovich tip made of diamond with highly precise three-sided pyramid geometry, or a spherical tip made of diamond or sapphire as demonstrated as sketches in Fig. 3-4.

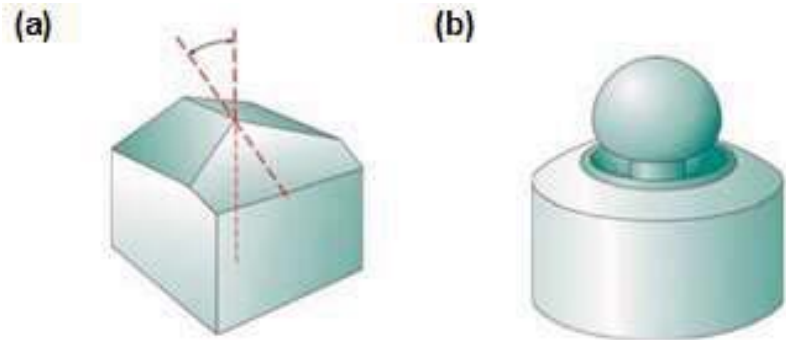

*Fig. 3-4 Illustrations of (a) a Berkovich tip; (b) a spherical tip [11].*

The elastic modulus and hardness of the bulk and film LSCF6428 samples were measured using nanoindentation, which is a static and microscopic method, in contrast to IET. All tests were made on the NanoTest platform at room temperature (~25 °C) and in a clean-air environment. Two indenter tips were used in this study depending on the sample forms under investigation; a Berkovich diamond tip and a spherical diamond tip of 50 μm diameter. The Young's moduli and Poisson's ratios for the diamond tips are $E_i$ = 1141 GPa and $v_i$ = 0.07, respectively. Note that compared to sharp indenter tips, such as the Berkovich tip, spherical tips facilitate the distinction from elastic to plastic deformation of materials during indentation due to their less drastic variation of stress (or stress singularity) under loading and hence the unlikelihood of plasticity induced immediately upon contact [12]. The spherical tip is particularly appropriate for porous materials as the deformation zone can be arranged to be of much greater length scale than the typical length scale of the porous microstructure (e.g. the average pore diameter). They thus can give a result that characterises the long range properties of the porous material. Other benefits of using spherical tips include less sensitivity to surface condition and more accurate resultant hardness [13].

At least 20 measurements were conducted in different locations for each sample in order to measure the variability of the mechanical response of the sample. Prior to nanoindentation tests, the NanoTest platform was precisely calibrated to establish the system frame compliance using a standard fused silica sample (Young's modulus $E_{fs}$ = 72 GPa, Poisson's ratio $v_{fs}$ = 0.17). Normally, the measurements were performed using the following parameters: indenter tip loading and unloading rate 1 mN/s; contact velocity 0.05 μm/s; maximum load allowed 500mN; maximum depth allowed ranging from 100-5000 nm according to the actual sample thickness. The routine also involved two periods of dwell time at peak load and at the end of unloading, respectively, depending on the testing samples, to accommodate the creep effect and thermal drift. 2 seconds and 20 seconds of dwell time were applied for those two periods in preliminary nanoindentation tests. In addition, the vibrations transmitted to the device, atmospheric temperature and pressure fluctuations during the course of an experiment might cause significant errors. To minimise these influences, the instrument was housed in an anti-vibration table-based environmental isolation cabinet, which reduced air turbulence upsetting the NanoTest pendulum, provided a thermally controlled environment, and also provided sound-proofing to reduce acoustic disturbance.

During a typical nanoindentation process, mechanical load and penetration depth (i.e. displacement) are recorded as the indenter tip is pressed into the test material's surface with a predefined loading/unloading rate. A record of these values can be plotted to create a load-displacement curve, which is often referred to as the $P$-$h$ curve, such as Fig. 3-5 (b) depicts. The shape of the $P$-$h$ curve differs from one material to another. The elastic moduli and hardness of the as-sintered LSCF6428 dense pellets and porous films were calculated by applying the Oliver-Pharr method [8] based on the $P$-$h$ data obtained.

Fig. 3-5 (a) shows the geometry of the indented sample during nanoindentation, assuming that an elastic process takes place with a recovery depth of $h_e$ upon unloading. For a porous film, the elastic modulus measured with a spherical tip is therefore the long range composite modulus which is affected by its porous microstructure. Indenting to a maximum depth less than 10% of the thickness of a film (namely $h_{max}/t_f < 0.1$) has been empirically considered as a safe condition to avoid effects from substrate and extract intrinsic film properties in routine nanoindentation tests [8]. However, in the current study we show later that the effect of the substrate is negligible for these porous films for penetration up to $h_{max}/t_f \cong 20\%$. The substrate effect, as well as the effect of surface roughness on nanoindentation measurement, will be further discussed later.

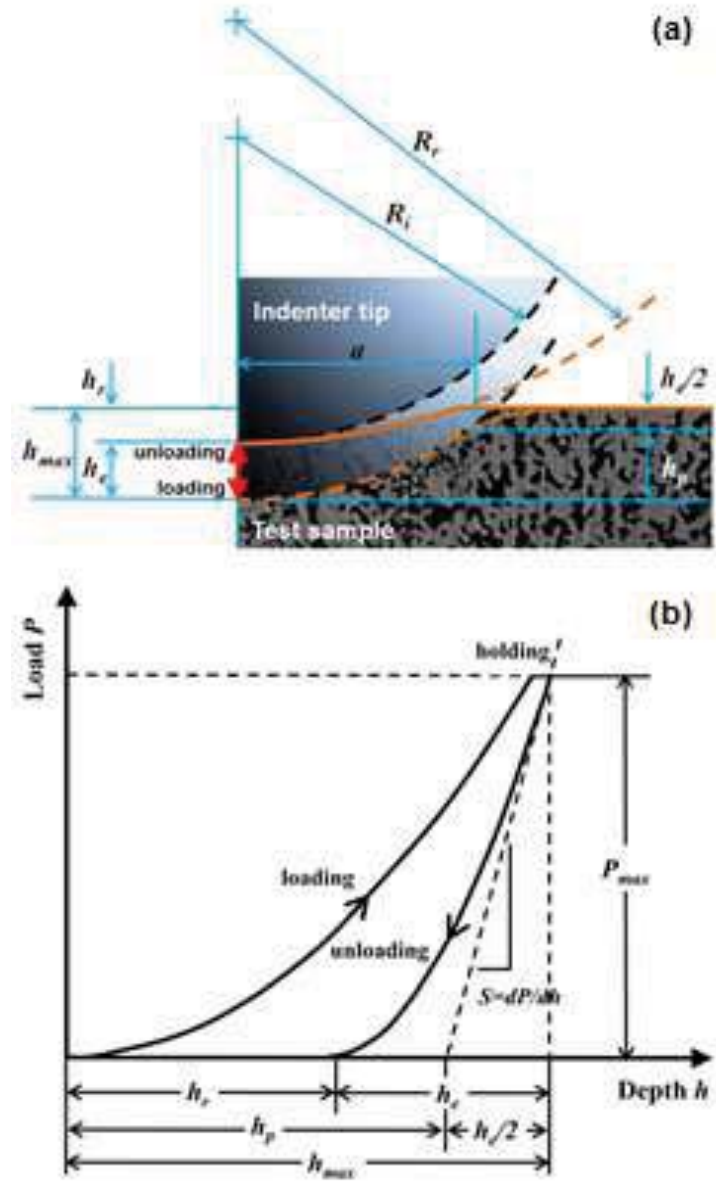

*Fig. 3-5 (a) Profile and geometries of sample under loading and unloading with a holding period at peak load $P_{max}$ using a rigid spherical indenter tip with radius of $R_i$; (b) response curve of load versus indent depth and the corresponding geometries. Notice that there is a residual depth $h_r$ with a cross-sectional profile radius $R_r$.*

In Fig. 3-5 (b), $P_{max}$ is the peak indentation load, $h_{max}$ is the indenter depth at peak load, $h_r$ is the residual depth of the impression after unloading and $S$ is the initial unloading stiffness, i.e. the slope of the unloading part of the curve (d$P$/d$h$). According to the Oliver-Pharr method, the indentation hardness $H$ (i.e. physically the effective contact pressure) and the reduced elastic modulus $E_r$ of the sample are given by:

$$H = \frac{P_{\max}}{A} \qquad\qquad 3.3$$

$$E_r = \beta \frac{\sqrt{\pi}}{2} \frac{S}{\sqrt{A}} \qquad 3.4$$

where $A$ is the projected contact area related to the plastic contact depth $h_p$, (which only accounts for plastic deformation and thus the hardness measured is load-independent plastic hardness), and $\beta$ is a geometry correction factor that depends on the shape of the indenter.

Regarding Equations 3.3 and 3.4, for a Berkovich indenter tip $\beta = 1.034$ and the area function is $A = 24.5h_p^2$, where $h_p$ is the contact depth. Hence Equation 3.5 for the Berkovich tip becomes:

$$E_r = 0.328 \frac{S}{h_p} = 0.328 \frac{1}{h_p} \frac{dP}{dh}\Big|_{P=P_{max}}^{unloading} \qquad 3.5$$

For a spherical indenter tip, $\beta = 1$ and $A = \pi a^2$, where $a$ is the radius of the contact circle at maximum depth $h_{max}$ and is given by: $a = \sqrt{2R_i h_p - h_p^2}$, where $h_p = h_{max} - \frac{h_e}{2}$ with reference to Fig. 3-5.

Hence for the spherical tip the equations become:

$$H = \frac{P_{max}}{\pi a^2} \qquad 3.6$$

$$E_r = \frac{1}{2a} \frac{dP}{dh}\Big|_{P=P_{max}}^{unloading} \qquad 3.7$$

The Young's modulus $E_s$ of the sample can then be calculated from the reduced modulus $E_r$ of the indenter-sample system through the following relationship [8]:

$$\frac{1}{E_r} = \frac{1 - v_i^2}{E_i} + \frac{1 - v_s^2}{E_s} \qquad 3.8$$

Here, the subscripts $i$ and $s$ indicate respectively a property of the indenter material and the test sample. The $E_i$ and $v_i$ of the diamond Berkovich and spherical indenter tips used in this study were given earlier. A sensitivity study showed that the indentation results depend little on the variation of Poisson's ratio, i.e. no more than 8% of change in sample elastic modulus when varying the Poisson's ratio in the range of 0.2 to 0.4 in the calculation [14]. As a result, the Poisson's ratio $v_s = 0.3$ of the fully dense LSCF6428 material [15] was used in this study for calculation based on nanoindentation.

### 3.2.7.3 Indentation for Fracture Toughness Measurement

Trials of fracture toughness $K_{Ic}$ measurement of both types of samples (i.e. porous films and bulk pellets) were also conducted using indentation microfracture (IM) methods (by both nanoindentation and microindentation where applicable). The comparison of the measurements of $K_{Ic}$ with Vickers and Berkovich indenters has been extensively discussed in the work of Dukino and Swain [16]. Their observations also revealed that the Berkovich indenter offers more advantages and gives more consistent estimates of toughness than using a Vickers indenter. Therefore, in the current study a Berkovich indenter tip was used to induce cracks. Samples with sound surface quality were prepared prior to indentation: bulk pellets were surface polished as when $E$ and $H$ were measured; films were fabricated in a crack-free form with very flat surface as a result of the film surface quality improvement detailed in Chapter 4. For each sample, at least 10 indentations were performed to average the measurement, by applying suitable loads at up to 500 mN or from 1 to 20 N.

Various formulae [17-20] have been proposed to connect the fracture toughness to the indenter type, crack geometries, load and material properties. One of the most widely used empirical formulae for the calculation of fracture toughness with mode I cracking using Vickers indentation was proposed by Lawn *et al.* [20], as shown below:

$$K_{Ic} = \alpha \left( \frac{E}{H} \right)^{1/2} \frac{P}{c^{3/2}} \qquad\qquad 3.9$$

where $\alpha$ is an empirical constant depending on the elastic/plastic deformation behaviour of the test material and the indenter geometry. It equals to 0.015 for a Vickers indenter. $E$ and $H$ are elastic modulus and hardness of the test sample, respectively. Fig. 3-6 (a) shows the schematic of the indent and radial cracks produced by indentation using Vickers tip, showing $c$ the crack length from the indent centre to crack tip and $a$ the half diagonal length of the indent area, which can be estimated using an optical microscope or SEM.

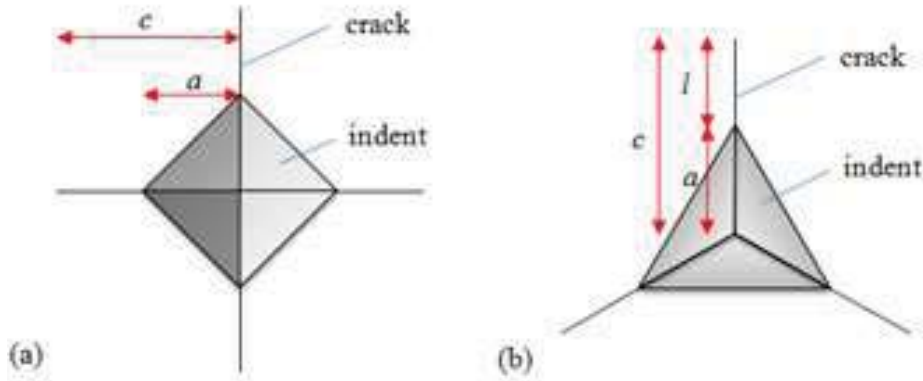

*Fig. 3-6 Schematic of indent and radial cracks generated by (a) Vickers indenter, and (b) Berkovich indenter*

On the other hand, for a Berkovich tip, which normally results in an indent as shown in Fig. 3-6 (b), Equation 3.9 was more specifically altered by Dukino and Swain [16], by applying a relationship of $\alpha$ between Berkovich and Vickers indenters proposed by Ouchterlony [21], who distinguished the star-shaped cracks with the number of crack paths, from which $\alpha_b$ and $\alpha_v$ have the relationship of: $\alpha_b = 1.073\alpha_v = 0.016$, with $\alpha_v = 0.015$ as described earlier. The fracture toughness with a Berkovich tip could then be obtained using the following equation [16, 19],

$$K_{Ic} = 0.016 \left( \frac{a}{l} \right)^{1/2} \left( \frac{E}{H} \right)^{2/3} \frac{P}{c^{3/2}}$$

3.10

where $P$ is the maximum indentation load, $c$ is the radial crack length measured by SEM from indent centre to the radial crack tip (i.e. crack end), $a$ is the diagonal length of the indent from centre to the indent corner and $l$ is the crack distance measured from the indent corner to the crack tip and was calculated as $l = c\text{-}a$, as shown in Fig. 3-6 (b). From a phenomenological point of view, Equation 3.10 is more relevant to the case of indentation-induced radial/Palmqvist crack morphology which is more likely to happen with the Berkovich indentation in the present study and is different from half-penny-shaped/median cracks as discussed in Chapter 6.

Although Equation 3.10 is applicable to the IM of thin films, it is worth noting that similar to the measurements of $E$ and $H$, effects on the film fracture process from the substrate, the film densification and residual stresses in the film are inevitable [22]. This can complicate the fracture toughness measurement of the film and effort must be made to minimise such effects. These, particularly the substrate effect, have only been acknowledged by some previous work on fully dense coatings without any in-depth investigation [22-25]. Moreover, little information can be found in the literature regarding such effects on porous thin films.

### 3.2.8 Microstructure Characterisation

### 3.2.8.1 Scanning Electron Microscopy

Scanning electron microscopy (SEM) was used to investigate the surface microstructure of sample surfaces and to characterise the change in microstructure after nanoindentation. A SEM can operate in secondary electron imaging (SEI) mode or in backscattered electron imaging (BEI) mode.

The specimen preparation process generally involves coating a thin layer of gold or carbon covering the specimen with a small or no electrical conductivity (such as the CGO in this study) to allow an electron conductive path to be created by connecting the surface of the specimen with the holder (i.e. electrically grounded) [26]. As the LSCF6428 samples all exhibited a sufficient electrical conductivity at room temperature, it was not necessary to use a conductive coating for them. Therefore, these samples were simply mounted onto aluminium stubs using carbon tape. Cross-sectional thickness measurement of sintered LSCF6428 films on CGO substrates required further preparation. The aluminium stub was orthogonally cut from the middle, and the specimen was then mounted on its side perpendicular to the bottom of the stub with carbon tape, which allowed the cross-section surface to be parallel to the SEM holder surface. Again the sample was coated with gold before imaging. Ag paste was deposited around the sample edge connecting to the stub to ensure conduction.

The SEM employed in this study was a JSM-5610LV (JEOL, Japan). The SEM operating parameters varied depending upon the materials used but in this study, generally the working distance was set to be 15 mm, accelerating voltage was 20 kV and spot size 30 nm (smaller size was required, such as 26 nm for higher magnification (> 6000) observation.

### 3.2.8.2 Optical Microscopy

Optical microscopy was used as a precursor to SEM characterisation. For example, the indents produced by nanoindentation were difficult to locate and distinguish with SEM due to the grey-coloured image effect while the optical microscope was used in bright field. The use of this microscope also allowed better identification of the cracking network in the films. The live images were displayed, captured, stored and analysed in a computer using a digital camera connected to the microscope. However, compared to SEM the optical microscope was of poor resolution. Therefore, SEM was extensively used to carry out precise observations and analysis.

### 3.2.8.3 FIB/SEM Dual Beam Tomography

To date, SEM is widely used as a non-destructive plan surface imaging technique capable of investigating 2D material microstructures, particularly the morphology, which are important links between certain materials properties and the processing methods applied. However, one of the most critical drawbacks of this technique is that it is impossible to examine the interconnected nature of regions within 3D space, such as tortuosity, which is directly related to the electrochemical performance of materials. The recently developed dual beam system, i.e. FIB/SEM (focused ion beam/scanning electron microscope), has overcome the above-mentioned disadvantage. As the dual beam system can be well controlled in size and position to nanometre resolution, and the signals are strong enough to be detected without excessive noise, it is a powerful tool for analysing samples in great detail over a wide range of magnifications, allowing the visualisation of 3D microstructures.

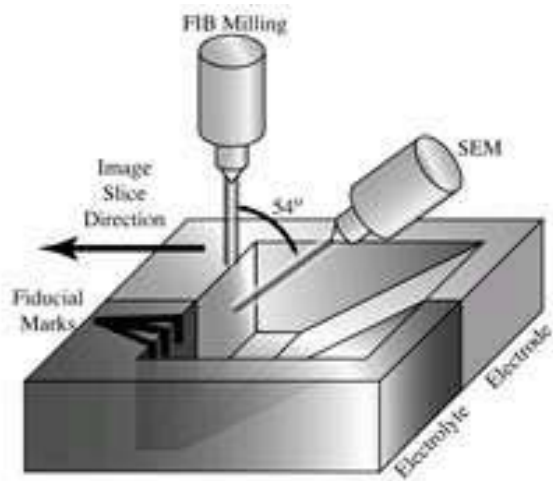

*Fig. 3-7 Schematic of FIB/SEM dual beam system working on an interface region of a SOFC [27].*

In the dual beam system, the FIB is used to mill the cross-section of the sample into slices of desired thickness, each of which is visualised by the SEM (as shown in Fig. 3-7). To start the process, the sample with finely polished surface is placed onto the holder in the chamber using silver paste and tilted to allow the sample surface to be normal to the FIB beam direction. After the chamber is evacuated, the beam coincident point (i.e. the working point on the sample) is then located. In order to prevent the remaining material from being destroyed by the high energy ion beam and to ensure the quality of image, a thin layer (~1 μm) of platinum is often deposited onto the material's upper surface using the incorporated Gas Injection System (GIS) before milling starts. When the sample and the system are ready, the gallium ion beam begins milling the material slice-by-slice with predefined area and depth. In the meantime, the SEM images the FIB-machined surface before the next slicing begins. As the FIB slicing process carries on, the surface information is acquired for each slice and the images are collected successively. In the final step, the images are combined in sequence and the material's 3D microstructure is ready to reconstruct. Next, with the aid of the reconstruction module of the software, the real-image-based microstructure was converted into a digital matrix of voxels for further processing; for example, visualising characteristic features and applying selective transparency to specific features based upon image or feature contrast, such as material phases and pores.

The FEI Helios NanoLab 600 FIB/SEM was used in this project for a simultaneous site-specific FIB cross-sectioning and high resolution SEM imaging. It is equipped with two GIS deposition facilities (Pt and C), which can be used for surface protection, patterning, welding, and so on. The resolution along the direction normal to the slice depended upon the predefined slice thickness, which could be on the nanometre scale. The resolution of the instrument is about 4.5 nm at 30 kV using the statistical method and 2.5 nm at 30 kV using the selective edge method [28]. The typical beam voltage and current used in this study are shown in Table 3-2. The rest working parameters are: working distance = 4 mm; imaging scanning speed = 300 ns - 10 μs and tilting angle = 52˚.

*Table 3-2 Typical beam voltage and current used of the FIB/SEM in this study*

| Beam type | Voltage | Current |
|-----------|---------|---------|
| SEM | 5-15 kV | 0.17 nA |
| FIB | 30 kV | Imaging: 93 pA<br>Patterning: 9.3-21 nA<br>Deposition: 0.92 nA<br>Slicing: 0.92 nA |

### *3.2.9 Thermal Analysis*

### *3.2.9.1 Thermogravimetric Analysis*

Thermogravimetric analysis (TGA) of the original commercial ink (i.e. INK-A) was performed in air using a JUPITER simultaneous DSC/TGA instrument (NETZSCH, Germany) at a heating rate of 10 °C/min with temperature heating up to 1000 °C in order to determine the thermal decomposition range of the organic components in the ink.

### *3.2.9.2 Dilatometric Analysis*

Dilatometric analysis on uniaxial-pressed green bodies of LSCF6428 powders were carried out using a push-rod dilatometer DIL-402E (NETZSCH, Germany) to investigate the sintering activity and shrinkage of the LSCF6428 material. The measurements were set from room temperature to 1250°C, with four constant heating rates of 3, 5, 10 and 20 °C/min.

## *3.3 3D Reconstruction and Finite Element Modelling*

### *3.3.1 Sample Preparation and Image Acquisition*

The real digital 3D microstructures of LSCF6428 films sintered at different temperatures from 900-1200 °C were reconstructed based on the serial images collected by SEM imaging of the FIB-prepared cross-sections of the films, as described earlier. It is worth noting that the accuracy of the 3D reconstruction relies on the resolution, contrast and noise level of the acquired SEM images. In general, images with higher resolution and contrast and lower noise level enable more accurate 3D microstructures to be generated. Therefore, before FIB machining, the LSCF6428 films (on the substrates as shown in Fig. 3-8) were impregnated with low viscosity epoxy resin under vacuum to enhance the grayscale contrast and edge definition between the pore phase and solid phase, and to ensure that the highly porous structures outside the milling section remained intact. The sample was then coated with a thin layer of gold to provide good electrical conductivity in the case of CGO substrates. A 1-2 μm thick protective platinum layer was then deposited on the top surface of the film after the sample was put into the FIB/SEM vacuum chamber (Fig. 3-8 (a)), in order to protect the upper surface and to prevent charging during slicing and imaging.

Prior to FIB slicing, it was necessary to machine trenches surrounding the three edges of the surface of interest so that a peninsula-shaped volume of interest (typically 10-20 microns wide and 10 microns deep) for sectioning could be made (Fig. 3-8 (a)). A cross-shaped fiducial mark was then cut on one side of the volume of interest to facilitate image registration during sectioning. Once the volume of interest was prepared and the sectioning parameters were defined, the serial cross-sectional milling by FIB and imaging of the resulting exposed surface by SEM was performed automatically by the instrument. Artefacts affecting the image quality such as tilting, drifting and curtaining arising during the image acquisition process could be corrected or minimised by adjusting the working parameters [29], as described later in Chapter 7.

The working parameters used for the FIB/SEM were listed earlier. In order to facilitate the subsequent meshing and FEM simulation processes, a suitable voxel size, preferably in cubic shape, was highly desired. This could be achieved by setting the slice thickness in Z direction between two adjacent images close to the pixel size of the SEM images acquired. The pixel size of a SEM image could be deduced by linking the horizontal field of view (HFW) and the resolution of the image, namely pixel size = HFW/Horizontal resolution. For instance, for an image obtained with 25.6 nm HFW and 2048×1768 (pixel×pixel) resolution, the pixel size of the image is 25.6 $\mu$m/2048 pixel = 12.5 nm/pixel. In this case the slice thickness should be set to be 12.5 nm so that a cubic voxel could be generated with the voxel size being 12.5×12.5×12.5 nm$^3$. Note that the preferable image format for the subsequent processing was TIF 8 bit encoded files.

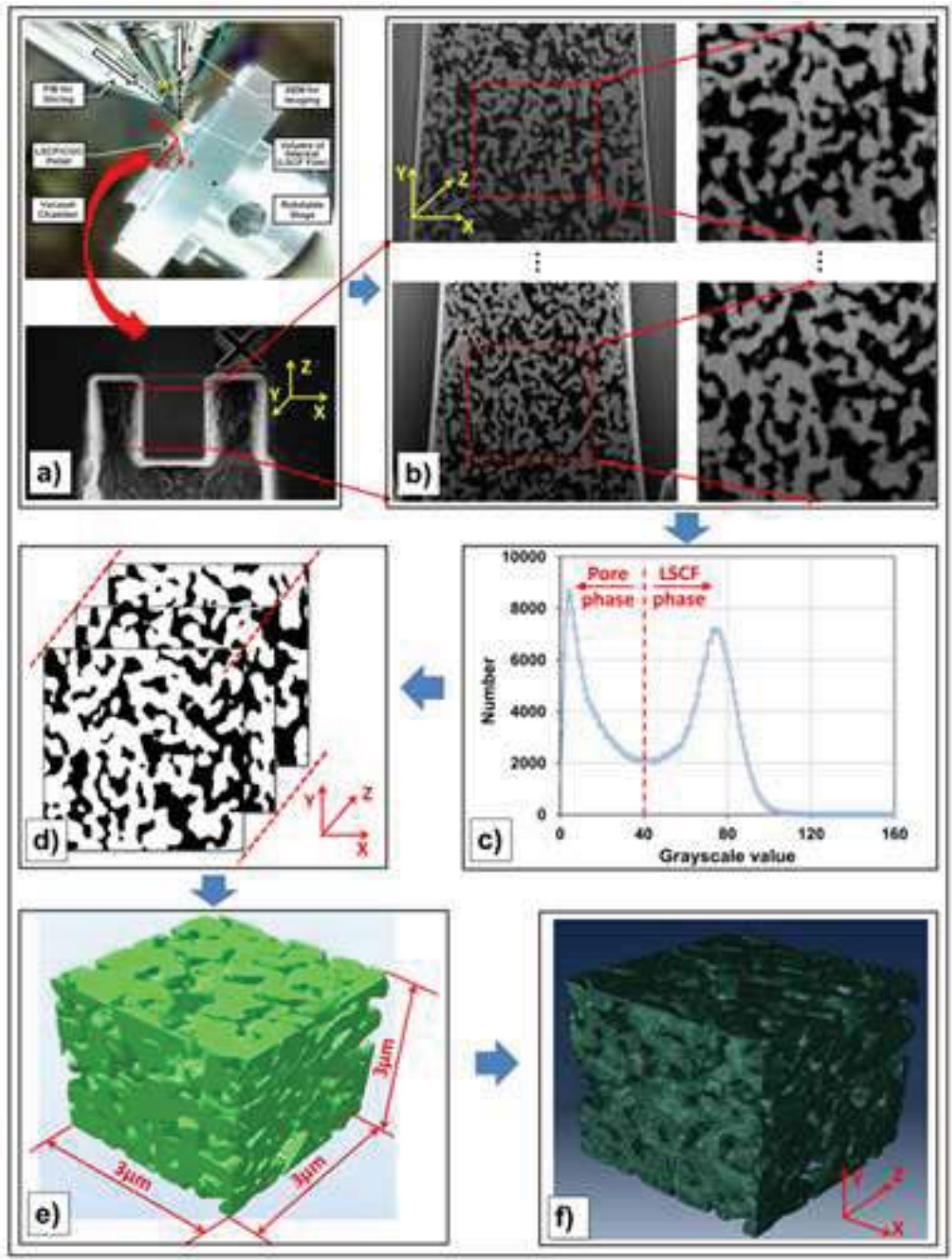

*Fig. 3-8 Steps in the 3D microstructural reconstruction of the LSCF6428 cathode using FIB/SEM. The example shown is a film sintered at 900 °C. (a) Schematic side-view of FIB/SEM set-up during operation and top-view of film under slice and view process; (b) Extraction and alignment of the acquired consecutive images; (c) Thresholding (i.e. segmentation) of the images; (d) Segmented image stack for reconstruction; (e) 3D reconstructed microstructure of the actual volume of interest of the film; (f) Meshing of the digital microstructure for the subsequent FEM.*

### *3.3.2 Image Processing and 3D Reconstruction*

In the current study the samples in both bulk and film forms were composed of two components, i.e. air-filled pores impregnated with resin and solid LSCF6428 material, which were normally represented by dark grey (or black) and light grey (or white) pixels in the FIB-milled SEM images obtained (Fig. 3-8 (b)). The 8 bit images could contain grayscale values ranging from 0 to 255 with 0 and 255 representing black and white pixels, respectively. A grayscale histogram was thus obtained to visualise the amount of pixels of a certain greyscale values (Fig. 3-8 (c)) and to be used for segmentation during the image data processing.

The ensuing image data processing followed a number of steps:

(i) *filtering* were applied for potential smoothing, sharpening of objects and for noise removal in the grayscale images;

(ii) *alignment* of the acquired consecutive images along the sectioning direction;

(iii) *segmentation* of the image stack by thresholding of the grayscale to accurately identify and partition the image into multiple segments which were assigned the same visual characteristics corresponding to relevant material components (or phases). This step would binarise the images and turn background pixels (pores) into black (value 0), foreground pixels (LSCF6428) white (value 1);

(iv) *resampling* of the dataset to generate cubic voxels and thus 3D reconstruction;

(v) *mesh generation* for the finite element model.

The threshold value for binarising the images (i.e. segmentation into pores and solid) was set to be the minimum value between the two peaks in the grayscale histogram, as shown in Fig. 3-8 (c). Applying this value to segmentation of the initial images (e.g. Fig. 3-8 (b)) resulted in the black and white images such as that in Fig. 3-8 (c). Once this was done for each 2D image, area the pore space and solid space were individually interconnected through the third dimension to generate the 3D reconstruction.

In the current study, the working volume of each LSCF6428 film was approximately $15 \times 20 \times 15 \ \mu m^3$, with SEM image pixel dimensions of 12.5 nm and slice thickness of 12.5 nm. The acquired images were then aligned and segmented to identify solid material and porous phase using Avizo 8.0 image processing software (VSG Co., USA). Smaller volumes of size ranging from $3 \times 3 \times 3$ to $10 \times 10 \times 10 \ \mu m^3$, consisting of around 240 - 800 serial 2D images, were extracted from the initial volumes for 3D reconstruction. The resulted 3D microstructural datasets with voxel size of $12.5 \times 12.5 \times 12.5 \ nm^3$ were then processed using ScanIP 6.9 software (Simpleware, UK) to generate high quality smooth hybrid hexahedral and tetrahedral meshes for the solid phase, to be used in the FEM simulation (Fig. 3-8 (f)). Artefacts (such as image intensity gradient that affected reconstruction and regions of isolated solid material that led to FEM calculation abortion) could be manually removed in the semi-auto segmentation process. More details on the image processing, 3D reconstruction and FEM will be found in Chapters 7 and 8.

### 3.3.3 Mechanical Simulation to Calculate Elastic Modulus and Poisson's Ratio for Films

FEM was performed using Abaqus CAE 6.10 (Dassault Systemes, USA) to calculate the effective elastic modulus and Poisson's ratio of the actual reconstructed 3D microstructures of the as-sintered LSCF6428 films. The elastic modulus of dense LSCF6428 was determined by nanoindentation of high density LSCF6428 pellets as described previously. A sensitivity study using FEM for the porous material showed that the results are not sensitive to Poisson's ratio of the solid phase. Varying the Poisson's ratio in the range of 0.2 to 0.4 in the simulation resulted in only 8% change in the calculated elastic modulus of the porous material [14]. As a result, the Poisson's ratio of fully dense LSCF6428 material reported in literature [15], $v_s = 0.3$, was chosen for use in all FEM simulations.

The process of FEM simulation is briefly described here, and the detailed related information will be found in Chapter 6. The simulation was run for each microstructure corresponding to a particular sintering temperature using the Abaqus Standard FE solver based on the assumption that the LSCF6428 solid was isotropic, linear and elastic, irrespective of the sintering temperature. A small displacement was applied on one free surface normal to the $Y$ direction (the film normal, as indicated in Fig. 3-8 (f)) so that the model deformed linearly along $Y$. The opposite face was constrained to have no displacement in this direction. Boundary conditions were also applied to constrain the degree of freedom of the normal displacement for the nodes on the model's other surfaces parallel to $Y$. Such settings allowed these surfaces to contract or extend freely once the displacement was applied. The resultant normal force on the displaced $Y$ surface was obtained from the model so that the effective elastic modulus of the 3D microstructure could be determined by simply dividing the resultant force by the total area (solid plus pore) of the displaced surface. Following the same way, elastic moduli in the other two directions could also be determined. As a result, the Poisson's ratio in each direction could be deduced as well by calculating the absolute ratio of the corresponding transverse strain over the axial strain. Details of the calculation and analysis process can be found in Chapter 7.

## Chapter 3 References

1.  Larson AC, Dreele RBV: **General Structure Analysis System (GSAS)**. *Los Alamos National Laboratory Report LAUR 86-748* 1994.

2.  Toby BH: **EXPGUI, a graphical user interface for GSAS**. *Journal of Applied Crystallography* 2001, **34**:210-213.

3.  Rietveld HM: **Line profiles of neutron powder-diffraction peaks for structure refinement.** *Acta Cryst* 1967, **22**:151-152.

4.  Rietveld HM: **A Profile Refinement Method for Nuclear and Magnetic Structures**. *Journal of Applied Crysallography* 1969, **2**:65-71.

5.  Fletcher DA, McMeeking RF, Parkin D: **The United Kingdom Chemical Database Service**. *J Chem Inf Comput Sci* 1996(36):746-749.

6.  **Advanced Technical Ceramics—Mechanical Properties of Monolithic Ceramics at Room Temperature—Part 2: Determination of Young's Modulus, Shear Modulus and Poisson's Ratio**. In.: British Standard Institute; 2006: 32-36.

7.  Bulychev S, Alekhin V, Shorshorov M, Ternovskii A, Shnyrev G: **Determining Young's modulus from the indentor penetration diagram**. *Ind Lab* 1975, **41**(9):1409-1412.

8.  Oliver WC, Pharr GM: **An improved technique for determining hardness and elastic modulus using load and displacement sensing indentation experiments**. *Journal of Materials Research* 1992, **7**(6):1564-1583.

9.  Oliver WC: **Measurement of hardness and elastic modulus by instrumented indentation: Advances in understanding and refinements to methodology**. *Journal of Materials Research* 2004, **19**(1):3.

10. Singh S, Smith J, Singh R: **Characterization of the damping behavior of a nanoindentation instrument for carrying out dynamic experiments**. *Experimental Mechanics* 2008, **48**(5):571-583.

11. **http://cp.literature.agilent.com/litweb/pdf/5990-4907EN.pdf**

12. Clément P, Meille S, Chevalier J, Olagnon C: **Mechanical characterization of highly porous inorganic solids materials by instrumented micro-indentation**. *Acta Materialia* 2013, **61**(18):6649-6660.

13. Swain MV, Menčík J: **Mechanical property characterization of thin films using spherical tipped indenters**. *Thin Solid Films* 1994, **253**(1–2):204-211.

14. Mesarovic SD, Fleck NA: **Spherical indentation of elastic–plastic solids**. *Proceedings of the Royal Society of London Series A: Mathematical, Physical and Engineering Sciences* 1999, **455**(1987):2707-2728.

15. Chou Y-S, Stevenson JW, TArmstrong TR, LPederson LR: **Mechanical Properties of La1-xSrxCo0.2Fe0.8O3-δ Mixed-Conducting Perovskites Made by the Combustion Synthesis Technique**. *Journal of American Ceramic Society* 2000, **83**(6):1457-1464.

16. Dukino RD, Swain MV: **Comparative Measurement of Indentation Fracture Toughness with Berkovich and Vickers Indenters**. *Journal of the American Ceramic Society* 1992, **75**(12):3299-3304.

17. Anstis G, Chantikul P, Lawn BR, Marshall D: **A critical evaluation of indentation techniques for measuring fracture toughness: I, direct crack measurements**. *Journal of the American Ceramic Society* 1981, **64**(9):533-538.

18. Niihara K: **A fracture mechanics analysis of indentation-induced Palmqvist crack in ceramics**. *Journal of materials science letters* 1983, **2**(5):221-223.

19. Laugier M: **New formula for indentation toughness in ceramics**. *Journal of materials science letters* 1987, **6**(3):355-356.
20. Lawn B, Evans A, Marshall D: **Elastic/plastic indentation damage in ceramics: the median/radial cracks system**. *J Am Ceram Soc* 1980, **63**:574.
21. Ouchterlony F: **Stress-Intensity Factors for the Expansion Loaded Star Crack**. *Engineering Fracture Mechanics* 1976, **8**(2):447-448.
22. Volinsky AA, Vella JB, Gerberich WW: **Fracture toughness, adhesion and mechanical properties of low-k dielectric thin films measured by nanoindentation**. *Thin Solid Films* 2003, **429**(1):201-210.
23. Schiffmann KI: **Determination of fracture toughness of bulk materials and thin films by nanoindentation: comparison of different models**. *Philosophical Magazine* 2011, **91**(7-9):1163-1178.
24. Vella J, Adhihetty I, Junker K, Volinsky A: **Mechanical properties and fracture toughness of organo-silicate glass (OSG) low-k dielectric thin films for microelectronic applications**. *International journal of fracture* 2003, **120**(1):487-499.
25. Scharf TW, Deng H, Barnard JA: **Mechanical and fracture toughness studies of amorphous SiC–N hard coatings using nanoindentation**. *Journal of Vacuum Science & Technology A: Vacuum, Surfaces, and Films* 1997, **15**(3):963-967.
26. Goodhew PJ, Humphreys FJ: **Electron Microscopy and Analysis**, 2nd edn: Taylor & Francis; 1988.
27. Wilson JR, Kobsiriphat W, Mendoza R, Chen H-Y, Hiller JM, Miller DJ, Thornton K, Voorhees PW, Adler SB, Barnett SA: **Three-dimensional reconstruction of a solid-oxide fuel-cell anode**. *Nature Materials* 2006, **5**(7):541-544.
28. **http://www.fei.com/uploadedFiles/DocumentsPrivate/Content/Helios-NanoLab-600i-ds-web.pdf**
29. Giannuzzi LA: **Introduction to focused ion beams: instrumentation, theory, techniques and practice**: Springer; 2004.

# 4 Characterisation of LSCF6428 and CGO Specimens

This Chapter is intended to summarise the results from the experimental work carried out for the characterisation of LSCF6428 and CGO materials, in the forms of powder, bulk and films. Particular attention is paid to the preparation of improved crack-free and smooth films suitable for nanoindentation. Some interesting findings and problems in the experiments will then be discussed and addressed.

## 4.1 Densities of Bulk Specimens

The LSCF6428 pellets were fabricated and sintered as described in Chapter 3. Their bulk densities were measured, and compared to the theoretical densities, in order to check that a relative density of approximately 95% could be achieved. The theoretical density of the LSCF6428 material obtained was 6.309 g/cm$^3$, as detailed in the previous Chapter. Table 4-1 shows the bulk and relative densities obtained for the pellets of LSCF6428 sintered at temperatures from 800-1200 °C.

*Table 4-1 Densities of LSCF6428 pellets sintered at varying temperatures*

| Sintering Temperature (°C) | Bulk Density (g/cm$^3$) | Relative Density (%) |
|---|---|---|
| 800 | 3.38 | 53.57 |
| 900 | 3.48 | 55.15 |
| 1000 | 4.02 | 63.72 |
| 1100 | 4.50 | 71.33 |
| 1200 | 5.96 | 94.47 |

It can be seen that there is an increasing trend in these results with the increasing sintering temperature, which shows that higher sintering temperatures result in a higher relative density. The maximum relative density of 94.47% was obtained for a sintering temperature of 1200 °C. The bulk density of the CGO pellet sintered at 1400 °C was measured to be 6.98 g/cm$^3$ compared with a theoretical density of 7.225 g/cm$^3$ resulting in a relative density of 96.61%, which ensured the CGO pellet was sufficiently dense for the use as a substrate for LSCF6428 films.

### 4.2 2D Microstructural Characterisation of Bulk Specimens

The CGO (sintered at 1400 °C) and LSCF6428 (sintered at 800-1200 °C) pellets' top surface and fracture surface microstructures were examined using SEM.

Fig. 4-1 (a) and (b) show the CGO pellet top surface and fracture cross-sectional surface, respectively. The images suggest a very fine distribution of grains under dense compaction.

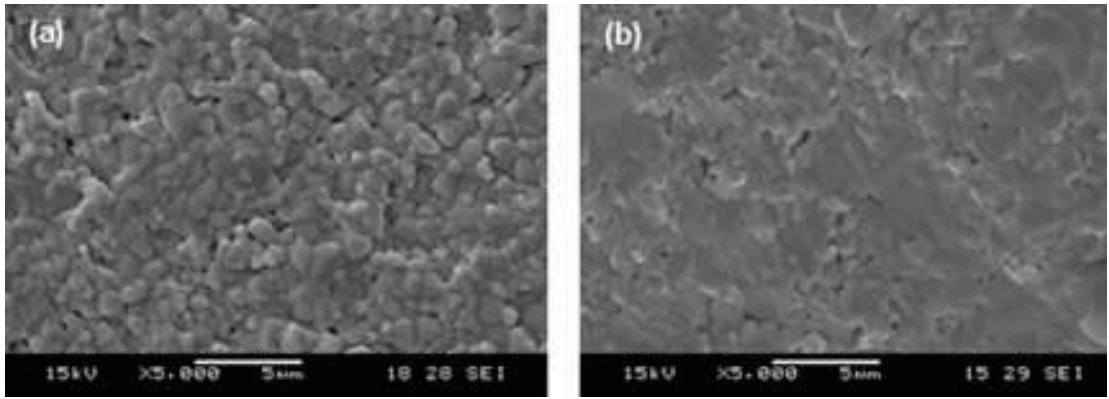

*Fig. 4-1 Microstructures of CGO sample (a) top surface (b) fracture surface.*

Fig. 4-2 displays the top surface microstructures of LSCF6428 pellets made from powders, with sintering temperature of (a) 800 °C, (b) 900 °C, (c) 1000 °C, (d) 1100 °C, and (e) 1200 °C.

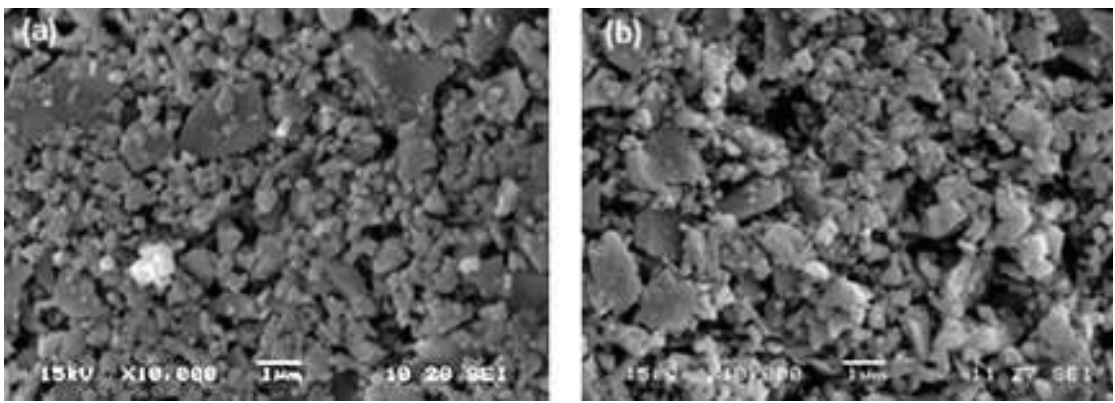

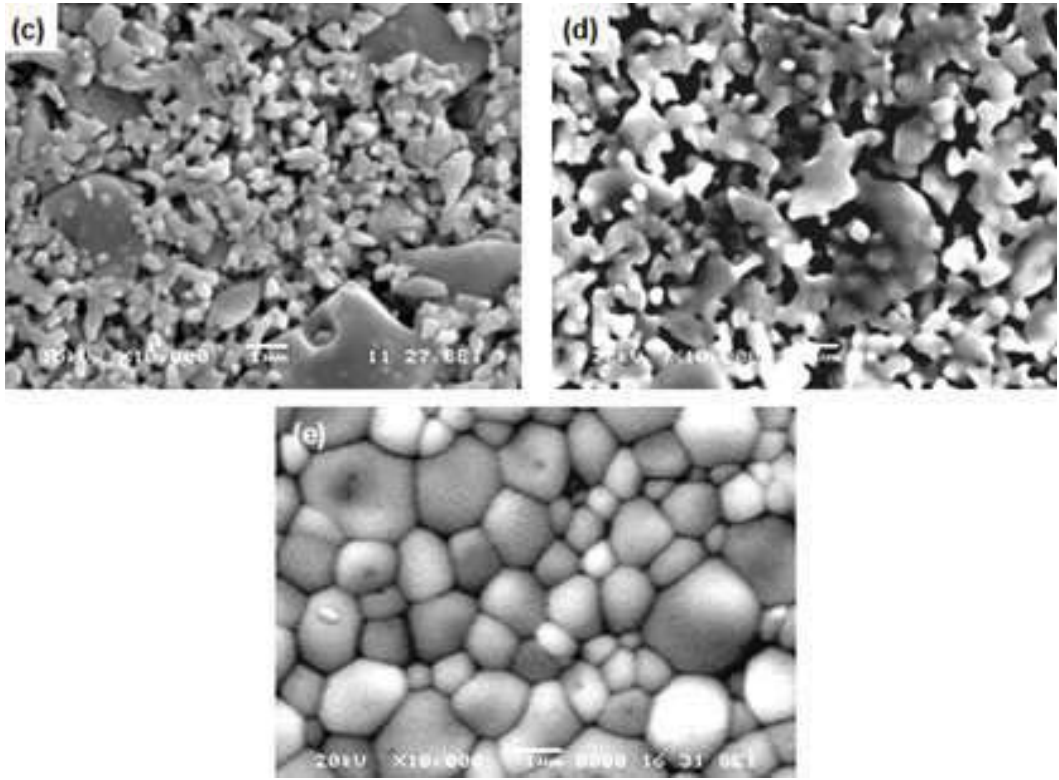

*Fig. 4-2 LSCF6428 pellets' top surface microstructures after sintering at (a) 800 °C, (b) 900 °C, (c) 1000 °C, (d) 1100 °C, and (e) 1200 °C.*

From Fig. 4-2 (c), it is evident that the particles began to bond together, which suggests that from 1000 °C the grains started growing; the degree of necking between particles generally increased with increasing sintering temperature; hence the open porosity reduced. For example, at 1200 °C, there was extensive sintering with substantial reduction in porosity, as indicated in Fig. 4-2 (e), which resulted in a very dense pellet. Moreover, no cracking appeared in those samples. However, it should be noted that although LSCF6428 powder specification provided by the manufacturer showed a fine particle size distribution (PSD) with single peak and $D_{50}$ = 0.669 µm, $D_{90}$ = 0.997 µm, it is obvious that there were quite a few large particles (particle size >> 1 µm) which were definitely not agglomerates. This study requires LSCF6428 powders with fine and uniform PSD in order to achieve representative microstructures from small sample volumes. This was a material supply issue, which will be addressed later in the study. Fig. 4-3 (a)-(e) are fracture cross-sectional SEM images of the samples corresponding to sintering temperature of 800-1200 °C.

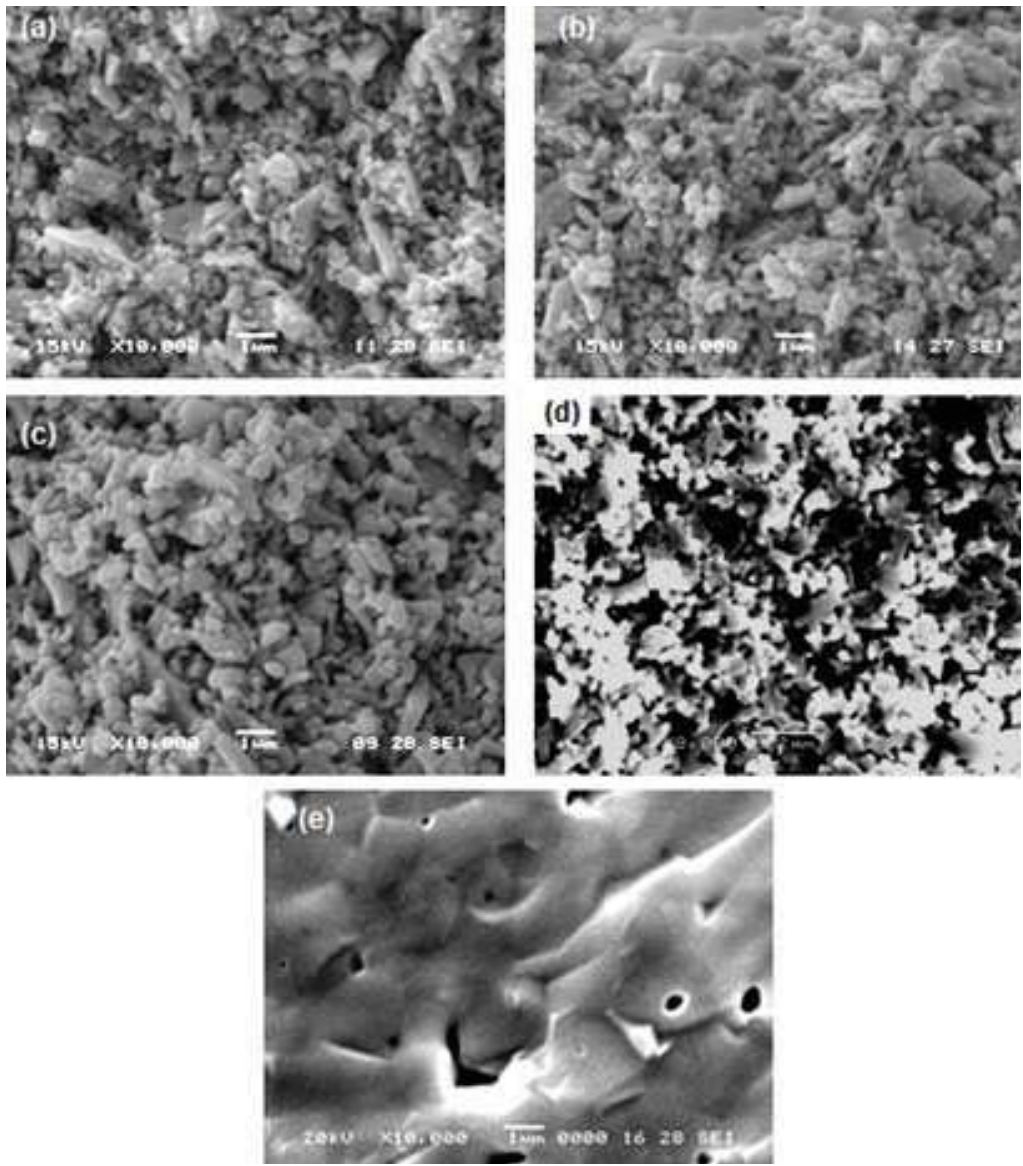

*Fig. 4-3 LSCF6428 pellet fracture surface microstructures sintered at (a) 800 °C, (b) 900 °C, (c) 1000 °C, (d) 1100 °C, and (e) 1200 °C.*

At 1200 °C, the extensive sintering is also evident in Fig. 4-3 (e) by the excellent bonding of the particles and few pores retained in the body.

## *4.3 Lumps of Abnormally Large Size in LSCF6428 Powder*

As described above, there were particles of very large size appearing in the LSCF6428 pellets, which was not consistent with the manufacturer provided specification. Therefore, a further study was carried out to investigate the original powders microstructure and the PSD.

The micrographs of LSCF6428 powders are presented in Fig. 4-4. Fig. 4-5 shows the FIB-milled cross-section of a film made from INK-B, which was prepared using the abovementioned powders as described in Chapter 3.

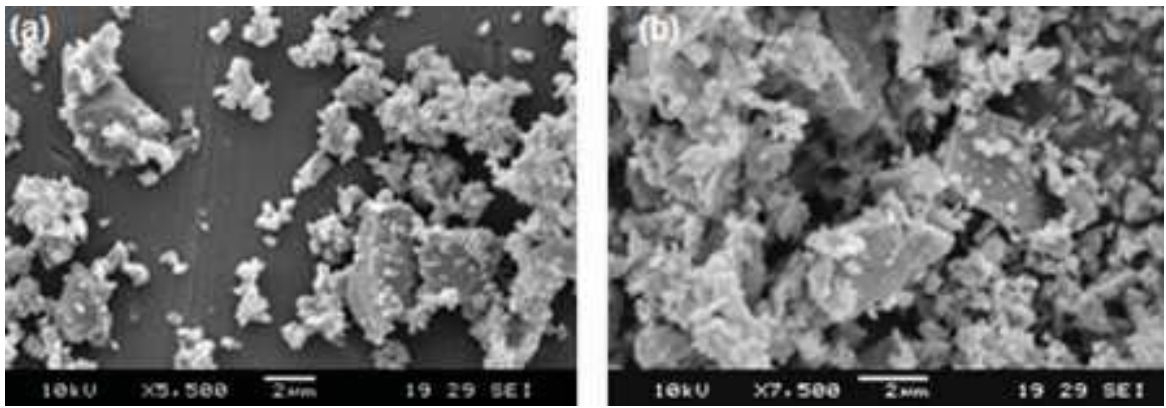

*Fig. 4-4 Large particles appeared in the LSCF6428 powders.*

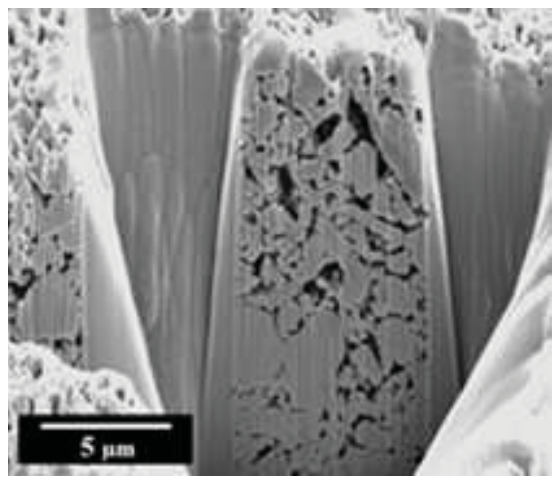

*Fig. 4-5 Large particles appeared in the LSCF6428 films after sintering at 1000 °C*

From the above figures, it is clear that large grains with particle size larger than 2 μm were present in the powders. They were not likely to be agglomerates since they appeared to have clear edges to be all single particles in the SEM images. The powders' chemical analysis with EDX in SEM also showed that they were not contaminations but pure LSCF. The PSD analysis result is shown in Fig. 4-6. Again, as expected, the dual peak result confirms the appearance of a large volume of 2-4 µm particles. Therefore, rather than using this batch of LSCF6428 powder to make inks for screen printing, it was decided that a ready-made LSCF6428 ink (i.e. INK-A, as described in Chapter 3) with a finer particle size distribution should be chosen to produce LSCF6428 films in order to guarantee the sample quality in terms of the uniformity of particle size for the later FIB/SEM 3D microstructural characterisation and nanoindentation measurements.

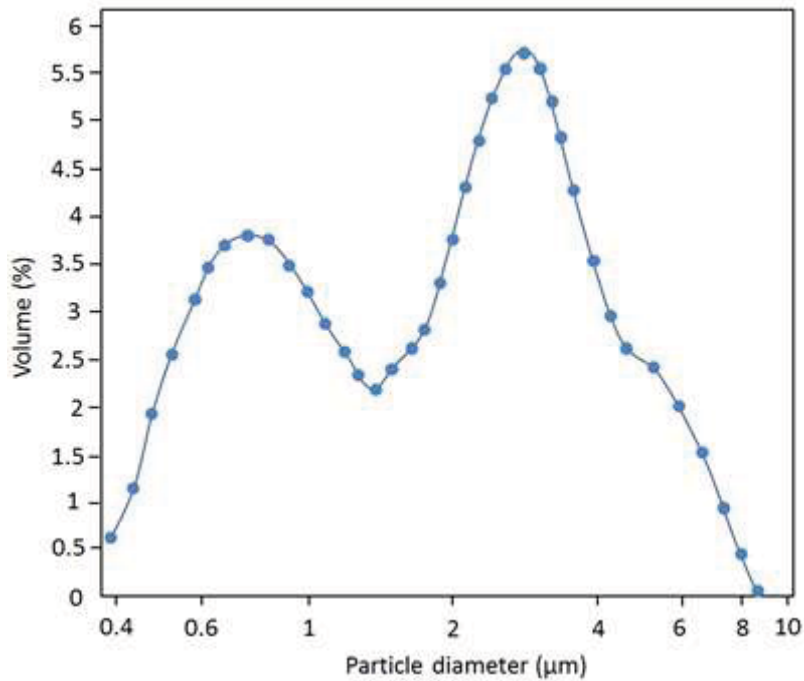

*Fig. 4-6 Particle size distribution measurement of LSCF6428 powders*

### 4.4 Measurement of LSCF6428 Film Thickness

In the research, a relatively thick sintered LSCF6428 film could help reduce the interference arising from the substrate during nanoindentation. Hence, the thickness of LSCF6428 films screen printed with a series of sequential passes was measured with SEM after sintering. The result is summarised in Table 4-2 and plotted in Fig. 4-7.

*Table 4-2 Sintered LSCF6428 film thickness vs. printing pass times*

| Number of Passes | 1 | 2 | 4 | 6 | 8 | 10 |
|---|---|---|---|---|---|---|
| Thickness (µm) | 5.9 | 6.0 | 14.0 | 13.4 | 7.6 | 6.0 |

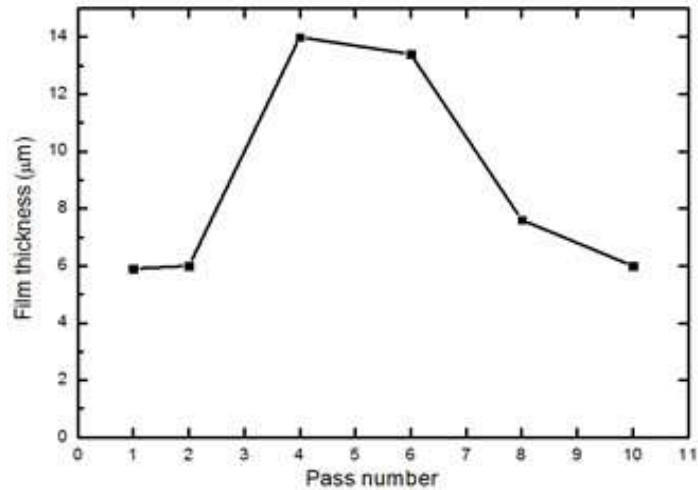

*Fig. 4-7 Relationship between LSCF6428 film thickness and printing pass number*

It is interesting that the film thickness first increased with the number of printing passes, then reached a maximum value of about 14 μm for 4 passes, after which it decreased. The minimum thickness was about 6 μm for 1 pass and 10 passes. However, the thickness had been assumed to have linear relationship with the number of passes, as the screen printing machine parameters were not changed. In order to check the reliability of the result, another method was used. After each pass finished, the sample's mass was quickly recorded using an electronic balance to find the total mass of the ink deposited on the substrate. Since the working parameters and the mesh geometry stayed unchanged, the weight gain should be proportional to the thickness of the film, which would result in the similar trend as the above thickness measurement. The result is displayed in Fig. 4-8, which is consistent with the above result.

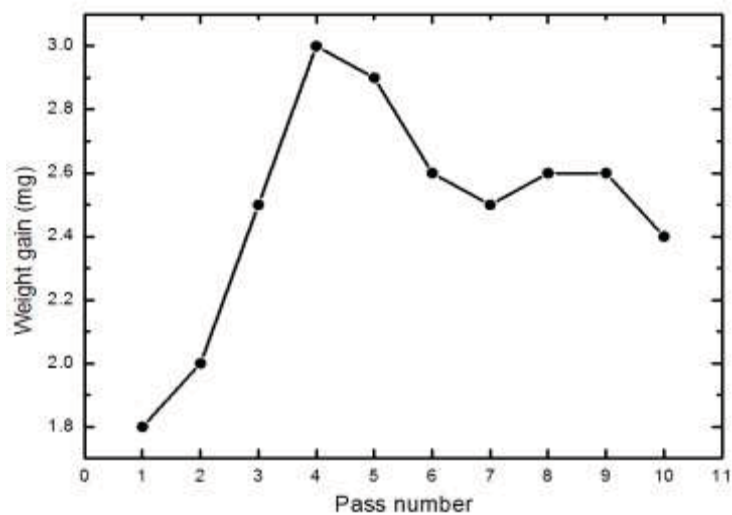

*Fig. 4-8 Weight gain of ink deposition vs. pass number*

Therefore, for the unchanged parameters and mesh in the screen printing machine, the maximum thickness of sintered LSCF6428 on CGO substrate was achieved as 14 µm.

Fig. 4-9 demonstrates the cross-sectional microstructures of sintered samples with 2 and 4 passes.

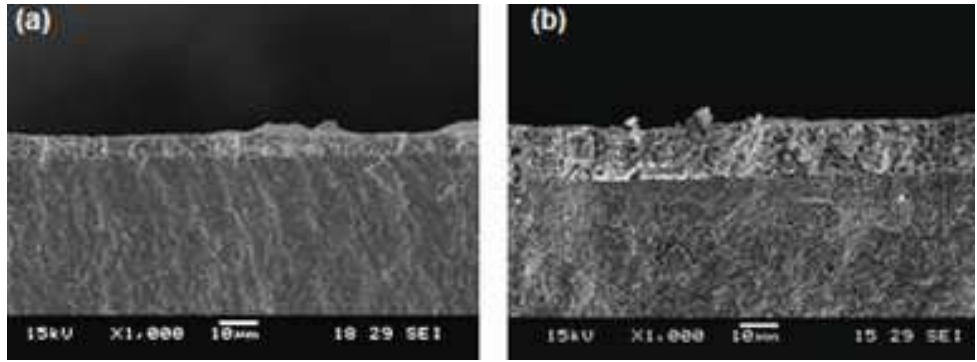

*Fig. 4-9 LSCF6428 film thickness measurements using SEM for 2 and 4 passes of printing*

## 4.5 Improvement of Film Quality Suitable for Nanoindentation

### 4.5.1 Defects in the As-sintered Films

As explained in the Introduction of this Thesis, the reliability of LSCF6428 to be used as cathode parts in SOFCs also relies on its mechanical properties, of which the measurements were mainly carried out by using nanoindentation in this study. Therefore, adequate surface conditions (i.e. flat and crack-free) are desirable when the as-sintered porous thin films are subjected to nanoindentation. However, these ideal surface features could not be readily generated in the preliminary work and therefore investigation of the origins of these defects was conducted and hence measures were taken to eliminate the defects.

In the preliminary study, severe crack networks and a large number of surface asperities (i.e. agglomerates) were generated in the as-sintered films deposited by screen printing using INK-A, which was an LSCF6428 ink prepared and provided by ESL, UK, as describe in Chapter 3. The surface and cross-section microstructural features of a film after sintering at 1000 °C for 4 h are shown in Fig. 4-10. The width of crack openings and the agglomerated islands were measured to be 1-4 µm and 20-40 µm, respectively. It should be noted, however, that there was a good adhesion between the film and the substrate. Fig. 4-11 shows a magnified cross-section sliced by FIB, revealing vertical cracks penetrating through the film thickness and a dense agglomerate present under the film surface.

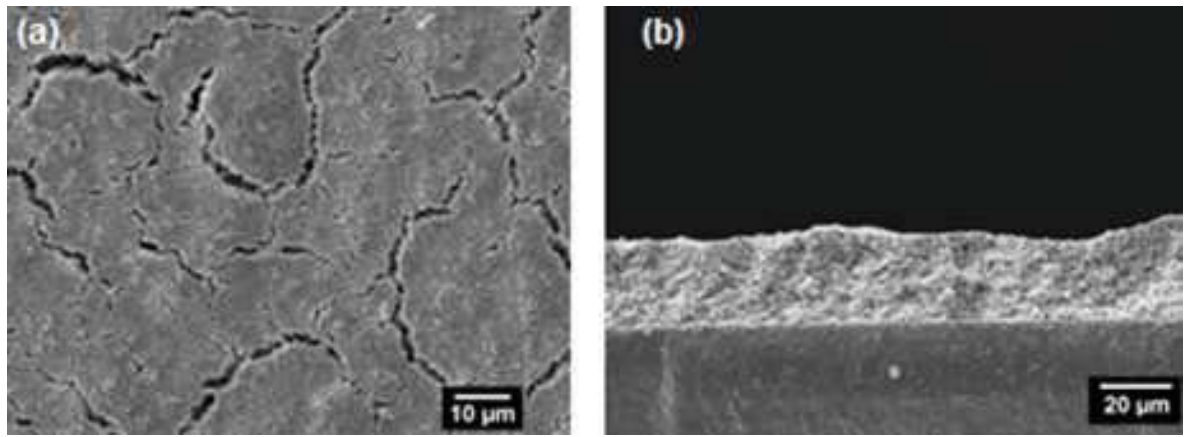

*Fig. 4-10 SEM micrographs of a LSCF6428 film after sintering at 1000 °C: (a) top surface view of crack networks and agglomerates; (b) fracture cross-sectional view of the poor surface smoothness*

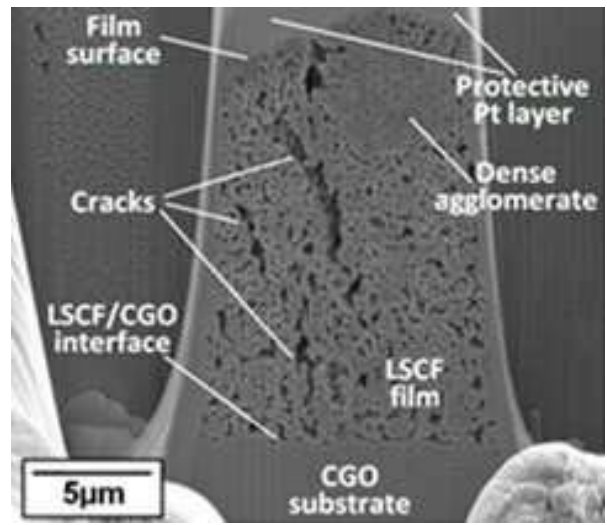

*Fig. 4-11 FIB-sliced cross-section showing cracks penetrating through a film sintered at 900 °C and a large asperity close to the film surface*

The fabrication of cathode components of SOFCs has been reported using various methods, such as screen printing, tape casting, dip casting, chemical vapor deposition, atmospheric plasma spray, colloidal spray deposition, pulsed-laser deposition, sputtering and painting, as reviewed by Fergus *et al.* [1] and Taroco *et al.*[2]. In Table 4-3 we summarise the techniques reported to be used for the deposition of LSCF6428 films on CGO substrates [3-8], and the resulting possible surface defects are shown in Fig. 4-12.

*Table 4-3 Summary of observations from the literature on the deposition of LSCF6428 films on CGO substrates*

| Reference | Deposition method | Heat treatment conditions | Film thickness (μm) | Crack opening width (μm) | Surface asperities |
|---|---|---|---|---|---|
| Lee *et al.* [3] | Screen printing | Sintering at 1000°C for 0.5h | < 40 | < 30 | Yes |
| Baque *et al.* [4][a] | Dip coating | Dried at 130 °C; Sintering at 900 °C for 6 h | < 28 | < 20 | Yes |
| Marinha *et al.* [5] | Electrostatic spray deposition | Substrate temperature 250-450 °C; Sintering at 900 °C for 2h | < 25 | < 5 | Yes |
| Hsu *et al.* [6] | Electrostatic-assisted ultrasonic spray pyrolysis | Deposition temperature 350 °C; Sintering at 1000°C for 2 h | -- | < 2 | Yes |
| Santillan *et al.* [7] | Electrophoretic deposition | Sintering at 950°C in air for 2 h | < 25 | No detectable crack | Yes |
| Wang *et al.* [8] | In-situ solid-state reaction sintering | Sintering at 1000 °C in air for 6 h | -- | No detectable crack | Yes |

[a]$La_{0.4}Sr_{0.6}Co_{0.8}Fe_{0.2}O_{3-\delta}$ was used as film material in this case.

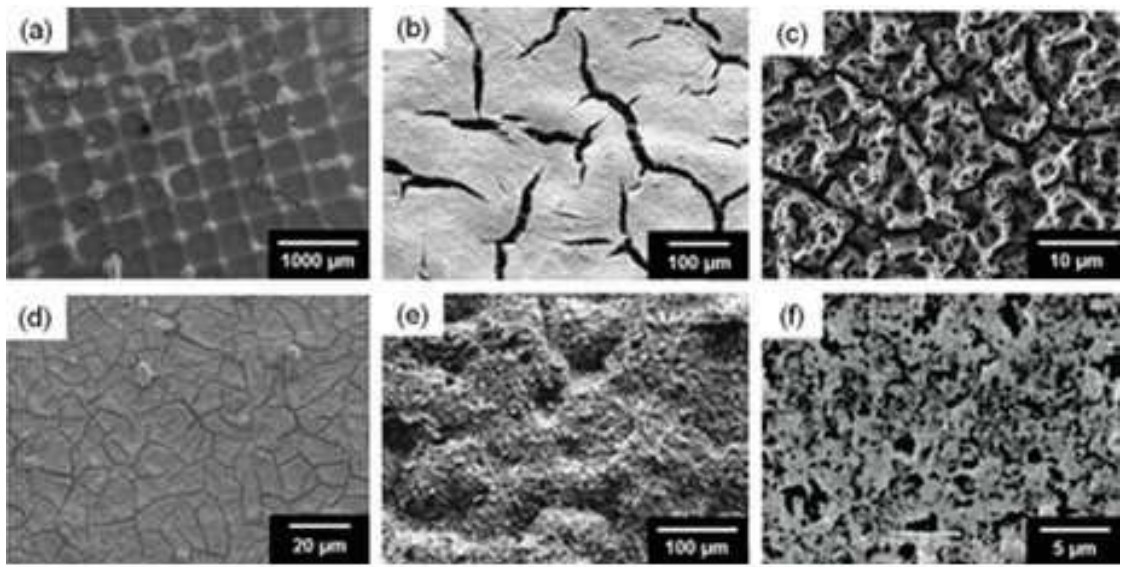

*Fig. 4-12 Micrographs showing surface defects on the as-sintered LSCF6428 films, which are reproduced from the references cited in Table 4-3: (a): [3]; (b): [4]; (c): [5]; (d): [6]; (e):[7]; (f):[8]. Note these images were in different magnifications.*

From the above literature survey, it can be concluded that most LSCF6428 films fabricated by different laboratories possessed both cracks and surface asperities after the final heat treatment process, irrespective of the deposition method applied. Although no detectable cracking was present in the films for [7] and [8], the surface asperities would remain a great concern when it comes to applying nanoindentation tests.

As a common issue in the research field and the current study, such defects not only were reported in the literature, but could also appear in some commercially available LSCF6428 films. An example is shown in Fig. 4-13, severe cracks and agglomerates were found in the screen-printed anode-supported LSCF6428 cathode provided by HTceramix-SOFCpower, Switzerland, which was sintered at 1000 °C. As can be seen in the micrographs, the agglomerates were approximately 100 μm in diameter, generating crack openings measured to be as wide as 20 μm, which seemed to be much more severe than in our case.

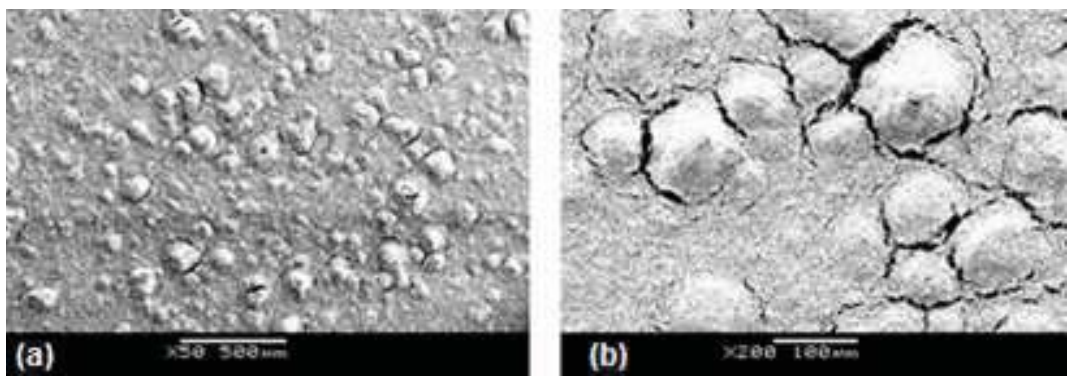

*Fig. 4-13 Severe surface defects in the commercial available LSCF6428 cathode films provided by HTceramix-SOFCpower. Images (a) and (b) are shown at different magnifications*

Adjusting the working parameters may to some extent reduce film cracking, but those deposition methods still struggle with preparing crack-free and flat LSCF6428 films containing highly porous structures. This implies that the existence of cracking is prevalent in many LSCF6428 films irrespective of their deposition techniques. Therefore, the challenge remained to be the successful fabrication of crack-free and smooth LSCF6428 films. Although many have observed such defects in their LSCF6428 films, they often attributed them to the densification and shrinkage of films during sintering, or the TEC mismatch between films and substrates during cooling, without any further experimental validation [3, 6]. Although these defects might not be of major importance for studies of electrochemical performance, they are, however, crucial for the present study, as their presence in the films would certainly cause errors in the nanoindentation response. Furthermore, they are likely to induce damage to cells in actual application.

### 4.5.2 Importance of Film Surface Quality to Nanoindentation

In order that nanoindentation tests be reproducible and thus the results be reliable, cracks and surface asperities (i.e. agglomerates) in the as-sintered films must be reduced to an acceptable level or be completely avoided, which means that the size of such defects should be much smaller to the size of indented features such as indentation depth, which in the current study was less than 2 μm. However, as can be seen above and later, such defects were found constantly in the sintered LSCF6428 films used initially, which severely scattered the nanoindentation response and thus the resulting measurements were considered invalid.

Although such defects in the as-sintered LSCF6428 films were also observed in some other studies, the problem has not yet been considered critical in many researches which essentially involved the study of electrochemical performance rather than mechanical properties of the SOFC components, as discussed later. Due to the nature of the fabrication process and the application in SOFCs, the LSCF6428 cathode films are difficult to be surface treated to obtain relatively smooth surfaces, and the cracking remains a problem. As a result, the relevant technological challenges could be related to the successful fabrication of thin, porous, flat and crack-free LSCF6428 cathode films.

To mitigate the cracking and surface asperity problems in LSCF6428 cathode films, in the current study various aspects of film fabrication were investigated, including: the comparison of deposition methods, the improvement of drying and sintering processes, and of the ink formulation. Results (which are described later) revealed that a sufficiently low viscosity of the ink could result in crack-free LSCF6428 films with smoother surfaces and a more homogeneous microstructure, while an alternative deposition method and different thermal treatments applied helped little to solve the problems. In other words, the surface quality of the films was significantly improved by using a reformulated less viscous ink which ensured more ability to self-level and allow particle rearrangement in the early wet state upon deposition. By using these improved films, effective nanoindentation measurements were achieved with high reproducibility and hence reliable results were obtained as described below.

### 4.5.3 Nanoindentation Response of "Defective" Films

Nanoindentation using a spherical indenter on the films similar to those shown in Fig. 4-10 and Fig. 4-11 was conducted and the response curves, namely the load vs. indentation depth curves, generated are shown in Fig. 4-14. Note that the determination of intrinsic elastic modulus of the porous thin ceramic films can be complicated, for the film-only results to be extracted detailed methodology based on nanoindentation data analysis will be addressed in Chapter 5. Here only some relevant preliminary results are shown to illustrate the reproducibility of the raw indentation data and elastic modulus from the initial unloading.

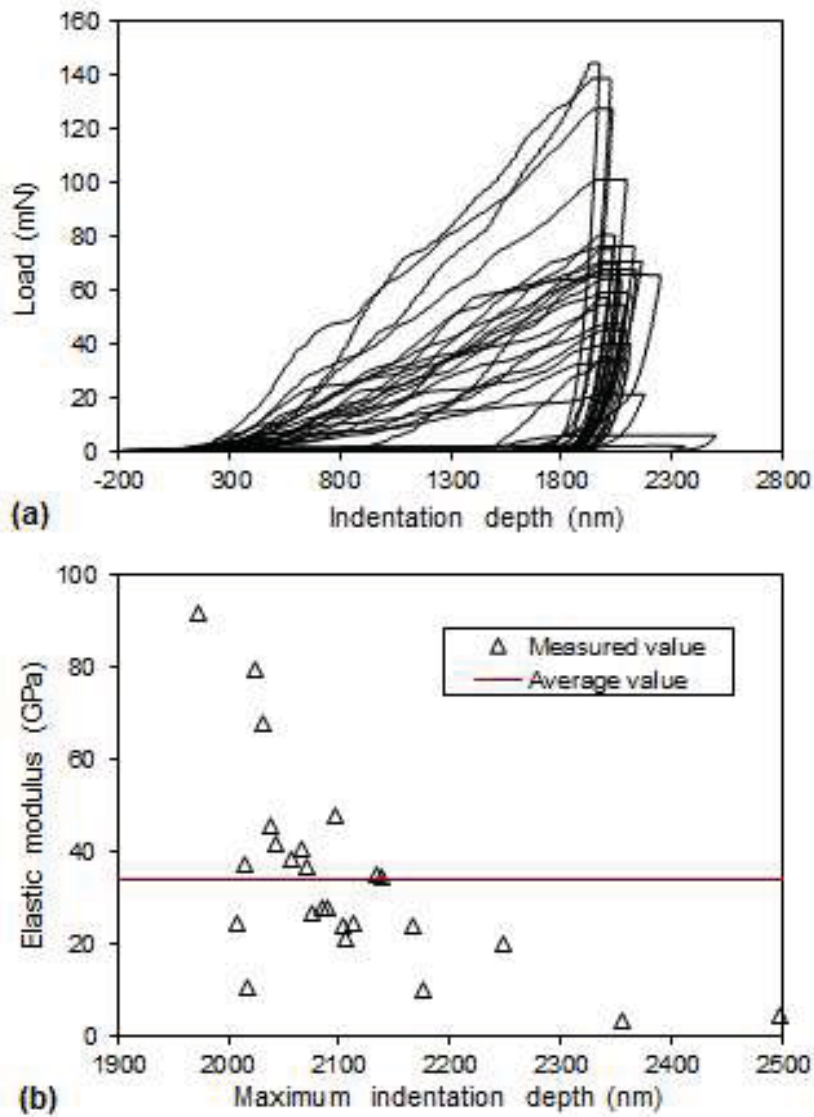

*Fig. 4-14 Nanoindentation load vs. indentation depth curves (a) and the corresponding elastic modulus values calculated (b) for the defective films similar to those in Fig. 4-10 and Fig. 4-11*

The extremely irregular shapes and wide variability of the response curves seen in Fig. 4-14 were generated by the indenter tip touching defective locations on the films during nanoindentation tests. The presence of cracks and poor surface flatness due to large asperities also induced significant variability in the measured elastic modulus values. In the current case, in addition to the extensive cracks, it was found that the average surface roughness $R_a$ for the defective film was measured to be 1.86 µm, which was very close to the indentation depth of approximately 2 µm, resulting in an enormous relative error as much as 63.2% for the apparent elastic modulus calculated (34±21 GPa), as shown in Fig. 4-14 (b).

Examples of three typical nanoindentation load-depth curves collected in this study are plotted in Fig. 4-15, which demonstrates the existence of three different contact processes of indenter tip with the sample surface and are related to the variation in characteristic surface features described earlier.

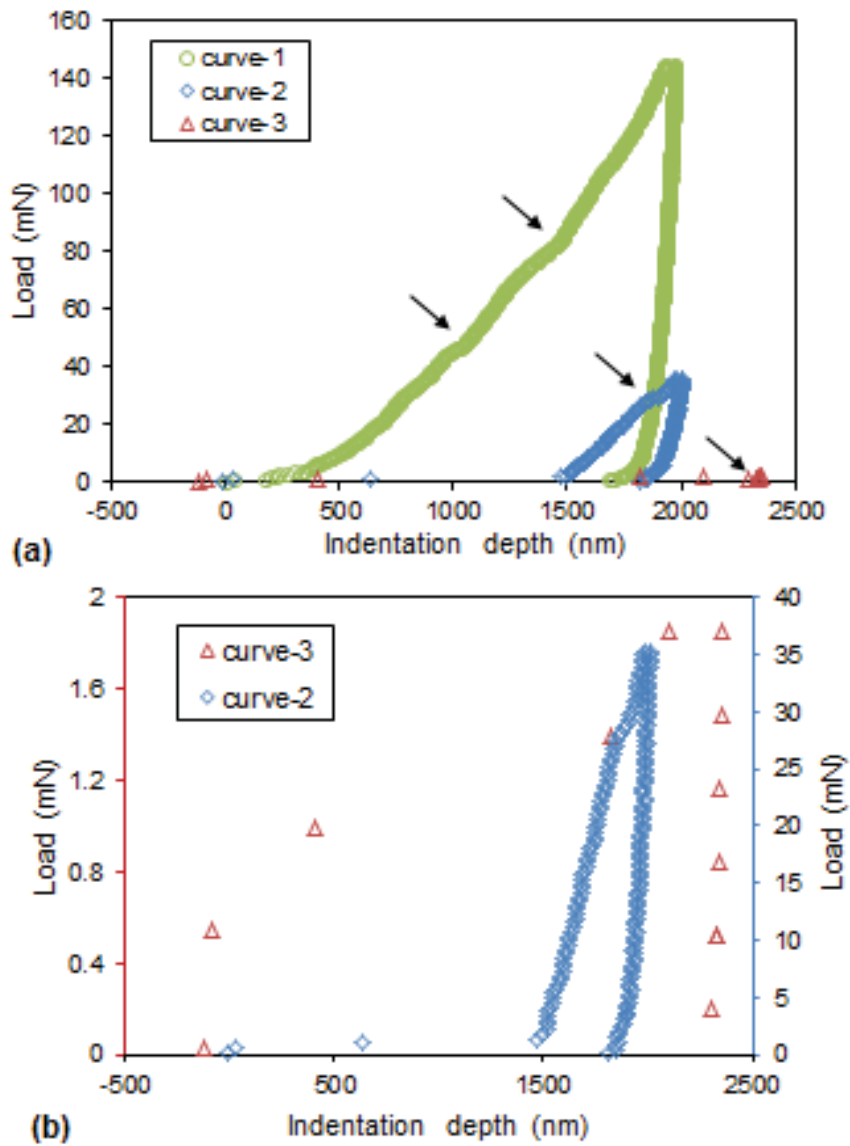

*Fig. 4-15 Examples of load-depth curves for nanoindentation using spherical indenter tip corresponding to the likely features at the indent locations. (b) plots curve-2 and curve-3on an expanded load scale.*

Curve-1 is typical of a relatively smooth and un-cracked surface location, it being less influenced by these defects. However, there are still two noticeable irregular responses during loading (such as the "pop-in" events, as indicated by arrows), which are attributed to the fact that the LSCF6428 film was not smooth enough. In contrast, curve-2 and curve-3 show extremely unreliable responses. Curve-2 exhibits a deep "pop-in" attributable to the cracks in the contact location, or to the poor bonding of the porous microstructures. The existence of the initial negative depth in curve-3 indicates that during loading an asperity on the indent surface was first touched by the tip. For a spherical indenter tip, the characteristic surface roughness of the test sample must be negligible compared to the maximum indentation depth. While for a Berkovich or other types of sharp tip, the indentation results are even more sensitive to the surface quality and presence of cracks. This was also the reason that in this work a spherical indenter tip was used for nanoindentation, particularly as the surface of the films could not be polished.

### 4.5.4 Improvement of Film Surface Quality

Given the analysis above, it was imperative that the cracking and surface asperities be reduced for the properties of the LSCF6428 films to be reliably and reproducibly measured using nanoindentation. The improvement of the film surface quality required the investigation of the possible factors influencing defect formation, such as the TEC mismatch between LSCF6428 films and CGO substrates, the drying, sintering and cooling processes applied, the deposition methods used as well as the formulation of the ink, which will be discussed below in more detail.

#### 4.5.4.1 TEC Mismatch between LSCF6428 Films and CGO Substrates

One reason that LSCF6428 is chosen as a promising cathode material is because its TEC ($15.3 \times 10^{-6}$ K$^{-1}$, 100-600 °C) [9] is not too different from that of CGO ($13.5 \times 10^{-6}$ K$^{-1}$) [10], which means there exists only a small thermal expansion mismatch. Nevertheless, even this small difference would induce a tensile thermal stress on cooling which could facilitate crack formation in the film during cooling after sintering. To avoid this possible effect caused by TEC mismatch, substrates of LSCF6428 were prepared by sintering at 1200 °C for 4 h with a heating rate of 300 °C/h. The LSCF6428 film was applied by screen printing the as-received

INK-A and the resulting sample was then sintered under the same condition as the previous samples on CGO substrates. This would certainly exclude any crack formation caused by the TEC mismatch between films and substrates. From Fig. 4-16, which shows SEM microstructures of the sintered LSCF6428 film, it can be seen that the film surface was full of asperities. Again, a cracking network was observed on the surface with crack openings of 1.5 µm wide. Hence little dependence was shown of the cracking on the thermal expansion coefficient difference between LSCF6428 and substrate materials since cracking still occurred for LSCF6428 films on LSCF6428 substrates, which involved zero TEC difference.

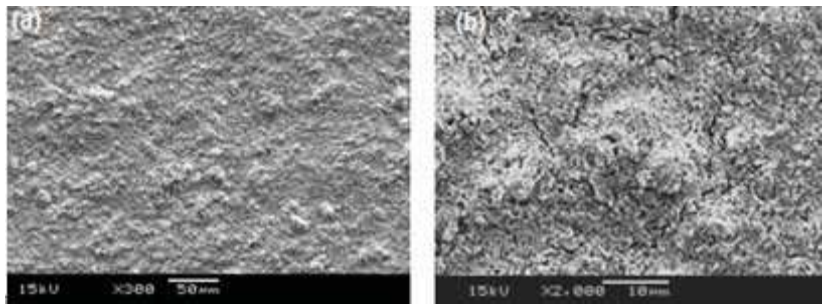

*Fig. 4-16 SEM images of sintered porous LSCF6428 films screen printed on dense LSCF6428 substrates. Images (a) and (b) are shown at different magnifications.*

### 4.5.4.2 Drying, Sintering and Cooling Process

Cracking might also be attributed to the shrinkage of the films themselves during the drying and constrained sintering processes. Therefore, an investigation of the influence of thermal treatments on defect formation was carried out.

### 4.5.4.2.1 Thermogravimetric Analysis

Thermogravimetric analysis (TGA) of the as-received LSCF6428 INK-A was performed in air atmosphere in order to determine the thermal decomposition range of the organic constituents. Fig. 4-17 shows the TGA curve of the ink at a heating rate of 10 °C/min.

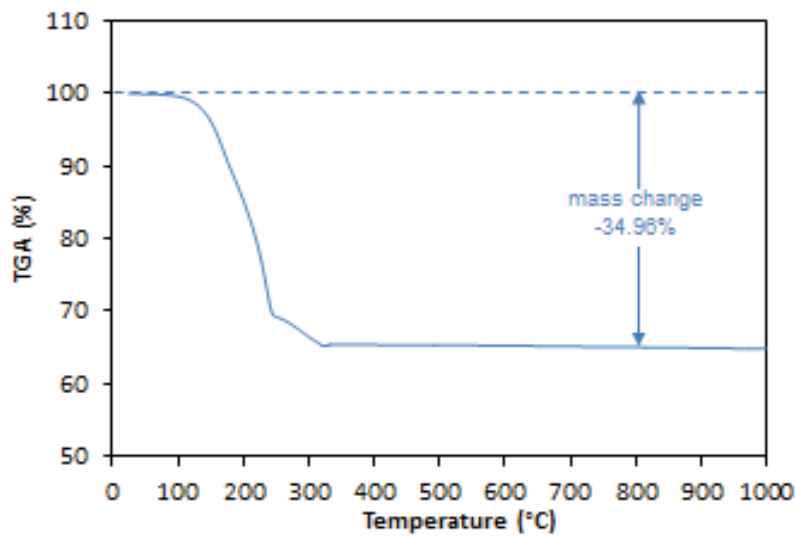

*Fig. 4-17 TGA curve of the as-received LSCF6428 INK-A in air atmosphere.*

As can be seen from the TGA curve, the thermal decomposition of the ink underwent three steps: the volatile organic solvent such as terpineol started volatilization at around 100 °C, followed by exothermic oxidation/decomposition of a majority of non-volatile organic content when the temperature increased to 220 °C. The organic residues continued to be burnt-out until the temperature reached 320 °C, after which the weight remained almost constant. (The oxygen desorption from the LSCF6428 perovskite lattice with increasing temperature [11] was too small to be observable on this scale.) The decomposition completed with a total mass loss of 34.96%.

### 4.5.4.2.2 Dilatometric Analysis

Dilatometry measurements on uniaxial-pressed green bodies of LSCF6428 powder were performed to investigate the sintering activity and shrinkage of the powder. The heating dilatometric curves are shown in Fig. 4-18 at four constant heating rates: i.e. 3, 5, 10, 20 °C/min from room temperature to 1250 °C.

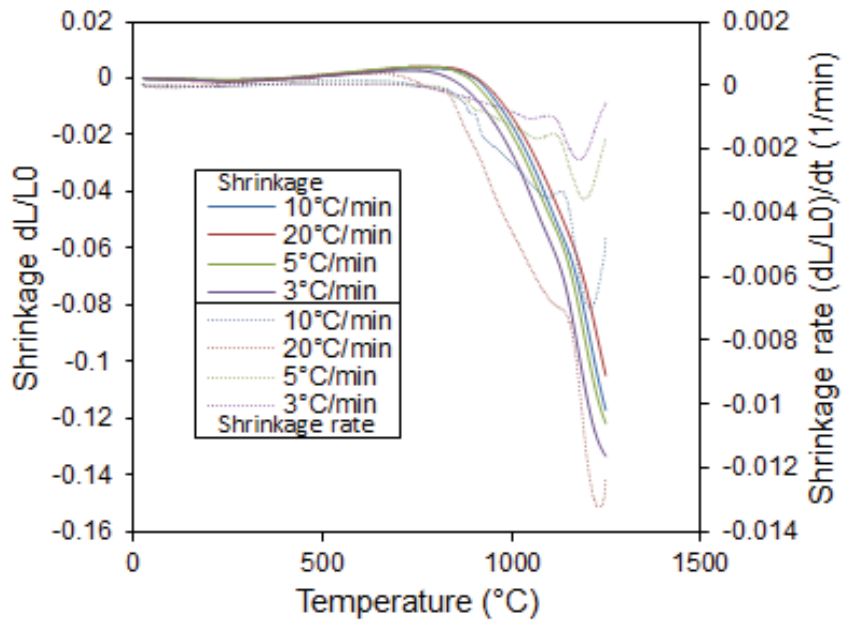

*Fig. 4-18 Shrinkage and shrinkage rate vs. temperature during constant heating rate sintering of LSCF6428 at 3, 5, 10, 20 °C/min heating rates.*

No significant dimensional changes were observed up to 500 °C, after which thermal expansion was observable until noticeable shrinkage started at around 800 °C. The densification process continued steeply up to 1250 °C. The final shrinkage ranged from 10% at a 20 °C/min heating rate to 14% at a 3 °C/min heating rate, suggesting that smaller heating rates allowed more time for shrinkage and resulted in a higher degree of densification. The shrinkage rate (dL/dt) depended on the heating rate applied. For example, above 1000 °C the shrinkage rate for a 20 °C/min heating rate was several times greater than that for a heating rate of 3 °C/min, indicating that a film would be more likely to crack when sintered at a 20 °C/min heating rate.

An excellent sintering activity was found for the LSCF6428 powder used in the present study. This was due to its high specific surface area and relatively homogeneous particle size distribution. The shrinkage rate reached a maximum at approximately 1200 °C. Considering a porosity approximately of 35 vol% was required for a cathode film, an appropriate sintering temperature for the LSCF6428 films should fall in the range 800-1000 °C.

### *4.5.4.2.3 Modified Sintering Programmes*

If the cracks were generated by the constrained shrinkage of the films, then it was considered that the cracking problem might be avoided by optimising the burn-out and sintering schedule (i.e. stepwise slowing down the heating process) based on the thermal analyses described above. The sintering programme was therefore modified as follows:

- For the organics removal step between room temperature and 350 °C, a very slow heating rate was used (3 °C/min or 1 °C/min, holding for 2 h at 350 °C);

- For the stable period between 350 and 800 °C, a medium heating rate was employed (5 °C/min) and a holding time was applied at 800 °C (4 h) at the start of sintering to strengthen the particle network;

- Between 800 °C and the target maximum sintering temperature, a much lower heating rate was used (1 °C/min) and a much longer holding time (48 h) was used at the top temperature; finally cooling to room temperature was done at 5 °C/min.

For the particular case of a top temperature of 1000 °C, the sintering programme is plotted in Fig. 4-19, compared with the initial sintering schedule which used a three-step sintering, at a rate of 5 °C/min and holding time of 4 h.

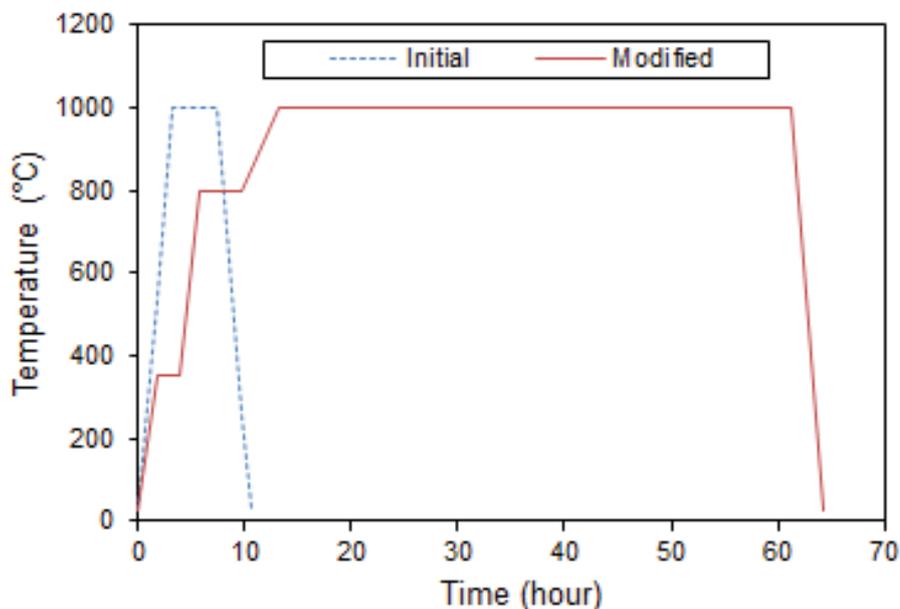

*Fig. 4-19 Comparison of the initial and modified sintering schedules.*

The micrograph of the sintered film using the modified sintering programme is shown in Fig. 4-20. Unfortunately, this shows little improvement of surface quality compared to that using the initial one. There are still many cracks typically 1.5 µm wide and asperities at least 10 µm in diameter, showing little effectiveness of the modified sintering conditions.

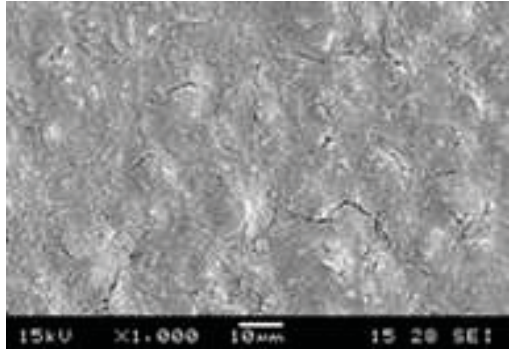

*Fig. 4-20 Micrograph of the surface features of a film sintered using the modified sintering programme*

Other sintering programmes with much slower heating rates or longer holding time were also explored to see if they could prevent the cracking. Fig. 4-21 are the top surface micrographs of a film sintered using these alternative sintering programmes:

- Sintering temperature 900 °C, holding for 4h, extremely slow heating rate of 5 °C/h between 800-900 °C, heating and cooling rate 5 °C/min for other segments;

- Sintering temperature 800 °C, holding for 4h, extremely slow heating rate 3 °C/h between 700-800 °C, heating and cooling rate 5 °C/min for other segments;

- Sintering temperature: 800°C, holding for 20h, heating and cooling rate 6 °C/h.

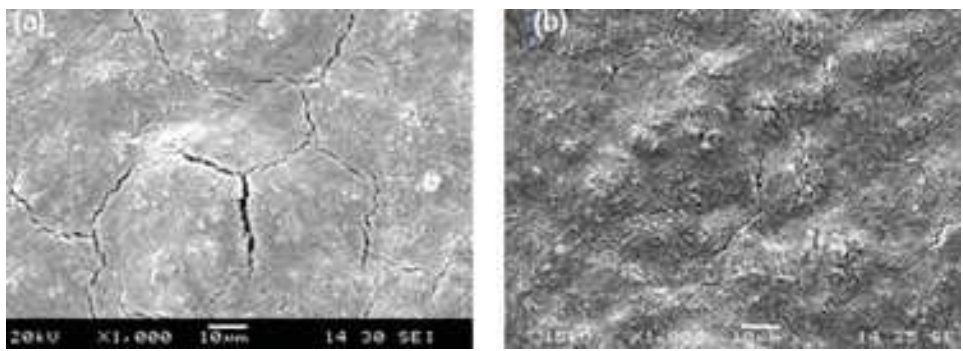

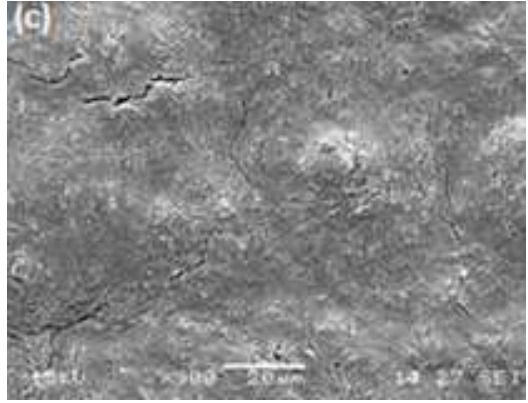

*Fig. 4-21 Micrographs of LSCF6428 film top surfaces after different sintering schedules: (a): Sintering temperature 900 °C, holding for 4h, slow heating rate of 5 °C/h between 800-900 °C, heating and cooling rate 5 °C/min for other segments; (b) Sintering temperature 800 °C, holding for 4h, slow heating rate 3 °C/h between 700-800 °C, heating and cooling rate 5 °C/min for other segments; (c) Sintering temperature: 800C, holding for 20h, heating and cooling rate 6 °C/h.*

Extensive cracks and asperities could still be found in all sintered films in the above experiments, indicating that the prevention of cracking was not achieved by lowering sintering temperatures or slowing down heating/cooling rates. This also implies that cracking in the films does not originate during the sintering process. However, the results do show that a lower sintering temperature results in a smaller crack opening width. Fig. 4-22 shows the cracks in LSCF6428 films after sintering at 1100 and 1200 °C. Compared with the film sintered at 1000 °C, the crack opening width increased to 6 and 10 μm, respectively, resulting in islands of approximately 30 μm diameter on the surface.

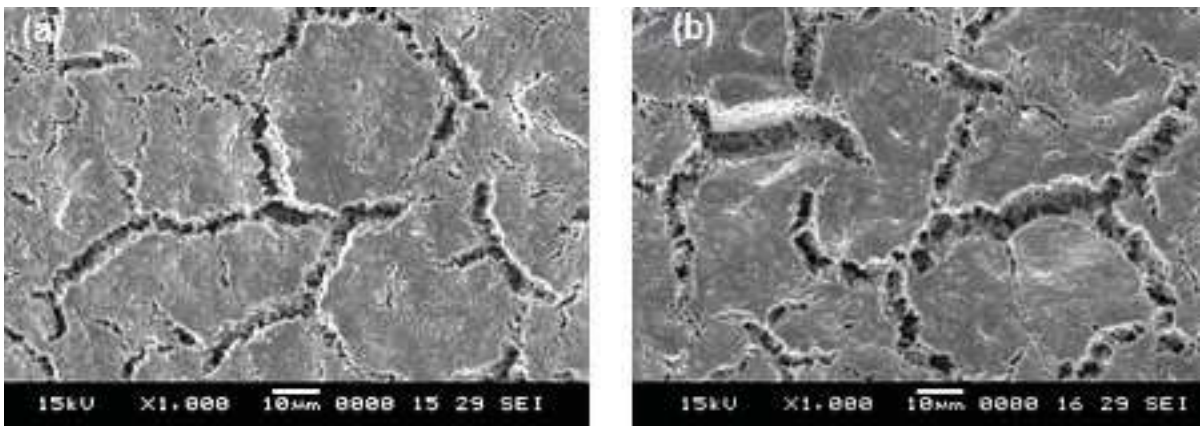

*Fig. 4-22 The surface cracks on LSCF6428 films sintered at (a) 1100 °C and (b) 1200 °C.*

A recent study [12] of crack formation in 8 mol% yttria stabilised zirconia (8YSZ) films showed that cracks already present in the dried films cannot be healed during constrained sintering, suggesting that cracks found after sintering are possibly initiated during the drying stage. However, it can be very difficult to directly observe drying cracks because their opening width is often as small as the particle size. Fig. 4-23 shows top surface micrographs of the as-dried LSCF6428 films, in which it is difficult to identify any cracks with certainty, although typical asperities, probably due to inhomogeneous packing of particles, are evident at this early stage.

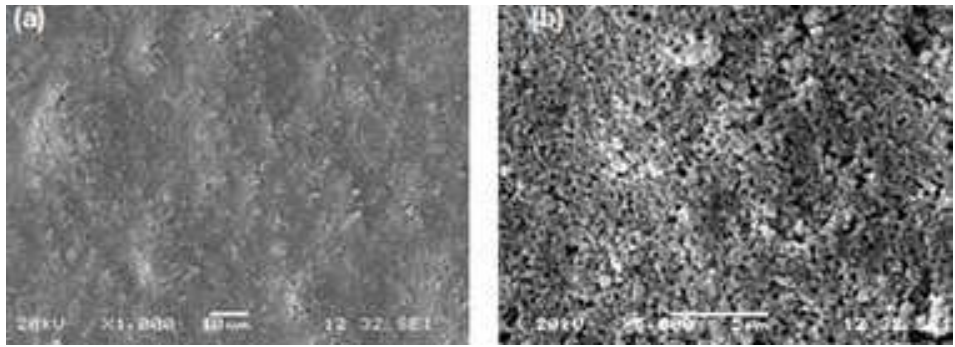

*Fig. 4-23 Micrographs of dried LSCF6428 films screen printed on CGO substrates. Images (a) and (b) are shown at different magnifications*

### 4.5.4.3 Film Deposition Methods

Considering that the non-uniform pressure distribution due to the wire mesh of the screen printer could be a reason for the presence of cracking and surface asperities, we investigated an alternative deposition method to deposit the ink, i.e. tape casting with a doctor blade. The CGO pellet was partially covered by a perimeter adhesive mask of a controlled thickness and a doctor blade coated with LSCF6428 INK-A was pressed under constant pressure and swept smoothly across the pellet surface by hand. Drying and sintering programmes, which were identical to the ones used for the screen-printed samples in the beginning of the study, were applied. The SEM micrographs of the sample surface after sintering are shown in Fig. 4-24. The surface is smoother and has finer cracks although the network remained visible. It is speculated that tape casting eased the homogeneous packing of the particles in the ink after deposition, so the presence of agglomerates was reduced, resulting in a flatter film surface, but the cracking was not prevented, although reduced in severity.

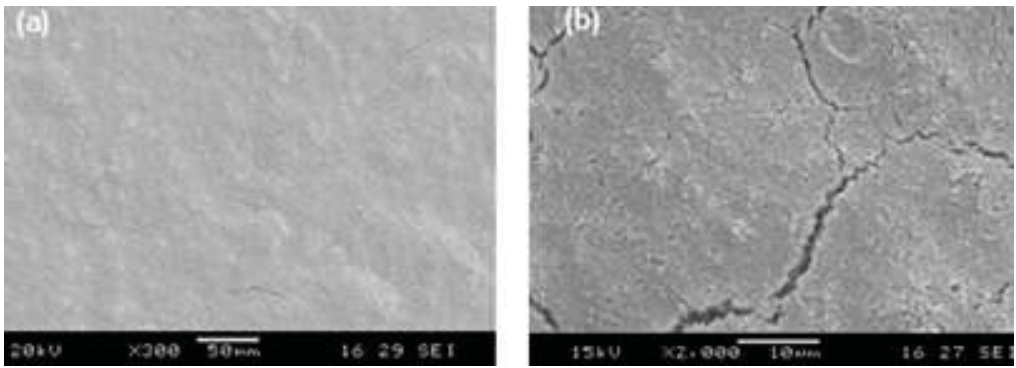

*Fig. 4-24 Micrographs of a sintered LSCF6428 films fabricated by tape casting method,
images (a) and (b) are shown at different magnifications*

### 4.5.4.4 Ink Formulation

In the light of the above experiments, the wetting of the ink and the dynamic interaction
of the particles in the ink, which were largely controlled by the ink viscosity, were thought to
play an important role in the packing of the particles in the films after deposition and
subsequent drying and sintering. Therefore, reformulated inks were prepared as described in
Chapter 3. The as-received LSCF6428 INK-A was diluted 1:1 (i.e. INK-C1) and 1:2 (i.e.
INK-C2) by volume with terpineol and then homogeneously ball-milled for deposition on
CGO substrates by tape casting. The initial drying and 3-step sintering programmes were
applied to sinter the as-deposited films at 1000 ºC. The surface and fracture cross-sectional
morphology of these two films were examined under SEM, as shown in Fig. 4-25.

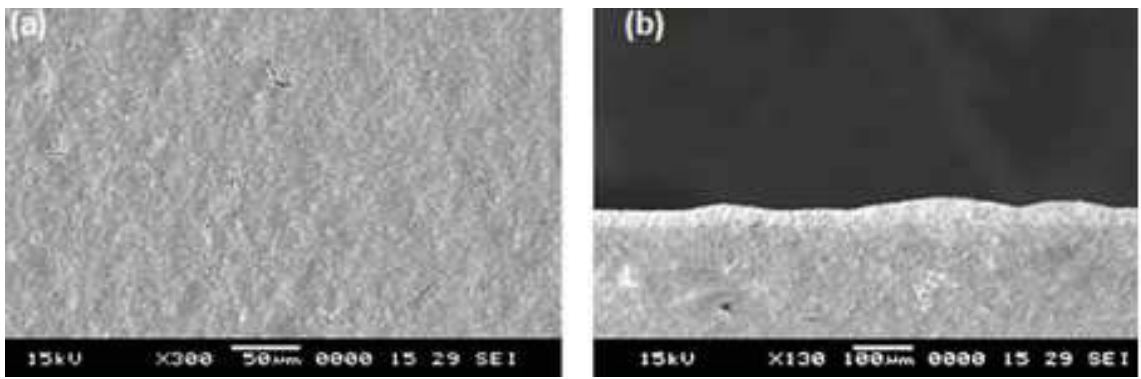

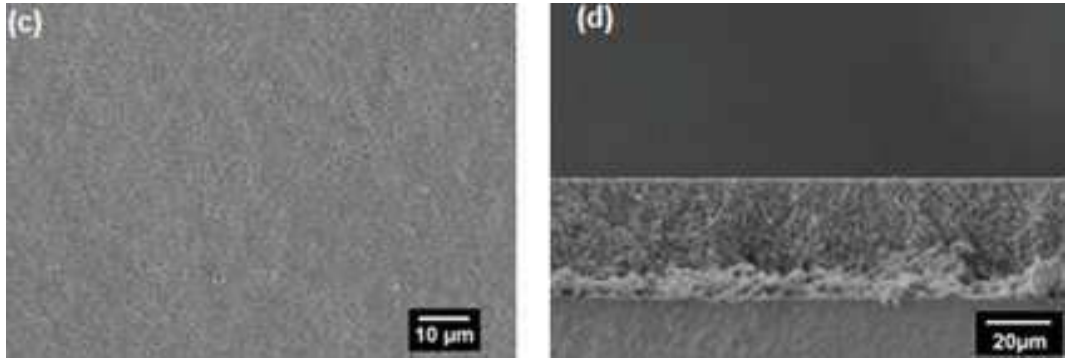

*Fig. 4-25 Micrographs of top surfaces and fracture cross-sections of as-sintered LSCF6428 films made by tape casting reformulated inks: (a, b): 1:1 diluted ink; (c, d): 1:2 diluted ink*

Compared with films (Fig. 4-10) made from as-received INK-A, Fig. 4-25 shows a drastic decrease in the number of cracks and surface agglomerates. Some fine cracks and surface asperities could still be observed in the film made from INK-C1, whereas for the film made from INK-C2, cracks are not seen and the film maintained a very flat surface. A more detailed observation using FIB cross-sectioning and SEM (Fig. 4-26) showed that the cracks were completely eliminated in the latter film, showing that by using the 1:2 diluted INK-C2 acceptable non-defective films could be fabricated. It is worth noting that compared with the microstructure shown in Figures 1 and 2 for the film made from the original ink, Fig. 4-25 (c), (d) and Fig. 4-26 demonstrate a much more homogeneously interconnected microstructure for the film made from INK-C2, as a result of the lower ink viscosity.

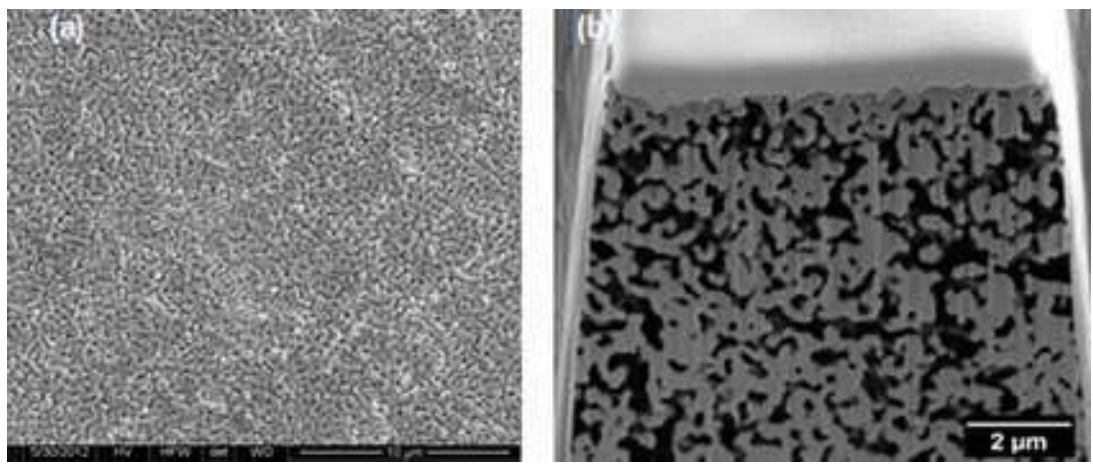

*Fig. 4-26 (a) High magnification surface and (b) FIB sliced cross-sectional micrographs at of a film made from the 1:2 diluted ink after sintering at 1000 ℃. Note that the black phase in (b) is epoxy resin impregnated prior to FIB slicing.*

Nanoindentation tests using the spherical indenter were performed subsequently on this crack-free and flat film. The response curves and the resulting elastic modulus are plotted in Fig. 4-27 (a) and (b), respectively.

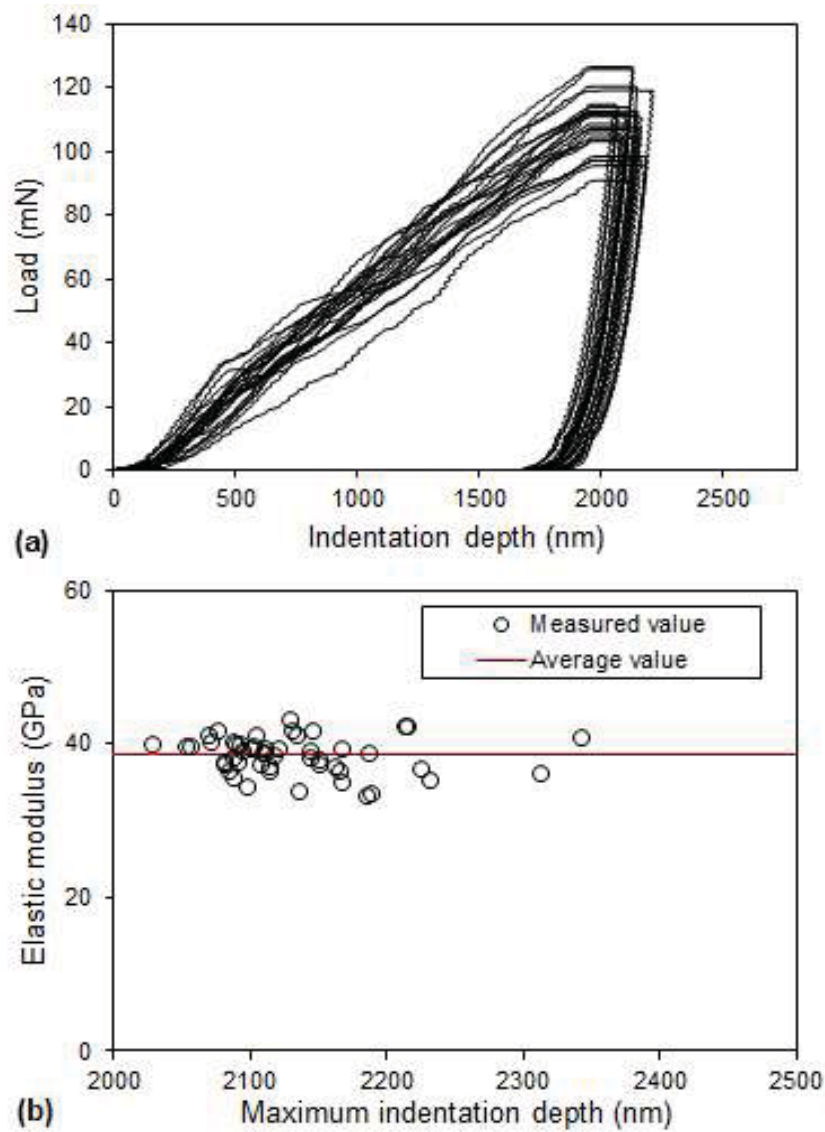

*Fig. 4-27 (a) Nanoindentation load vs. indentation depth curves and (b) the corresponding elastic modulus values calculated for the crack-free and flat film made from 1:2 diluted ink.*

The results above show that the measurements for the non-defective film were significantly less scattered than for the defective ones (Fig. 4-14), as the unloading response curves plotted in Fig. 4-27 (a), from which the elastic modulus was determined, were highly consistent.

Fig. 4-28 shows the viscosity measurements of as-received ink and the diluted inks and the surface roughness profiles performed using Zygo for the corresponding as-sintered films made from them.

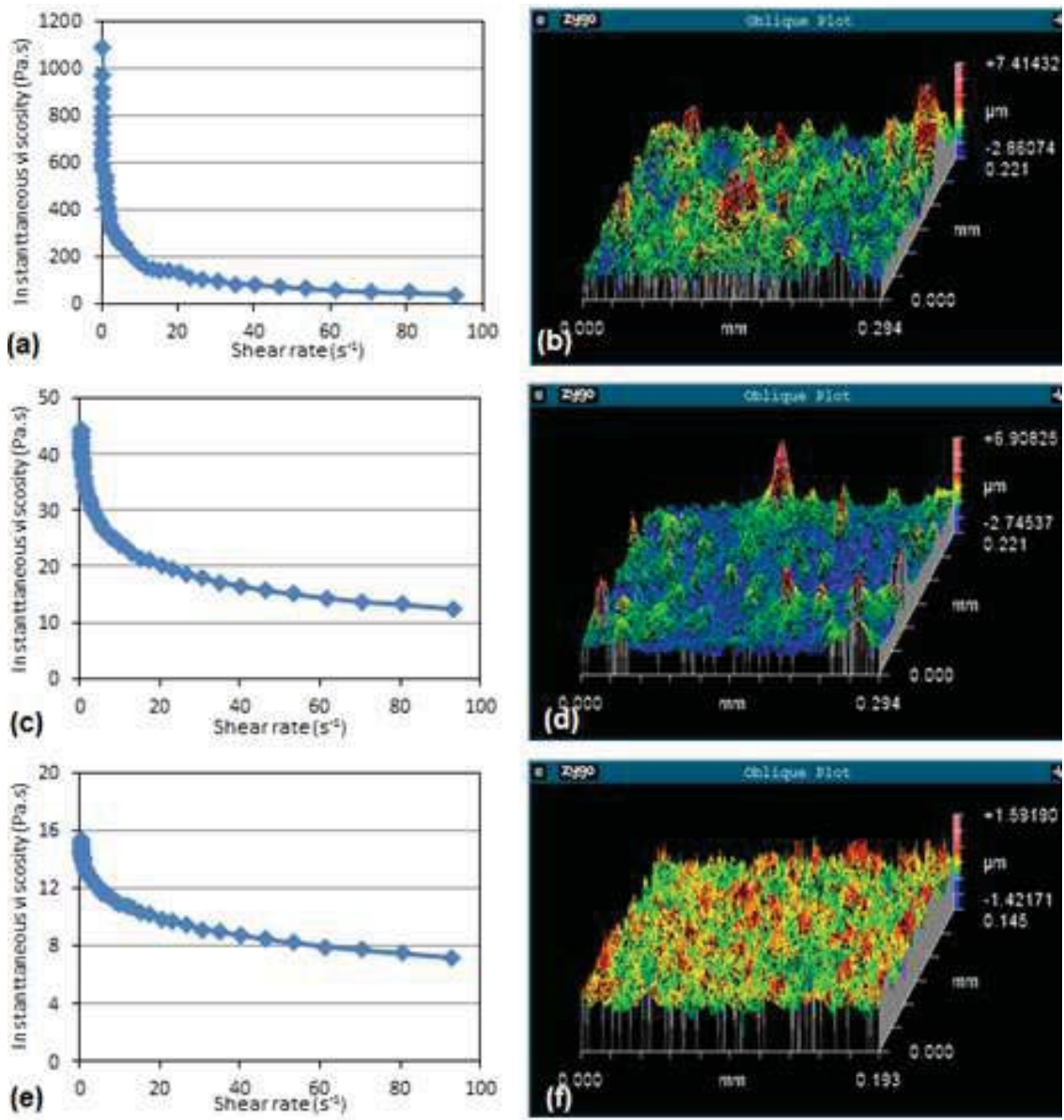

*Fig. 4-28 Images (a), (c) and (e) are viscosity measurement results for INK-A, INK-C1 and INK-C2. Images (b), (d) and (f) are surface roughness profiles of the as-sintered films made from INK-A, INK-C1 and INK-C2 measured by Zygo.*

Obviously, the viscosity reduced repidly when the inks are diluted. The Zygo profiles plotted in the figure also readily show that much smoother surfaces were generated using diluted inks. Table 4-4 further compares the properties of the three inks as well as the characteristics of the films fabricated from them. Note that in the table the viscosity values are compared at a shear rate of 95 s$^{-1}$.

*Table 4-4 Comparison between properties of the as-received and diluted inks and their corresponding films sintered at 1000 °C.*

| Properties | INK-A | INK-C1 | INK-C2 |
|---|---|---|---|
| Viscosity (*Pa.s*) | 41.51 | 12.36 | 7.19 |
| | Corresponding films fabricated | | |
| Average surface roughness $R_a$ (*μm*) | 1.82 | 0.89 | 0.20 |
| Elastic modulus (*GPa*) | 34±21 | 35±11 | 39±2 |

Lower ink viscosities resulted in remarkable decrease of surface roughness and standard deviation of elastic modulus, implying a significantly reduced number of surface defects in the corresponding films. As can be seen in Table 4-4, the surface roughness for the films made from INK-C2 was reduced to merely 10% of that of films made from the as-received INK-A. This is much smaller than the indentation depth, resulting in a steep reduction of relative standard deviation of measured elastic modulus from 63.2% to 6.2%.

The experiments above revealed that rather than shrinkage during sintering, or differential contraction during cooling as proposed in many studies [6, 13], the more critical factor for obtaining crack-free and flat films in the current study was the ability of the ink to be self-levelling in the early wet state. Cracking is most likely to initiate at the drying stage if the particles are prevented from packing more effectively as the liquid content was removed. Lowering ink viscosity by adding more terpineol solvent in this case, could effectively keep the ceramic particles in a stable suspension in the ink and flocculated agglomerates could be avoided. It also promoted the dispersion of the particles in a more stable and homogeneous way as reported by Maiti and Rajender in YSZ suspensions [14]. As a result, cracking and asperity formation could be reduced to an acceptable level, or even be completely avoided. Consistent and reliable nanoindentation measurements could thus be conducted with high reproducibility.

## 4.6 Comparison of Porosities and Microstructures of LSCF6428 Bulk and Films

The porosities measured for LSCF6428 in both film and bulk forms sintered at 900-1200 °C are summarised in Table 4-5. Note that the films were made from the reformulated INK-C2. It is found that sintering at 900 °C tended to generate approximately 45 % porosity for both forms of samples. After sintering at 1000 and 1100 °C, both the films and bulk samples had similar porosities. However, the bulk porosity experienced a huge drop to only 5 % after sintering at 1200 °C, compared to 15 % for films. This significant difference might be attributed to the different sintering constraints for bulk and films, as the films deposited on substrates underwent constrained sintering in which lateral shrinkage was prevented by the substrate. For bulk samples much more homogeneous and faster densification occurred along all three directions and hence much less porosity was generated. These trends in the evolution of porosity as a function of sintering temperature can be readily seen in the micrographs shown later.

*Table 4-5 Porosity vs. sintering temperature for LSCF6428 films and bulk specimens*

| Sintering Temperature (°C) | Film Porosity (%) | Film average grain size (nm) | Bulk Porosity (%) |
|---|---|---|---|
| 900 | 46.9±2.2 | 200 | 44.85±0.32 |
| 1000 | 39.7±2.6 | 270 | 36.28±1.12 |
| 1100 | 24.1±1.8 | 450 | 28.67±0.95 |
| 1200 | 15.2±1.2 | 690 | 5.22±0.01 |

The surface features of the as-sintered films are shown in Fig. 4-29. The average grain sizes of these films were estimated based on the SEM micrographs obtained and are shown in Table 4-5, indicating that higher sintering temperatures resulted in coarser grains.

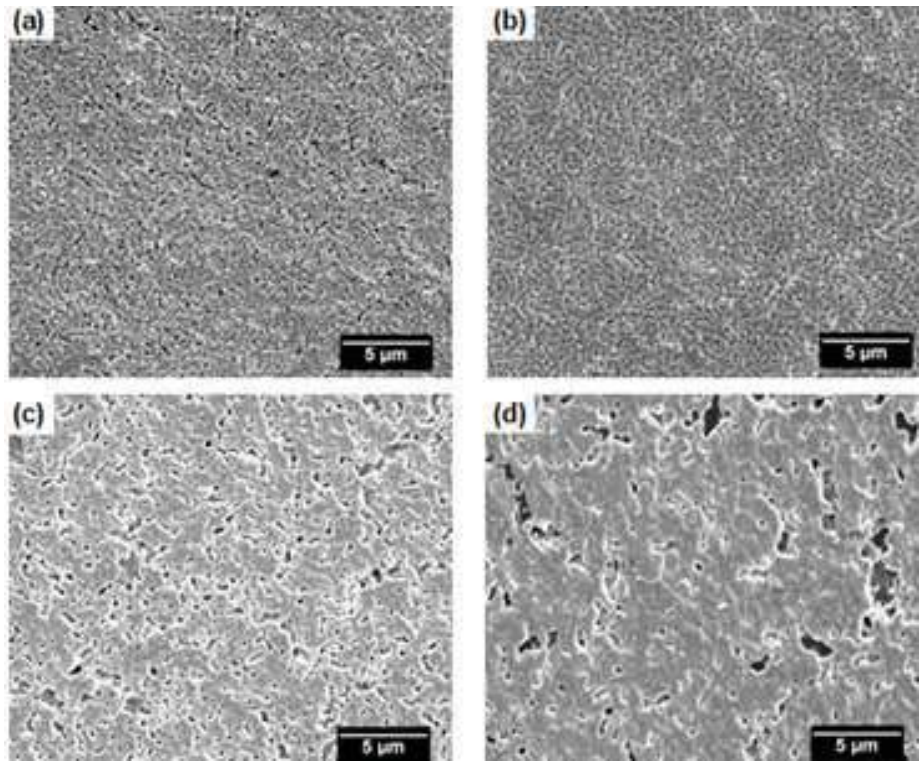

*Fig. 4-29 Micrographs of top surface of LSCF6428 films after sintering at (a) 900; (b) 1000; (c) 1100; and (d) 1200 °C*

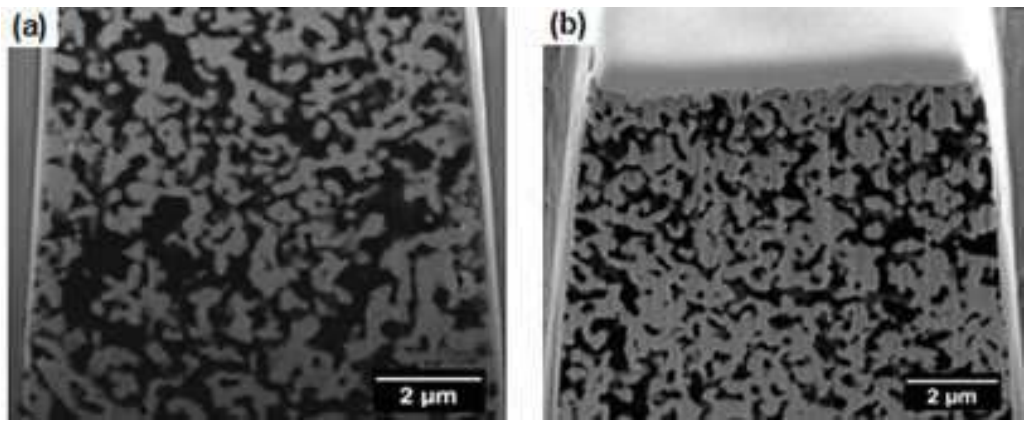

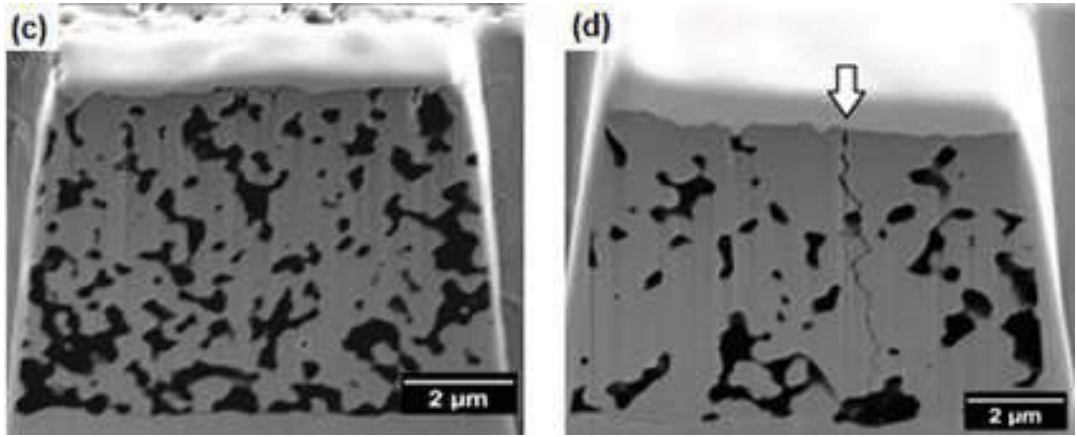

*Fig. 4-30 FIB-milled cross-sectional micrographs of LSCF6428 films after sintering at (a) 900; (b) 1000; (c) 1100; and (d) 1200 °C (black represents porosity and gray is LSCF6428). The bright layer at the top is sputtered Pt to protect the surface.*

As can be seen in Fig. 4-29, little densification or grain growth took place during sintering at 900 and 1000 °C, consistent with the large porosity shown in Table 4-5. Whereas, during the sintering at 1100 and 1200 °C, the continuous growth of grains could lead to the coalescence of neighbouring pores, thus the average pore size increased, even if the porosity decreased [15]. These increasingly large pores observed on the surface with higher sintering temperature are typical of films formed by constrained sintering. It is also apparent that there are no detectable surface cracks in these films for all sintering temperatures. The evolution of the corresponding cross-sectional microstructures shown in Fig. 4-30 is consistent with the images from the top surfaces. A very narrow micro-crack penetrating through the film can be seen in Fig. *4-30* (d) (shown by an arrow). The narrowness of the crack indicates that it was formed after sintering, probably due to thermal contraction mismatch on cooling. Such a crack might cause errors if the nanoindentation test was conducted nearby, but these cracks are rare and any individual indentations affected by them would be apparent in the distribution of measured values.

In the same way, the SEM micrographs of the polished top surfaces (which are different from the unpolished pellet surfaces shown in the beginning of this Chapter) and FIB-milled cross-sections of each sintered bulk sample are shown in Fig. 4-31 and Fig. 4-32. Note that resin impregnation was not performed for these samples.

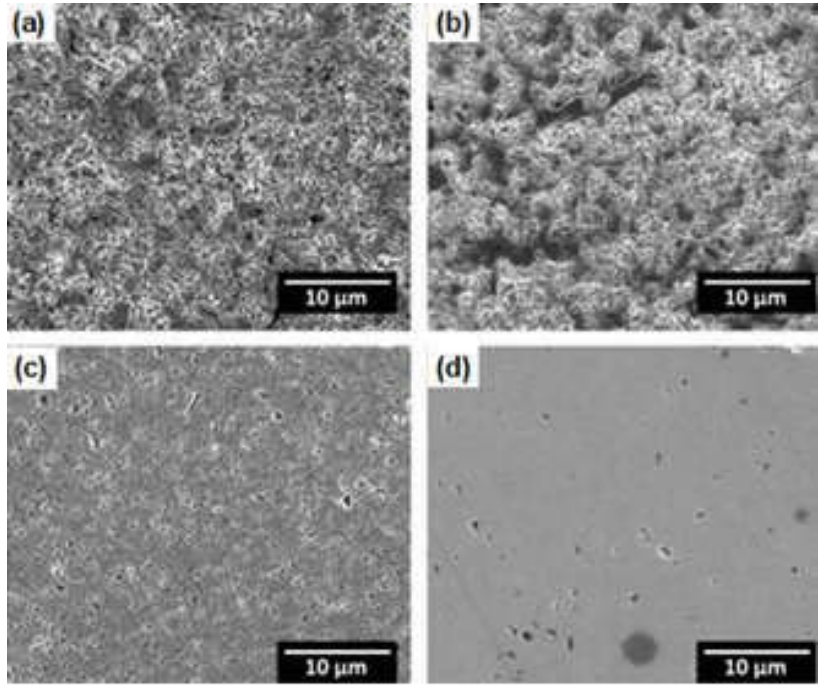

*Fig. 4-31 Polished surface SEM micrographs of bulk samples after sintering at (a) 900; (b) 1000; (c) 1100; and (d) 1200 °C*

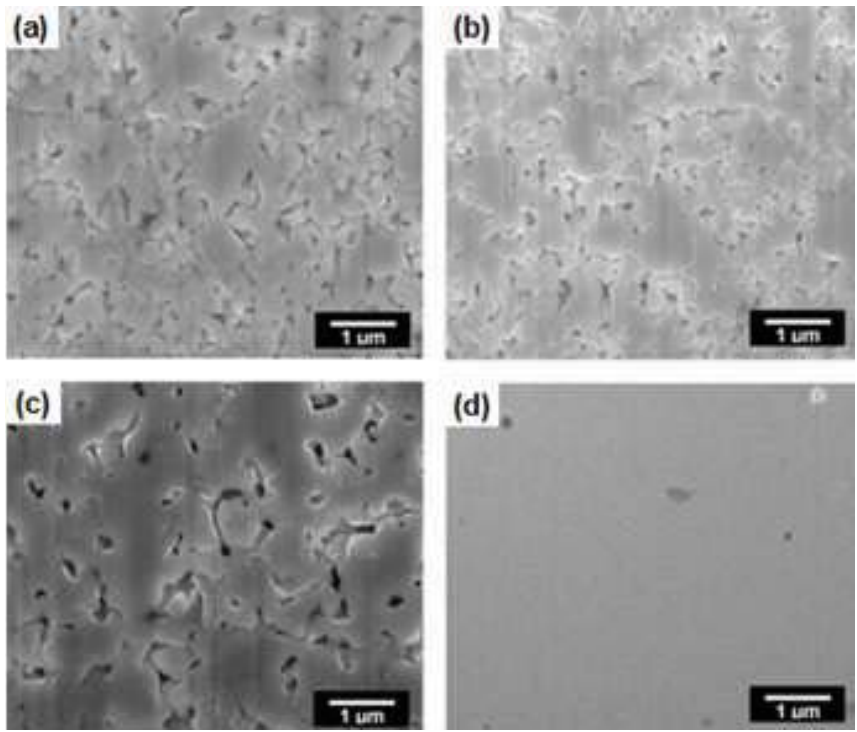

*Fig. 4-32 FIB-milled cross-sectional micrographs of bulk samples after sintering at (a) 900; (b) 1000; (c) 1100; and (d) 1200 °C*

These micrographs show that bulk samples after sintering at 900 and 1000 °C inherited similar porous features, while for samples after sintering at 1000 and 1200 °C, the solid content significantly increased, particularly for the one sintered at 1200 °C, for which only few pores can be identified in the micrograph. The observation was confirmed by the porosity measurement shown in Table 4-5, the porosities for 900 and 1000 °C-sintered samples were very close, while as the sintering temperature increased from 1000 to 1200 °C, the porosity was reduced from 36% to 5%, implying almost complete solid content in the 1200 °C-sintered sample as shown in the SEM micrographs. In addition, similar microstructural features revealed by the SEM images for films and bulk samples sintered at the same temperature below 1200 °C readily confirm the comparable porosity levels presented in Table 4-5. The porosity-dependent mechanical properties of both film and bulk samples, which are the main concern of the current research, will be discussed later in Chapters 5 and 6. 3D microstructural characterisation using FIB/SEM and the subsequent finite element modelling using the actual microstructures will be shown in Chapters 7 and 8.

## *Summary*

In this Chapter, important physical properties such as density, porosity, and 2D microstructures of the LSCF6428 films and bulk as well as CGO bulk were investigated. Results show that high density LSCF6428 and CGO pellets are suitable to be substrates for the deposition of LSCF inks. Abnormally large lumps were found in the commercially provided LSCF6428 powder which was therefore subjected to further mechanical granulation to generate finer particle size suitable to be used in the study, particularly to make bulk samples rather than inks for screen printing.

A ready-made LSCF6428 ink (i.e. INK-A) was chosen for film deposition due to its finer and homogeneous particle size distribution. The thickness of sintered LSCF6428 films was measured to reach a maximum as a function of the number of screen printing passes. More importantly, sintered porous thin LSCF6428 films fabricated using this typical screen-printing LSCF6428 ink were full of cracks and surface asperities, which caused extremely inconsistency and errors in attempts to measure the film elastic modulus by nanoindentation. Various processing parameters were investigated in order to eliminate these defects. Different film deposition methods and thermal treatments showed little effect on the prevention of the defect formation. Furthermore the cracks were shown not to arise from thermal expansion mismatch between film and substrate and it was concluded that the defects initiated during drying of the ink and then were made more severe by the sintering process. Thus, sintered porous thin LSCF6428 cathode films were successfully fabricated without any crack or surface asperities by using a much less viscous ink. Results suggested that when a sufficient amount of terpineol solvent was added to the as-received ink, it eased the wetting and self-levelling of the ink upon deposition and thus a more homogeneous packing of LSCF6428 particles in the films was achieved. As a result there was more homogeneous shrinkage in the film during the thermal treatments that followed and lower local stresses in the films. With the non-defective films, consistent and reliable loading and unloading data could then be obtained using nanoindentation.

# *Chapter 4 References*

1.   Fergus JW: **Solid oxide fuel cells: materials properties and performance**, vol. 1: CRC PressI Llc; 2009.

2.   Taroco H, Santos J, Domingues R, Matencio T: **Ceramic Materials for Solid Oxide Fuel Cells**. 2011.

3.   Lee S, Chu C-L, Tsai M-J, Lee J: **High temperature oxidation behavior of interconnect coated with LSCF and LSM for solid oxide fuel cell by screen printing**. *Applied Surface Science* 2010, **256**(6):1817-1824.

4.   Baque L, Serquis A: **Microstructural characterization of La0.4Sr0.6Co0.8Fe0.2O3–δ films deposited by dip coating**. *Applied Surface Science* 2007, **254**(1):213-218.

5.   Marinha D, Dessemond L, Cronin JS, Wilson JR, Barnett SA, Djurado E: **Microstructural 3D Reconstruction and Performance Evaluation of LSCF Cathodes Obtained by Electrostatic Spray Deposition**. *Chemistry of Materials* 2011, **23**(24):5340-5348.

6.   Hsu C-S, Hwang B-H: **Microstructure and Properties of the La0.6Sr0.4Co0.2Fe0.8O3-δ Cathodes Prepared by Electrostatic-Assisted Ultrasonic Spray Pyrolysis Method**. *Journal of The Electrochemical Society* 2006, **153**(8):A1478-A1483.

7.   Santillán MJ, Caneiro A, Quaranta N, Boccaccini AR: **Electrophoretic deposition of La0.6Sr0.4Co0.8Fe0.2O3−δ cathodes on Ce0.9Gd0.1O1.95 substrates for intermediate temperature solid oxide fuel cell (IT-SOFC)**. *Journal of the European Ceramic Society* 2009, **29**(6):1125-1132.

8.   Wang S, Awanob M, Matsuda K, Maeda K: **Porous La0.6Sr0.4Co0.2Fe0.8O2.8 electrodes prepared by in-situ solid-state reaction sintering**. In: *Power Sources for the New Millennium: Proceedings of the International Symposium: 2001*: The Electrochemical Society; 2001: 118.

9.   Tai LW, *et al.*: **Structure and Electrical Properties of La1-xSrxCo1-yFeyO3. Part 2: The System La1-xSrxCo0.2Fe0.8O3**. *Solid State Ionics* 1995(76):273-283.

10.  Jacobson AJ: **Materials for Solid Oxide Fuel Cells†**. *Chemistry of Materials* 2009, **22**(3):660-674.

11.  Zhang HM, Shimizu Y, Teraoka Y, Miura N, Yamazoe N: **Oxygen sorption and catalytic properties of La1−xSrxCo1−yFeyO3 Perovskite-type oxides**. *Journal of Catalysis* 1990, **121**(2):432-440.

12.  Wang X, Chen Z, Atkinson A: **Crack formation in ceramic films used in solid oxide fuel cells** *Journal of European Ceramic Society* 2013, **in press**.

13.  Lai BK, Xiong H, Tsuchiya M, Johnson AC, Ramanathan S: **Microstructure and Microfabrication Considerations for Self-Supported On-Chip Ultra-Thin Micro-Solid Oxide Fuel Cell Membranes**. *Fuel Cells* 2009, **9**(5):699-710.

14.  Maiti A, Rajender B: **Terpineol as a dispersant for tape casting yttria stabilized zirconia powder**. *Materials Science and Engineering: A* 2002, **333**(1):35-40.

15.  Rahaman MN: **Ceramic processing**: CRC Press; 2007.

# 5 Characterisation of Mechanical Properties: Elastic Modulus and Hardness

This Chapter shows how reliable measurements on porous ceramic films can be made by appropriate indentation (i.e. nanoindentation and microindentation) experiments and analyses. Room-temperature mechanical properties including elastic modulus and hardness of LSCF6428 were investigated by spherical nanoindentation of bulk samples and porous films which were sintered at 900-1200 °C. The elastic moduli measured by nanoindentation were compared with those measured by the resonance method (i.e. IET) on bulk samples. The mechanical properties were evaluated with regard to the influence of porosities. FIB/SEM slice and view technique was applied to examine the subsurface microstructural changes, such as plastic deformation, crack morphologies, and film/substrate interface condition.

## 5.1 Elastic Modulus and Hardness of Bulk Samples at RT

### 5.1.1 Nanoindentation Results

The elastic modulus and hardness of bulk samples measured using spherical nanoindentation are shown in Fig. 5-1 and Fig. 5-2 plotted as a function of the maximum indent load, which is related to the maximum indentation depth.

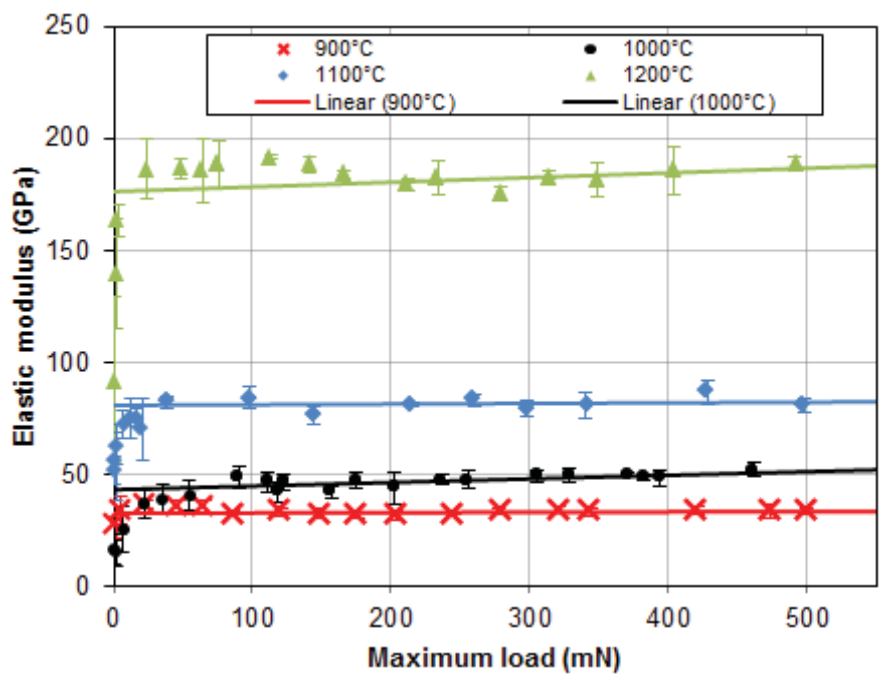

*Fig. 5-1 Elastic modulus vs. maximum indentation load for bulk samples after sintering at 900 - 1200 °C*

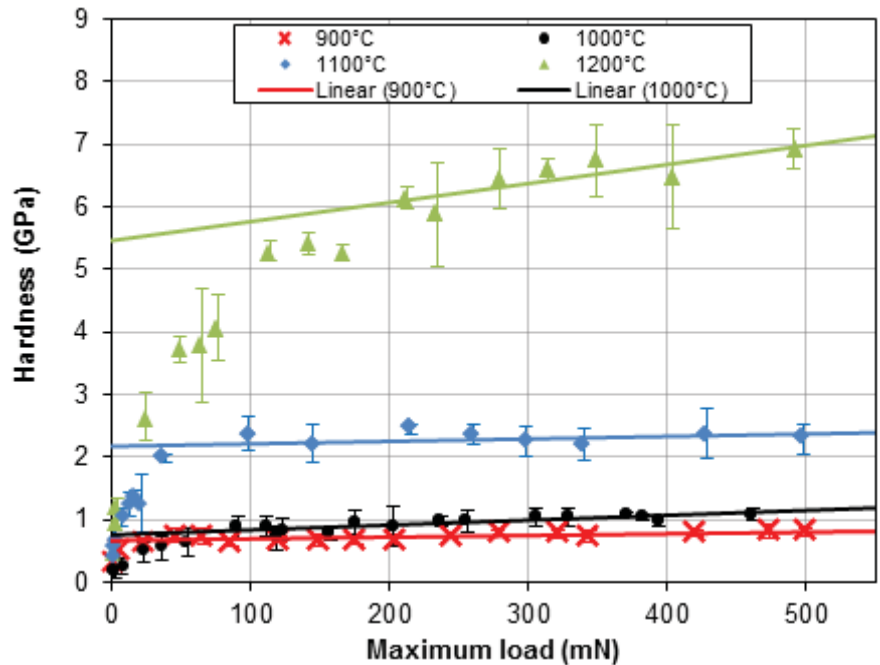

*Fig. 5-2 Hardness vs. maximum indentation load for bulk samples after sintering at 900 - 1200 °C*

Fig. 5-1 clearly shows that, in all the bulk samples, the apparent elastic modulus exhibits an initial dramatic increase with increasing maximum indentation load below approximately 50 mN, followed by a levelling off at higher loads, although there is a high level of scatter between 50-200 mN for 1200 °C data. The variation of elastic modulus at loads lower than 50 mN is probably due to the inevitable surface roughness which was comparable to the indent depth at these low loads. Therefore in our case, the best estimate of true elastic modulus was obtained by extrapolating the data points from the load range 50 – 500 mN to zero load for 900-1100 °C data and 200-500 mN for 1200 °C data, as shown by the solid lines in Fig. 5-1. The same method was used to obtain the best estimates of the microstructure hardness. The estimated elastic modulus and hardness are summarised in Table 5-1.

*Table 5-1 Elastic moduli and hardness of the sintered bulk samples obtained by nanoindentation*

| Sintering Temperature (°C) | Elastic modulus (GPa) | Hardness (GPa) |
|---|---|---|
| 900 | 34.2±2.1 | 0.69±0.09 |
| 1000 | 44.5±3.2 | 0.86±0.20 |
| 1100 | 80.2±1.9 | 2.35±0.14 |
| 1200 | 174.3±2.8 | 5.76±0.12 |

The larger scatter in the results for the bulk sample after sintering at 1000 °C (Fig. 5-1, Fig. 5-2 and Table 5-1) was attributed to its relatively rougher surface (see Fig. 4-31 in Chapter 4) compared to other samples despite fine surface polishing. The observation that for loads > 50 mN there is only a weak dependence of apparent modulus on load, indicates that densification (crushing) of the porous bulk material under the indenter had a negligible influence on the elastic response. In principle, densification of the material generated by crushing under the indenter could significantly increase the local elastic modulus. This lack of sensitivity to densification is presumably because the densified zone is small compared with the longer range of elastic deformation of the rest of the material. This phenomenon will be examined in more detail later in the data analysis of the porous films, for which the factors influencing the measurements are more complicated. However, it can be seen from Fig. 5-2 that the hardness is more sensitive to indenter load, implying that the hardness was more influenced by densification. This is to be expected since the hardness (permanent plastic deformation) of these materials is controlled by crushing rather than the more usual plastic deformation (viscous flow or dislocation motion) seen in other non-porous materials.

### 5.1.2 IET Results and Comparison with Nanoindentation

The elastic modulus of nominally dense LSCF6428 samples was also measured by impulse excitation test (IET) to compare with the nanoindentation result after the accuracy of the IET measurements was calibrated. The elastic modulus measured by IET for the nominally dense pellet was 147±3 GPa, which is close to, but slightly lower than the results reported by Chou *et al.* [1] (152±3 GPa) and Kimura *et al.* [2] (164 GPa), as shown in Table 5-2. Nanoindentation tests using a spherical tip tended to generate slightly smaller values of both elastic modulus and hardness, compared to the use of the Berkovich tip. Vlassak and Nix [3] also reported that modulus measured with a triangular indenter was typically 5-6% higher than that obtained using an axisymmetric indenter, possibly due to the indenter geometry effect on the material plastic flow upon indentation. Surprisingly, Li *et al.* [4], using nanoindentation, reported much lower elastic moduli (155±4 GPa) for even higher relative density (98.3%) compared to our nanoindentation result. In addition, the surprisingly low value (76 GPa) reported by Huang *et al.* [5] might be related to some creep during the ring-on-ring test which greatly decreased the apparent elastic modulus. The reliability of both these latter values remains highly questionable and the origins of such discrepancies need to be identified. The results in our study confirm that using a spherical tip, the standard deviations were less scattered, compared with the Berkovich tip, which might be due to it being less sensitive to the sample surface condition. Nevertheless, it is clear that the indentation method gave significantly higher modulus than the other methods and we therefore considered whether non-elastic behaviour might be responsible for this.

*Table 5-2 Comparison of Young's modulus and hardness measurements for nominally dense LSCF6428 samples*

| Reference | Sintering Conditions in Air (sintering T/holding time/heating rate) | Relative Density (%) | Main Grain Size (μm) | Measurement Technique | Young's Modulus at RT (GPa) | Hardness (GPa) |
|---|---|---|---|---|---|---|
| Kimura *et al.* [2] | 1300°C/6h/106°C·h⁻¹ | 98 | 5.0 | IET | 164 | n/a |
| Chou *et al.* [1] | 1250°C/4h/300°C·h⁻¹ | 95.4±0.2 | 2.9 | Ultrasonic method | 152±3 | n/a |
| Li *et al.* [4] | 1200°C/2h/300°C·h⁻¹ | 98.3 | 0.8 | Nanoindentation (Berkovich tip, $P_{max}$=490mN) | 155±4 | 8.6±0.3 |
| Huang *et al.* [6] | 1200°C/3h/300°C·h⁻¹ | 96.6±0.2 | 0.6 | Ring-on-ring bending test | 76±5 | 6.3±0.4 |
| This work | 1200°C/4h/300°C·h⁻¹ | 94.78±0.01 | 1.6 | IET | 147±3 | n/a |
| | | | | Nanoindentation (Berkovich tip, $P_{max}$=500mN) | 180±10 | 7.0±0.2 |
| | | | | Nanoindentation (Spherical tip, $R$=25μm, $P_{max}$=500mN) | 174±3 | 5.3±0.1 |

LSCF6428 has a rhombohedral perovskite structure at room temperature [7, 8], with a transition to cubic perovskite occurring at approximately 773 K [8]. Therefore, any phase transition of the material due to temperature difference can be ruled out as a source of the discrepancy, as the IET and nanoindentation tests were performed in air at room temperature. We therefore considered other possibilities such as: (1) ferroelastic behaviour induced by the indentation stresses; (2) inhomogeneous distribution of oxygen vacancies across the sample; (3) micro-cracking in the sample; (4) nanoindentation creep; and (5) residual porosity. These were investigated successively as described below.

### 5.1.2.1 Ferroelastic Behaviour

Ferroelastic behaviour in crystals was first presented and summarised by K. Aizu in 1969 [9]. Materials possessing ferroelasticity are often characterised to exhibit strain vs. stress hysteresis, which is an analogue with polarisation of a ferroelectric material in an electric field and magnetisation of a ferromagnetic material in a magnetic field [9]. The origin of this phenomenon is considered to be domain switching induced by the application of mechanical stress below a critical temperature above which the material transforms to a paraelastic phase showing the strain is a single-valued function of stress.

Ferroelasticity in both rhombohedral and orthorhombic lanthanum-based perovskite materials has been observed, such as $LaFeO_3$ [10], $LaAlO_3$ [11], $La_{1-x}Sr_xMnO_3$ [12] and $LaCoO_3$ [13]. More recently, ferroelastic behaviour of $La_{0.5}Sr_{0.5}Co_{1-x}Fe_xO_{3-\delta}$ was reported by Lein *et al.* [14] and ferroelasticity in $La_{0.58}Sr_{0.4}Co_{0.2}Fe_{0.8}O_{3-\delta}$ was reported by Huang *et al.* [15] and Araki *et al.* [16]. Most of the ferroelastic behaviour in the abovementioned materials was experimentally observed as a nonlinear pop-in/pop-out response in the load-displacement curves or hysteresis loop in the stress-strain curves generated from experiments such as uniaxial compression, four-point bending and indentation. Fig. 5-3 illustrates an example of the ferroelastic hysteresis observed by Orlovskaya *et al.* [17] in the perovskite material $La_{0.8}Ca_{0.2}CoO_3$ under contact loading using a conical indenter. It was confirmed that ferroelastic domain switching in this perovskite material caused by the applied stress and unloading led to the hysteresis behaviour. In addition, it has been reported [18] that the critical stress decreases from lower temperatures to higher temperatures for ferroelastic materials, which was explained by the reduced mechanical energy threshold triggering domain switching at higher temperatures. This, in other words, means that a ferroelastic material at room temperature may require a larger applied stress to exhibit the ferroelastic behaviour, compared with that at an elevated temperature.

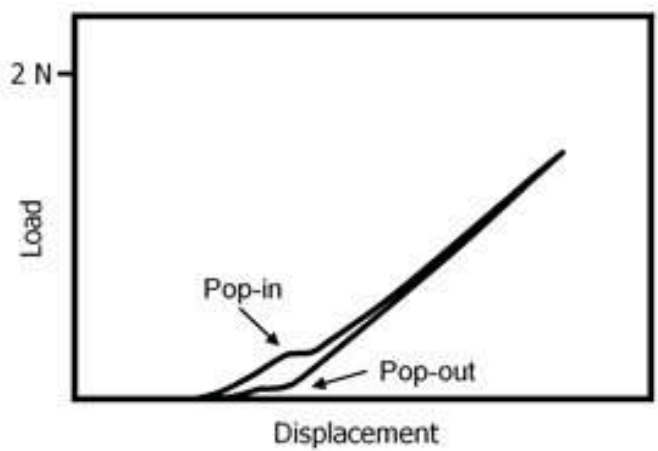

*Fig. 5-3 Ferroelastic hysteresis loop in the load-displacement curves for $La_{0.8}Ca_{0.2}CoO_3$ under contact loading using a conical indenter (after Orlovskaya et al. [17]).*

In the current study, however, the so-called nonelastic pop-in and pop-out was not found during indentation tests. Loading-partial-unloading tests with the 50 μm diameter spherical indenter, even after the maximum indentation load was increased to 20 N, showed no hysteresis in the elastic region as can be seen from Fig. 5-4. It was revealed by SEM observation that the small potential pop-in and pop-out events arrowed in the figure were, in effect, due to cracking induced at high load ( > 12 N).

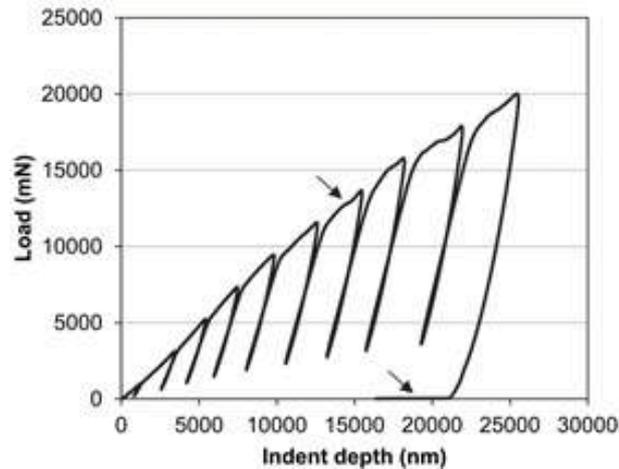

*Fig. 5-4 Load-displacement curve of LSCF6428 dense sample under loading-unloading cycles of microindentation using a spherical tip*

Fig. 5-5 also shows a set of typical load vs. depth curves suggesting very smooth nanoindentation response for bulk samples sintered at varying temperatures.

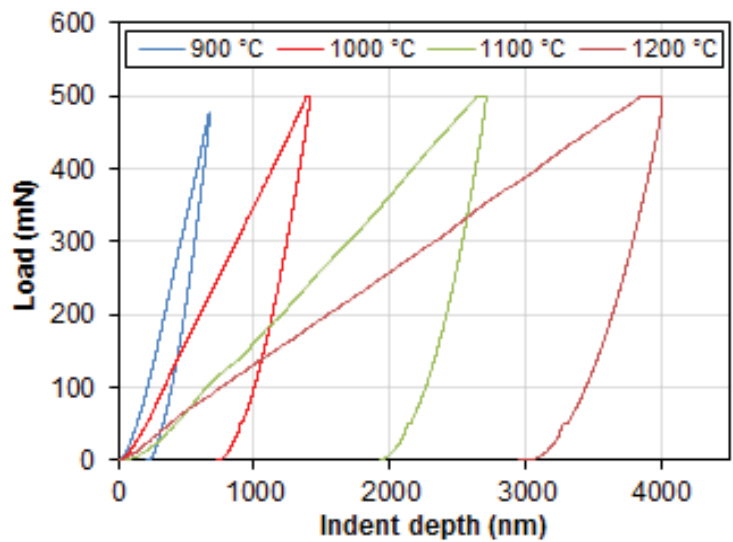

*Fig. 5-5 Typical nanoindentation curves of bulk samples sintered at varying temperatures showing smooth response. Note that the pop-out at 50 mN was due to the 20 s holding time to check system drift.*

The reason for ferroelasticity not being observed in this study may be due to the domain dimension being larger than the grain size, or the stress induced by the applied load being below the critical stress for domain wall motion, possibly because the use of spherical tip led to small change in stress with the change of load (although this seems unlikely at these high loads). Further study is required to clarify this. Nevertheless, it would appear that ferroelasticity is not responsible for the difference between Young's modulus measured by indentation and resonance methods.

### 5.1.2.2 Oxygen Vacancy Gradient

As explained in some studies [19-22], absorption of oxygen, and hence loss of oxygen vacancies from outer surface regions of a sample, might be expected when perovskite materials with oxygen deficiency are cooled. This would result in a gradient of oxygen vacancies from inside out across the thickness of the samples. Since oxygen vacancies increase the lattice parameter this could lead to differential contraction between the surface and the central core of the sample and a residual tensile stress and higher modulus (due to lower vacancy concentration) in the surface region. Therefore, the local elastic modulus across the thickness of a cut specimen should have a symmetric distribution and a minimum would ideally be reached in the centre. If observed, this could possibly account for the sample's overall modulus measured by dynamic methods being lower than that measured on the surface using nanoindentation.

In order to verify the above assumption, nanoindentation tests were performed along the thickness of a cut sample from the original outer surface to the opposite face of the 2 mm thick pellet. The measurement result is shown in Fig. 5-6.

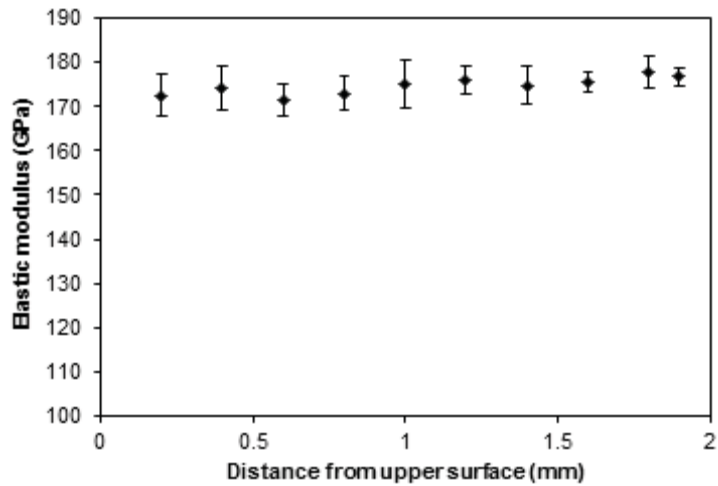

*Fig. 5-6 Distribution of the local elastic modulus measured across the thickness on the cut surface of the bulk sample*

Surprisingly the local elastic modulus across the thickness turned out to be constant, within experimental error, and close to the previous nanoindentation result. There is no obvious decreasing trend with distance from outer surface into the centre and no aforementioned elastic modulus gradient was found. Therefore, any gradient in oxygen vacancy concentration could be ruled out as being the cause of the significant discrepancy between results measured by dynamic (resonance method) and static methods (nanoindentation).

### 5.1.2.3 Micro-cracking

No evidence of micro-cracking in the present specimens used in IET experiments could be detected by SEM even at high magnification as shown in Fig. 5-7 (a). A load-depth curve of nanoindentation on such a well-polished dense LSCF6428 pellet using the spherical tip is shown in Fig. 5-7 (b), which further demonstrates no noticeable influence from intrinsic cracks.

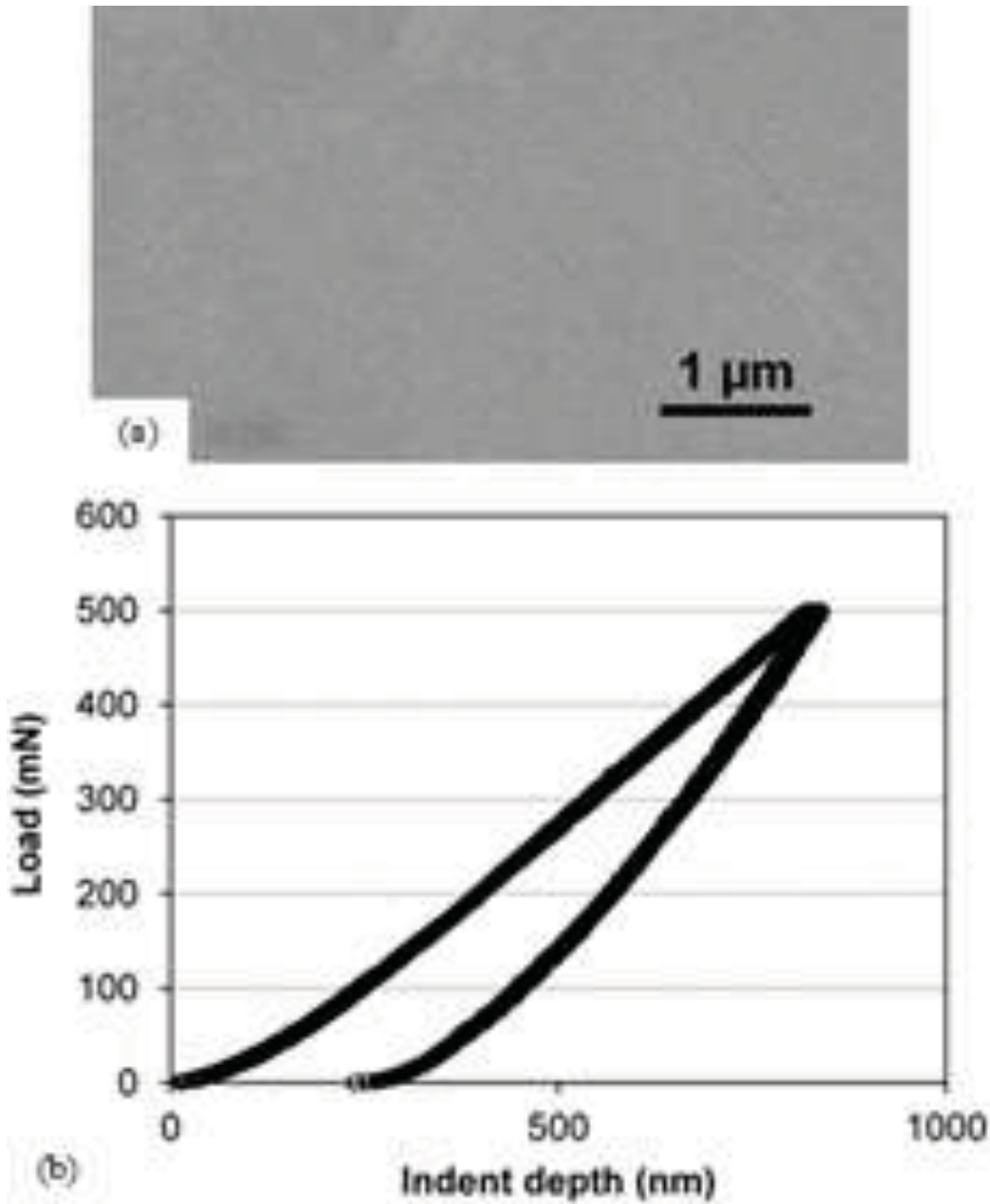

*Fig. 5-7 Spherical nanoindentation on high density LSCF6428 pellet, (a) SEM micrograph of the original top surface, (b) a load-depth curve obtained.*

### 5.1.2.4 Nanoindentation Creep Effect

Depending on the tested materials, creep during nanoindentation may also have varying degrees of influence on the mechanical property measurements, as demonstrated in some reports [23, 24]. For the elastic modulus closest to the true elastic value to be obtained, it is therefore necessary to introduce a sufficient period of dwell time at peak load during nanoindentation to accommodate the influence of the creep effect which led to a negligible influence on the elastic modulus calculation. Too short a dwell time can even lead to bowing in the initial part of the unloading curve due to the time-dependent creep deformation in an extreme case [23, 25], as an example illustrated in Fig. 5-8.

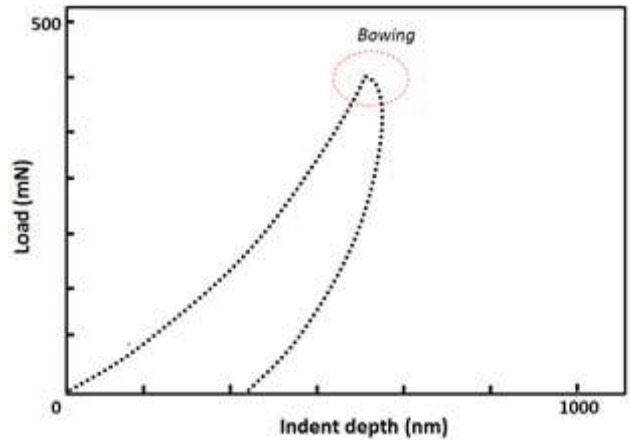

*Fig. 5-8 Example of a bowing event in the initial part of the unloading curve due to creep*

Even when bowing is not evident, creep can still cause overestimation of the sample's elastic modulus calculated using the Oliver-Pharr method, as the slope of the initial portion of the unloading curve (see Fig. 3-5 (b) in Chapter 3) becomes steeper (hence larger stiffness obtained). Chudoba *et al.* [23] found that the apparent elastic modulus of aluminium measured by indentation increased by up to 50% with zero dwell time compared to that with a holding time of 45 seconds or more, in which case a negligible contribution of creep effect was introduced. However, although a much longer dwell time might be favourable to the measurement results, compromise should be made between the productivity of the tests and the precision of the results.

In order to assess the influence of creep on the measured results, namely the elastic modulus and hardness of LSCF6428 bulk samples, spherical nanoindentation tests with varying dwell times ranging from 0 s to 400 s at maximum load applied (500 mN) were performed. Fig. 5-9 only shows the load-displacement curves for dwell times of 5 s, 20 s and 400 s for comparison. The dependence of the calculated elastic modulus and indentation hardness on the varying dwell time is displayed in Fig. 5-10.

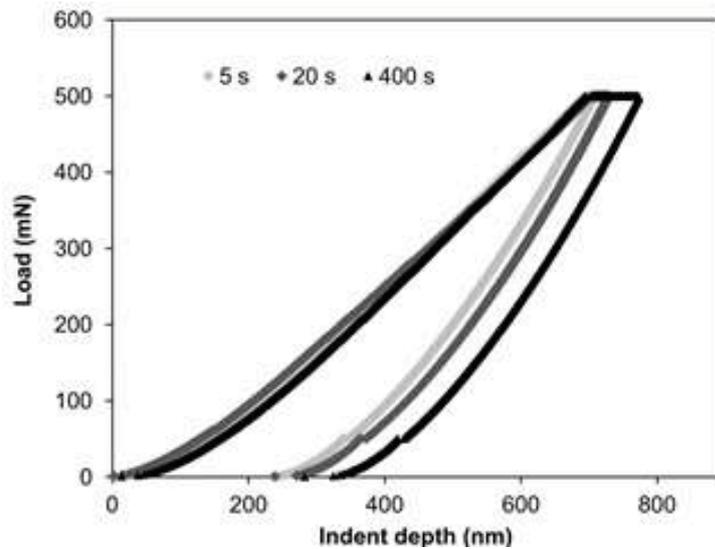

*Fig. 5-9 Load-displacement curves for nanoindentation tests with 5, 20 and 400 s of dwell time at maximum load*

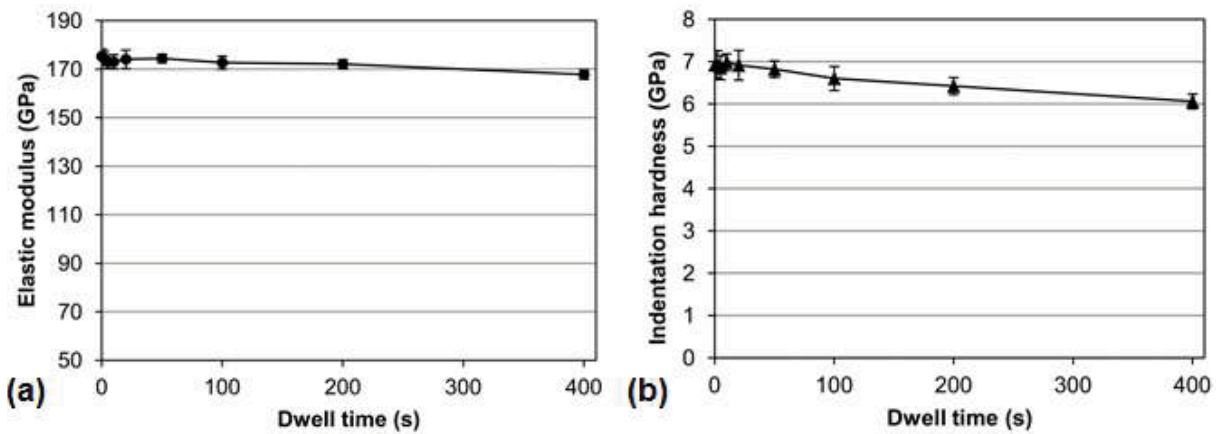

*Fig. 5-10 (a) Elastic modulus and (b) indentation hardness as a function of dwell time for the highest density specimens.*

The result in Fig. 5-10 (a) reveals that the elastic modulus of the sample almost remained constant, with the mean value in the range of 175 GPa to 168 GPa, over the change of dwell time. This was because the stiffness in the initial unloading part of each curve experienced little change (Fig. 5-9). From this perspective, compared to the porosity effect, such a creep event had a minor or negligible contribution towards the calculation of the high elastic modulus measured with nanoindentation. Therefore the LSCF6428 dense sample in the current study could be regarded as having high creep resistance compared to the aforementioned aluminium, which is the opposite. However, indentation hardness was more sensitive to the creep as it dropped from about 7 GPa to 6 GPa over the interval studied. This could be possibly due to the fact that the longer the dwell time lasted the deeper the indent depth would be, so that a larger contact area was reached.

As a result, the overall creep effect on the elastic modulus calculation could be neglected and thus any period of dwell time could be applied to conduct the nanoindentation tests. Nonetheless, it was reasonable to apply a dwell time of 2 s in order that the accuracy of measurements could be reached with minimised creep effect.

### 5.1.2.5 Effect of Residual Porosity

However, there was approximately 5% of residual open porosity in the nominally dense LSCF6428 samples which would affect the elastic modulus measured by macroscopic methods. The effect of this porosity was estimated by applying the model proposed by Ramakrishnan and Arunachalam [26] (denoted hereafter as RA model) using a composite sphere method for solids with randomly distributed isolated pores, namely:

$$E_0 = \frac{E_p(1 + b_E p)}{(1 - p)^2} \hspace{3cm} 5.1$$

In Equation 5.1, $E_p$ is the modulus for porous material having porosity, $p$, and $E_0$ is the modulus of the fully dense material. $b_E$ is a parameter depending on Poisson's ratio $v_0$ of fully dense material, with $b_E = 2 - 3v_0$. The values of $E_p$ measured by the IET method and the corresponding porosities were extrapolated to fully dense material using Equation 5.1 and the results are shown in Table 5-3.

*Table 5-3 Elastic moduli ($E_0$) of high density LSCF6428 extrapolated using Equation 5.1 from the experimental dynamic test results*

| Reference | Porosity (%) | Measured Elastic Modulus $E_p$ (GPa) | Extrapolated Elastic Modulus $E_0$ (GPa) |
|---|---|---|---|
| Kimura *et al.* [2] | 2 | 164 | 175 |
| Chou *et al.* [1] | 4.64±0.22 | 152±3 | 176±3 |
| This work | 5.22±0.01 | 147±3 | 173±3 |

The extrapolated values for fully dense material are all in agreement to within experimental error. Furthermore, the extrapolated results are also in agreement with the indentation results to within experimental error. Therefore we can conclude that the reason for the discrepancy between modulus measured by IET and modulus measured by indentation is residual porosity in the nominally dense bulk specimens. The IET method samples the entire specimen volume and is affected by residual porosity, whereas the indentation method possibly samples smaller volumes. Nevertheless, it should be acknowledged that the elastic region could be much larger than indentation size and therefore than the scale of the microstructure, so that the nanoindentation measurements could also be influenced by the residual porosity. Since the porosity for the current nominally dense sample was quite low, the distribution of pores might be non-uniform. This would possibly result in high values of elastic modulus when indentation sampled regions with less pores present even though the elastic volume could be very large, so that the results were less influenced by the porosity effect. However, the indentation method might be approximate in the present conditions, which might also contribute to the discrepancy measured. In any case, according to the analysis above, the elastic moduli reported by Li *et al.* [4] and Huang *et al.* [6] as described at the beginning of this Chapter would appear to be unreliable.

In fact, available literature also indicates the common existence of a large discrepancy between measured elastic moduli of polycrystalline solids using the IET method and nanoindentation. It is likely that this difference is caused by residual porosities as found in the present study. Rodavic *et al.* [27] investigated and compared various experimental techniques for determination of elastic properties of different solids, including $Al_2O_3$, 7075 aluminum, 4140 steel and Pyrex glass. Nanoindentation moduli were always higher than IET results and a major discrepancy was found for $Al_2O_3$ samples. These had approximately 4.3% porosity and elastic moduli measured using nanoindentation were 10% higher than that measured by IET. Wachtel *et al.* [28] reviewed and compared the elastic moduli of pure and doped ceria using dynamic and static methods including IET and indentation, revealing the static moduli were significantly larger (up to 25%) than the dynamic moduli for samples of the same composition with up to 5% porosity for $Ce_{0.9}Gd_{0.1}O_{1.95}$. To sum up, in the current study, the order of the degree of influence of each factor on elastic modulus of the bulk nominally dense LSCF6428 samples measured using nanoindentation would be: porosity > indentation creep effect > oxygen vacancy gradient / ferroelastic behaviour.

## 5.2 Elastic Modulus and Hardness of Porous Films at RT

As mentioned earlier, nanoindentation measurements of porous films deposited on substrates are much more complicated to analyse than for bulk samples due to more factors affecting the indentation behaviour. Although steps had been taken to minimise roughness, it was still inherent to the processing method of our films. Since these films would be used as-processed in SOFCs, it was imperative to obtain their properties without any surface treatment such as polishing. Moreover, there would be an effect from the substrate as indentation approached the film/substrate interface, which would be expected to affect the experimental data. Consequently the true elastic modulus and hardness of the sample could not be obtained straightforwardly from the nanoindentation data as the measured apparent modulus and hardness varied with the indentation depth, but could be obtained only by removing those effects.

The following Fig. 5-11 and Fig. 5-12 show the nanoindentation data of apparent elastic modulus and hardness as a function of maximum indentation depth relative to film thickness ratio (i.e. $h_{max}/t_f$) of films after sintering at 900-1200 °C. Further data analysis will be discussed afterwards.

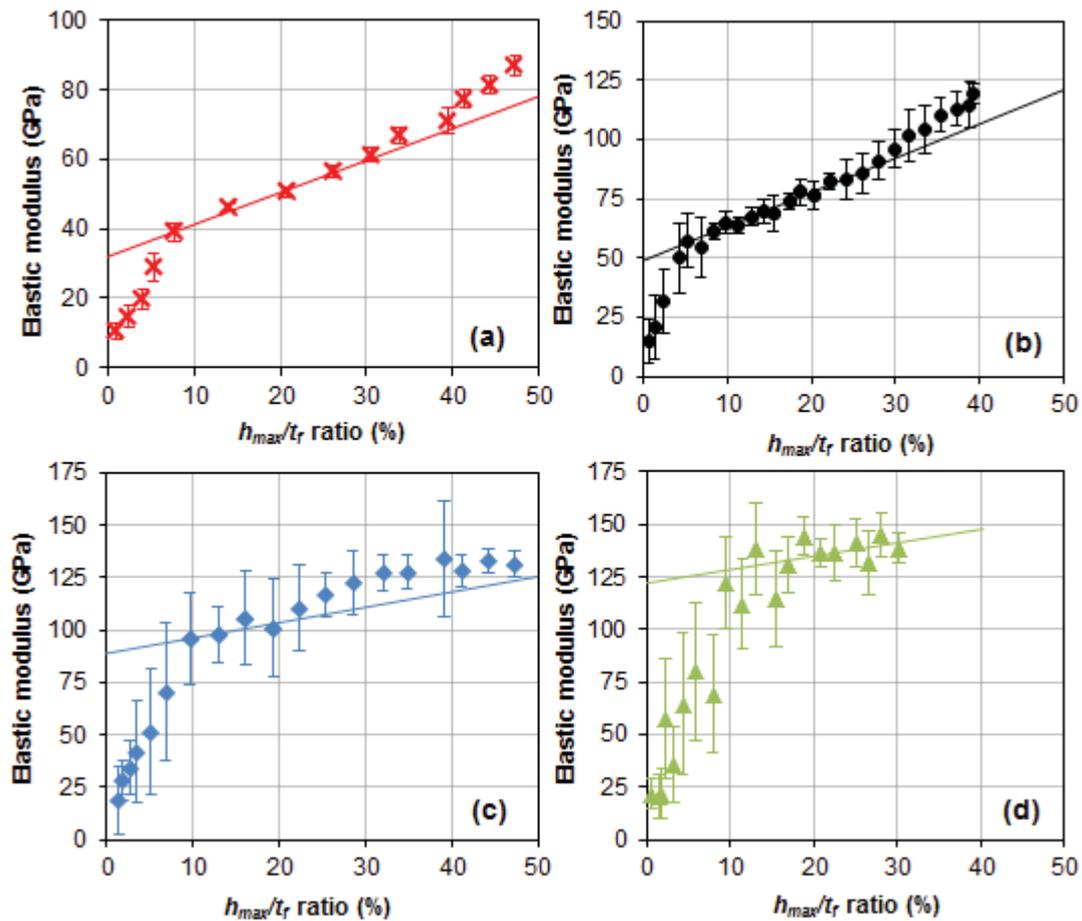

*Fig. 5-11 Apparent elastic modulus vs. $h_{max}/t_f$ ratio for porous films after sintering at (a)-(d): 900, 1000, 1100 and 1200 °C*

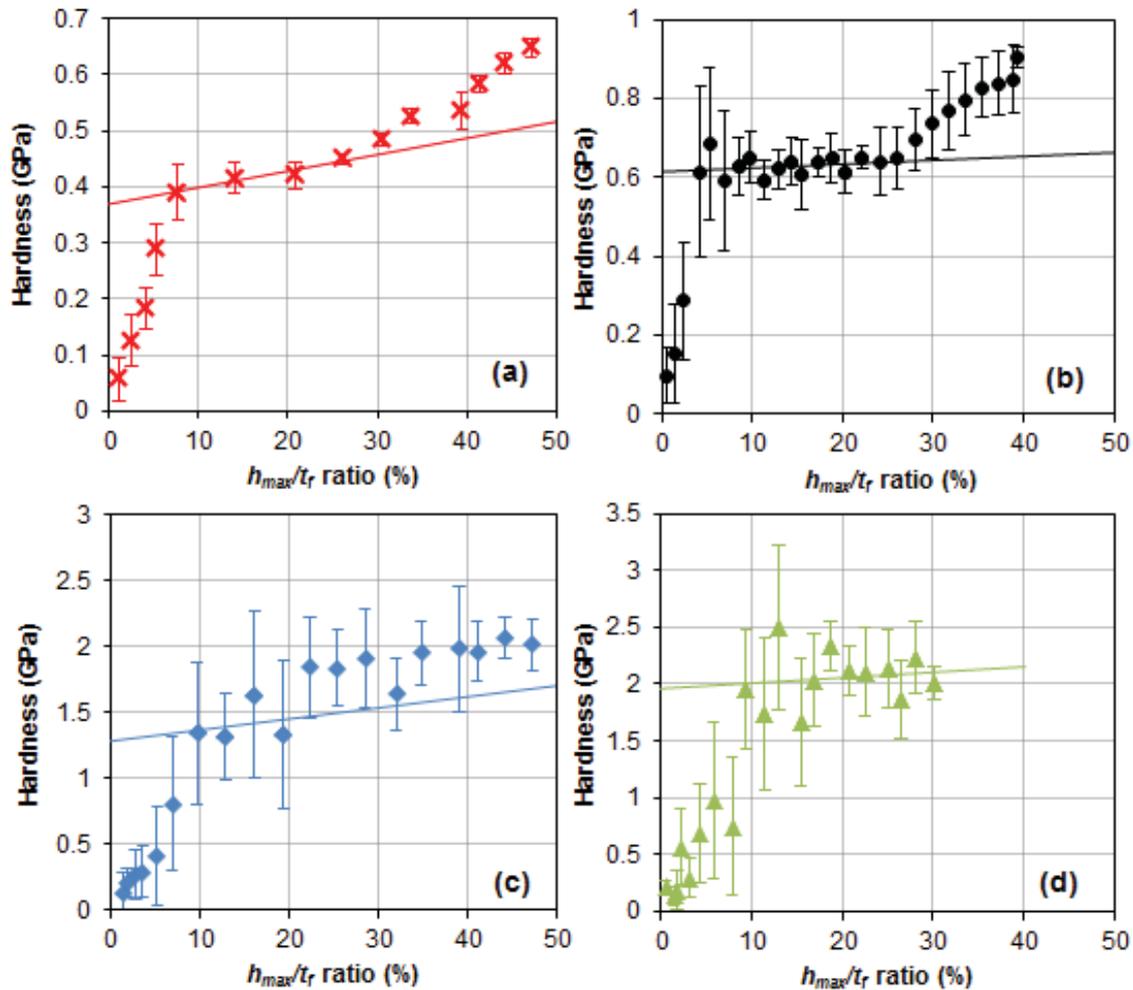

*Fig. 5-12 Apparent hardness vs. $h_{max}/t_f$ ratio for porous films after sintering at (a)-(d): 900, 1000, 1100 and 1200 °C*

The results show a clear dependence on normalised indentation depth for apparent elastic modulus and hardness, which were largely controlled by the combined effects from surface roughness at shallow depth, densification under the indenter and substrate interference at deep penetration.

### 5.2.1 Surface Roughness Effect

It can be seen from the results above that, the scatter for shallow indent depths less than 10% of the film thickness is particularly large. This is due to the surface roughness. Take 1000 °C sintered films for example, the corresponding shallow indentation depths are comparable to the grain size (270 nm) and the surface roughness ($R_a$ = 202 nm) and result in variability in the initial stages on indentation. A cross section through a shallow indent is shown in Fig. 5-13 together with the outline of the indenter tip where it contacts the surface. It can be seen that $h_{max}$ in this case is of similar magnitude to the surface roughness and that the contact is at asperities on the specimen surface. The stresses at these points are much greater than for ideal contact between two smooth surfaces and the displacements are consequently greater.

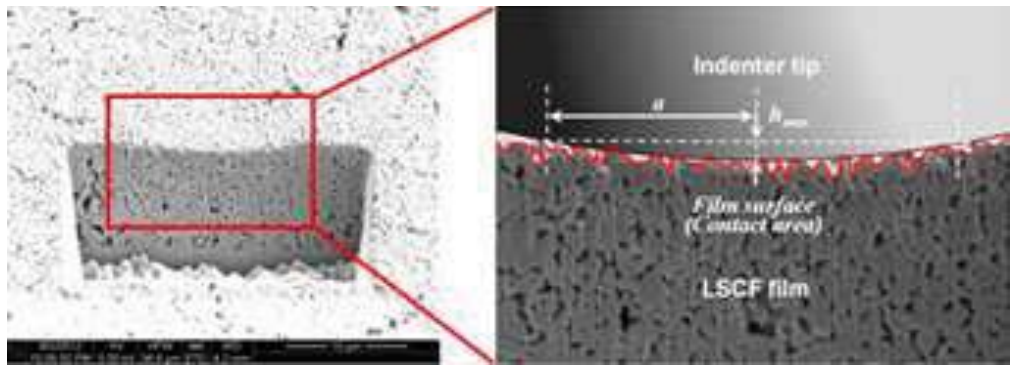

*Fig. 5-13 Effect of surface roughness on the real contact area with the indenter tip, leading to errors in indentation results at shallow penetration depth. Here the sample shown was sintered at 1000 °C*

Due to the facts presented above, surface roughness might have two main effects on the measured results. One was the underestimation of $E$ and $H$ at shallow indentation depth, as shown in Fig. 5-11 and Fig. 5-12. This underestimation was due to the overestimated contact area, provided that the indentation depth was comparable to the roughness, as also reported by other investigations [29, 30]. The measured apparent $E$ and $H$ increased with the indentation depth as the surface roughness effect gradually diminished when an increasing number of the untouched valleys close to the asperities on the surface under compression were touched by the indenter tip. Such an effect can be eliminated in data analysis by simply discarding the data for indentation depths smaller than a critical value (which should be linked to the $R_a$ value), for which the results are less reliable, as suggested by Mencik and Swain [31]. In order to reduce such an effect induced by surface roughness on the measurements, indentation depths larger than a critical value ($h_{cr}$) should be considered. Such an $h_{cr}$ relies on the tested specimen's $R_a$ in and in the current study $h_{cr}$ of each type of specimen was deduced to be approximately 5 times the corresponding $R_a$ value according to the Fig. 5-11 and Fig. 5-12. In other words, the undue underestimation due to the roughness could be avoided by indenting a depth $h_{max} > 5R_a$, corresponding to indentation depths greater than 10% of the film thickness for these LSCF6428 films. Besides, surface roughness could also be responsible for the scatter in the measured data over the whole range of indentation depth as shown in Fig. 5-11 and Fig. 5-12. Nevertheless, it appears that the standard deviations of both the elastic modulus and hardness data for the indentation depth in the range of 0.1-0.2$t_f$ were more consistent and reproducible than that for the other depths. This was particularly true for films sintered at 900 and 1000 °C. While for the films sintered at 1100 and 1200 °C, the errors were much larger, which were attributed to the much coarser surface features after sintering at higher temperature, as large pores exhibited in micrographs Fig. 4-29 (c) and (d).

For many other materials the indentation hardness increases at shallow indentation depths (referred to as the indentation size effect) and is related to constraints on the generation of dislocations within small plastic deformation zones [32, 33]. However, the present results show that for porous materials the opposite is true due to the roughness of the contact at shallow displacements. Nevertheless, it should be noted that the former is a plastic effect while ours is effectively structural fractures.

### 5.2.2 Plastic Deformation under the Indented Surface

The effect of densification upon indentation results could be neglected since it was shown earlier that it had no influence on the elastic modulus measurement for porous bulk samples with the same material composition. Nevertheless, further insights into microstructural changes near the indents in the LSCF6428 films were obtained using FIB machined cross-sections through the indents. Here a film after sintering at 1000 °C is taken as an example. SEM micrographs of its top surface (inset) and cross-section through the middle section of an indent are presented in Fig. 5-14. The spherical tip generates an axisymmetric volume of indented material with its axis at the indent centre and parallel to the direction of indentation. Therefore these images are representative of the whole indented volume. Compressive crushing of the porous structure generated an approximately parabolic "plastic" zone in which the relative density was increased by the crushed debris filling some of the original pore space.

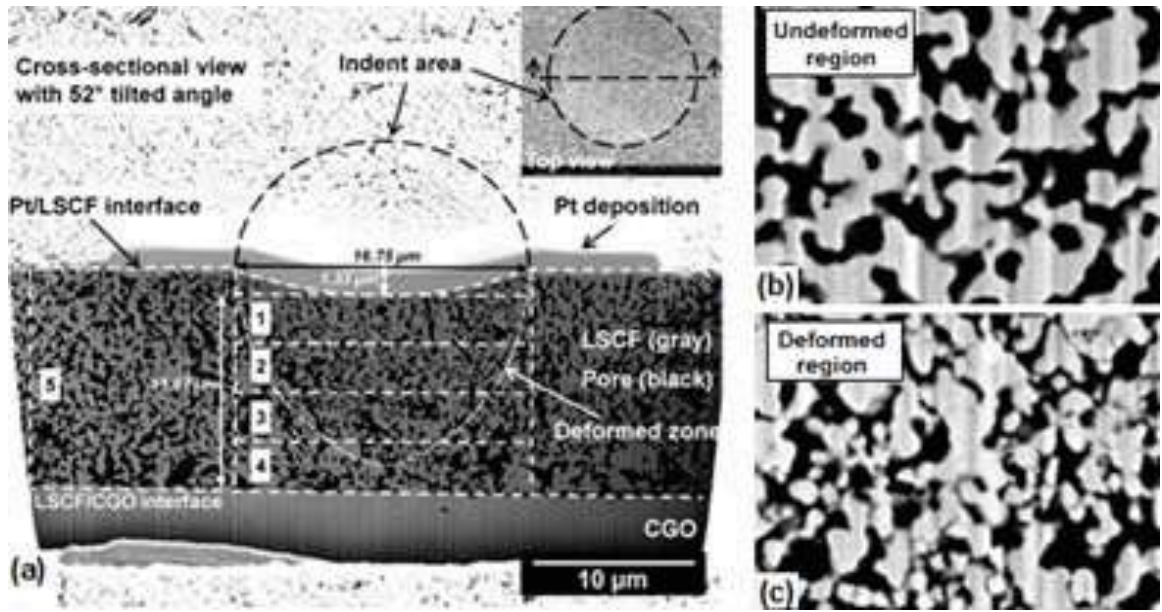

*Fig. 5-14 SEM images of the (a) top surface and cross-section of the indent for a LSCF6428 film sintered at 1000 °C; (b) higher magnification image showing undeformed region and (c) shows deformed region*

Nevertheless, no significant cracks were found outside this zone so that the original microstructure was preserved. Another observation which can be made from the SEM images is that the common pile-up and sink-in effects (as illustrated in Fig. 5-15) of the film material [34] are not found here.

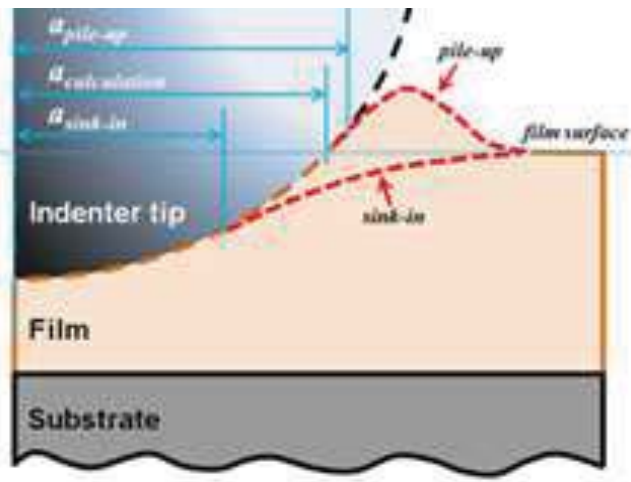

*Fig. 5-15 Schematic of half-space cross-sectional view of the effect of pile-up and sink-in on the actual contact area for nanoindentation on films with a spherical indenter*

Soft films deposited on hard substrates often exhibit pile-up upon indentation due to the change of the films' plastic deformation behaviour attributed to the constraint stress enforced by the relatively harder substrate, whereas in the opposite case (hard film on soft substrate) the films tend to sink-in. As a result, the former case could erroneously increase the apparent mechanical properties calculated using the Oliver-Pharr method because according to the method $H \propto A^{-1}$ and $E \propto A^{-1/2}$ where $A$ is extracted from nanoindentation load-depth data and not from direct contact area measurement, while the piled-up material would accommodate an extra amount of load but causes underestimation of the real contact depth and hence the contact area. On the other hand, the sink-in effect causes overestimation of the real contact area and hence decreases of the calculated mechanical properties. Both of these cases would cause problems in the calculation using the Oliver-Pharr method, which was first established for monolithic materials use only. Therefore, in the absence of the pile-up and sink-in effects, the calculation in this study using the Oliver-Pharr method could generate estimates of $E$ and $H$ free from these effects.

In order to gain a further understanding of the deformation, the deformed area (in 2D) underneath the indenter was divided into four rectangular shapes with identical height and width equivalent to the diameter of the projected circular area of the indent. The area ratio of LSCF6428 material in gray for each rectangle was measured as a matrix "density" using the segmentation module in Avizo software. Fig. 5-16 shows the variation of this densification with distance from the bottom of the indenter tip.

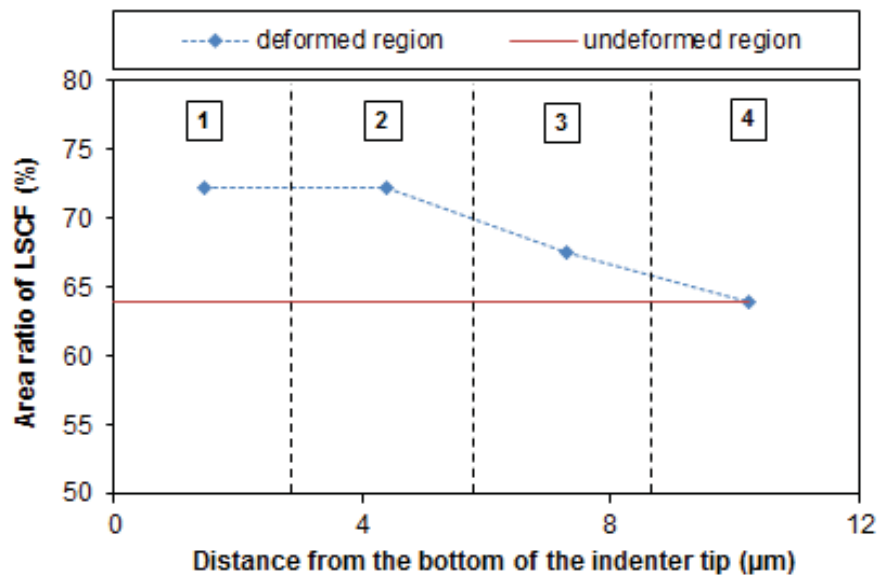

*Fig. 5-16 Densification variation across the film thickness. Note that the numbers in boxes refer to the numbered areas in Fig. 5‑13(a).*

In this specimen the relative density in the undeformed region was measured in area 5 of Fig. 5-14 (a) to be 63.9%. The density gradually decreased from 72.3% in the area close to the indent, to the average value of 63.9% in the area beyond the influence of the indentation.

Thus, in this particular case the "plastic" zone was a crushed region having higher density and extending to a depth of approximately 10 μm.

### *5.2.3 Substrate Effect*

In nanoindentation tests of thin films, one of the key difficulties is to avoid errors in measuring film properties caused by the influence of the substrate. To do so, an appropriate indentation depth ($h_{max}$) and/or an indent load ($P_{max}$) are desired by taking into account the test film's thickness and the nature of the substrate (whether it has higher or lower elastic modulus than the film). Indenting to a depth less than 10% of the thickness of a film (namely $h_{max}/t_f < 0.1$) has been empirically considered as a safe condition to avoid effects from substrate and extract intrinsic film properties in routine nanoindentation tests [35]. This was however not applicable for the current LSCF6428 films as (i) the surface roughness effect took place when indenting less than 10% of the film thickness and (ii) it can be seen in our case that the effect from the substrate was yet much less significant for indentation depth up to $0.2t_f$, as shown in Fig. 5-11 and Fig. 5-12. Moreover, the influence of the substrate was not as marked for films sintered at 1100 and 1200 °C, as the plots stabilised as plateaus after indentation depth increased even beyond $0.2t_f$. This was most probably because the stiffness of these films was closer to that of the substrate.

Fig. 5-17 (a) shows the "plastic" zone (higher density caused by crushing deformation) under an indent performed at a relatively deep penetration ($h_{max}/t_f = 0.25$) in a LSCF6428 film after sintering at 1100 °C reaching as far as the interface with the substrate. Measurements of the local "density" using the method described earlier are presented in Fig. *5-17* (b), which shows that the local density close to the substrate was 79.6%, greater than the density of the material away from the indent (74.4%). This indicates that the substrate had interfered with the progression of the "plastic" zone and, by inference, the elastic zone beyond. Therefore in this particular case the ratio $h_{max}/t_f = 0.25$ was too large for the results not to be influenced by the substrate and, since the substrate possessed a higher modulus (221 GPa) than the film, the measured apparent film modulus $E = 115$ GPa would be larger than the true value. To avoid the substrate effect, a lower penetration of around $0.15t_f$ was applied as shown in Fig. 5-18. In this case the "plastic" zone had less densification and did not reach the substrate. The apparent modulus of 93 GPa obtained is therefore significantly smaller.

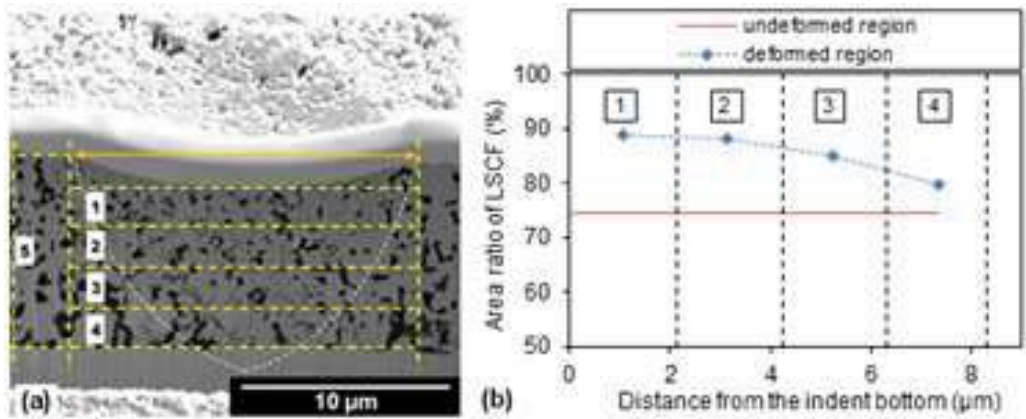

*Fig. 5-17 Indentation of an 1100 °C-sintered film: (a) cross-sectional SEM through a relatively deep indent; (b) variation of solid area ratio. Note that the numbers in boxes refer to the numbered areas in (a).*

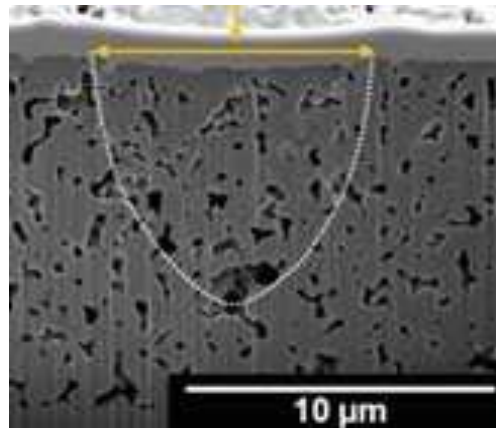

*Fig. 5-18 A shallower indent in the same film as in Fig. 5-17 showing a less pronounced "plastic" zone that did not reach the substrate*

Further evidence of the substrate effect was observed as described below. Fig. 5-19 shows representative SEM images of the indentation imprints and their cross-sections at indentation depths from 800 to 3600 nm in an 1100 °C-sintered film ($t_f \approx 10$ μm) obtained using the FIB/SEM instrument. The imprints for indentation depth lower than 800 nm (i.e. $h_{max}/t_f < 8\%$) could not be identified with SEM because the residual depths were too shallow. To simplify the image acquisition process, no surface protection coating or impregnation was applied to this sample.

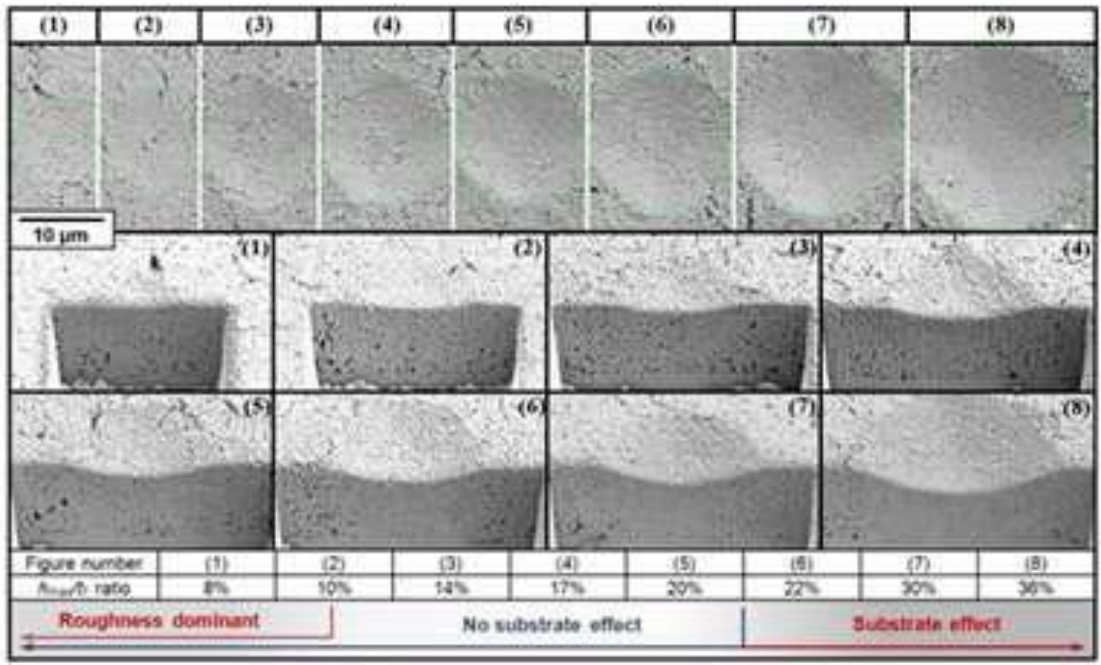

*Fig. 5-19 Top surface and cross-sectional SEM images of indents corresponding to maximum indentation depths from 800 to 3600 nm (i.e. 8% < $h_{max}/t_f$ < 36%) in a film sintered at 1100 °C.*

The results clearly show no delamination of the films from the substrates, indicating good interfacial adhesion. Increasing compaction of the film against the substrate can be found at large indentation depth. In this film a critical indentation depth of approximately 2000 nm ($h_{max}/t_f$ = 20%) can be deduced above which the "plastic" deformation region reached the film/substrate interface and the substrate had a significant effect on the apparent modulus of the film. It should be noted, however, that the elastic region underneath the indented area was even larger and might already exceed the interface.

Saha and Nix [36] studied the substrate effect on nanoindentation for eight different film/substrate systems: Al and W films on aluminium, glass, silicon and sapphire substrates. They concluded that for soft film/hard substrate system the substrate hardness had negligible effect on the film hardness because the plastic deformation was contained within the film and the plastic deformation occurred only when the indenter penetrated the substrate. However, this was not the case in the current study (i.e. structural fracture) which shows significant impact from the substrate for the film hardness once the indentation depth exceeds 20% of the film thickness. Two significant differences might account for this. First, in the present study a spherical (blunt) indenter was used which resulted in a larger deformation volume than a sharp indenter for a given penetration depth. Second, the films in this study were highly porous and the "plastic" zone, formed by crushing, was denser than the original microstructure of the films. Thus the harder densified material was compressed onto the substrate even when the penetration depth only reached 20% of the film thickness and, for greater indentation depths the substrate had an increasing effect on the measured hardness of the film. In the literature, most studies have adopted the approximate guideline of limiting the indentation depth to less than one tenth of the film thickness in order to obtain reliable values for the elastic modulus of the film [35]. In the present study we have shown that, for the LSCF6428 films investigated here, indentation depths of up to 20% of the film thickness could be used before the influence of the substrate became much noticeable, particularly for the calculation in the case of lower specimen sintering temperatures. It should be emphasised that although the elastic region under the plastic deformation region cannot be readily seen and its size was not easy to measure, the effect of substrate on it had primarily link to the elastic modulus measured. As a result the substrate effect could be further minimised by increasingly reducing the indentation depth. However, complete elimination of surface roughness effect or substrate effect could never be possible and for such thin porous films a balance between deep penetrations to reduce surface roughness effect and shallow penetrations to reduce substrate effect should be achieved.

### 5.2.4 Elastic Modulus and Hardness Estimations

It can be concluded from the analyses above that indenting shallower than $h_{cr} = 5R_a$ (which is converted to be $0.1t_f$ in our films) the results were unduly influenced by surface roughness and above $0.2t_f$ they were increasingly influenced by the substrate. Therefore, the most reliable data points for the films to be extrapolated to estimate true properties in this study are those with the ratio $h_{max}/t_f$ in the range 0.1 to 0.2, where much less significant influence from surface roughness and substrate were introduced. Consequently, extrapolations were carried out to zero $h_{max}/t_f$ by ignoring these most affected data ranges, as shown in Fig. 5-11 and Fig. 5-12, particularly for the films sintered at lower temperatures (i.e. 900 and 1000 °C). The estimated results for true film values were obtained as listed in Table 5-4. It is evident from the table that both the elastic modulus and hardness increased dramatically with sintering temperature as further densification took place at higher temperatures. For example, the elastic modulus rose nearly four-fold and the hardness rose over six-fold across the whole range of sintering temperatures.

*Table 5-4 Estimated true values of elastic modulus and hardness for LSCF6428 films measured by nanoindentation*

| Sintering temperature (°C) | Elastic modulus (GPa) | Hardness (GPa) |
|:---:|:---:|:---:|
| 900 | 32.4±1.2 | 0.37±0.08 |
| 1000 | 48.3±4.6 | 0.61±0.11 |
| 1100 | 90±6.4 | 1.28±0.14 |
| 1200 | 121.5±7.2 | 1.97±0.20 |

### 5.2.5 Elastic Modulus and Hardness vs. Porosity for LSCF6428 Films and Bulk Samples

Fig. 5-20 (a) displays the relationship between the elastic modulus and porosity results obtained from aforementioned analysis, for both porous films and bulk samples. Note that in this figure the elastic modulus for the 95% dense bulk sample (i.e. 174 GPa) was plotted at zero porosity because as revealed by the previous nanoindentation analysis, this was a local measurement of fully dense material (while the IET modulus should be at 5% porosity, which is also shown in the figure).

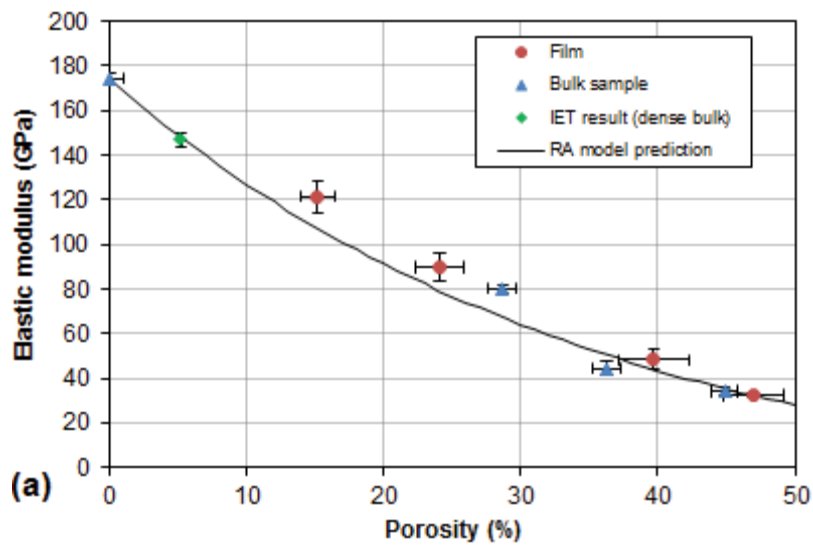

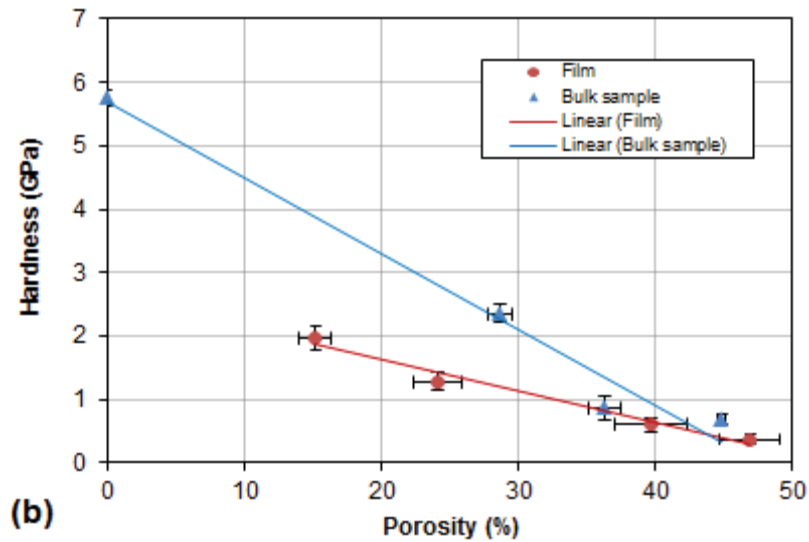

*Fig. 5-20 Comparison of (a) elastic modulus and (b) hardness vs. porosity between sintered LSCF6428 films and bulk samples*

Although there could be some other factors controlling the elastic modulus besides porosity (as discussed in the next chapters) and the relationship of elastic modulus and porosity might not be necessarily exponential, the consistency between the two datasets shown in Fig. 5-20 (a) clearly implies that the porosity almost played an identical role in LSCF6428 films and bulk samples regarding its effect on the sample's elastic modulus, particularly for samples of high porosity. It should be noted that for a film sintered on a substrate, the microstructure produced by constrained sintering could differ from that of an as-sintered bulk sample for a given porosity (e.g. by being anisotropic, or having a different ratio of pore size to grain size) and this might account for the small difference between the films and bulk samples (i.e. slightly higher moduli for films). Nevertheless, the effect of porosity on the elastic modulus of both films and bulk can be regarded as being similar. Fig. 5-20 (a) also compares the plot (curve in black) of the prediction of elastic modulus as a function of porosity using the RA model as explained by Equation 5.1, with $E_0$ and $v_0$ chosen to be 174 GPa and 0.3, respectively. It is found that this model fit very well with the measured data for both films and bulk samples.

On the other hand, as shown in Fig. 5-20 (b), hardness appears to be linearly related to the porosity for both types of samples. However, the validity of such remains highly questionable due to the limited number of experimental data points, although the hardness does increase with rising relative density as normally behave in other ceramic materials [37, 38]. Moreover, the different slopes for these two types of samples imply that a universal empirical function to describe the hardness vs. porosity relationship of both samples is unlikely, which deserves further work because of the complicated permanent plastic deformation mechanism in the porous samples. This also suggests that the applicability of the empirical single variable relationships found in the literature for relating hardness and porosity [39, 40], is very limited and material dependent and thus invalid for our specimens. Although the interrelationship of the mechanical properties for both types of samples largely stemmed from their porosities, the distinguishable behaviours of plots shown in Fig. 5-21 for elastic moduli versus hardnesses also suggest that apart from porosity, there should be other factors controlling the mechanical properties, such as particle neck size, which require further studies.

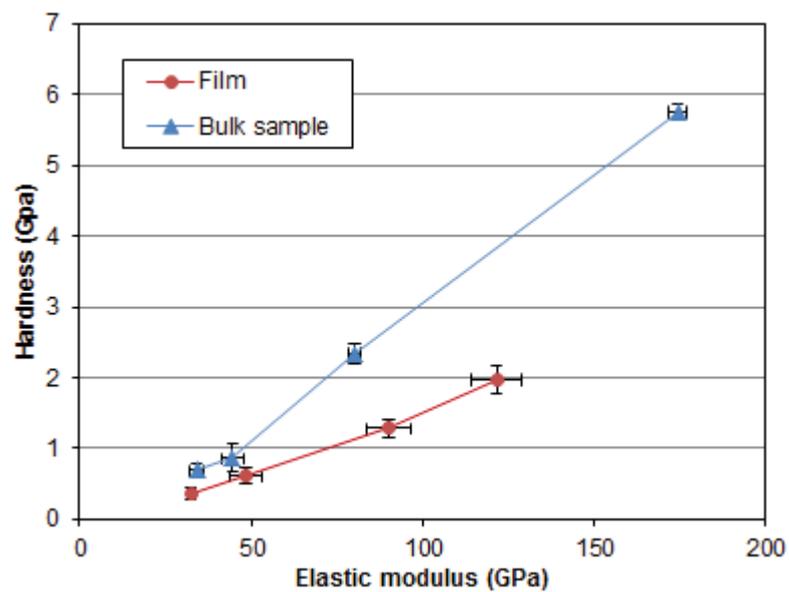

*Fig. 5-21 Linear relationship between elastic moduli and hardnesses of bulk and films, implying that the interrelated mechanical properties may have several influencing factors.*

## *Summary*

In this Chapter, the measurements of important mechanical properties including elastic modulus and hardness of LSCF6428 in both bulk and film forms were investigated after sintering at 900-1200 °C.

For elastic modulus and hardness measurements, a spherical indenter was used since the contact area was then much greater than the scale of the porous microstructure. The elastic modulus of the bulk samples was found to increase from 34-174 GPa and hardness from 0.64-5.3 GPa as the porosity decreased from 45-5% after sintering at 900-1200 °C. The results were found to be sensitive to surface roughness at shallow indentation depth, while stable values were obtained at larger depths. Densification under the indenter was found to have little influence on the measured elastic modulus. However, the elastic modulus measured by indentation of "dense" bulk specimens sintered at 1200 °C, 174 GPa, was significantly greater than that measured by impulse excitation, 147 GPa. This was shown to possibly be due to residual porosity of approximately 5% in the nominally dense specimens which influenced the long range elastic modulus measured by impulse excitation. Therefore the higher value is characteristic of fully dense material. No evidence of a ferroelastic contribution to the load-deflection indentation response was found.

Crack-free LSCF6428 films of acceptable surface roughness for indentation were also prepared by sintering at 900-1200 °C. The porosities of the films were in the range 15-47%. The influence of surface roughness was due to the granular nature of the porous films, while the influence of the substrate was due to formation of a "plastic" zone of crushed, higher density, material under the indent which touched the substrate if the indentation was too deep. The experiments in this study showed that for this type of porous film and using a spherical indenter, relatively reliable measurements of the true properties of the films were obtained by data extrapolation provided that the indentation depth ($h_{max}$) fell in the range of $5R_a < h_{max} < 0.2t_f$, which in the current case corresponded to the ratio of indentation depth to film thickness being $0.1 < h_{max}/t_f < 0.2$, acknowledging the fact that the substrate effect persisted but at a lower level in this depth range. Thus it was found that the elastic modulus of the films increased from 32 to 121 GPa and hardness from 0.37 to 1.97 GPa as the sintering temperature increased from 900 to 1200 °C. Comparison with bulk specimens clearly showed that the porous films behaved very similarly to the porous bulk specimens in terms of the dependency of elastic modulus on porosity. Microstructures obtained by FIB/SEM slice and view of the film specimens, after indentation revealed the nature of the "plastic" deformation zone and how this affected the measurement of elastic modulus when it reached the substrate.

## Chapter 5 References

1.  Chou Y-S, Stevenson JW, TArmstrong TR, LPederson LR: **Mechanical Properties of La1-xSrxCo0.2Fe0.8O3-δ Mixed-Conducting Perovskites Made by the Combustion Synthesis Technique**. *Journal of American Ceramic Society* 2000, **83**(6):1457-1464.
2.  Kimura Y, Kushi T, Hashimoto S-i, Amezawa K, Kawada T: **Influences of Temperature and Oxygen Partial Pressure on Mechanical Properties of La0.6Sr0.4Co1−yFeyO3−δ**. *Journal of the American Ceramic Society* 2012, **95**(8):2608-2613.
3.  Vlassak JJ, Nix WD: **Measuring the elastic properties of anisotropic materials by means of indentation experiments**. *Journal of the Mechanics and Physics of Solids* 1994, **42**(8):1223-1245.
4.  Li N, Verma A, Singh P, Kim J-H: **Characterization of La0.58Sr0.4Co0.2Fe0.8O3−δ–Ce0.8Gd0.2O2 composite cathode for intermediate temperature solid oxide fuel cells**. *Ceramics International* 2013, **39**(1):529-538.
5.  Huang BX, Malzbender J, Steinbrech RW, Singheiser L: **Mechanical properties of La0.58Sr0.4Co0.2Fe0.8O3-δ membranes**. *Solid State Ionics* 2009, **180**(2-3):241-245.
6.  Huang B, Chanda A, Steinbrech R, Malzbender J: **Indentation strength method to determine the fracture toughness of La0.58Sr0.4Co0.2Fe0.8O3-delta and Ba0.5Sr0.5Co0.8Fe0.2O3-delta**. *Journal of Materials Science* 2012, **47**(6):2695-2699.
7.  Tai LW, et al.: **Structure and Electrical Properties of La1-xSrxCo1-yFeyO3. Part 2: The System La1-xSrxCo0.2Fe0.8O3**. *Solid State Ionics* 1995(76):273-283.
8.  Wang S, Katsuki M, Dokiya M, Hashimoto T: **High temperature properties of La0.6Sr0.4Co0.8Fe0.2O3-δ phase structure and electrical conductivity**. *Solid State Ionics* 2003, **159**(1-2):71-78.
9.  Aizu K: **Possible Species of "Ferroelastic" Crystals and of Simultaneously Ferroelectric and Ferroelastic Crystals**. *Journal of the Physical Society of Japan* 1969, **27**(2):387-396.
10. Abrahams SC, Barns RL, Bernstein JL: **Ferroelastic effect in lanthanum orthoferrite**. *Solid State Communications* 1972, **10**(4):379-381.
11. Kim CH, Jang JW, Cho SY, Kim IT, Hong KS: **Ferroelastic twins in LaAlO3 polycrystals**. *Physica B: Condensed Matter* 1999, **262**(3–4):438-443.
12. Déchamps M, de Leon Guevara AM, Pinsard L, Revcolevschi A: **Twinned microstructure of La1-x SrxMnO3 solid solutions**. *Philosophical Magazine A* 2000, **80**(1):119-127.
13. Kleveland K, Orlovskaya N, Grande T, Moe AMM, Einarsrud M-A, Breder K, Gogotsi G: **Ferroelastic Behavior of LaCoO3-Based Ceramics**. *Journal of the American Ceramic Society* 2001, **84**(9):2029-2033.
14. Lein HL, Andersen ØS, Vullum PE, Lara-Curzio E, Holmestad R, Einarsrud M-A, Grande T: **Mechanical properties of mixed conducting La0.5Sr0.5Fe1−xCoxO3−δ (0≤x≤1) materials**. *Journal of Solid State Electrochemistry* 2006, **10**(8):635-642.
15. Huang BX, Malzbender J, Steinbrech RW, Wessel E, Penkalla HJ, Singheiser L: **Mechanical aspects of ferro-elastic behavior and phase composition of La0.58Sr0.4Co0.2Fe0.8O3−δ**. *Journal of Membrane Science* 2010, **349**(1-2):183-188.

16. Araki W, Malzbender J: **Ferroelastic deformation of La0.58Sr0.4Co0.2Fe0.8O3−δ under uniaxial compressive loading**. *Journal of the European Ceramic Society* 2013, **33**(4):805-812.

17. Orlovskaya N, Gogotsi Y, Reece M, Cheng B, Ion G: **Ferroelasticity and hysteresis in LaCoO3 based perovskites**. *Acta Materialia* 2002, **50**(4):715-723.

18. Leist T, Webber KG, Jo W, Granzow T, Aulbach E, Suffner J, Rodel J: **Domain switching energies: Mechanical versus electrical loading in La-doped bismuth ferrite–lead titanate**. *Journal of Applied Physics* 2011, **109**(5):054109-054109-054109.

19. Hilpert K, Steinbrech RW, Boroomand F, Wessel E, Meschke F, Zuev A, Teller O, Nickel H, Singheiser L: **Defect formation and mechanical stability of perovskites based on LaCrO3 for solid oxide fuel cells (SOFC)**. *Journal of the European Ceramic Society* 2003, **23**(16):3009-3020.

20. Lee S: **Mechanical properties and structural stability of perovskite type oxygen-permeable, dense membranes**. *Desalination* 2006, **193**:236-243.

21. Kuhn M, Hashimoto S, Sato K, Yashiro K, Mizusaki J: **Oxygen nonstoichiometry, thermo-chemical stability and lattice expansion of La0.6Sr0.4FeO3-δ**. *Solid State Ionics* 2011, **195**(1):7-15.

22. Atkinson A: **Chemically-induced stresses in ceramic oxygen ion-conducting membranes.pdf**. *Solid State Ionics* 2000, **129**:259-269.

23. Chudoba T, Richter F: **Investigation of creep behaviour under load during indentation experiments and its influence on hardness and modulus results**. *Surface and Coatings Technology* 2001, **148**(2–3):191-198.

24. Fischer-Cripps AC: **A simple phenomenological approach to nanoindentation creep**. *Materials Science and Engineering: A* 2004, **385**(1–2):74-82.

25. Feng G, Ngan A: **Effects of creep and thermal drift on modulus measurement using depth-sensing indentation**. *Journal of Materials Research* 2002, **17**(03):660-668.

26. Ramakrishnan N, Arunachalam VS: **Effective elastic moduli of porous solids**. *Journal of Materials Science* 1990, **25**(9):3930-3937.

27. Radovic M, Lara-Curzio E, Riester L: **Comparison of different experimental techniques for determination of elastic properties of solids**. *Materials Science and Engineering: A* 2004, **368**(1–2):56-70.

28. Wachtel E, Lubomirsky I: **The elastic modulus of pure and doped ceria**. *Scripta Materialia* 2011, **65**(2):112-117.

29. Rico A, Garrido MA, Otero E, Rodriguez J: **Roughness Effect on the Mechanical Properties of Ceramic Materials Measured from Nanoindentation Tests**. *Key Engineering Materials* 2007, **333**:247-250.

30. Walter C, Mitterer C: **3D versus 2D finite element simulation of the effect of surface roughness on nanoindentation of hard coatings**. *Surface and Coatings Technology* 2009, **203**(20):3286-3290.

31. Mencik J, Swain M: **Errors associated with depth-sensing microindentation tests**. *Journal of Materials Research* 1995, **10**(6):1491-1501.

32. Fischer-Cripps AC, : **Nanoindentation**, 3rd edn. New York: Springer; 2011.

33. Gao H, Huang Y, Nix WD, Hutchinson JW: **Mechanism-based strain gradient plasticity— I. Theory**. *Journal of the Mechanics and Physics of Solids* 1999, **47**(6):1239-1263.

34. Tsui TY, Vlassak J, Nix WD: **Indentation plastic displacement field: Part I. The case of soft films on hard substrates**. *Journal of Materials Research* 1999, **14**(06):2196-2203.

35.   Oliver WC, Pharr GM: **An improved technique for determining hardness and elastic modulus using load and displacement sensing indentation experiments**. *Journal of Materials Research* 1992, **7**(6):1564-1583.

36.   Saha R, Nix WD: **Effects of the substrate on the determination of thin film mechanical properties by nanoindentation**. *Acta Materialia* 2002, **50**:23-38.

37.   Kim SW, Khalil KA-R: **High-Frequency Induction Heat Sintering of Mechanically Alloyed Alumina–Yttria-Stabilized Zirconia Nano-Bioceramics**. *Journal of the American Ceramic Society* 2006, **89**(4):1280-1285.

38.   Chen D, Jordan EH, Gell M, Ma X: **Dense Alumina–Zirconia Coatings Using the Solution Precursor Plasma Spray Process**. *Journal of the American Ceramic Society* 2008, **91**(2):359-365.

39.   Tricoteaux A, Rguiti E, Chicot D, Boilet L, Descamps M, Leriche A, Lesage J: **Influence of porosity on the mechanical properties of microporous β-TCP bioceramics by usual and instrumented Vickers microindentation**. *Journal of the European Ceramic Society* 2011, **31**(8):1361-1369.

40.   Luo J, Stevens R: **Porosity-dependence of elastic moduli and hardness of 3Y-TZP ceramics**. *Ceramics International* 1999, **25**(3):281-286.

# 6 Characterisation of Mechanical Properties: Fracture Toughness

This Chapter shows the measured results and analyses of room-temperature mode I fracture toughness ($K_{Ic}$) of the tested specimens using both Berkovich indentation and an adapted *in situ* DCB testing method. The resulting fracture and crack morphologies were investigated by the SEM.

## 6.1 Indentation Result for Bulk Samples

### 6.1.1 Microstructural Observation

As explained in Chapter 2, accurate fracture toughness estimations require well-defined cracks which should not have sizes comparable to grain sizes. In the current study, as found earlier according to the surface micrograph observations, the average grain sizes for the bulk samples were estimated to be approximately 0.5, 0.8, 1.0 and 1.5 µm, much smaller than the corresponding crack sizes shown later, with decreasing porosities of 44.9%, 36.3%, 28.7% and 5.22%, respectively after sintering at 900-1200 °C.

Although obvious pop-in events did not occur in the recorded indentation load-displacement curves, the indentation-induced cracks were observable in SEM micrographs of the bulk pellets sintered at 900-1200 °C. Fig. 6-1 shows clear cracks in a nominally dense pellet (porosity=5.22%) sintered at 1200 °C and indented at 5 N load.

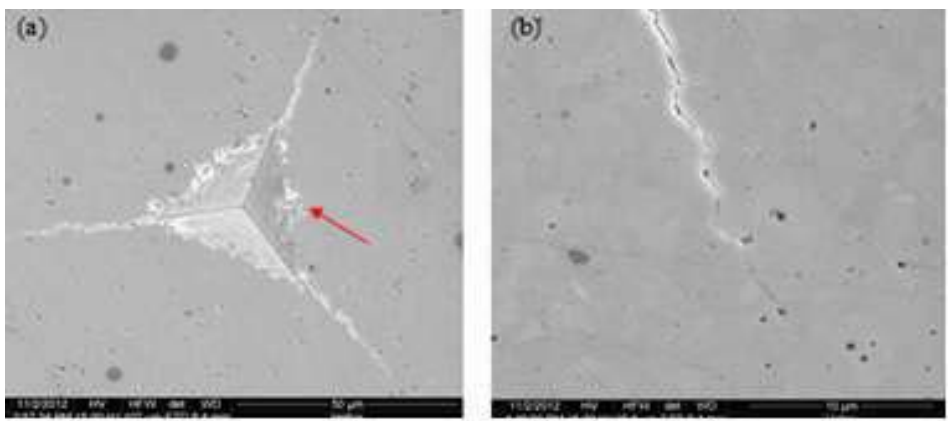

*Fig. 6-1 SEM micrographs of a pellet surface sintered at 1200 °C after indentation at 5 N: (a) top-surface showing radial star-shaped cracks around the Berkovich indented area, (b) a crack tip showing intergranular crack propagation*

It can be seen from Fig. 6-1 (a) that the brittle cracks induced by indentation of a Berkovich tip emanated linearly on the sample surface along the radial diagonals from the three corners of the indented area. The arithmetic mean of the total crack length ($c$) measured from the indent central apex was approximately 50 µm and the indent half diagonal was 20 µm. Some granular chippings on the periphery of the pyramidal indented area can be found as indicated by a red arrow. A higher magnification SEM image in Fig. *6-1* (b) further shows one of the crack tips with distinguishable grain boundary cracks, suggesting that the induced cracks were propagating intergranularly with the surrounding grain size being approximately 1.5 µm. The average total crack length was estimated to be 57±4 µm and half diagonal 22±2 µm, resulting in a *c/a* ratio equal to 2.6.

While conventional studies have investigated cracks underneath the indented area using a bonded interface technique [1] or by simply polishing the cross-sections [2], of which the main drawback would be either change of stress state or damage of materials, the much more straightforward FIB/SEM slice and view technique was used in the current study to have a close insight of the underlying deformation features. Three locations across one of the radial diagonals of the residual imprint were chosen for sequential FIB milling and SEM imaging, as presented in Fig. 6-2. Note that he stripes on the indented area in Fig. 6-2(b) were due to the ion-beam scanning.

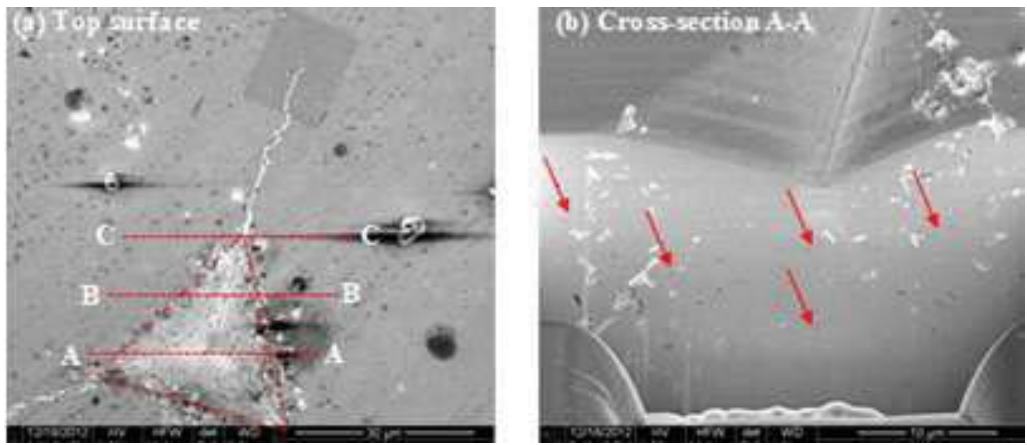

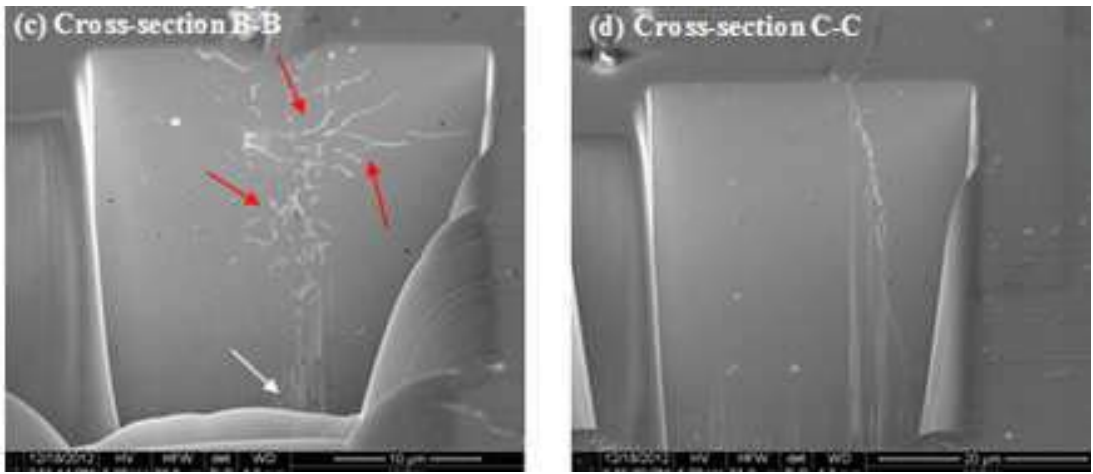

*Fig. 6-2 FIB-sliced cross-sectional features of the deformed locations under the indented area of a pellet sintered at 1200 °C and indented at 5 N: (a) top surface showing the three chosen location for cross-sectioning, (b) A-A cross-section through the residual imprint centre, showing a homogeneous distribution of flaws/microcracks (c) B-B cross-section through the half way of a diagonal edge, showing more concentrated interconnected flaws/microcracks, (d) C-C cross-section through the residual imprint corner, showing a well-developed crack.*

Several interesting features can be identified in the above SEM figures. First of all, obvious cracking is not likely to be observed immediately below the contact area of the residual imprint. Rather, very different from the so-called "rosette" shear flow lines that can evolve into lateral cracks and have large opening found in soda-lime glasses after indentation reported by Hagan and Swain [3], and different from the cone cracks induced by Berkovich tips found by Pharr *et al.*[4], here a great number of small segments of interconnected microcracks of submicron opening size are found confining to the plastic deformation zone with wider and more homogeneous distribution towards the area under the apex, as shown by arrows in Fig. 6-2 (b). The presence of such microcracks indicates the existence of residual stress field produced by internal strains due to non-uniform plastic deformation under the contact area [5]. In addition, the orientation dependence of these flaws/microcracks is noticeable as they were preferentially developed to be parallel to the sample surface. The reason for this appears to be the predominant subsurface tensile residual stress orthogonal to the surface plane induced by recovery of the elastically deformed material upon unloading of the indenter tip. It is important to emphasise that the sectioned C-C face in Fig. 6-2 (d) confirms the immediate emanation of the well-developed radial crack (vertical penetration and horizontal propagation) from the corner of the residual imprint. The cracks penetrated under the sample surface to a depth of several times the indentation depth without extending into or beneath the deformed zone for joining, although in Fig. 6-2 (c) a segment of crack pointed by white arrow under the plastic zone can be clearly seen. Based on the above observations, the crack shape induced by Berkovich indentation here is the Palmqvist radial crack type [6, 7] (see Fig. 2-3 (a)), as the cracks generated from each extremity of the corner are independent, rather than joined to form the median or half-penny shaped cracks beneath the deformed zone [8, 9], as shown in Fig. 2-3 (b). Nevertheless, the development of Palmqvist cracks into median cracks remains possible once a critical load is surpassed [10, 11]. Note that the morphologies found in the present study with the FIB-milled cross-sections are adequate to be representative of the features in three-dimension.

The initiation of cracks can only start from flaws which are either pre-existing or induced by indentation itself, as claimed by Lawn *et al.* [12]. For the current study, in the absence of any detectable pre-existing cracks in the bulk LSCF6428 samples as described in previous microstructural analysis, the above observations further confirm that the accumulated interconnected flaws induced by the indentation process at the deformed zone boundary (i.e. plastic-elastic boundary) immediately below the imprint corners where stress concentration took place gave rise to the radial and Palmqvist cracks. They emanated from the corners during the residual stress relaxation upon unloading [7], as can be speculated from Fig. 6-2 (b) to (d). These cracks were mainly nucleated and propagated by the residual tensile unloading stresses, which were generated due to the material being unable to recover fully elastically as a result of the presence of the plastic deformation, as suggested in [3, 13].

A quantitative study of the crack morphology evolution is out of the scope of the current study, since it requires a detailed knowledge of the elastic-plastic contact stress fields, which could be hard to deduce. Some complicated models have been proposed [8, 14-16], for axisymmetric indentation, but are not strictly applicable to an asymmetrical Berkovich indenter.

Fig. 6-3 shows the residual indent and a crack on a pellet sintered at 900 °C, after indentation at 20 N. It appears that the crack morphology is far more difficult to identify because the surface is quite different from that of a highly dense pellet's smooth surface. Hence the highly porous microstructure is not favourable for crack identification because even at a relatively low magnification the crack features might be disguised, as shown in Fig. 6-3 (a). This is possibly due to the fact that the porosity of the pellet sintered at 900 °C is close to 45%, with a small average grain size of 0.5 μm, which is comparable to the pore size and larger than the crack opening size.

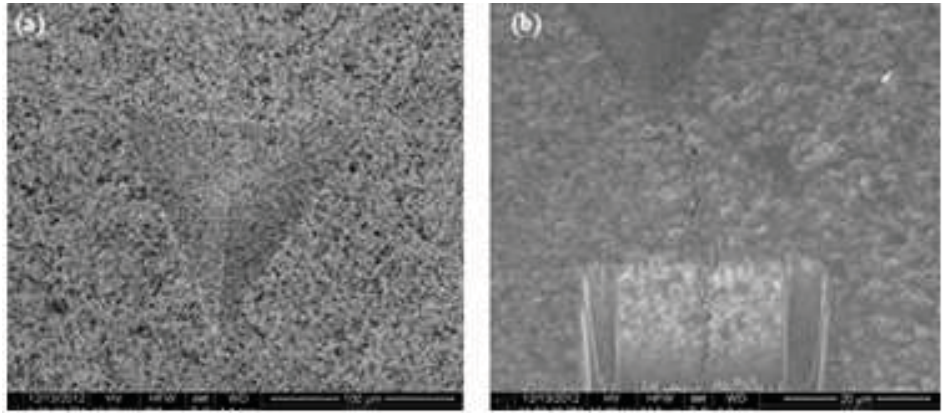

*Fig. 6-3 Crack morphologies of a pellet sintered at 900 °C: (a) the residual indent, (b) cross-sectional face of the crack.*

However, with careful observation at higher magnifications, the average total crack length was measured to be approximately 99±5 μm, with indent diagonal length of 50±4 μm, with a *c/a* ratio of 2.0, based on the observation of 20 indents. The observation at high magnification also indicates that the crack was intergranular due to low particle bonding strength.

Fig. 6-4 shows the SEM photos of the indentation-induced crack morphologies after microindentation at 2 N for a pellet sintered at 1000 °C. FIB sequential slicing was also used in three locations of the residual indent (Fig. 6-4 (a)) for closer insight of the microstructures under the cracks and the indent. Penetrating cracks are indicated by red arrows.

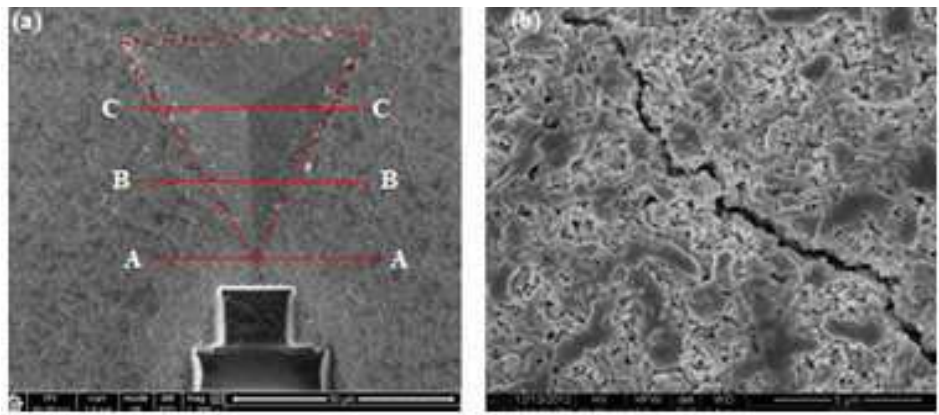

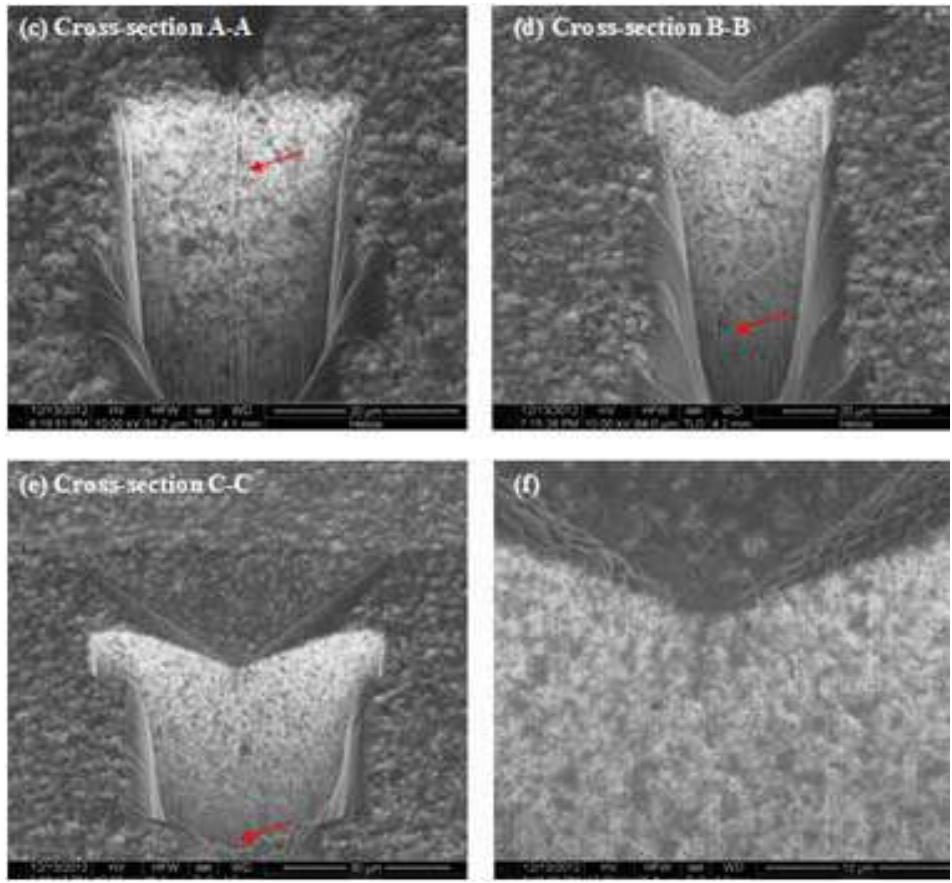

*Fig. 6-4 Micrographs of the crack after indentation at 2 N of a pellet sintered at 1000 °C: (a) residual indent surface,(b) high magnification image showing intergranular cracking, (c-e) A-A, B-B, C-C cross-sectional micrographs, (f) a closer view of C-C section.*

Apart from the similar intergranular cracks as found in the 1200 °C as-sintered pellet shown earlier, it is interesting to notice in Fig. 6-4 (e) the appearance of a vertical crack, as arrowed, on the bottom of the sectioned C-C face through the residual indent apex. This existence indicates the possible joining of the Palmqvist cracks underneath the deformed region for the porous pellet (36.3% porosity), for which the critical load was small and easy to surpass during indentation, resulting in the development of half-penny cracks from Palmqvist cracks by joining under the deformed region, as demonstrated in Fig. 6-5.

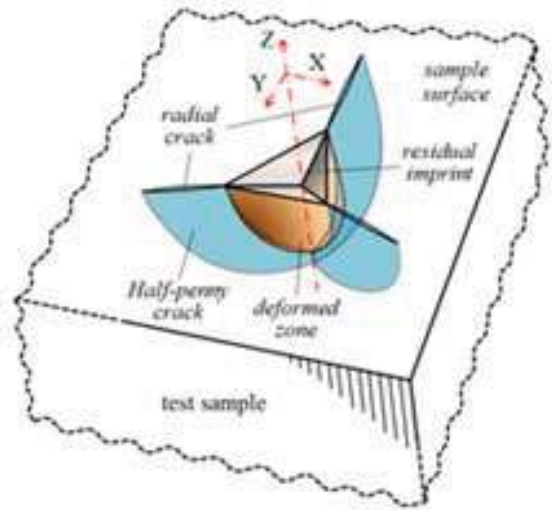

*Fig. 6-5 Isometric schematic of half-penny cracks joining under deformed region*

This type of subsurface cracks joining under the plastic deformation zone induced by a three-face indenter recently has been visualised using FIB tomography by Cuadrado *et al.* [17] and Rueda *et al.* [18], in a soda-lime glass sample and in dental porcelain, respectively, as shown in Fig. *6-6*, although they used cube-corner indenters.

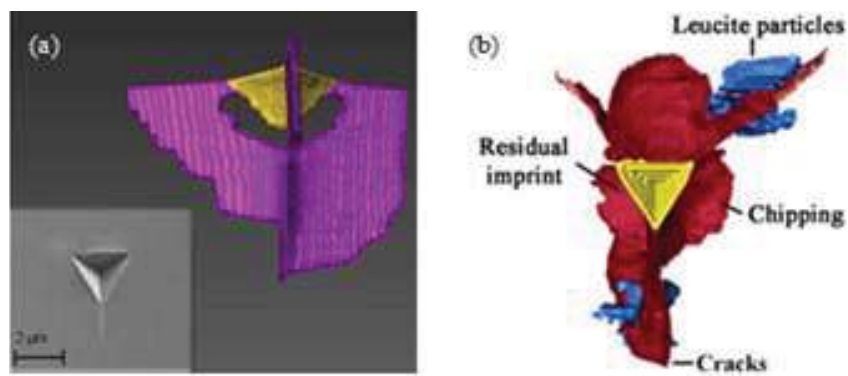

*Fig. 6-6 3D reconstructions of crack morphologies induced by cube-corner indenter for (a) soda-lime glass [17], and (b) dental porcelain [18]*

Regarding the dimensions of the cracks of our film, the average total crack length *a* was measured to be 49±2 μm and indent diagonal *c* 34±2. μm, the ratio of c/a, is 1.5, also suggesting that cracks emanating from an indentation corner would deeply penetrate into the pellet, as shown in Fig. 6-4 (c). Another important feature worth noticing in this specimen is that microcracks and collapsed particle networks cannot be readily seen in the deformed regions, as shown in Fig. 6-4 (f), in contrast to the clear porosity gradient found in the indented films as shown in Fig. 5-14. This difference might not be due to the use of a sharp indenter because similar feature was also observed when a spherical indenter was used, as shown in SEM micrographs in Fig. *6-7*. This might indicate that the particle bonding in the bulk samples is stronger than in the films, despite possessing similar porosities.

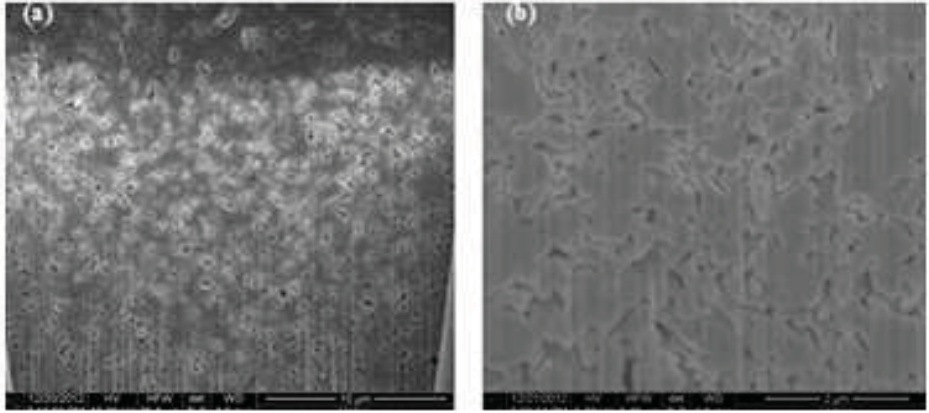

*Fig. 6-7 Micrographs of (a) a cross-sectional surface of the residual area after spherical indentation at 2 N of a pellet sintered at 1000 °C,(b) close view at high magnification*

Fig. 6-8 illustrates the crack morphologies generated by indentation at 500 mN load on a pellet sintered at 1100 °C, showing readily the occurrence of transgranular crack propagation. The average total crack length measured was $16\pm2$ µm, and the indent half diagonal $11\pm1$ µm, with a *c/a* ratio of 1.5.

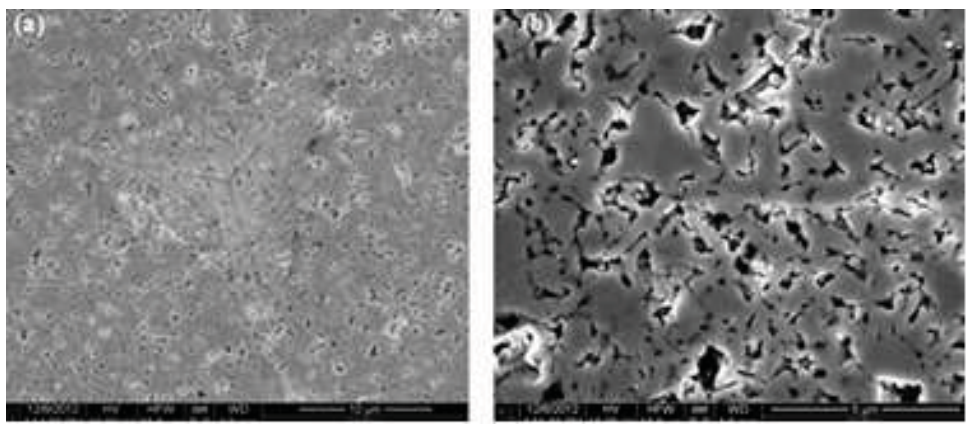

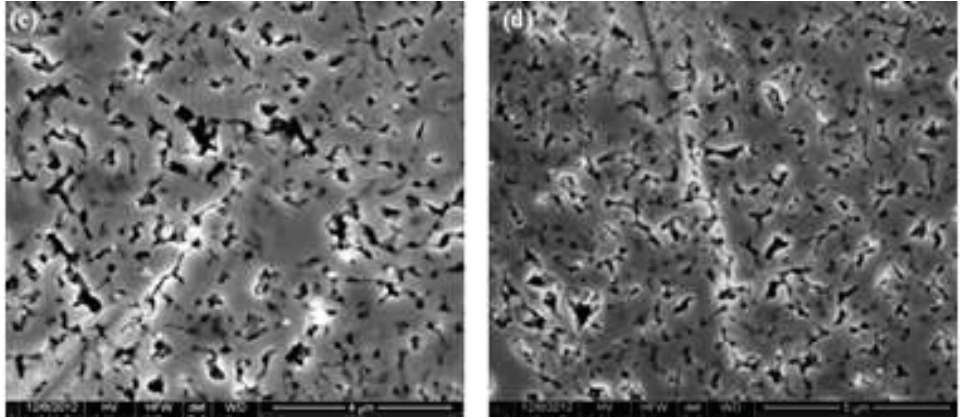

*Fig. 6-8 Crack morphologies on the surface of a pellet indented at 500 mN after sintering at 1100 °C: (a) SEM image of the residual indent with radial cracks emanating from the corners, (b), (c) and (d) high magnification SEM images of the cracks.*

It is important to emphasise that although cracking took place in the above samples, the indentation load vs. displacement curves did not show any relevant pop-in or pop-out events, which suggests that the cracking was probably induced after complete removal of load, as revealed by Lawn *et al.* [9].

### 6.1.2 Fracture Toughness Derived from Indentation

Based on the previously determined parameters, i.e. crack dimensions, elastic moduli and hardnesses, Equation 3.10 in Chapter 3 was used to estimate the apparent fracture toughnesses for the bulk samples sintered at 900-1200 °C, as shown in Fig. 6-9.

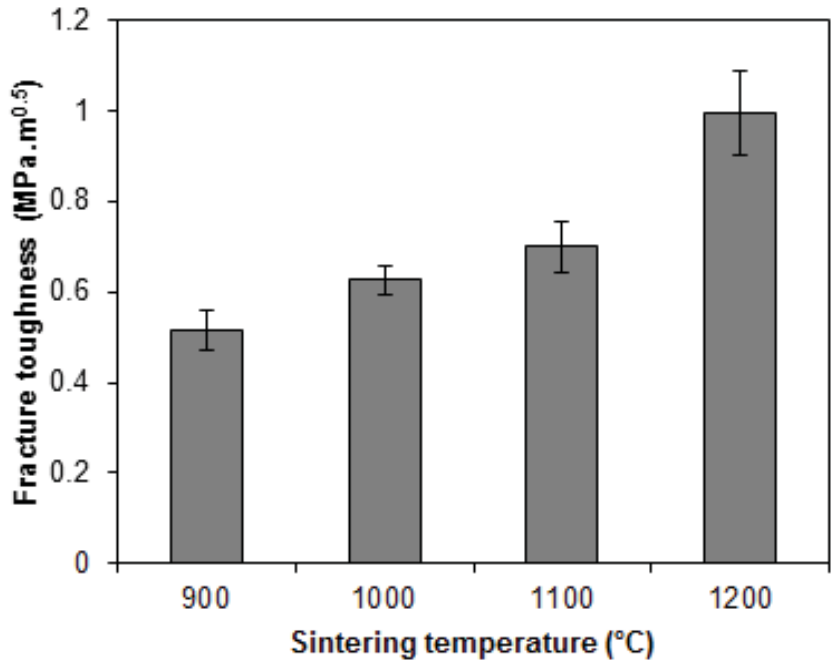

*Fig. 6-9 Fracture toughness measurement by Berkovich indentation for bulk specimens after sintering at different temperatures*

It can be seen that $K_{Ic}$ increased with increasing sintering temperature as expected. All these values are equal to or less than 1 MPa·m$^{1/2}$ which is regarded as a typical fracture toughness value for untoughened ceramics. The estimated toughness for the pellet sintered at 1200 °C (0.99±0.10 MPa·m$^{1/2}$) falls between the literature values reported by Chou *et al.*[19] (1.10±0.05 MPa·m$^{1/2}$) and Huang *et al.* [20] (0.91±0.05 MPa·m$^{1/2}$), as compared in Table 6-1.

*Table 6-1 Comparison of fracture toughness measurements for nominally dense LSCF6428 samples*

| Reference | Sintering Conditions (in air) | Relative Density (%) | Main Grain Size (μm) | Measurement Technique | Fracture Toughness at RT (MPa·m$^{1/2}$) |
|---|---|---|---|---|---|
| Chou *et al.* [19] | 1250°C/4h/300°C·h$^{-1}$ | 95.4±0.2 | 2.9 | Vickers Indentation | 1.10±0.05 |
| Huang *et al.* [20] | 1200°C/4h/300°C·h$^{-1}$ | 96.6±0.2 | 0.6±0.2 | Vickers Indentation | 0.91±0.05 |
| Li *et al.* [21] | 1200°C/2h/300°C·h$^{-1}$ | 98.3 | 0.8 | Vickers Indentation | 1.75±0.25 |
| This work | 1200°C/4h/300°C·h$^{-1}$ | 94.78±0.01 | 1.6 | Berkovich Indentation | 0.99±0.10 |

If we further plot the fracture toughness values against the porosity of each as-sintered pellet, a fairly linear relationship can be obtained, as illustrated in Fig. 6-10, showing that an increasing porosity causes the reduction of fracture toughness.

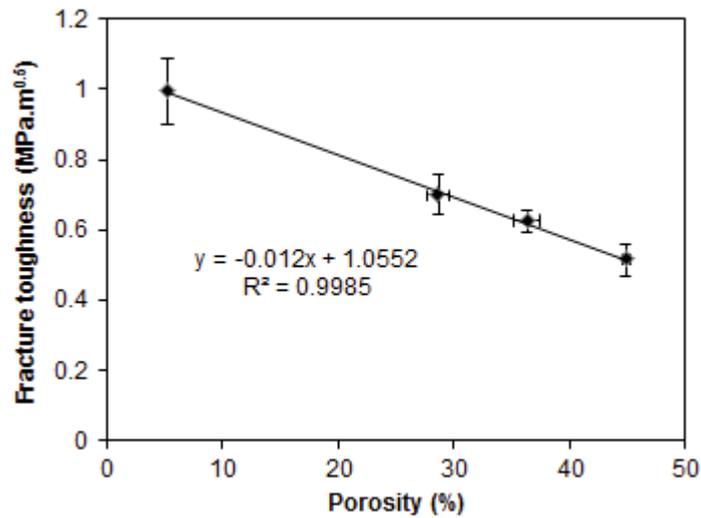

*Fig. 6-10 Fracture toughness of bulk samples as a function of bulk porosity and the linear fit parameters*

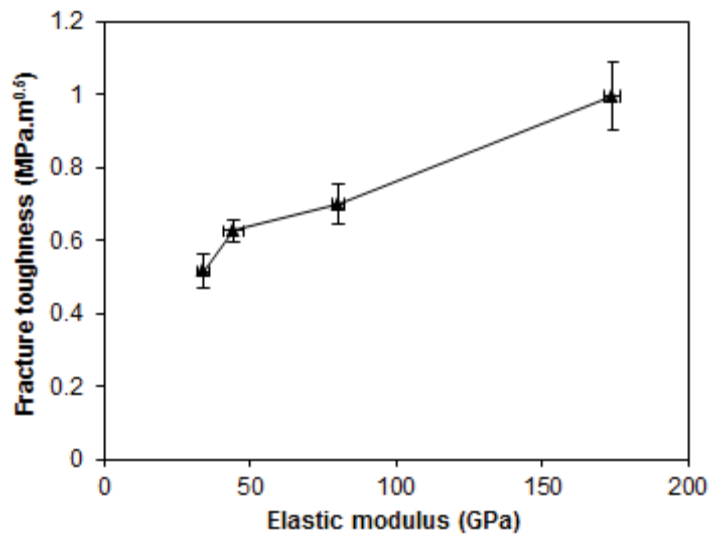

*Fig. 6-11 Plot of fracture toughness as a function of elastic modulus*

Although Fig. 6-11 reflects some proportionality between fracture toughness and elastic modulus, this is very unlikely as the dependencies of $E$ and $K_{Ic}$ on porosity are different (one is exponential and the other is linear).

## 6.2 Indentation Results for Films

### 6.2.1 Microstructural Observation

Fracture toughness measurements in the case of porous thin films need acceptable indents that show radial cracks and no interfacial delamination cracks. Table 6-2 summarises the indentation loads applied and the resulting crack detectability using SEM for the as-sintered LSCF6428 films during the fracture toughness measurements.

*Table 6-2 Indentation loads applied for toughness measurements and the resulting crack detectability using SEM for the LSCF6428 films*

| Sintering Temperature (°C) | Load without Detectable Crack (mN) | Load with Detectable Crack (mN) |
|---|---|---|
| 900 | 50/100/200/300/400/500/1000/2000/3000 | -- |
| 1000 | 50/100/200/300/400/500/1000/2000/3000 | -- |
| 1100 | 50/100/200/300/400/500/1000/2000/3000 | -- |
| 1200 | 50/100 | 200/300/400/500 |

The table above shows that detectable cracks were not found at any loads in the tested films sintered at 900-1100 °C, either in the nanoindentation load range (0-500 mN) or beyond (> 500 mN). Note that higher loads (deeper indenter penetration) led to more noticeable effect from substrates [22], which resulted in erroneous toughness measurements. Nevertheless, the SEM results revealed distinguishable and clear indentation cracks at loads ranging from 200-500 mN for the films sintered at 1200 °C.

Examples of top-surface and cross-sectional SEM images of the films' residual imprints resulting from 500 mN indentation are shown in Fig. 6-12. Note that visible and measurable cracks were only possible to be found in films sintered at 1200 °C, whereas for films sintered at 900 °C indentation on the surface did not result in any detectable residual imprint, even in very high magnification, thus the picture is not shown in the figure.

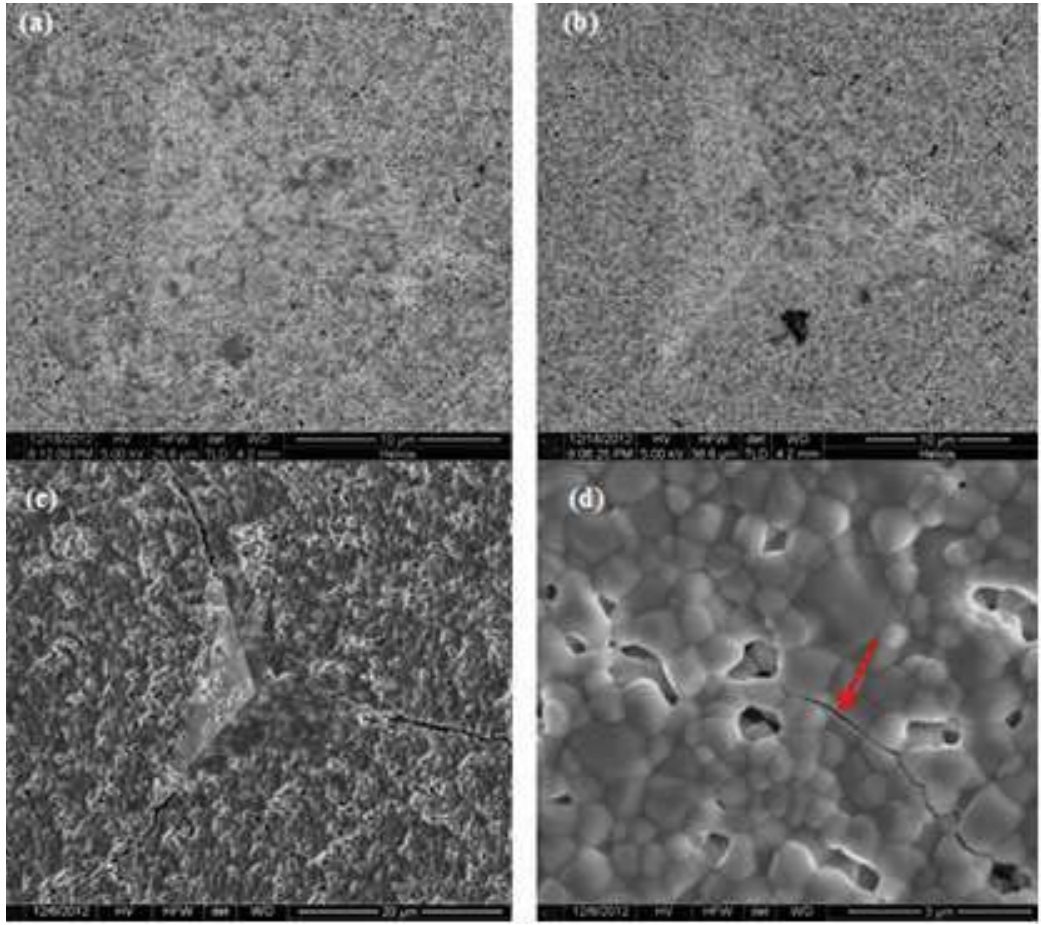

*Fig. 6-12 Micrographs of surface morphologies of Berkovich-indented films at 500 mN load, which were sintered at (a)1000 °C, (b) 1100 °C, (c) 1200 °C, (d) is a much magnified SEM picture of a crack tip of sample in (c).*

The above figures show that Berkovich indentation in the highest density film developed well-defined crack morphologies without any chipping or secondary cracks. However, cracks in films sintered at temperatures lower than 1200 °C tended to be difficult to detect. It can be seen from Fig. 6-12 (c) that cracks emanated from the extremities of the imprint corners, as observed earlier for the bulk samples. As for bulk samples, sequential cross-sectional imaging was also made for the films using FIB/SEM slice and view to see the deformation below the residual indent, as shown in Fig. 6-13.

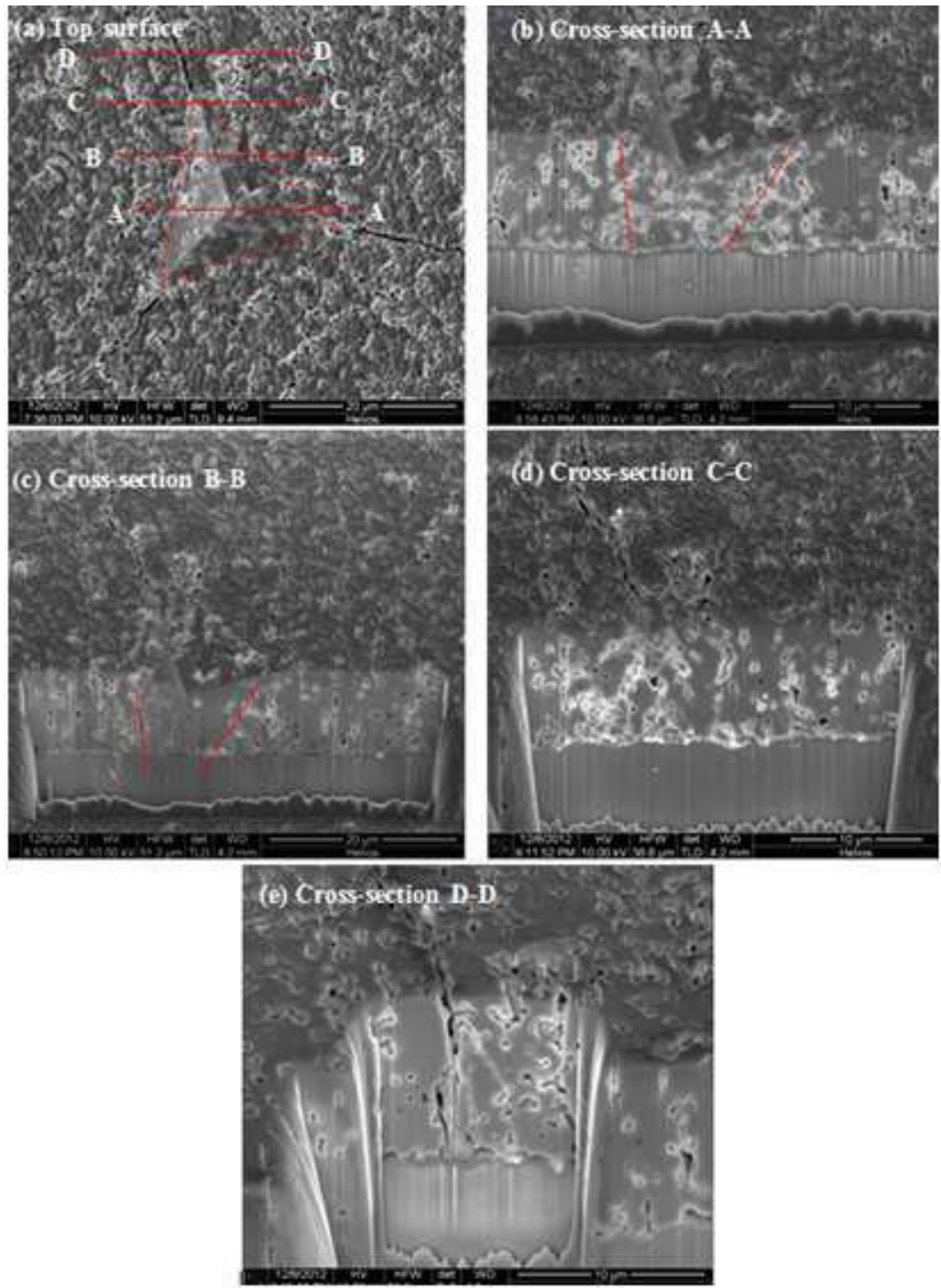

*Fig. 6-13 FIB-sectioned cross-sectional micrographs in the residual indent: (a) Top surface showing the three chosen locations for sectioning, (b) A-A cross-section through the imprint centre, showing plastic deformation (i.e. densification) below the imprint, (c) B-B cross-section through the half way of a diagonal edge, showing little densification, (d) C-C cross-section through the imprint corner, showing a well-developed crack, and (e) D-D cross-section through the radial crack.*

Fig. 6-13 shows some characteristics of deformation below the residual imprint for the film which are similar to those of bulk samples shown earlier. In addition, subsurface delamination cracks were not found evident. The densification in the deformed "plastic" zone immediately underneath the residual imprint did not extend to the region beneath the imprint corner. The densification is similar to that discussed earlier for spherical indentation. In Fig. 6-13 (e), a radial crack propagated through the film thickness and arrested on the interface with the substrate, above which one or two smaller secondary cracks can be found branching off the main large crack. Therefore, the crack system in the indented film could be regarded as approximating a radial/Palmqvist crack. The effect of the densification on the initiation and propagation of cracks and hence on the estimation of fracture toughness should be further studied in the future.

Despite the potential influence from the substrate, which needs further studies, the highest density film was subjected to indentation at a series of maximum loads of up to 500 mN. Cracks were not detected using SEM for loading at maximum load below 200 mN, which implies that the cracking threshold with a Berkovich indenter was not less than 200 mN for this film. (Note that for the previous $E$ and $H$ measurements described in Chapter 5, there were no such fractures over all specimen sintering temperatures from 900-1200 ˚C and nanoindentation load range from 0-500 mN because a spherical indenter was used.) The average total crack length $c$, indent diagonal $a$ and the ratio of $c/a$ are plotted as a function of the maximum load $P$, as shown in Fig. 6-14.

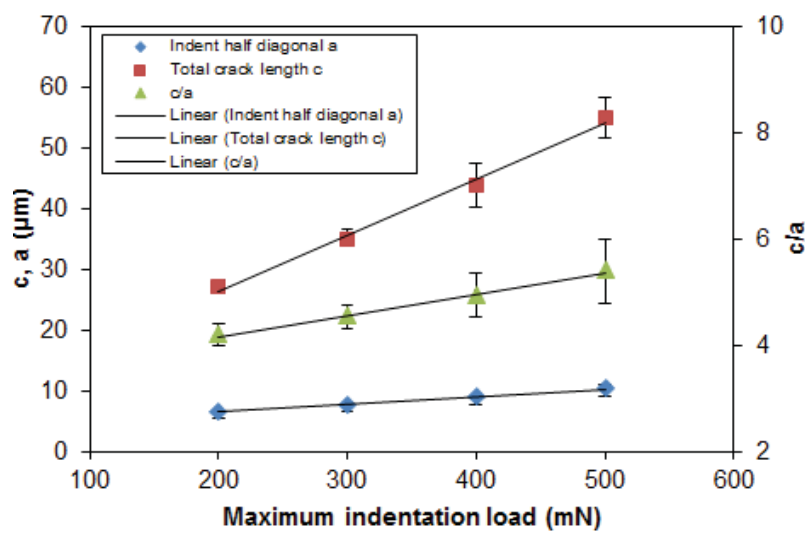

*Fig. 6-14 Relationships between indent half diagonal a, total crack length c, the c/a ratio and maximum indentation load P for a film sintered at 1200 °C*

Fig. 6-14 demonstrates increasing crack dimensions with increasing maximum indentation load applied. High $c/a$ ratios greater than 4 are found, suggesting that the test film showed well-defined crack patterns in the range of applied load. Schnerder *et al.* [23] have also reported that half-penny morphology with underneath joining is prevailing in Vickers-indented bulk brittle materials with $c/a$ larger than 3, while radial/Palmqvist cracks are developed at lower indentation loads or/and in materials with higher toughness, regardless of the indenter type, and with $c/a$ smaller than 2.5. This is reasonably consistent with the observations in the current study for both bulk and films.

### 6.2.2 Fracture Toughness Derived from Indentation

In Fig. 6-15 the plot of the maximum indentation load as a function of $(l/a)^{1/2}c^{3/2}$ reveals a fairly linear proportionality as expected from Equation 3.10. Therefore the value $P/((l/a)^{1/2}c^{3/2})$ could be regarded constant (i.e. the slope of the linear fit) and independent of the maximum load applied, suggesting that Equation 3.10 might be applicable to the current films. By applying this equation $K_{Ic}$ of 0.16±0.02 MPa.m$^{0.5}$ was estimated for the film sintered at 1200 °C. This value appears to be very low and is much smaller than that of the bulk specimens with even larger porosity as shown earlier. Such a low fracture toughness value obtained is possibly related to the effect from the substrate. Nevertheless, the method used could have a typical level of accuracy (or uncertainty) of 40% (as proposed by Harding *et al.* [24]), within which the toughness measured from the crack lengths were found to be accurate, as also found by Volinsky *et al.* [25] for the fracture measurement using nanoindentation on a the low-$k$ film.

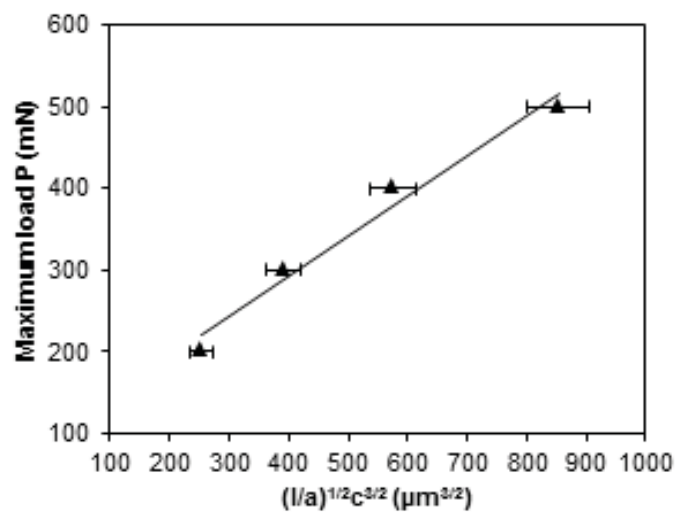

*Fig. 6-15 Maximum indentation load P, as a function of $(l/a)^{1/2}c^{3/2}$*

Fracture toughness measurements using the IM method on partially sintered highly porous ceramic bulk specimens and films are rare. As mentioned in Chapter 2, apart from characterisations of hardness and elastic modulus, considerable emphasis has been put on determining fracture toughness of dense (or nano-porous) thin coatings. Despite the existence of the similar crack patterns exhibited in the LSCF6428 bulk and films, as found earlier, it is worth noting that the toughness equation was developed for bulk materials and therefore its applicability for assessing film fracture toughness remains disputable [25]. Nevertheless, it has been directly used without necessary verification to assess the thin layer fracture toughness by several workers, as reported in [26-29]. Further work may be required on the reliability of such measurements in this study.

## *Summary*

In this Chapter, the measurements of fracture toughness of LSCF6428 in both bulk and film forms were investigated after sintering at 900-1200 °C.

Based on crack length measurements from micrographs obtained for Berkovich-indented specimens, the fracture toughness of bulk LSCF6428 was determined to be approximately 0.51-0.99 MPa·m$^{1/2}$, after sintering at 900-1200 °C. Experimental results for films showed that the generation of observable indentation-induced cracks was very difficult for porous thin films sintered at temperatures below 1200 °C. Cracks were only detectable in the films sintered at 1200 °C, which gave a toughness of 0.16 MPa·m$^{1/2}$, much smaller than that of the bulk specimens with even higher porosity. The surface and subsurface crack morphologies of the test samples, which were investigated using SEM and FIB/SEM slice and view, revealed the presence of radial and half-penny crack systems induced by Berkovich indentation.

## *Chapter 6 References*

1.    Tarrés E, Ramírez G, Gaillard Y, Jiménez-Piqué E, Llanes L: **Contact fatigue behavior of PVD-coated hardmetals**. *International Journal of Refractory Metals and Hard Materials* 2009, **27**(2):323-331.

2.    Ceseracciu L, Anglada M, Jiménez-Piqué E: **Hertzian cone crack propagation on polycrystalline materials: Role of R-curve and residual stresses**. *Acta Materialia* 2008, **56**(2):265-273.

3.    Hagan J, Swain MV: **The origin of median and lateral cracks around plastic indents in brittle materials**. *Journal of Physics D: Applied Physics* 1978, **11**(15):2091.

4.    Nastasi MA, Parkin DM, Gleiter H: **Mechanical properties and deformation behavior of materials having ultra-fine microstructures**, vol. 233: Kluwer Academic Pub; 1993.

5.    Lei Y, O'dowd N, Webster G: **Fracture mechanics analysis of a crack in a residual stress field**. *International journal of fracture* 2000, **106**(3):195-216.

6.    Palmqvist S: **Indentation hardness and fracture toughness in single crystal**. *Jernkontorets Ann* 1957, **141**:300-306.

7.    Cook RF, Pharr GM: **Direct Observation and Analysis of Indentation Cracking in Glasses and Ceramics**. *Journal of the American Ceramic Society* 1990, **73**(4):787-817.

8.    Lawn B, Evans A, Marshall D: **Elastic/plastic indentation damage in ceramics: the median/radial cracks system**. *J Am Ceram Soc* 1980, **63**:574.

9.    Lawn B, Swain M: **Microfracture beneath point indentations in brittle solids**. *Journal of Materials Science* 1975, **10**(1):113-122.

10.   Liang KM, Orange G, Fantozzi G: **Evaluation by indentation of fracture toughness of ceramic materials**. *Journal of Materials Science* 1990, **25**(1):207-214.

11.   Chicot D, Pertuz A, Roudet F, Staia MH, Lesage J: **New developments for fracture toughness determination by Vickers indentation**. *Materials Science and Technology* 2004, **20**(7):877-884.

12.   Lawn B, Wilshaw R: **Indentation fracture: principles and applications**. *Journal of Materials Science* 1975, **10**(6):1049-1081.

13.   Hockey B, Rice R: **The science of ceramic machining and surface finishing II**. 1979.

14.   Lee JH, Gao YF, Johanns KE, Pharr GM: **Cohesive interface simulations of indentation cracking as a fracture toughness measurement method for brittle materials**. *Acta Materialia* 2012, **60**(15):5448-5467.

15.   Ponton CB, Rawlings RD: **Vickers indentation fracture toughness test Part 1 Review of literature and formulation of standardised indentation toughness equations**. *Materials Science and Technology* 1989, **5**(9):865-872.

16.   Zhang W, Subhash G: **An elastic–plastic-cracking model for finite element analysis of indentation cracking in brittle materials**. *International Journal of Solids and Structures* 2001, **38**(34–35):5893-5913.

17.   Cuadrado N, Casellas D, Anglada M, Jiménez-Piqué E: **Evaluation of fracture toughness of small volumes by means of cube-corner nanoindentation**. *Scripta Materialia* 2012, **66**(9):670-673.

18.   Rueda AO, Seuba J, Anglada M, Jiménez-Piqué E: **Tomography of indentation cracks in feldspathic dental porcelain on zirconia**. *Dental Materials* 2013, **29**(3):348-356.

19. Chou Y-S, Stevenson JW, TArmstrong TR, LPederson LR: **Mechanical Properties of La1-xSrxCo0.2Fe0.8O3-δ Mixed-Conducting Perovskites Made by the Combustion Synthesis Technique**. *Journal of American Ceramic Society* 2000, **83**(6):1457-1464.

20. Huang B, Chanda A, Steinbrech R, Malzbender J: **Indentation strength method to determine the fracture toughness of La0.58Sr0.4Co0.2Fe0.8O3-delta and Ba0.5Sr0.5Co0.8Fe0.2O3-delta**. *Journal of Materials Science* 2012, **47**(6):2695-2699.

21. Li N, Verma A, Singh P, Kim J-H: **Characterization of La0.58Sr0.4Co0.2Fe0.8O3−δ–Ce0.8Gd0.2O2 composite cathode for intermediate temperature solid oxide fuel cells**. *Ceramics International* 2013, **39**(1):529-538.

22. Scharf TW, Deng H, Barnard JA: **Mechanical and fracture toughness studies of amorphous SiC–N hard coatings using nanoindentation**. *Journal of Vacuum Science & Technology A: Vacuum, Surfaces, and Films* 1997, **15**(3):963-967.

23. Schneider GA, Fett T: **Computation of the stress intensity factor and COD for submicron sized indentation cracks**. *Journal of the Ceramic Society of Japan* 2006, **114**(1335):1044-1048.

24. Harding D, Oliver W, Pharr G: **Cracking during nanoindentation and its use in the measurement of fracture toughness**. In: *Materials Research Society Symposium Proceedings: 1995*: Cambridge Univ Press; 1995: 663-663.

25. Volinsky AA, Vella JB, Gerberich WW: **Fracture toughness, adhesion and mechanical properties of low-k dielectric thin films measured by nanoindentation**. *Thin Solid Films* 2003, **429**(1):201-210.

26. Morris DJ, Cook RF: **Indentation fracture of low-dielectric constant films: Part I. Experiments and observations**. *J Mater Res* 2008, **23**(9):2429.

27. Malzbender J, den Toonder JMJ, Balkenende AR, de With G: **Measuring mechanical properties of coatings: a methodology applied to nano-particle-filled sol–gel coatings on glass**. *Materials Science and Engineering: R: Reports* 2002, **36**(2–3):47-103.

28. Morris DJ, Cook RF: **Indentation fracture of low-dielectric constant films: Part II. Indentation fracture mechanics model**. *J Mater Res* 2008, **23**(9):2443-2457.

29. Üçisik AH, Bindal C: **Fracture toughness of boride formed on low-alloy steels**. *Surface and Coatings Technology* 1997, **94–95**(0):561-565.

# 7 3D Microstructure Reconstruction, Quantification and FEM Simulation

This Chapter describes the results of (i) 3D microstructural reconstruction based on acquisition of image stacks using the FIB/SEM slice and view technique, (ii) quantification of microstructural parameters and (iii) mechanical simulation to calculate the elastic modulus and the Poisson's ratio using FEM. Fig. 7-1 shows the workflow of the 3D microstructure reconstruction, mesh generation and FEM simulation process including applications and software used and the resulting output file types as well as the file extensions.

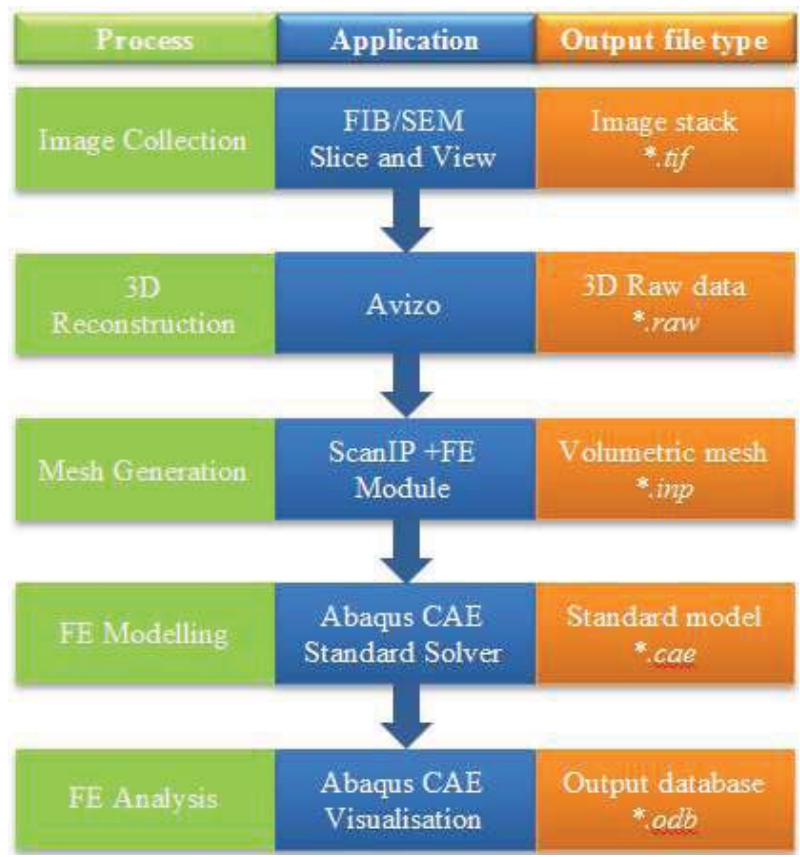

*Fig. 7-1 Workflow of dataset processing for image analysis and finite element modelling*

Efforts were made to prevent image artefacts arising from image acquisition process. The sensitivity to working parameters during image acquisition (with FIB/SEM slice and view), processing parameters during 3D reconstructions (with Avizo) and meshing methods during meshing (with ScanIP) of the microstructural properties such as porosity, necking size and surface areas of the films are discussed.

## *7.1 Artefacts Arising from Image Acquisition*

As briefly described earlier in Chapter 3, in this study consecutive high resolution cross-sectional images of the LSCF6428 films sintered at different temperatures were produced by the FIB/SEM slice and view technique. However, numerous artefacts can originate from the image acquisition process, such as the so-called "curtain effect", ion charging, low phase contrast, shadow effect (i.e. image grayscale gradient), redeposition effects and image drift. Their origins and possible influences on the quality of the images obtained, as well as the measures taken for their elimination are discussed below.

### *7.1.1 Curtain Effect*

For a highly porous material to be milled using FIB, a so-called "curtain effect" can frequently occur on the cross-section of the volume of interest (VOI), manifesting as vertical curtain-shaped strips of material on the surface, as in the severe case shown in Fig. 7-2.

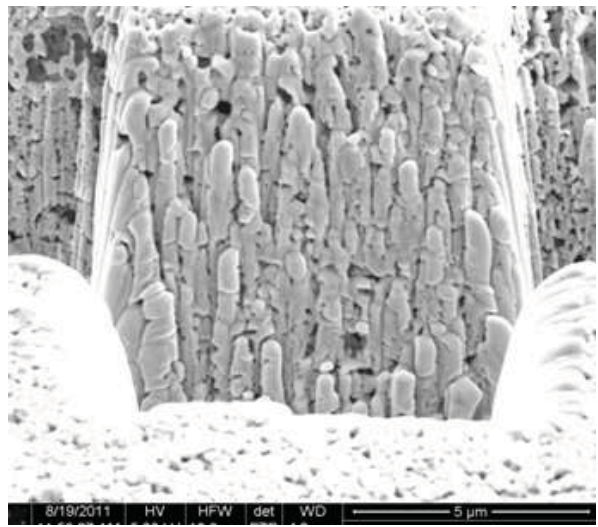

*Fig. 7-2 Severe curtain effect after ion milling of a 1000 ˚C sintered LSCF6428 film. Note that there was no deposition of a protective layer on the top of the sample and the specimen was not resin-impregnated.*

Several factors which can induce the effect, such as the non-uniform distribution of material on the surface, a material structure vulnerable to damage under ion beam erosion, ion beam current being too low or the milling depth being insufficient.

These factors can be mitigated by adjusting the ion beam current and milling depth. A thin layer of protective material (e.g. platinum (Pt) or carbon (C)) deposited on the surface of the VOI by the built-in gas injection system (GIS) prior to the automatic slice and view operation was effective to minimise the curtain effect caused by non-uniform surface in the present study. Compared with Fig. 7-2, Fig. 7-3 demonstrates a significant reduction of the curtain effect on the cross-section thanks to adequate adjustment of beam current and the protective Pt layer deposited on top of the VOI. Resin impregnation also helped to lessen the curtain effect and protect the interconnected highly porous microstructures, as explained later.

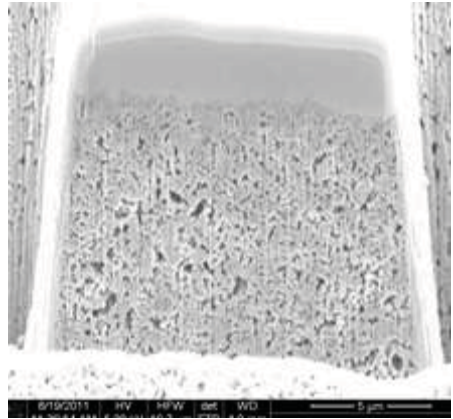

*Fig. 7-3 Curtain effect after ion milling was minimised by deposition of a protective Pt layer on the top of a 1000 ℃ sintered LSCF6428 film which was not resin-impregnated.*

### 7.1.2 Charging Effect

This effect is due to the build-up of the trapped electrons or ions on a poorly conductive material surface, due to the number of electrons/ions flowing to the sample being different from that escaping from within the sample. This is particularly severe with materials of large surface area to volume ratio (i.e. porous or loose materials such as particles). This may induce an electric field in the sample and influence the escape of secondary ions/electrons from the sample surface. Thus the image signal reflecting the local specimen structure will be significantly affected (e.g. image distortion, drift, partially bright or dark appearance and/or loss of image contrast) [1].

An example of extreme contrast reflecting a strong charging effect was observed on the milled cross-section of a sample, as shown in Fig. 7-4 (a), thereby prohibiting a good view of the cross-section. The bright white charging area overlapped a major part of both the solid and pore phases, which were hard to distinguish visually. Accordingly, Fig. 7-4 (b) reveals the disadvantageous grayscale histogram which was unsuitable for further segmentation of the image, and hence any images acquired with this type of charging effect were discarded.

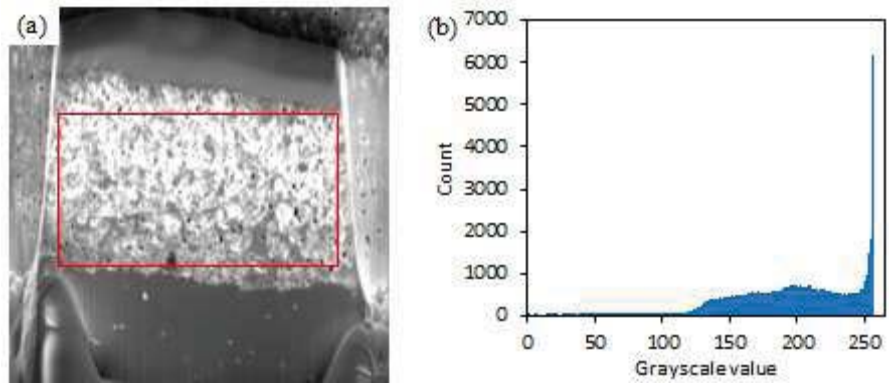

*Fig. 7-4 (a) Charging effect on the milled cross-section and (b) the resulting grayscale histogram of the selected region*

Although LSCF6428 is a good mixed electronic-ionic conductor, its deposition on an isolating CGO substrate drastically reduces the overall conductance when attached on an aluminium microscopy stub. Therefore, in order to minimise the charging effect, the sample was firmly attached to the stub using conductive Ag paste and the paste was also used to draw a low resistant path from the film edge to the stub, crossing the CGO substrate. In addition, the entire sample was subjected to sputter coating of a thin film of gold. If no measure was taken to provide electrical grounding, the charging effect could be even more noticeable. Apart from good electrical grounding and conductive contacting of the sample, another method tried to prevent charging was to use a lower accelerating voltage for the electron beam [2].

### 7.1.3 Low Phase Contrast

The grayscale contrast of the images acquired relied on the quantity of the emitted electrons, both back scattered and secondary, received by the detector. In the current study, there are only two phases in the films, namely the LSCF6428 solid phase and the pore phase. Low contrast (as an example shown in Fig. 7-5 (a)) resulted in a poor level of automatic discrimination between phases and was attributed to the charging effect.

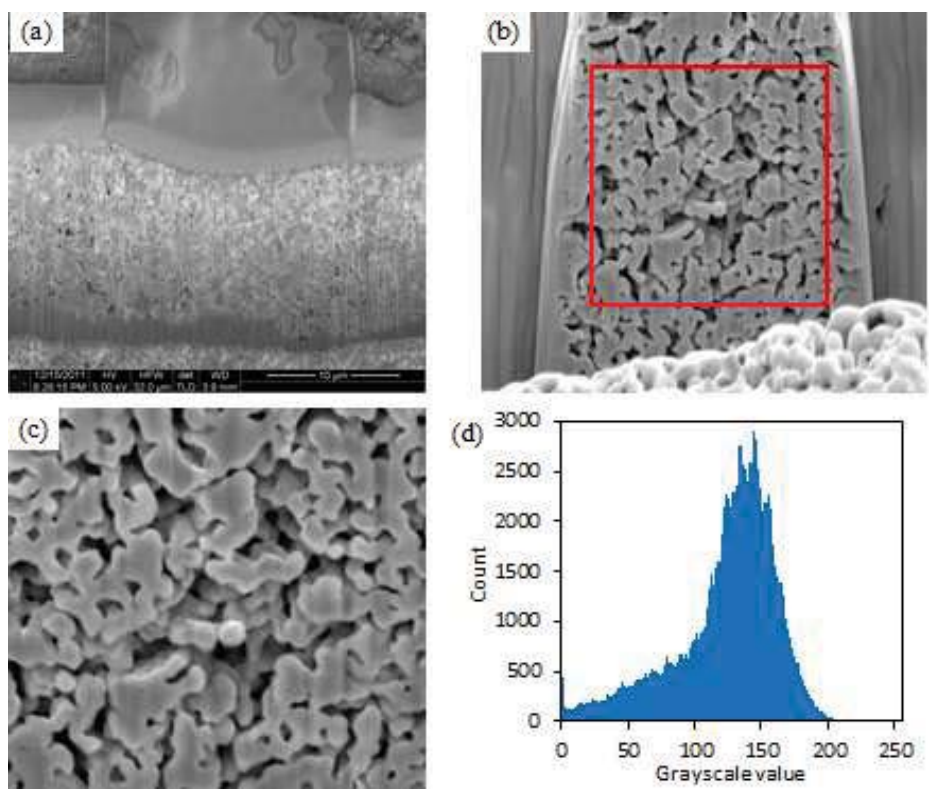

*Fig. 7-5 Example of low and unreliable contrast between porous phase and solid phase in specimens not impregnated with resin (a), (b) and (c), and histogram (d) corresponding to micrograph (c)*

Despite some enhancement of phase contrast achieved by lowering the accelerating voltage of the electron beam (from 15 kV to 5 kV), the poor differentiation between pores (especially the background pores below the milled cross-section) and solid in most areas remained unresolved in the example shown in Fig. 7-5 (b) and (c). Although the pore and solid features are still visually distinguishable upon close view, without a tedious manual correction the automated image processing (i.e. segmentation) was not possible because of the unimodal grayscale histogram shown in Fig. 7-5 (d).

Rather than devising more complicated image processing techniques, resin (e.g. epoxy) impregnation was found to be an effective way to increase the contrast between pore and solid phases for highly porous structures. Prior to sectioning, the sample was placed in a vacuum impregnation chamber and one or two drops of epoxy resin, which had been homogeneously mixed with curing precursor, were placed on the specimen and then held under vacuum for 2 hours. Fig. 7-6 (a) shows an example of a film after sintering at 900 °C in which the resin impregnation induced a very well defined contrast between the two phases as well as distinct phase boundaries. This gave a grayscale histogram (Fig. 7-6 (b)) which met the requirement for automated segmentation based on grayscale thresholding without manual correction. Lower grayscale values belong to pores ((i.e. darker regions around left hand peak) and higher values solid phase (brighter regions around right hand peak). Since the image grayscale value intensities in the histogram were clearly divided into two groups showing two distinct peaks, the threshold value could be chosen at the middle of two peaks in the histogram. Fig. 7-6 (c) therefore shows the corresponding segmented image by applying a threshold value of 38 according to the histogram in this example.

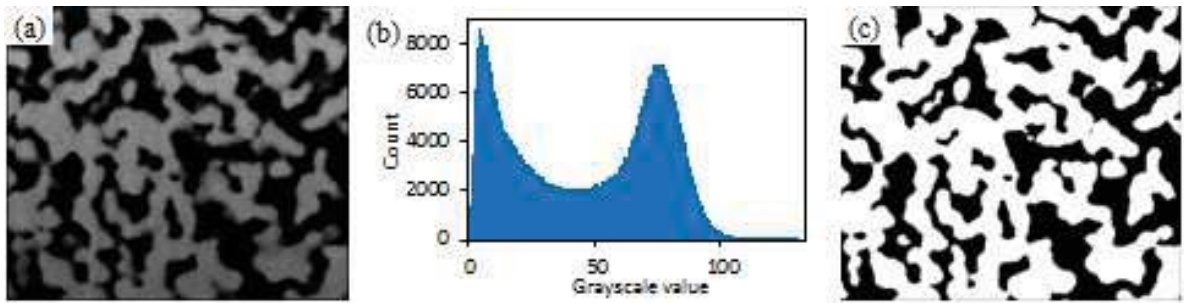

*Fig. 7-6 Sufficient phase contrast resulting from resin impregnation, (a) the SEM image; (b) the corresponding grayscale histogram; and (c) the segmented image with a threshold at the minimum between the two peaks in (b)*

It should be noted that in Fig. 7-6 the image and histogram show relatively dark grayscale values, which was found to be more favourable for the segmentation process. In practice it was not always the case that the brighter the images the better the image quality in terms of the subsequent image processing, unless the contrast is sufficiently strong. As shown in Fig. 7-7 and Fig. 7-8, sometimes a lower brightness is more favourable to better distinguish between porous phase (dark) and solid phase (bright). The relatively bright noise/charging inside the dark phase and at the phase boundaries, which would be improperly regarded as solid phase in a bright image, can be included with the dark phase, so that differentiation can be made easier for image segmentation.

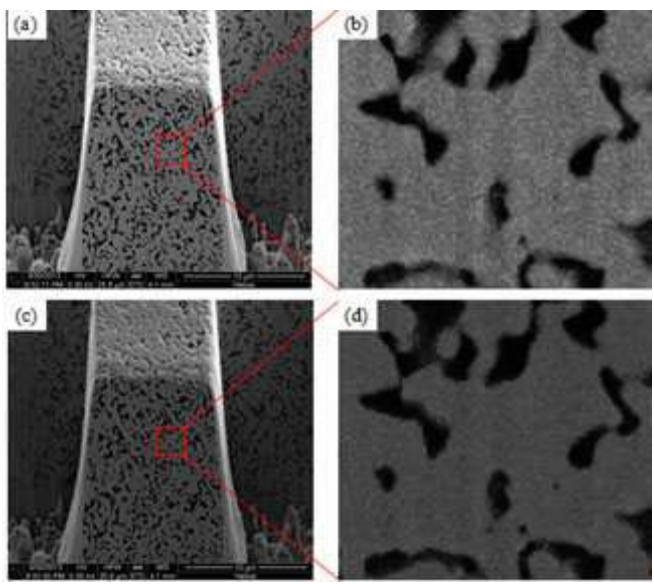

*Fig. 7-7 Comparison between noise in (a-b) a relatively bright image and (c-d) a dark image of the same porous microstructure*

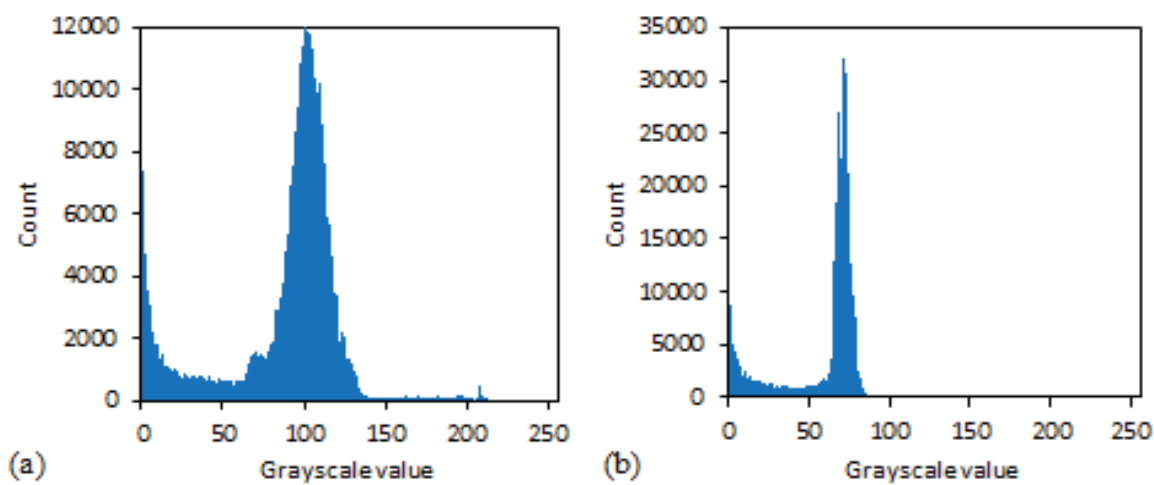

*Fig. 7-8 Histograms showing (a) more noise in the bright image than (b) in the relatively darker image from Fig. 7-7*

### 7.1.4 Shadow Effect

The shadow effect is a gradient effect, as illustrated in Fig. 7-9 (a), in which the image appears to be increasingly dark towards the bottom. This was because the trenches created around the VOI had a small volume. In this particular example the trench in front of the peninsula (i.e. the VOI) is short, as shown in Fig. 7-9 (b), so that the access of the ions/electrons flowing in and exiting from regions near the trench (e.g. the lower part of the front face of the VOI in this case) was restricted.

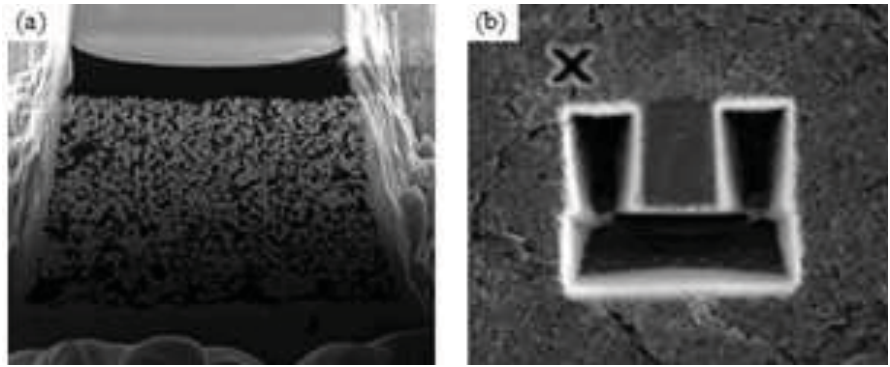

*Fig. 7-9 (a) Shadow effect on a SEM image and (b) the corresponding top view of the VOI.*

As the streams of charged particles were obscured, their quantities detected by the detector diminished so that the corresponding area in the image was shaded and non-uniform contrast was generated. This generated segmentation problems, as it was then hard to automatically segment the image using a simple thresholding method. Fig. 7-10 (a) shows the 2D SEM image corresponding to Fig. 7-9 (a), and its grayscale histogram is shown in Fig. 7-10 (b). Although there are two distinct primary peaks in Fig. 7-10 (b), the simple one-off thresholding method resulted in inaccurate segmentation as shown in Fig. 7-11 where the features for the lower part of the image appear to be lost.

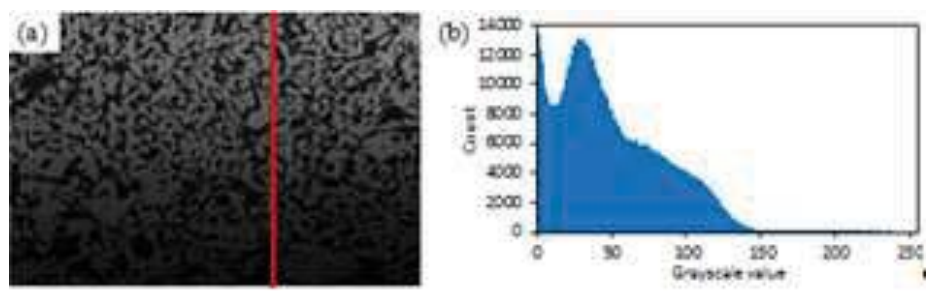

*Fig. 7-10 (a) 2D SEM image cut from Fig. 7-9 (a) showing a shadow effect and (b) the corresponding grayscale histogram*

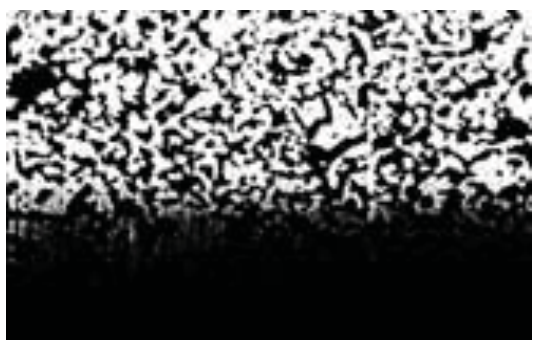

*Fig. 7-11 The resulting segmented image of Fig. 7-10 (a) based on simple thresholding using the histogram of Fig. 7-10 (b)*

For a more informative representation of the non-uniform contrast distribution, Fig. 7-12 (a) and (b) respectively show the 3D surface plot of the grayscale values across the image and the grayscale values across a line cut in the image shown in red. These figures readily show the gradient of grayscale value over the image from the top to the bottom, as it tended to decrease to zero (= black).

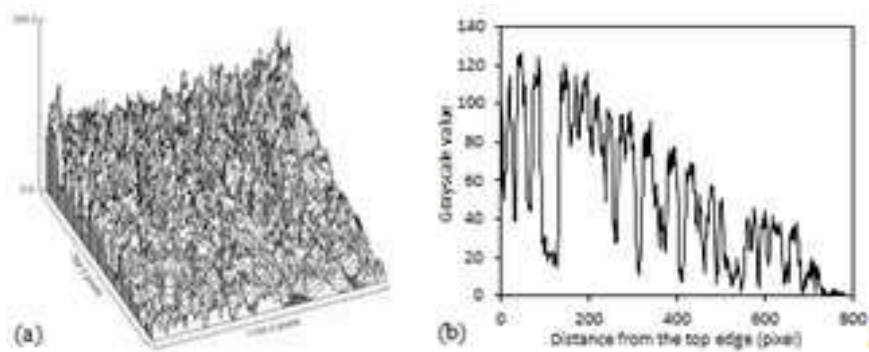

*Fig. 7-12 (a) A 3D surface plot and (b) a 2D plot of the grayscale values of the entire image and across the line cut in red in Fig. 7-10 (a)*

Measures taken to minimise the shadow effect were considered to be either conducting extra image processing to subtract the gradient information in the images or milling much wider and larger trenches during FIB slice and view process.

The removal of the grayscale gradient was investigated by using a number of image processing packages, such as ImageJ, Lispix and Matlab, based on background subtraction or filtering algorithms (such as median filter [3] and edge-preserving smoothing filter [4]). Fig. 7-13 compares the quality of an as-acquired image and the image after gradient removal by applying the "flatten background" function in Lispix. Compared with the original grayscale histogram shown in Fig. 7-13(b), the result after gradient removal in Fig. 7-13 (d) shows a highly improved histogram despite the presence of some trivial noise events. Fig. 7-14 (a) and (b) demonstrate the corresponding segmented binary images of Fig. 7-13 (a) and (c) by thresholding, respectively, and demonstrate the improvement of image quality by the aforementioned gradient removal.

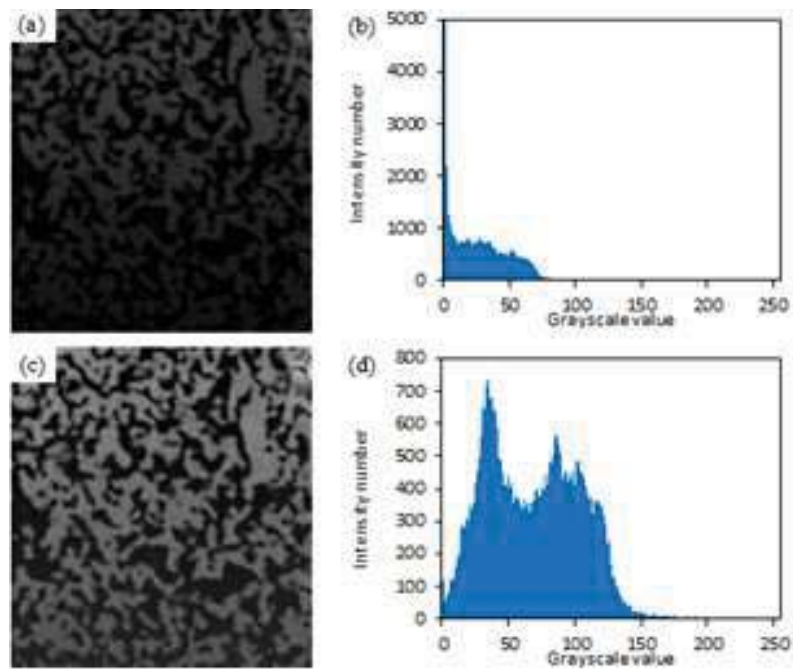

*Fig. 7-13 Removal of grayscale gradient by image processing using background subtraction: (a-b) before, and (c-d) after*

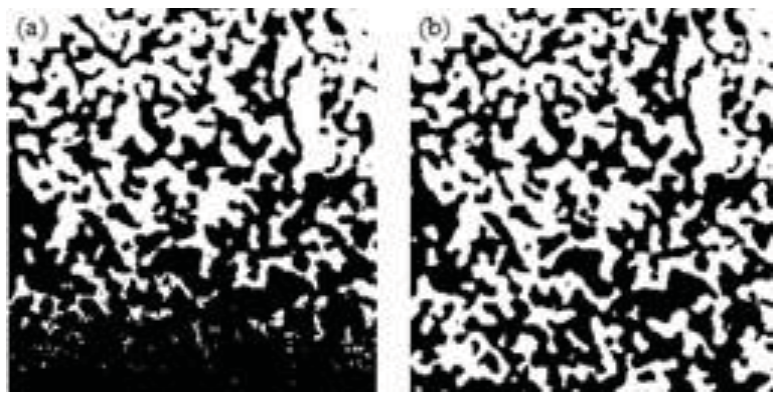

*Fig. 7-14 The resulting segmented binary images corresponding to Fig. 7-13 (a) as-acquired and (b) after gradient removal. (black=pore, white=LSCF6428)*

It was concluded that the cut area of the front trench should have a length at least twice the height of the VOI. Fig. 7-15 (a) shows a slice and view operation on a sample with a wider and larger U-shape trench. This resulted in a uniform brightness across the front face of the VOI (Fig. 7-15 (b)).

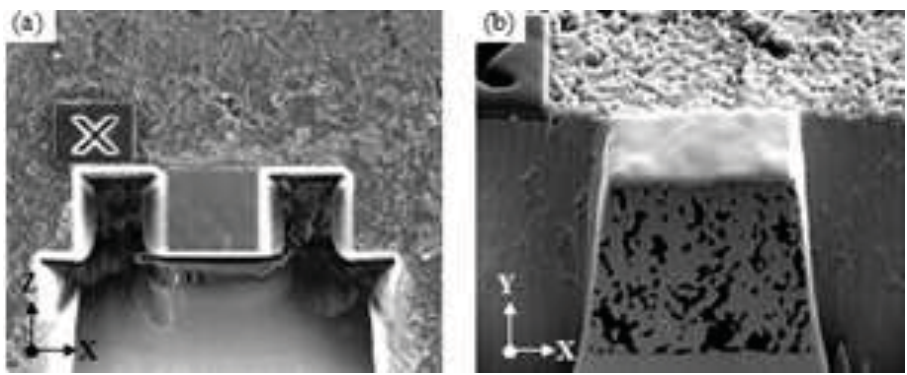

*Fig. 7-15 The shadow effect could be avoided by creating larger trenches.*

### 7.1.5 Redeposition of Sputtered Material

When a relatively high ion-beam current ( > 2.8 nA) was used to mill the cross-section of the sample, some of the material sputtered from the milled region redeposited on the cross-section (Fig. 7-16) as well as in the trenched area (Fig. 7-17). The redeposition on the cross-section could eventually obscure the true microstructural features of the face being analysed.

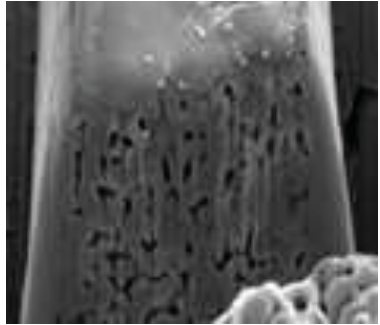

*Fig. 7-16 An extreme example of redeposition of sputtered material on the cross-section*

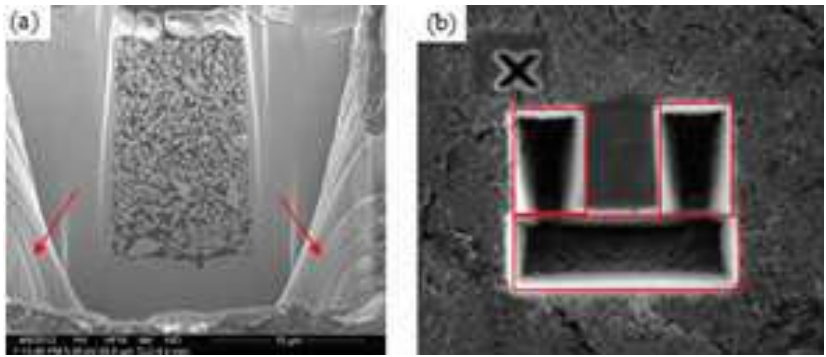

*Fig. 7-17 Two examples of redeposited material (bright area, as the (a) red arrows and (b) boxes show) in the trenches surrounding the cut face*

In addition to facilitating the access of the ion/electron flux, another advantage of creating U-shape trenches was also to accommodate the material sputtered during milling, so that the redeposition on the cross-section of the VOI could be minimised. Nevertheless, some redeposition could still be found on the cross-section. In order to remove this residual redeposition and to avoid further redeposition during milling, a relatively small ion beam current (e.g. 2.8 or 0.92 nA) was be used to finely polish each cross-section after milling, while keeping the real features of the section unaffected. This was very useful for the long-period automated slice and view runs. As shown in Fig. 7-18, the redeposition on a cross-section due to milling using a high beam current (9.3 nA) could be removed by finer polishing using a lower beam current (2.8 nA).

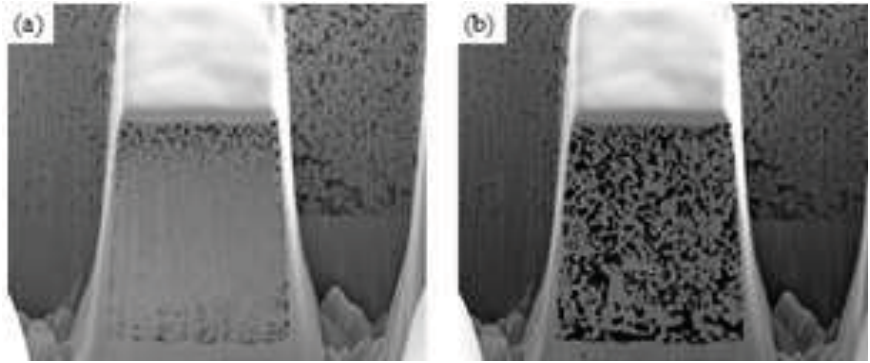

*Fig. 7-18 (a) Redeposition of material on the cross-section after milling using a high beam current of 9.3 nA, and (b) after fine polishing using a lower beam current of 2.8 nA.*

### 7.1.6 Image Drift

Significant image drift problems were very rarely encountered with the current advanced FIB/SEM system, as all the image shifts were compensated during the automated slice and view procedure. However, occasionally the system was not adequately stabilised for the thermal or turbulent drift in the working environment, so that uncorrected image drift could still be significant. Since the slice and view of several hundred images could take over 10 hours to finish, image drift could be noticeable. Illustrated in Fig. 7-19 is an example in which there was significant displacement of the images acquired by SEM before and after 500 milling slices (a–b), while the corresponding view from ion beam did not change much (c-d).

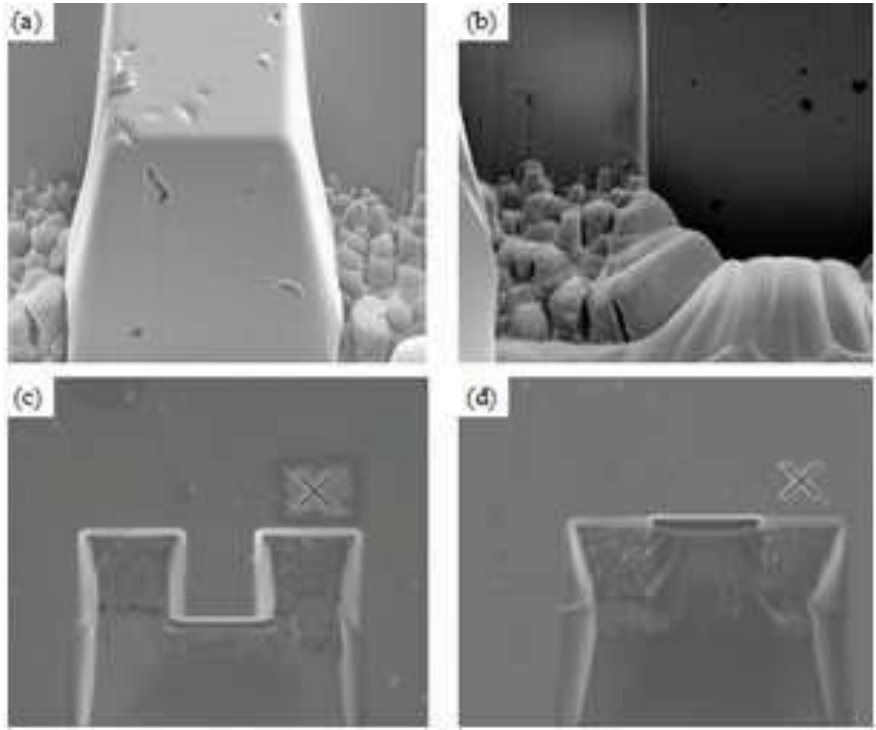

*Fig. 7-19 Example of image shift caused by electron beam drift (bulk sample of LSCF6428 sintered at 1200 °C), (a) and (b) show displacement in the SEM images after 500 slices and (c) and (d) show much smaller displacement in the corresponding ion beam images.*

If no drift correction was made afterwards, in this example severe distortion in the reconstructed 3D microstructures would be induced, as described in the next section. Furthermore, in the case as shown in Fig. 7-19 some of the later SEM images only recorded a part of the cross-sections. This would not generate a sufficiently larger VOI by 3D reconstruction unless the VOI was completely included in the stack of images.

## 7.2 3D Microstructural Reconstruction Using Avizo Software

As previously mentioned in Chapter 3, the 3D microstructures of LSCF6428 films sintered at different temperatures were reconstructed using Avizo software, by applying a series of image processing techniques. A typical process outline from a recorded image stack to a reconstructed 3D microstructure would include: (i) image preparation, which involves correction of image resolution (i.e. resampling) and cropping; (ii) alignment of the image stack, which involves drift and shear correction; (iii) image filtering, such as noise removal and island removal; (iv) segmentation (by grayscale thresholding and phase labelling) and (v) 3D model generation. Note that stages (i) to (iii) are not strictly sequential but can be inter-coordinated.

### *7.2.1 Image Preparation*

As found later at the meshing stage, image stacks of very high resolution (i.e. in this study the highest resolution was 12.5 nm/pixel) resulted in very large datasets ( > 1 GB) which Abaqus CAE would not be able to handle due to the limits on job capacity in the workstation used. Therefore, some reduction of image resolution was necessary, but without significant loss of key microstructural detail. This was been achieved by down-sampling the image by a factor of 2 slice by slice using common image processing tools, such as IrfanView, ImageJ, Matlab, Avizo and ScanIP, after which the resolution was reduced to 25 nm/pixel. In order to reduce the resolution across the direction normal to the image planes, the images were selected every other slice, by which the slice number was reduced to a half and the slice distance increased from 12.5 nm to 25 nm, to produce a new dataset with a new voxel size of $25 \times 25 \times 25$ nm$^3$. An example is shown in 2D in Fig. 7-20 (black=pores and blue=solid), which demonstrates down-sampling by a factor of 2 of a segmented image collected in this study using ScanIP based on linear interpolation. In this case the stack porosity was kept constant but the morphologies of the pore and solid phases were slightly altered with some small interconnected particle apertures being eroded by relatively large pores or *vice versa* as shown in the magnified images in Fig. 7-20 (c) and (d). It is worth emphasising that the change in the feature shapes was negligible and thus the physical properties were hardly affected while the dataset size was significantly reduced.

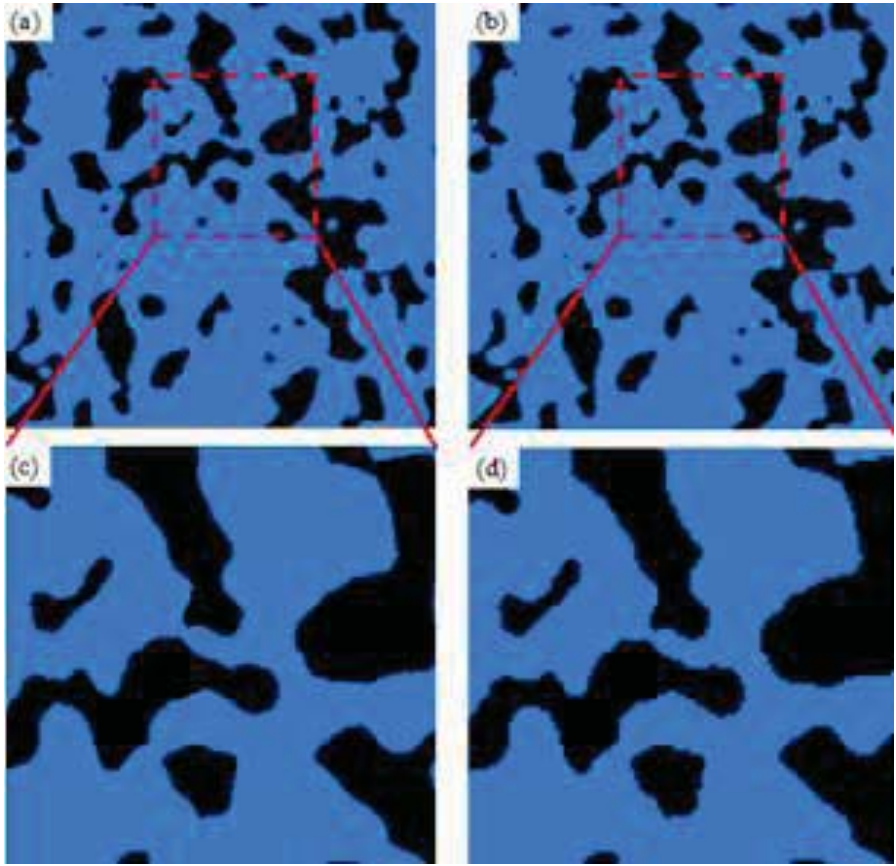

*Fig. 7-20 Down-sampling of a 2D image by a factor of 2: (a) before and (b) after, (c) and (d) are magnified images as a part of (a) and (b), respectively.*

### 7.2.2 Alignment of Image Stack

In this study, the aforementioned stack of images with relatively lower resolution was then imported into Avizo for further processing. Alignment of the image stack involved drift correction and shear correction, in order to remove any distortions of dimension and aspect ratio in the stack in all the three directions.

Tilt correction and dynamic focus correction were applied by the image acquisition procedure (to compensate the tilted slicing and image angle of 52° in the FIB/SEM), but the drift effect remained. When the desired cross-sectional images captured of the VOI were all within the field of view, the drift components in the X and Y directions were corrected using the following image alignment process in Avizo. This is illustrated in Fig. 7-21, where (a) represents a stack of sequential images comprising 200 slices, (b) and (c) are the side views along X axis before and after shear correction, respectively; (d) and (e) are the top views along Y axis before and after drift correction, respectively. The automated alignment process was made using a least squares algorithm based on image intensity. Note that slices of image were stacked in the Z direction (i.e. FIB slicing direction). Once the stack was aligned, the dataset was cropped into the desired size of the VOI. Fig. 7-22 (a) shows an example of the aligned 3D stack of the VOI, and (b) shows examples of the resulting segmented images in three orthogonal planes.

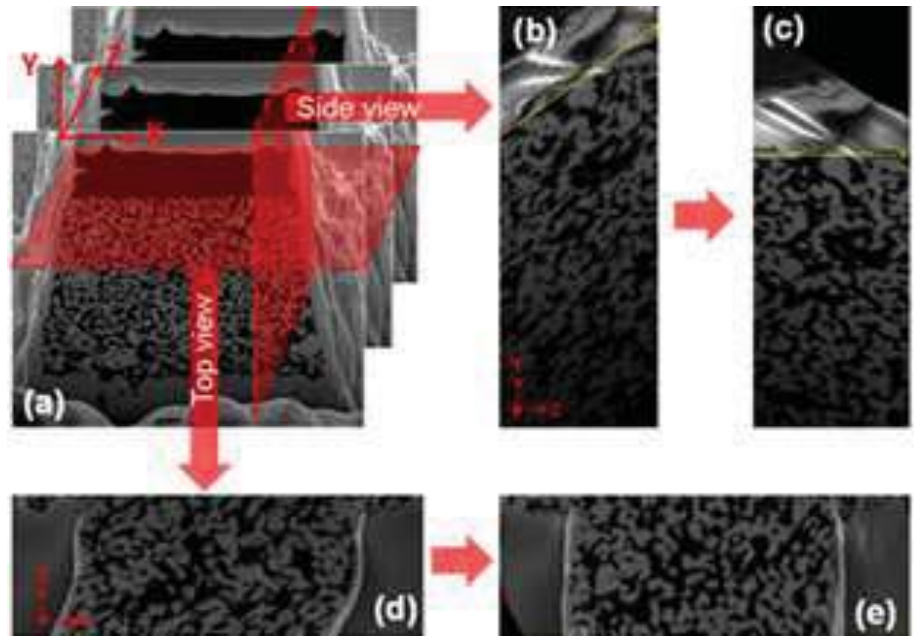

*Fig. 7-21 (a) represents the studied image stack, (b) and (c) are side views (along Y axis) before and after shear correction, (d) and (d) are top views (along X axis) of the stack before and after drift correction.*

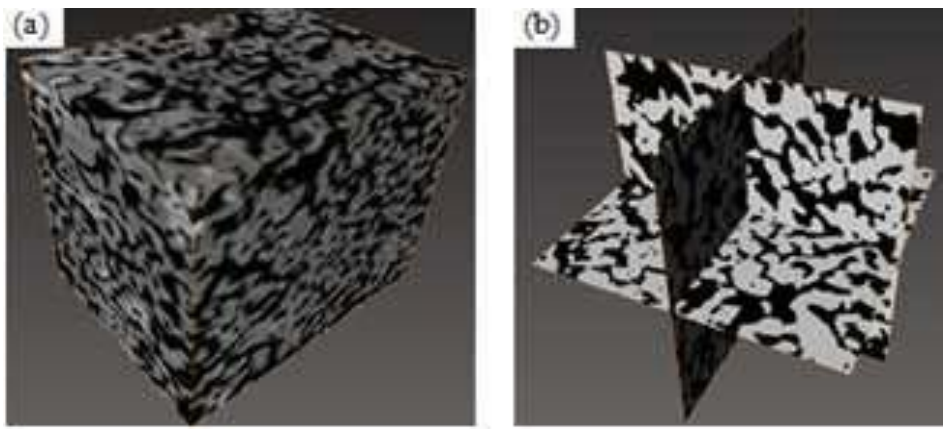

*Fig. 7-22 (a) Stack of cropped and aligned images in 3D, and (b) segmented slices of images in three orthogonal planes*

### *7.2.3 Image Filtering*

Various 2D and 3D filters are available in Avizo for adjusting contrast, removing noise and enhancing features in a single slice, all the slices, and the 3D volume. 2D smoothing based on a median filter algorithm was usually applied in 2D to reduce noise before and after segmentation. For most of the images collected in the current study, there was little to improve in terms of 2D image quality provided that the artefacts had been mitigated during image acquisition.

### *7.2.4 Image Segmentation*

As described previously, the most common method used for segmentation in tomographic images is histogram thresholding whereby a fixed threshold grayscale value is chosen based on the grayscale histogram of the analysed volume. During segmentation each pixel within a given grayscale value range in an 8 bit image acquired is assigned a corresponding material phase label. As in the current study, there were only two phases (i.e. solid LSCF6428 and pores) in the microstructures, the process was made simpler and easier: the grayscale images acquired were binarised with only interior (i.e. solid material) and exterior (i.e. pores) labels. This was performed based on the simple thresholding method, which chose a threshold according to the resulting grayscale histogram, above which all pixels' grayscale values were set to be 1 (white, solid) and the rest were set to be 0 (black, pores), as explained earlier. Problems arising during segmentation usually related to the quality of the images acquired by FIB/SEM slice and view. In other words, the automated segmentation was very successful provided that sound bimodal grayscale histograms (such as the example shown in Fig. 7-6 in the previous section) were generated after the removal of artefacts in the images, as described above.

### *7.2.5 3D Dataset Generation*

Once the aligned images of the VOI had been segmented into binary images, the 3D dataset was generated using Avizo which allows the visualisation in surface view, isosurface rendering or volumetric rendering, as examples respectively shown in Fig. 7-23 and Fig. 7-24.

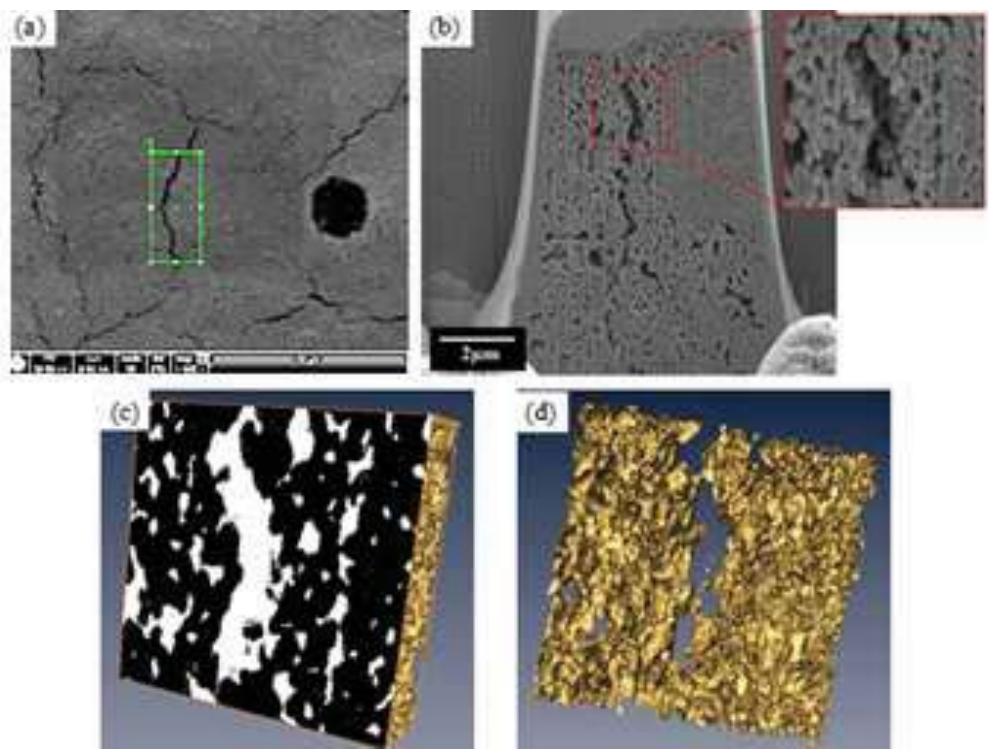

*Fig. 7-23 A LSCF6428 film sintered at 900 ℃ with surface cracks analysed by FIB/SEM slice and view and 3D reconstruction: (a) top surface with green box indicating the milling region; (b) one of the cross-sectional slices milled by FIB with a higher magnification of part of the VOI face; (c) isosurface rendering of the VOI with a front binarised image; (d) complete isosurface rendering of the VOI showing the crack feature.*

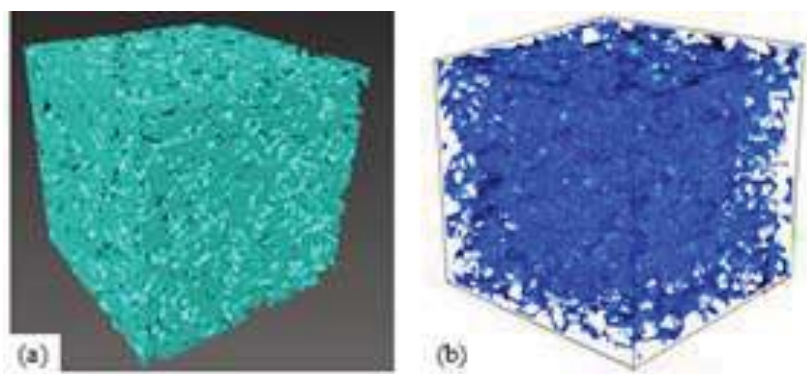

*Fig. 7-24 Example of 3D volumetric rendering showing (a) LSCF6428 solid phase and (b) pore phase*

However, the 3D microstructure could contain some apparently isolated solid regions (islands) that appeared not connected to the main body of the microstructure due to limited resolution. These features would eventually give rise to non-convergence of the subsequent FEM calculation, and therefore it was necessary to exclude them before importing the dataset into Abaqus CAE. Some such isolated islands were removed by applying the Avizo's "Island

Removal" filter which removed islands smaller than a user-defined threshold pixel size. Others that could be readily visually detected were removed during segmentation. Fig. 7-25 shows an example of a solid region of this sort which had to be excluded from further meshing and FEM. Fig. 7-26 illustrates the removal of it by manual labelling the region as pore phase during Avizo's segmentation process.

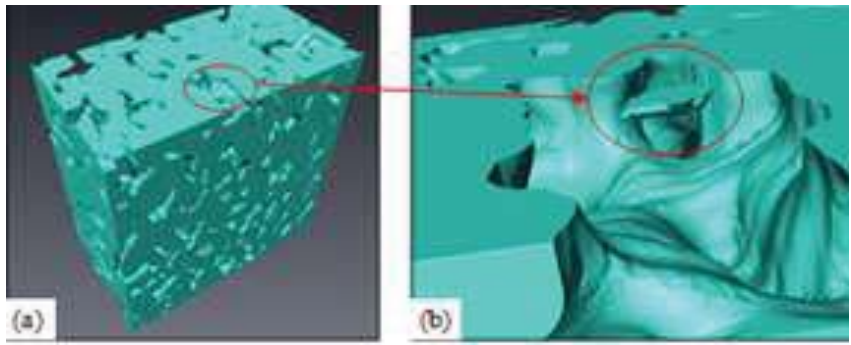

*Fig. 7-25 Example of an isolated solid region (island) "floating" in the pore region of a microstructure, for which removal was desired, (b) is the magnified view.*

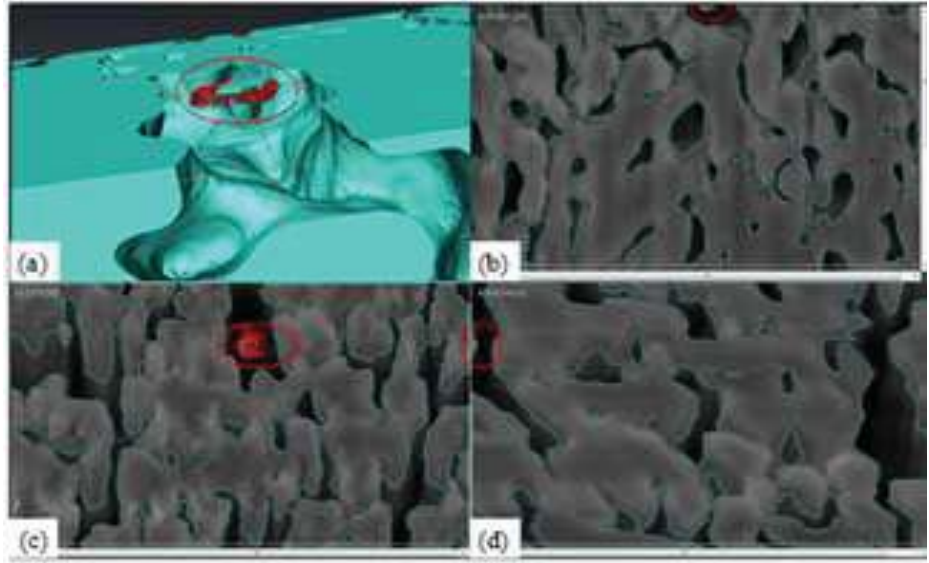

*Fig. 7-26 The "floating" solid region in (a) was removed by manually labelling it as pore phase during segmentation using Avizo, (b), (c) and (d) are of the same location viewed in three orthogonal directions.*

Islands that were not visually noticeable (e.g. hidden in a cluster) and of size slightly above the island removal threshold were moved using ScanIP software, as described in Section 7.3. Because all isolated islands were of very small size compared to the overall microstructure, the influence on the final FEM calculation could be regarded negligible.

### 7.2.6 Pore Volume Fraction Measurement

As explained earlier in Chapter 3, instead of using the conventional Archimedes' method whereby the porosity of porous thin films is almost impossible to measure, the quantification of volume fractions of the two phases were made using Avizo. The porosity was calculated as the ratio of the number of voxels segmented as pore phase over the total voxel number as shown in Fig. 7-27.

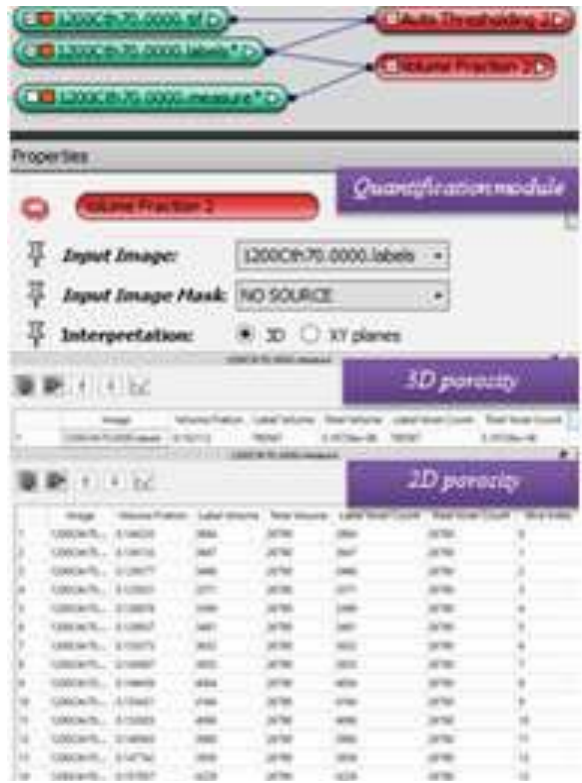

*Fig. 7-27 Data processing and outputs for calculating 3D or 2D volume fractions of solid or pores in a dataset*

The 2D surface area fraction of porous phase (called 2D porosity) in each as-acquired image of a dataset was also measured to assess the microstructure "homogeneity". In this case, the "XY planes" mode was selected. Other important microstructural parameters which may have influence on elastic modulus measurements also can be determined using Avizo and will be discussed in Section 7.5.

## 7.3 FE Model Meshing Using ScanIP Software

Once the 3D microstructures were successfully reconstructed, the Avizo-binarised image dataset was exported as "3D raw data (*.raw)" file and then imported into ScanIP for mesh generation. A typical process would include segmentation, mask filtering and mesh generation. The meshing algorithms as well as the other parameter settings for mesh generation can have a large influence on the mesh quality, meshing time and FEM time. Thus a trade-off between a desired fine mesh quality and meshing time was also desired.

### *7.3.1 Mask Filter Operation*

The volumetric meshing process was carried out using ScanIP software's +FE module, which provides a robust approach for the conversion of segmented serial 3D image data into multi-part volumetric and/or surface meshes. Because the binary data were composed of image slices rather than an individual reconstructed 3D microstructure, prior to the mesh generation process, a simple and quick threshold-based segmentation was applied to all slices for creating new masks (unlinked regions) for the two phases, with white (grayscale value=255) being material and black (grayscale value=0) being pores.

As mentioned earlier, after processing using Avizo, the microstructure (masks) might still contain some small size isolated "islands". Therefore these were removed to produce the final *.inp in the following ways. The whole assembly of solid masks was further split into separated connected regions (which were defined by the face-to-face connectivity of the voxels within each mask), by selecting the appropriate option under mask filtering operations. The isolated "islands" that were not connected to the main body were then successfully removed. By choosing the "General Statistics" option, information on the number and volume of masks could be shown. The FE model created from the remaining mask was then meshed and exported for FEM simulation.

### *7.3.2 Mesh Generation*

In a preliminary study, the default meshing algorithm (i.e. +FE Grid algorithm) which uses a mesh morphing technique, was applied for mesh generation (Fig. 7-28 (a) and (b)). Meshes of sufficiently high quality can be directly imported into a range of commercial FE packages, such as Abaqus CAE used in the current study. Due to the complicated distribution of the irregular interconnected material and pore shapes in the microstructures, mesh generation using +FE Grid could result in a *.inp data file (ready to import into Abaqus) larger than 1 GB, containing over 7 million elements. A compromise had to be established between the size of the executable files and the computational capability. In the current study, the Abaqus CAE run on the workstation could not handle a *.inp file over 500 MB due to the very large transitory memory usage required. Therefore a more effective meshing algorithm, i.e. +FE Free, which offers greater flexibility in the element creation process and allows more control over the number of elements that are generated than +FE Grid, was applied to reduce the number of elements for the meshed model in order to reduce the computational demands.

The +FE Free meshing algorithm allows the multipart conforming surfaces to be automatically adapted to features in the segmentation. Several steps are involved in volume mesh generation (this is the reason why +FE Free requires longer time than +FE Grid).The algorithm first creates the +FE Grid mesh of high quality, which is characterised by a consistent element edge length based on the voxel size. The surfaces of the +FE Grid mesh are then extracted (single and multipart) and are then remeshed using the +FE Free meshing algorithm. It preserves small features in the model with fine local mesh generation. The remeshed surfaces are then filled with tetrahedral elements using a Delaunay/Advancing front approach [5].

Parameter setting for +FE Free consists of the "Compound Coarseness" slider to control the global coarseness of the model (see Fig. 7-28 (b)). At zero setting the surface is not remeshed but simply taken from the +FE Grid mesh and the interior filled using the free-meshing approach. If the slider is moved to the left (negative values) the remeshing algorithm will be activated and allowed to increase the element edge lengths based on features. The rate of change and size of the edge lengths will both increase as the slider is moved further to the left. If the slider is moved to the right (positive values) the remeshing will again be activated and will attempt to densify the mesh. In the current study, -20 or -30 was typically chosen to reduce the element number. Local mesh refinement could be achieved by applying the "Mesh refinement volumes" option, which allows refining the volume mesh around a VOI, with both +FE Grid and Free meshing algorithms. When the number of elements was too large but down-sampling would lead to loss of potentially significant features, the mesh refinement tools were used to set a lower global mesh density, but preserve some features by using a higher resolution locally. A smoothed surface mesh was generated for all the models. For a smoother resampling "Linear" was found to be a good option to choose. The "Advanced Parameters" could be set to manually adjust target element edge lengths.

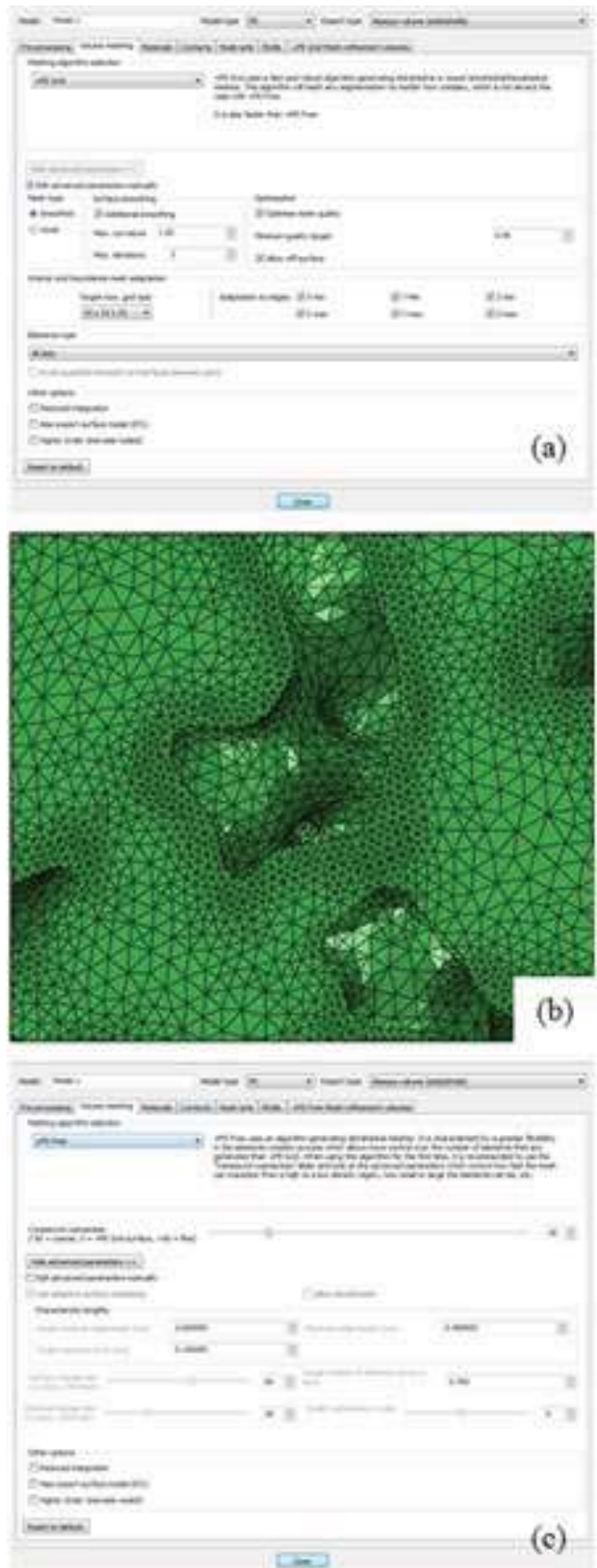

*Fig. 7-28 Mesh generation using the ScanIP +FE meshing module: (a) Parameter setting dialogue for +FE Grid, (b) a large number of hexahedral elements generated by using +FE Grid, (c) Parameter setting dialogue for +FE Free*

As a result, the number of elements was reduced to less than 1 million and hence the *.inp file size decreased to around 100 MB, which was adequate to be imported into Abaqus CAE for FEM simulation. However, it is worth noting that the time required for the +FE Free algorithm was dramatically increased compared to the use of +FE Grid on a 12-core workstation with 48 GB RAM (i.e. 10-30 mins for +FE Grid compared to several hours using +FE Free).

There were other different possible ways to reduce the number of elements, which included cropping and downsampling of the test models. It should be noted that in the current study, the meshing and modelling processes effectively treated the pores as holes within the microstructures, which means the pores (i.e. empty space) should not be meshed as in the case for computational simulation.

In the current study, the actual voxel size of the image data acquired by FIB/SEM slice and view process and treated using Avizo was 12.5 nm/pixel with 12.5 nm slice spacing. Taking a $10 \times 10 \times 10 \ \mu m^3$ VOI as an example, the voxel number would be $800 \times 800 \times 800 = 512$ million, which was too high for subsequent computation. Therefore, the data were first down-sampled to reduce the size of the model. Down-sampling, which also reduced the memory usage dramatically, was done with caution by using either image processing software (e.g. IrfanView or ImageJ) or by ScanIP with "Linear" as the background image interpolation method and the mask interpolation method. It is necessary to keep in mind that down-sampling (i.e. increasing the pixel spacing) by a factor of $r$ in all directions will decrease the models size by $r$ to the power 3. Thus if down-sampling by a factor of 2 in all three directions, a considerable difference in the memory used ($2 \times 2 \times 2 = 8$ times less) would be noticed. It is in general preferable to resample with the same resolution in $x$, $y$ and $z$ as this would provide with the best finite element qualities. In the current study down-sampling was made as much as possible while ensuring that the features of interest were preserved. The "ideal spacing" had to be chosen depending on the size of the smallest feature considered to be significant in the model. Therefore in the current study, the model was down-sampled by a factor of 2 in all directions, which gave a resolution of 25 nm/pixel and 25 nm for slice spacing. As a result the number of voxels was significantly reduced to 64 million.

In the ScanIP +FE module material properties could also be assigned. Node sets are useful during FEM as they can save time and effort when applying boundary conditions to the model in the FEM processor. To avoid time-consuming selection of node sets using Abaqus CAE and to facilitate the application of boundary conditions in FEM processor, the nodes on the six surfaces of the model were selected and defined as node sets in ScanIP.

Finally, a 4-node linear three dimensional stress tetrahedral element type (C3D4) was applied to generate meshes for the models. Fig. 7-29 (a) shows the meshed isoview of a microstructure after sintering at 1000 °C. Fig. 7-29 (b) shows an outer planar surface meshing (area in white), compared to the visible mesh inside the microstructure (area in green). Fig. 7-29 (c) represents in detail a magnified view of adaptive meshing with different mesh element size.

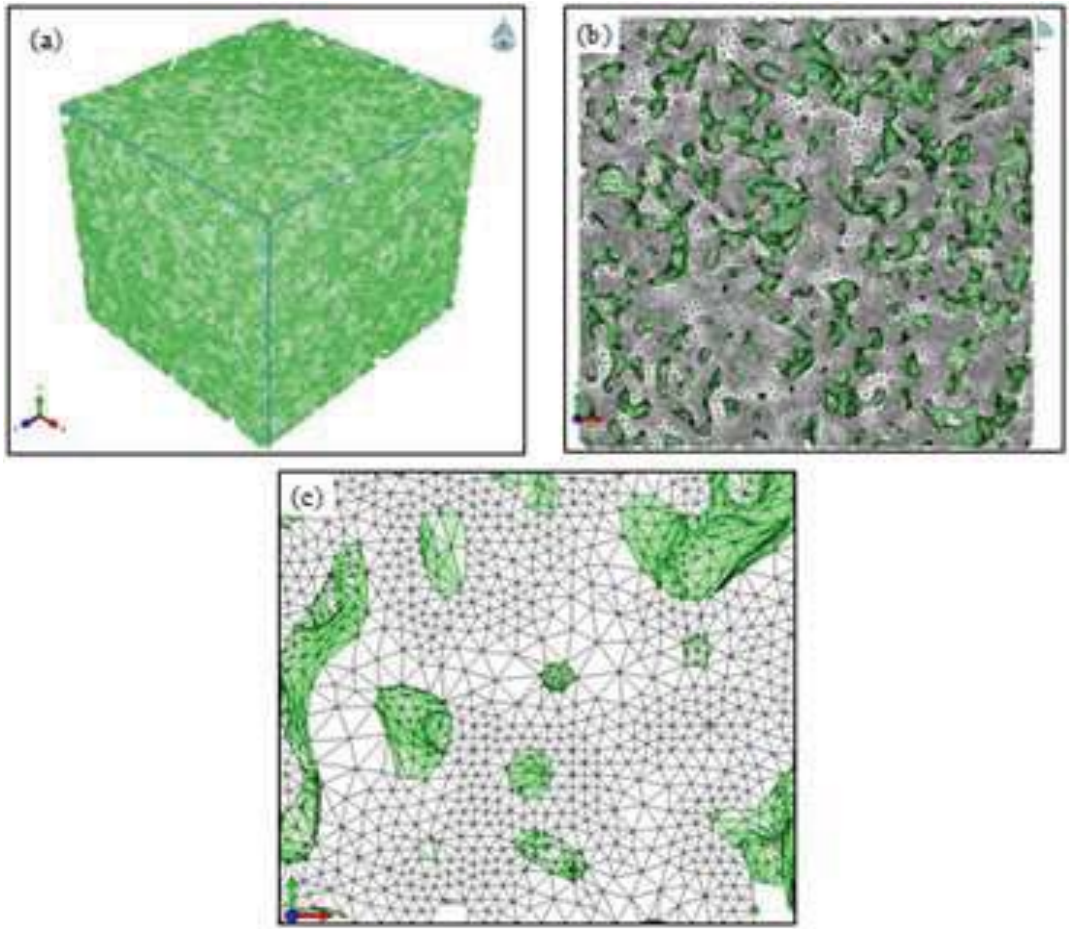

*Fig. 7-29 Mesh generation for a microstructure: (a) isoview of the meshed model; (b) surface view at Zmax; (c) adaptive mesh at higher magnification*

In the current study, at least three independent VOIs were sectioned in each as-sintered film to perform the reconstruction and the subsequent FEM. Approximately 2 million tetrahedral elements were generated for each VOI after the reconstructed 3D microstructures were meshed using ScanIP software.

## *7.4 Mechanical Modelling Using Abaqus CAE*

After the 3D microstructures were fully meshed, they were exported as FE compatible model with a file suffix *.inp and imported into Abaqus CAE for further finite element modelling, as described below.

### *7.4.1 Material Parameters*

In Abaqus CAE, the model imported possessed orphan mesh status, which means it was not generated by Abaqus and might be difficult to modify. The FEM simulation was run for each microstructure corresponding to a particular sintering temperature using the Abaqus Standard FE solver based on the assumption that the LSCF6428 solid was isotropic and linear elastic, irrespective of the sintering temperature. In Abaqus CAE, a Poisson's ratio of fully dense LSCF6428 material reported in literature [6], $v_s = 0.30$, was chosen for use in all FEM simulations. An elastic modulus of fully dense LSCF6428 material determined by nanoindentation in this study was chosen as the solid phase elastic parameter: $E_0 = 175$ GPa.

### *7.4.2  Boundary Conditions*

After the assignment of material properties, a new equation solver step named "step-1"of the general static type was created after "initial step" and "default time incrementation" were used. The boundary conditions which were created from "initial step" and propagated to "step-1" and displacement created from "step-1" applied on the node sets are shown in Table 7-1. It should be noted that mechanical simulation in the Y direction was not necessarily equivalent to the nanoindentation of the actual films in that direction because the nanoindentation-induced deformation of the films was also influenced by stresses in the other two directions. Therefore, normally each of the models was subjected to three simulation tests (i.e. Test 1, 2 and 3), corresponding to all three directions (X, Y and Z) to calculate the elastic moduli in the three directions. The corresponding three pairs of surfaces (node sets) were denoted as Xmin/Xmax, Ymin/Ymax and Zmin/Zmax and for the boundary conditions U1 means displacement in the X direction, U2 in the Y direction, U3 in the Z direction.

*Table 7-1 The boundary conditions as well as the compressive boundary displacement applied for each FEM model*

| Test No. | Direction | Boundary Conditions Applied | Displacement Applied (pixel) |
|----------|-----------|------------------------------|-------------------------------|
| 1 | X | Xmin: U1=0; Ymin: U2=0; Zmin: U3=0 | Xmax: U1=-1 |
| 2 | Y | Xmin: U1=0; Ymin: U2=0; Zmin: U3=0 | Ymax: U2=-1 |
| 3 | Z | Xmin: U1=0; Ymin: U2=0; Zmin: U3=0 | Zmax: U3=-1 |

An example is shown in Fig. 7-30 of the simulation in the Y direction. A small displacement (compression) was applied on one free surface normal to the Y direction (U2=-1 at Ymax), so that the model deformed linearly in the Y direction. The opposite face (node set Ymin) was constrained to have no displacement (U2=0 at Ymin) in this direction. Boundary conditions were also applied to constrain the degree of freedom of the normal displacement for the nodes on the model's other surfaces parallel to Y (U1=0 at Xmin; U2=0 at Ymin; U3=0 at Zmin). Such settings allowed these surfaces to move freely in the Y direction once the displacement was applied.

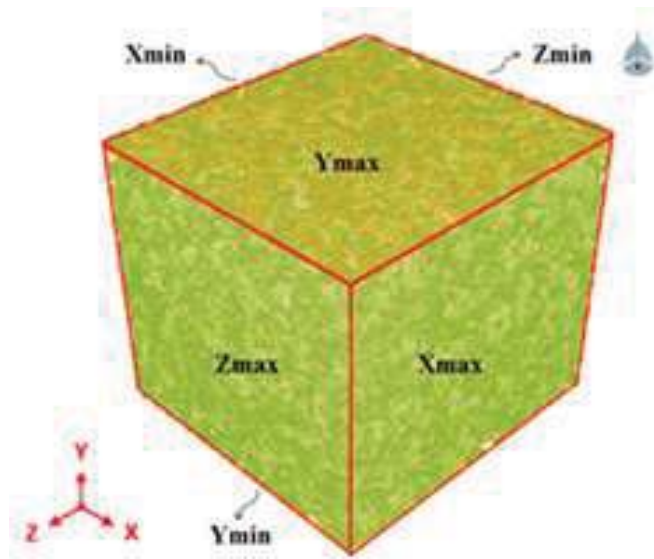

*Fig. 7-30 The meshed model under application of boundary conditions (spots in orange).*

It is worth noticing that, the boundary conditions for the three orthogonal node sets required that there existed one corner node shared by the three node sets Xmin, Ymin and Zmin (Fig. 7-31), which means that as a result this corner node had an encastre boundary condition (i.e. U1=U2=U3=0). Otherwise, an encastre boundary condition should be applied on a node on one of these node sets. If this constraint requirement was not applied, it would result in non-convergence of the FEM calculation.

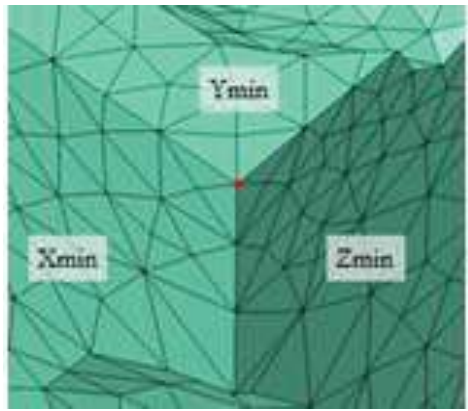

*Fig. 7-31 Corner node (red) shared by all three orthogonal node sets*

### 7.4.3 Modelling Output File Setting

New field output requests for the reaction forces (RF) of the Ymax node set and for the translational displacements (UT) of the Xmax and Zmax node sets at the last time increment were created (Fig. 7-32), before an analysis job was created for submission.

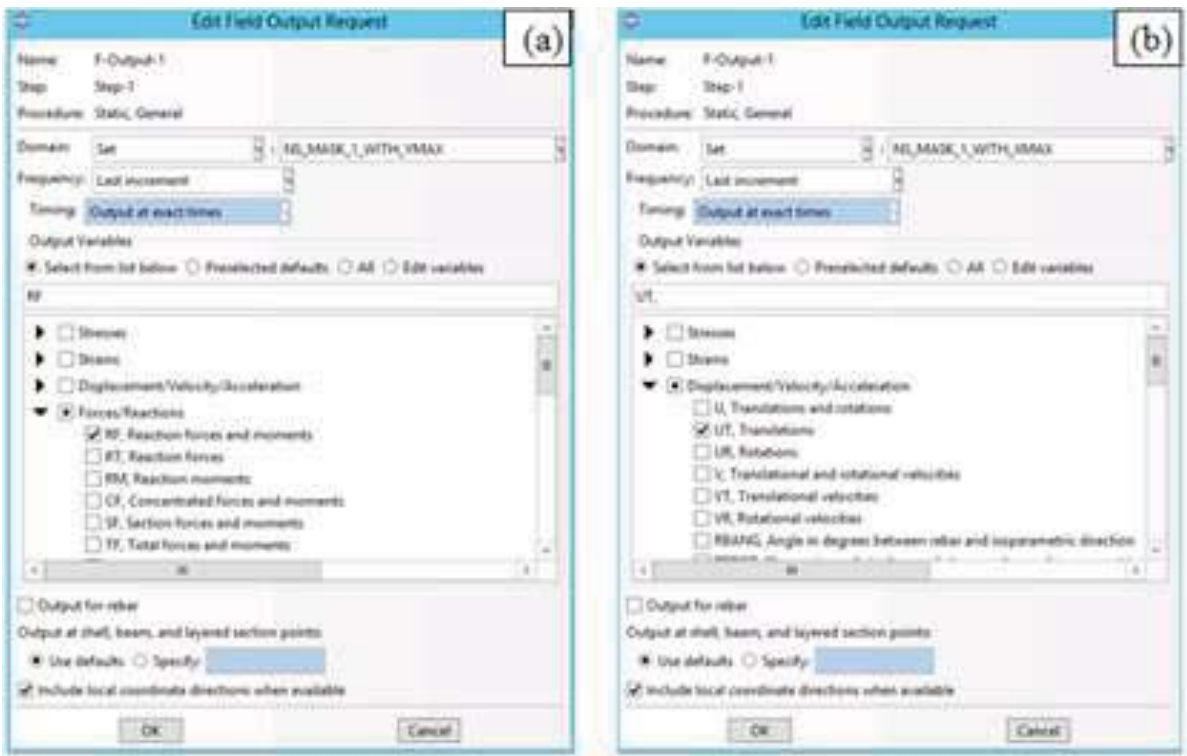

*Fig. 7-32 Setting of field output requests in Abaqus for FEM (a) force and (b) displacement*

After a successful analysis, the result could be obtained by first creating and saving the results for the node sets Ymax, Xmax and Z max, and then outputting the specific field variables (*RF2, U1* and *U3,* respectively) in a table report (Fig. 7-33) which included the integral values. *RF2* was then divided by the total surface area of Ymax surface ($A_{Ymax}$) to for the compression stress, $\sigma2 = RF2/A_{Ymax}$. The resulting strain in the Y axis $\varepsilon2$ was calculated as averaged displacement/height, i.e. $\varepsilon2 = U2/H_Y$. As a result, the effective elastic modulus in the Y direction was determined by $E2 = \sigma2/\varepsilon2$ and the corresponding Poisson's ratio was calculated by $v2 = (v21+v23)/2,$ where $v21= -\varepsilon1/\varepsilon2$ and $v23 = -\varepsilon3/\varepsilon2$ (both *v21* and *v23* should have the same value for an isotropic microstructure). In the same way, the effective elastic moduli *E1* and *E3* and the Poisson's ratios *v1* and *v3* in the other two axes were also calculated.

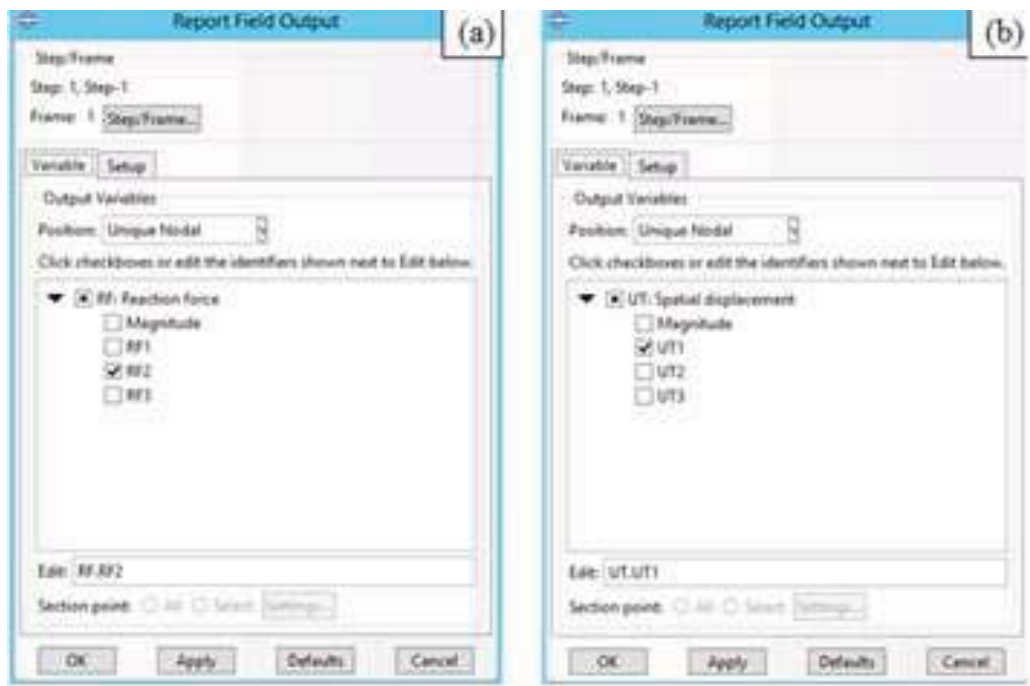

*Fig. 7-33 Field output report generation for (a) reaction force in the Y axis of node set Ymax and (b) translational displacement of node set Xmax in the X axis*

### 7.4.4 Mechanical Modelling Results

#### 7.4.4.1 Elastic Moduli

The average total time consumed to complete the simulation was less than 10 minutes for each microstructure model, on a Dell PRECISION T5500 workstation with a Linux OS and an Intel Xeon quad-core 64 bit 2.76 GHz processor with 46 GB RAM. Some of the detailed data on the segmentation, meshing, and FEM processes are presented in Table 7-2. Note that the different thresholding values for segmentation of different image stacks acquired are attributed to their respective grayscale distribution (brightness and contrast) caused by the image acquisition conditions of the time. From the table, it is also found that with the increasing sintering temperature (i.e. reducing porosity as shown earlier), the LSCF6428 voxel number increased while the element number after meshing decreased and hence the total simulation time was also diminished. This was because lower porosity could reduce the complexity of meshing by generating more regular mesh shape and adapting to larger mesh size.

*Table 7-2 Indicative data on the reconstruction and elastic simulation of the films*

| Sintering Temperature (°C) | Thresholding Grayscale Value | LSCF6428 Voxel Number ($10^6$) | Mesh Element Number ($10^6$) | FEM Simulation Time Used (min) |
|---|---|---|---|---|
| 900 | 40 | 7.34 | 2.15 | 8.2 |
| 1000 | 70 | 8.35 | 2.09 | 7.3 |
| 1100 | 35 | 10.49 | 1.86 | 6.8 |
| 1200 | 75 | 11.72 | 1.75 | 6.2 |

The reconstructed 3D microstructures are shown in Fig. 7-34 (a) - (d) in the order of increasing sintering temperature from 900 to 1200 °C. From these reconstructed microstructures, the gradual densification due to the increasing sintering temperature can be readily observed.

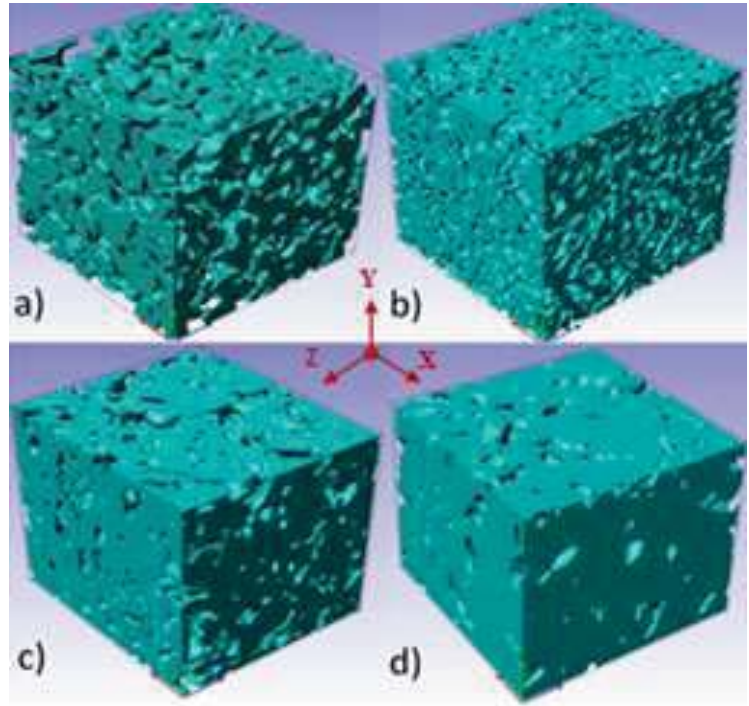

*Fig. 7-34 Examples of the reconstructed 3D microstructures of the 900-1200 °C as-sintered LSCF6428 films, with voids being pores*

Although assumptions of ideal properties of the dense solid (such as linear elasticity and isotropy) were made for the mechanical simulations of the 3D reconstructed microstructures, the actual microstructures might not be isotropic. One reason for possible anisotropy is that the films experienced constrained sintering which induced greater shrinkage along the direction normal to the surface than that in the other two directions. This was explored by comparing the elastic moduli calculated by FEM in different directions.

Table 7-3 shows the calculated elastic modulus in the three principal orthogonal directions for the microstructures of films sintered at different temperatures. Note that as explained earlier, although the primary displacement in nanoindentation test was in the Y direction normal to the film surface, due to the stress influence also coming from the X and Y directions, the averaged modulus over all the three directions would be the most appropriate to compare with the elastic modulus measured by nanoindentation.

*Table 7-3 Effective elastic moduli calculated of actual reconstructed 3D models*

| Sintering Temperature (°C) | E in X (E1) (GPa) | E in Y (E2) (GPa) | E in Z (E3) (GPa) |
|---|---|---|---|
| 900 | 26.2±1.5 | 36.2±1.2 | 22.9±2.6 |
| 1000 | 50.1±3.2 | 70.4±4.3 | 57.2±1.8 |
| 1100 | 95.4±5.2 | 109.1±6.4 | 90.6±3.4 |
| 1200 | 114.3±6.1 | 125.6±7.5 | 111.8±4.6 |

From the table, it can be seen that the elastic moduli in all the three directions increased with the increased sintering temperature, attributable to the overall densification (i.e. decrease of porosity) of the microstructures. More specifically, for all the samples the elastic modulus in the direction Y possessed a relatively larger value compared to that in the other two directions (i.e. X and Z), regardless of the sintering temperature. This anisotropy is expected from the aforementioned constrained sintering during which grain boundary diffusion and surface diffusion along the Y direction are inactivated due to the tensile stress in the film plane, while diffusion parallel to the film plane is active because there is no constraint in the Y direction. As a result, atoms of the solid phase are diffused to fill the pore area parallel to the film plane, resulting in the contraction of the pores along the X and Z directions and thus elongation normal to the film plane, although these differences cannot be readily distinguished in the 2D SEM image shown in Fig. 4-29 or Fig. 4-30 in Chapter 4, or in the reconstructed 3D microstructures shown in Fig. 7-34. Similarly the necks parallel to the film plane between particles aligned perpendicular to the film plane are relatively larger than those with other orientations, particularly in the initial stages of sintering or after sintering at lower temperatures. This is consistent with the observation found by Wang *et al.* [7] in the constrained sintering of YSZ films in which the pores are preferentially oriented in the direction normal to the film plane. Hence the modulus in direction Y is larger.

In order to effectively assess the variation of the directionally dependent elastic modulus, the results were normalised relative to elastic modulus along the Y direction, i.e. $E^* = E_i/E_2$ ($i$=1, 2 and 3), as shown in Fig. 7-35. The comparison and difference can be more readily seen with the normalised average values added. Because $E_1$ and $E_3$ for each type of model had similar values and were both smaller than $E_2$, the degree of anisotropy (DA) was thus be calculated and plotted for each sintering temperature: $DA = (E_2-(E_1+E_3)/2)/((E_1+E_3)/2)$.

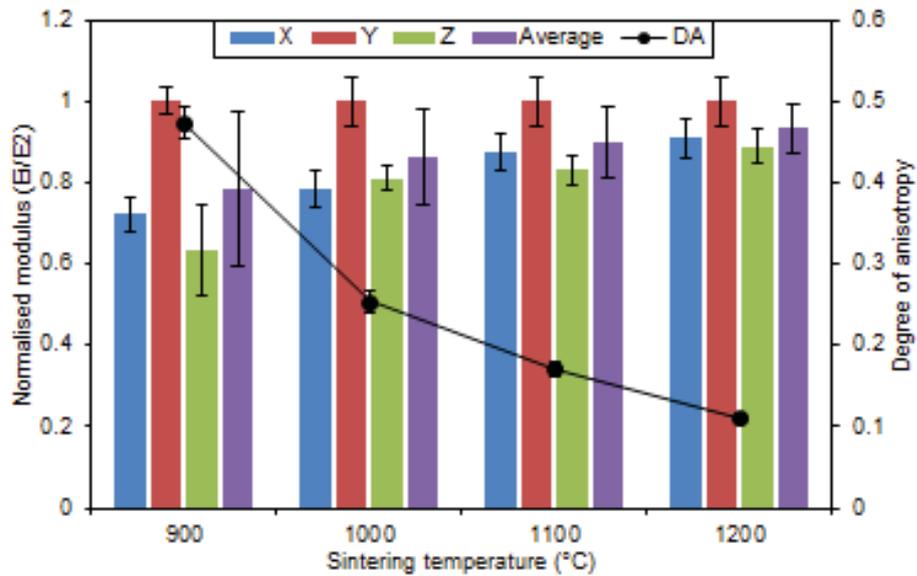

*Fig. 7-35 Normalised elastic modulus in three directions for films sintered at different temperatures.*

It can be seen from Fig. 7-35 that normalised elastic moduli in the X and Z directions have similar values but are both smaller than that in the Y direction, as found earlier with the cause explained. Besides this, another interesting aspect can also be found, the normalised elastic moduli in the X and Z directions experienced similar increases and thus tended to gradually approach the modulus in the Y direction with the rising sintering degree. This means that the variation of the calculated elastic moduli became increasingly smaller in all the three directions as the sintering temperature increased. Consequently, the degree of anisotropy decreases from about 0.5 to 0.1 with the increase of sintering temperature from 900-1200 °C. This decrease is also attributable to the enhanced surface diffusion along the Y direction with the increase of sintering degree. This resulted in the growth of necks aligned perpendicular to the X and Z directions, which accelerated the increase of elastic moduli in these two directions, so that the difference with the elastic modulus in the Y direction gradually diminished. Such a decrease can also be explained by taking highly dense samples for examples. Considering a sample of merely 1% porosity with rarely distribution of pores in the sample, elastic moduli along the three directions are thought to be the same so that the material is almost isotropic, irrespective of the pore distribution, shapes or orientations. In a word, the less porosity there is, the less difference it makes to elastic modulus, so the less the anisotropy it can introduce to the material.

### *7.4.4.2 Comparison with Nanoindentation Results*

The comparison of elastic moduli measured by nanoindentation with the elastic modulus derived from FEM is shown in Fig. 7-36. Note that *Enan* denotes the nanoindentation-derived elastic modulus and *Eavg* is the averaged value of *E1, E2* and *E3*.

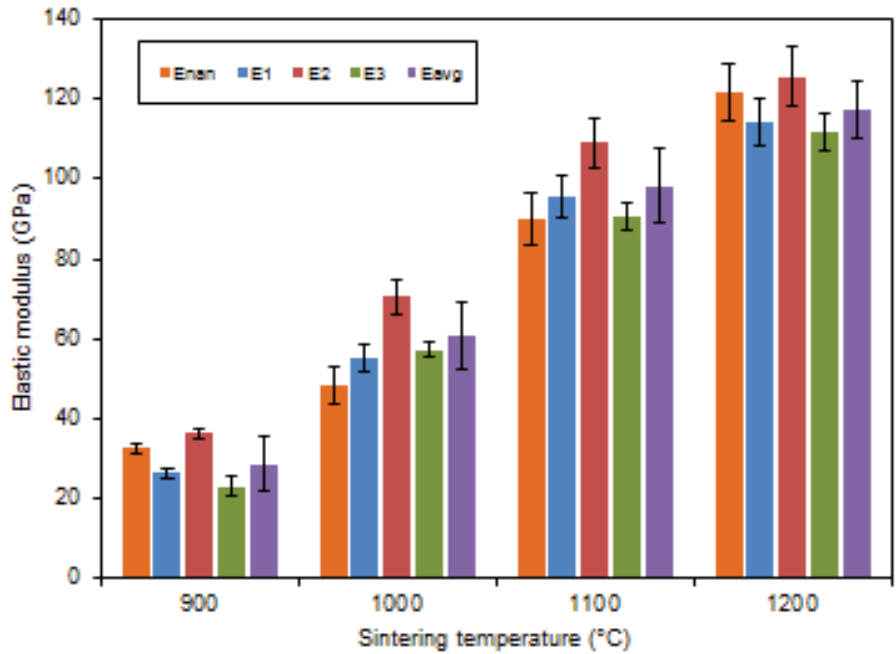

*Fig. 7-36 Comparison of elastic moduli measured by nanoindentation and calculated by FEM*

As explained earlier, the values for the X and Z directions in the film plane are very close to each other but smaller than that for the Y direction, which were due to the constrained sintering process. Although the samples exhibited different degrees of anisotropy, the average values were considered to better represent the true elastic modulus values of the porous thin films. Furthermore, the figure indicates that the elastic modulus measured by nanoindentation for each sintering temperature was reasonably consistent with the average of the elastic moduli in the three directions to within experimental error, although slightly lower values were derived from FEM than from nanoindentation for films sintered at 1000 and 1100 °C sintered films. The experimental errors of the elastic moduli calculated by FEM are caused by the porosity variation of the three different physical sampling locations for each type of model (i.e. VOIs).

### *7.4.4.3 Poisson's Ratios*

The Poisson's ratios of individual directions were calculated, as shown in Fig. 7-37 and Fig. 7-38. Note that the Poisson's ratios were preferentially oriented. Note that for the symbol "*vij*" shown in Fig. 7-37, *i* denotes the direction of the applied displacement and *j* the direction of the resulting displacement. While in Fig. 7-38 "*vi*" represents the averaged value of the corresponding two "*vij*" in the direction *i*. For example, in the case when the elastic modulus *E2* along the Y direction (i.e. direction 2) was simulated as described earlier, the direction 2 was applied a displacement, then the corresponding Poisson's ratio was calculated by *v2 = (v21+v23)/2*, where *v21= -ε1/ε2* and *v23 = -ε3/ε2* (both *v21* and *v23* should have the same value for an isotropic microstructure). The overall average shown in Fig. 7-38 was based on the three values (i.e. *v1, v2* and *v3*) for each specimen sintering temperature.

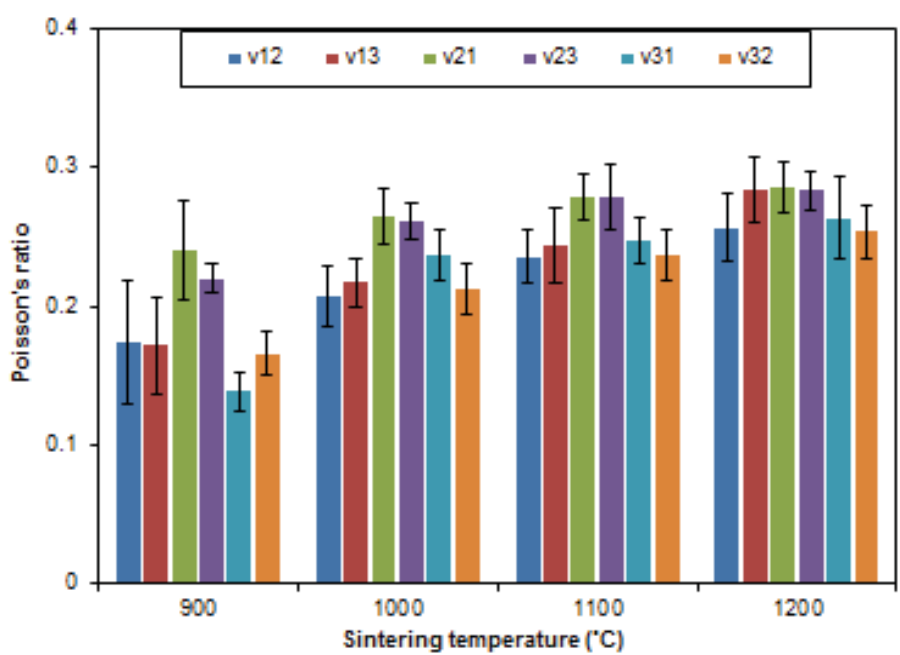

*Fig. 7-37 Comparison of Poisson's ratios for each individual directions calculated by FEM*

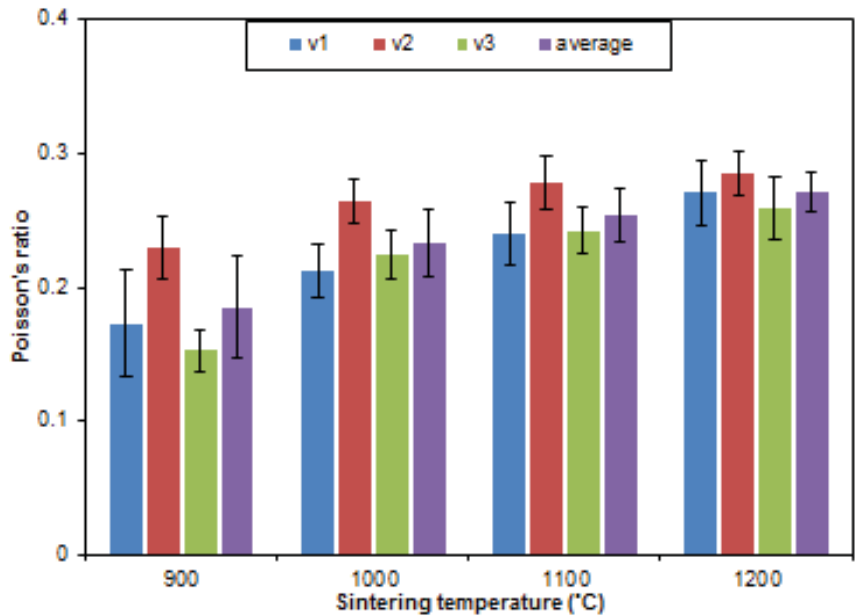

*Fig. 7-38 Comparison of the averaged Poisson's ratios calculated by FEM for each specimen sintering temperature*

It is found from Fig. 7-37 that the two Poisson's ratios (*vij*) in each direction had very close values. This is particularly true for *v21* and *v23*, due to the uniformity in the film plane being different from the direction normal to the plane (i.e. the direction 2 or Y), as result of the constrained sintering.

More clearly, Fig. 7-38 shows a trend fairly similar to that of the elastic moduli as shown earlier in Fig. 7-36. The overall averaged value increased as the sintering temperature increased, from 0.19 for 900 ˚C to 0.27 for 1200 ˚C. Meanwhile, the standard deviation decreased with the increased sintering temperature. For each specimen sintering temperature, *v2* remained to be larger than *v1* and *v3* (both of which also had similar values), while the difference became less obvious as the sintering temperature grew. The reasons for all the above behaviours regarding the Poisson's ratios are considered to be the same as found in the case of the elastic moduli described earlier.

## 7.5 Quantification of Microstructural Parameters and Their Influence on Elastic Modulus Simulation

It was concluded in Section 5.2.5 of Chapter 5 that apart from porosity there could be other factors controlling the elastic properties of the samples, which might include for example particle (or pore) size distribution, neck sizes between the separated particles and their orientations and so on, which require further consideration.

### 7.5.1 Application of Avizo's Separate Objects Module

As explained earlier in Section 7.2 of this Chapter, Avizo allows the quantification and analysis of other fundamental parameters of the reconstructed 3D microstructures such as volume fraction and surface area of constituent phases and tortuosity factor of the phase of interest. More specifically, Avizo's advanced *Separate Objects* module (which is based on watershed, distance transform and numerical reconstruction algorithms [8]) enables the separation of interconnected particles and pores into smaller units. Thus further quantification of detailed microstructural properties is possible, including the determination of particle (or pore) size distribution, neck sizes between the separated particles (or pores) and their orientations, all of which might have an influence on the mechanical properties of the films.

For automated separation of objects, the *Separate Objects* module involves the identification of interconnected regions by using a 3D watershed algorithm. Analogous to flooding of geographic watersheds of a basin, the watershed algorithm simulates the progressive immersion from a set of labelled regions in a 2D or 3D image, and expands the regions according to a distance map until the regions reach at the same final watershed lines [9]. The application of the module in a 3D image for *particle* or *pore* separation used the following sequence (this example is for "particles" of the solid phase) [10]. First, a segmentation of phases was performed by thresholding to assign each voxel as either pore or solid. Then a dilation algorithm was used to expand the surrounding pores to the centre of each possible "particle" to find the number of voxel steps needed. The centroid of each particle was filled last. Using the centroids and the number of dilation steps as a distance map, the 3D watershed algorithm was applied to divide the image into individual particles. Next, the interface between any two separated particles was identified as a particle "neck". Finally, the distributions of particle (or pore) size, neck sizes between the separated particles (or pores) and their orientations were determined.

### *7.5.2 Quantification Methods of Microstructural Parameters*

Besides the conventional volume fraction and surface area of particles and pores in the films, the determinations of some other important parameters are explained here.

### *7.5.2.1 Tortuosity Factor*

Tortuosity (denoted as $\tau$) is a parameter to describe the extent of twisted curves and is defined as the ratio of the length of a curve to the distance between the two ends of it. In the current study, tortuosity was measured by applying Avizo's *Centroid Path Tortuosity* module on the binary 3D dataset, based on a computed path formed by centroids of each interconnected region identified as same phase on each plane of the dataset along the Z axis as shown in Fig. 7-39. As a consequence, $\tau$ was calculated by dividing the path length through the centroids by the straight length between the two ends of the path.

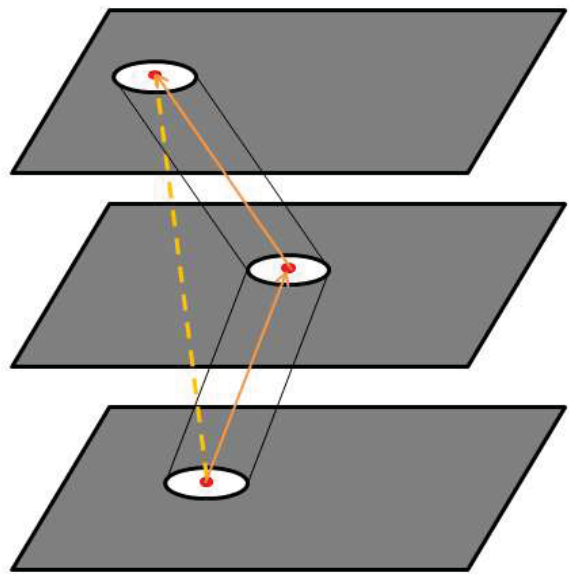

*Fig. 7-39 Schematic of computing tortuosity (in an interconnected pore structure) using the Centroid Path Tortuosity module of Avizo, after [8]*

It should be noted that for a volume of interest, the tortuosities of solid and porous phases may be different. In the current study both tortuosities were measured.

### *7.5.2.2 Particle and Pore Size Distribution*

Depending on which phase was of interest, the separation of the corresponding interconnected regions in a 3D dataset was performed by following the steps described earlier. After a solid/porous phase was separated into a group of individual particles (or pores), the volume of each particle/pore was converted to an *Equivalent Spherical Diameter (ESD)* for plotting the particle/pore size distribution histogram, which involved the frequency of a series of *ESD* ranges. *ESD* of a particle/pore in 3D can be obtained by $ESD = (6V/\pi)^{1/3}$, where *V* denotes the particle/pore volume.

### 7.5.2.3 Sphericity of Particles and Pores

Sphericity, which is the parameter describing how spherical a particle shape is, is defined as the surface area of a sphere possessing the same volume of the given particle divided by the surface area of the particle. As a result, a sphere's sphericity is 1 and any non-spherical particles should have sphericity less than 1. The derivation can be expressed as follows:

$$\psi = \frac{\pi^{1/3}\left(6V\right)^{2/3}}{A} \qquad 6.1$$

where *V* and *A* are the volume and surface area of the given particle, respectively. As examples, a cube's sphericity is 0.806, a 4-faces tetrahedron has a sphericity of 0.671 and a cylinder with a height two times its radius possesses a sphericity of 0.874.

### 7.5.2.4 Neck Size of Interconnected Particles

Necks are defined as a 2D interface between two interconnected particles when separation is performed. They are also known as minimum solid area (MSA) as proposed by Rice [11], in a porous microstructure. The surface area of a neck was converted into an *Equivalent Circular Diameter (ECD)* for neck size distribution analysis. The *ECD* of an interfacial neck can be derived by $ECD = 2(A/\pi)^{1/2}$, where *A* denotes the neck surface area.

The importance of necks is due to the fact that, as explained previously, they serve as load bearing areas when microstructures are subjected to mechanical constraints. Therefore, the size of the load bearing areas (i.e. neck size) is a key to the elastic modulus of the microstructures. Here a simple 2D illustration is demonstrated in Fig. 7-40 of microstructures at constant 2D porosity to emphasise the significance of the particle neck size to the resistance to mechanical constraints.

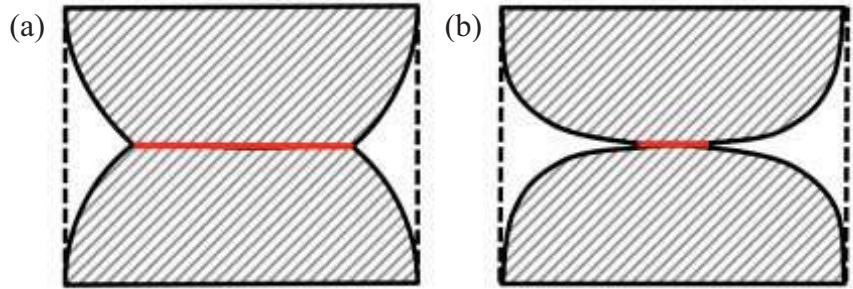

*Fig. 7-40 Necks (shown as lines in red) of two interconnected particles (cross-hatched areas) in simple microstructures at constant 2D porosity: (a) large neck size; (b) small neck size.*

As can be seen in Fig. 7-40 (a), large neck size resulting from a greater overlapping area of two interconnected particles would induce a stronger inter-particle connection and less stress concentration, and thus macroscopically a higher external vertical uniaxial load (compressive or tensile) has to be imposed to yield a same vertical strain when compared with the case where only a smaller neck present (Fig. 7-40 (b)). Therefore, it would be expected that in a porous single phase microstructure larger neck size generally leads to a higher effective elastic modulus.

### 7.5.2.5 Orientation of Individual Particle and Pore

The orientation of individual particles and pores was determined by measuring the direction of its major axis. It is given as the eigenvector of the largest eigenvalue of the inertia matrix [8]. The orientation of an enclosed shape in 3D space is expressed as *Orientation Phi* ($\phi$) relative to the Y axis and *Orientation Theta* ($\theta$) relative to the X axis, as depicted in Fig. 7-41. The resulting $\theta$ falls between -180˚ and +180˚, while $\phi$ falls between 0˚ and 90˚.

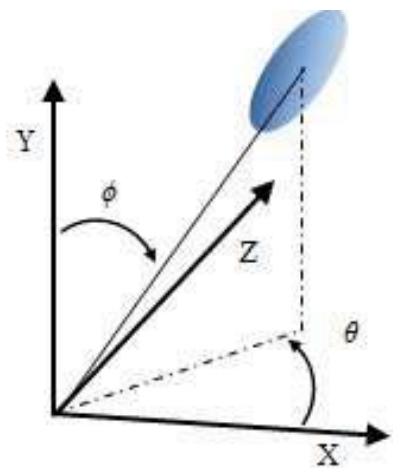

*Fig. 7-41 Schematic of orientation of an object in 3D space*

### *7.5.2.6 Typical Processing, Quantification and Analysis Route*

Here a typical route of processing, quantification and analysis using Avizo is demonstrated in Fig. 7-42 for solid phase (particles) in a LSCF-9 (i.e. LSCF6428 film sintered at 900 ˚C) dataset. Icons in green on the left are either datasets or quantitative results before and after applying the relevant module commands which are displayed on the right in red and attached to the left icons by blue lines.

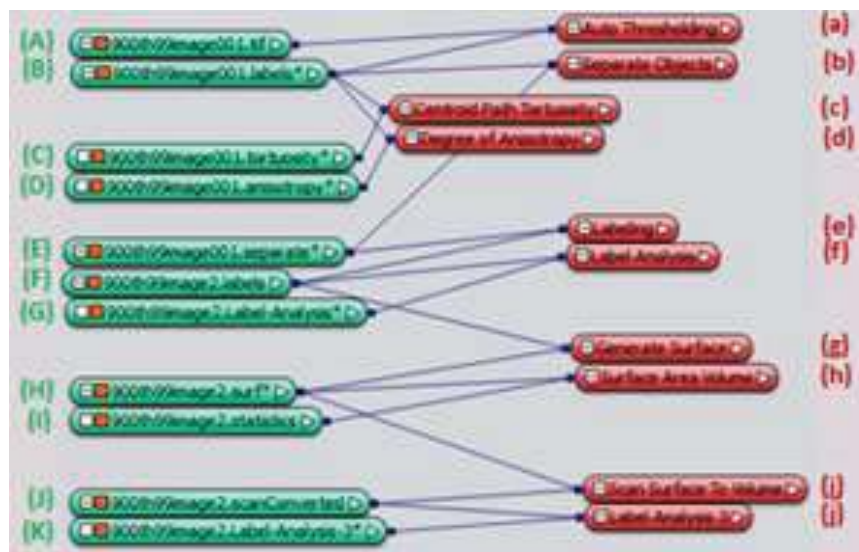

*Fig. 7-42 Processing, quantification and analysis route using Avizo for a dataset*

The route could be described as follows:

**(i)** A dataset containing a serial binary image stack (A) generated from image segmentation described in Section 6.2 was imported as the entry dataset;

**(ii)** *Auto Thresholding* module (a) was applied to segment (A) into gray level-labeled data of two phases (B) (i.e. solid=0 and pores=1);

**(iii)** Tortuosity (C) and degree of anisotropy (D) were directly measured by applying (c) and (d) modules on data (B);

**(iv)** *Separate Objects* module (b) was applied on dataset (B) for separation of particles of the interconnected solid phase, whereby the resulting dataset was (E);

**(v)** *Labeling* module was applied on (E) for assigning the same gray level value to all the pixels belonging to the same object (i.e. particle). As a result a different gray level value starting from 1 was assigned for each object (i.e. each particle) depending on the location of the objects in the image, resulting in dataset (F);

**(vi)** *Label Analysis* module (f) was used to compute a group of measures (such as surface area, volume, ESD and orientation of particles) on each separated or connected components of the input dataset. Note that customised measures could be created from the list of existing ones and the analysis results were read-only spreadsheets shown in (G) which could be imported into external spreadsheet programmes such as Microsoft Excel for further processing (e.g. value editing, chart plotting and quantitative analysis).

**(vii)** *Generate Surface* module (g) was used on (F) to compute a triangular approximation of the interfaces (H) between different material types (i.e. phases). Since individual particles were obtained based on the application of *Separate Objects* module, interfaces (i.e. necks) between previously connected particles were produced, of which the surface areas were then calculated in patch mode by *Surface Area Volume* module (h). The resulting spreadsheet data (I) were further processed to estimate the ECD of the 2D necks.

**(viii)** In order to retrieve the orientation data of necks, *Scan Surface to Volume* module (i) was applied to compute a volumetric representation of closed neck surface (J). This way the neck orientation (K) could be estimated using the *Label Analysis* module (j). During the above procedure, *Ortho Slice* and *Surface View* commands could be used on certain datasets to realise the visualisation of 2D and 3D features.

The processing, quantification and analysis of the pore phase followed exactly the same route as described for the solid phase, but the entry dataset was segmented into phases with inverse gray level value (i.e. solid=1 and pores=0) due to the fact that Avizo only takes into account the interior (gray level=0) as enclosed material for subsequent object separation.

### 7.5.3 Results for As-sintered Microstructures

Quantification of the particle phase was first performed following the above procedure for all datasets acquired from films sintered at 900-1200 ˚C. The resulting microstructural parameters are compared and analysed in this section.

### 7.5.3.1 Visualisation of the Processed Datasets

Fig. 7-44, Fig. 7-43, Fig. 7-45 and Fig. 7-46 respectively illustrate the 2D *Ortho View* and 3D *Surface View* of the *particle* datasets generated from the LSCF6428 films sintered at 900-1200 ˚C during and after processing. Once particle separation was performed based on the watershed algorithm, individual particles are readily seen in Fig. 7-44 (d), Fig. 7-43 (d), Fig. 7-45 (d), and Fig. 7-46 (d). One may assess the quality of the particle separation when looking at the watershed lines shown in the images, but note that as the separation was carried out in 3D mode, the watershed lines in the image were not as straightforward as separating a 2D image. Furthermore, it is noteworthy that all the separated particles and pores shown in these pictures have extremely irregular shapes, which represent the complexity of the actual 3D microstructures of the films.

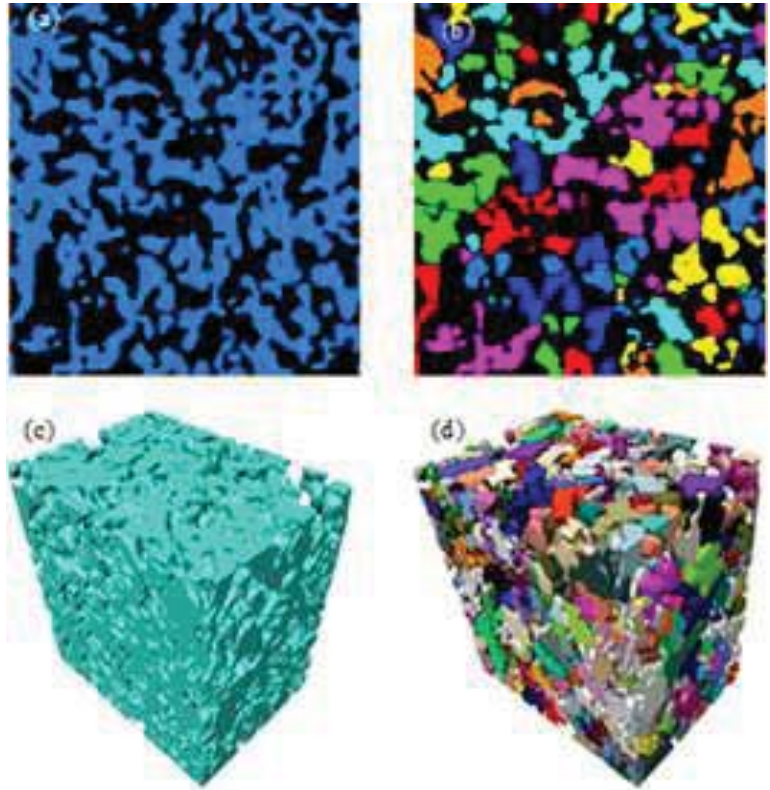

*Fig. 7-43 LSCF-9 particle dataset: Ortho view of an image (a) after "Auto Thresholding"; (b) after "Separate Objects" and colour labelling; 3D surface view of the dataset (c) before particle separation and (d) after particle separation. Individual particles are readily seen.*

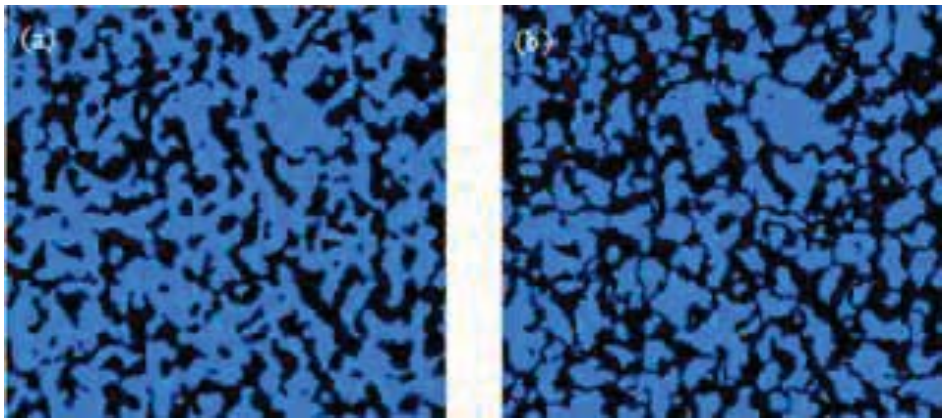

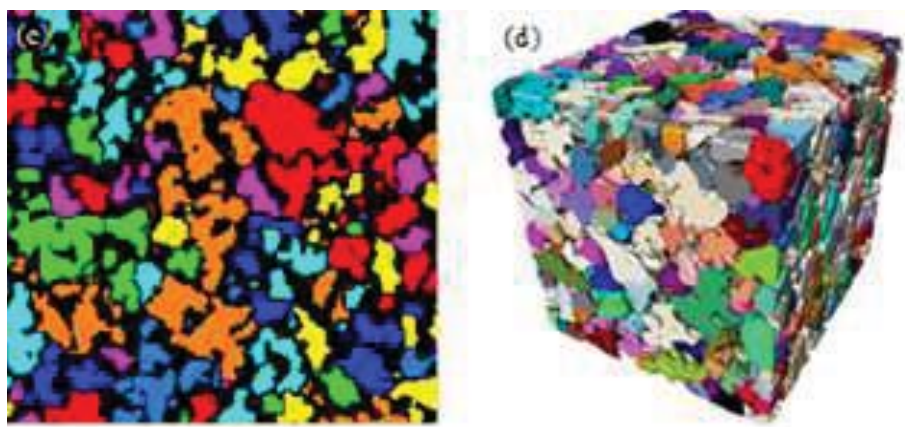

*Fig. 7-44 LSCF-10 particle dataset: Ortho view of an image (a) after "Auto Thresholding". Blue denotes particles and black pores; (b) after "Separate Objects", showing watershed lines cross the solid phase; and (c) after labelling with different colours; (d) 3D surface view of the dataset after particle separation.*

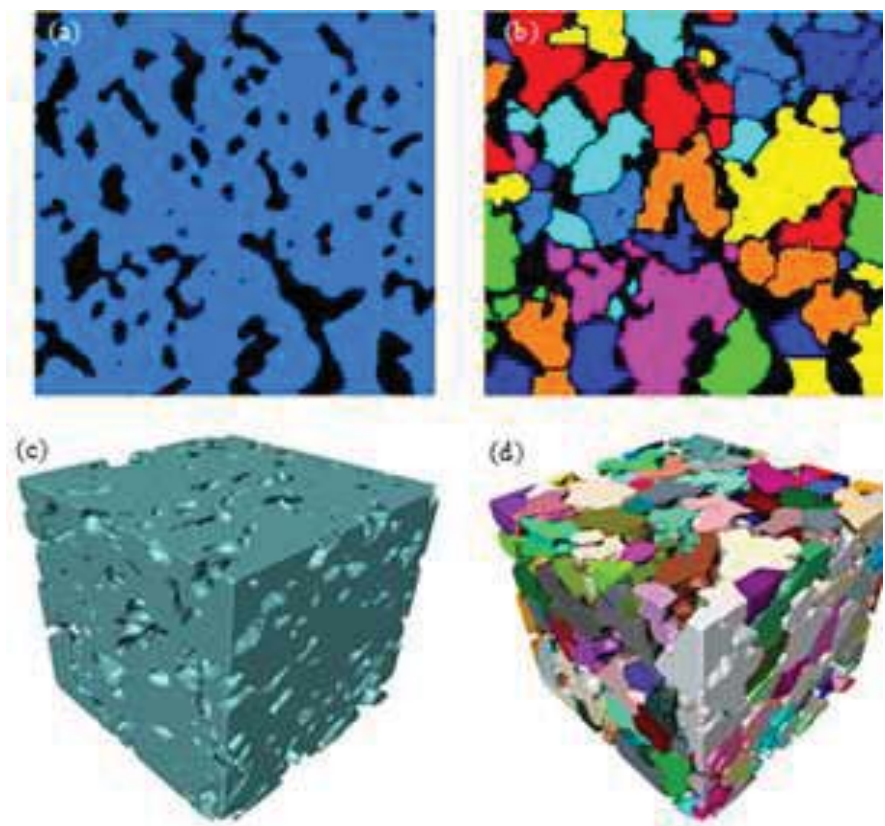

*Fig. 7-45 LSCF-11 particle dataset: Ortho view of an image (a) after "Auto Thresholding" and (b) after "Separate Objects" and colour labelling; 3D surface view of the dataset (c) before particle separation and after particle separation.*

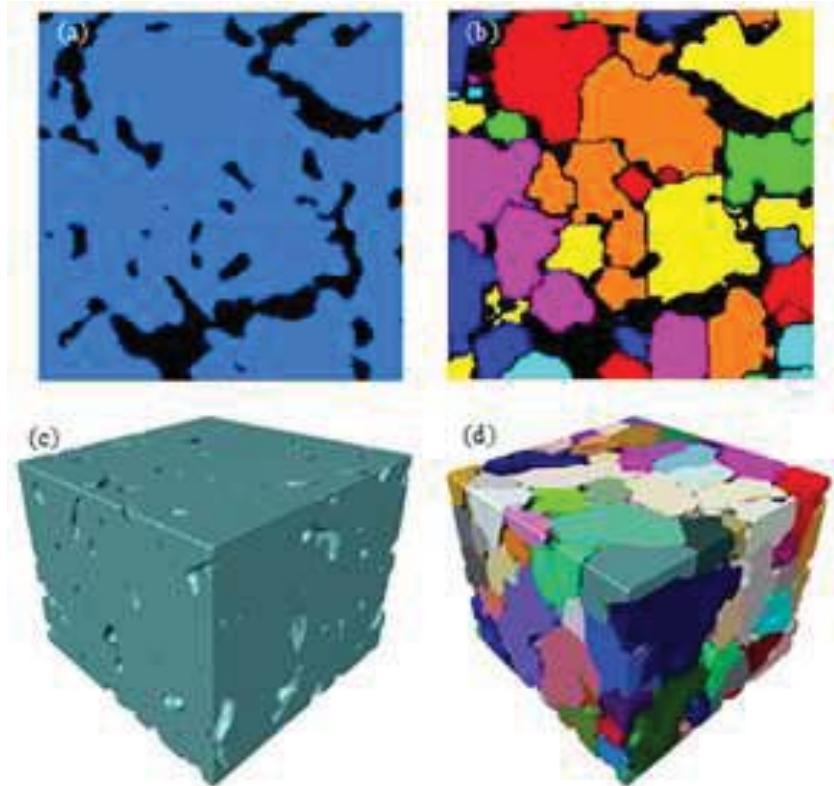

*Fig. 7-46 LSCF-12 particle dataset: Ortho view of an image (a) after "Auto Thresholding" and (b) after "Separate Objects" and colour labelling; 3D surface view of the dataset (c) before particle separation and after particle separation.*

In addition, the pictures in Fig. 7-47 illustrate respectively the 3D *Surface View* of the *pore* datasets generated after processing. Again, very irregular shapes are noticeable.

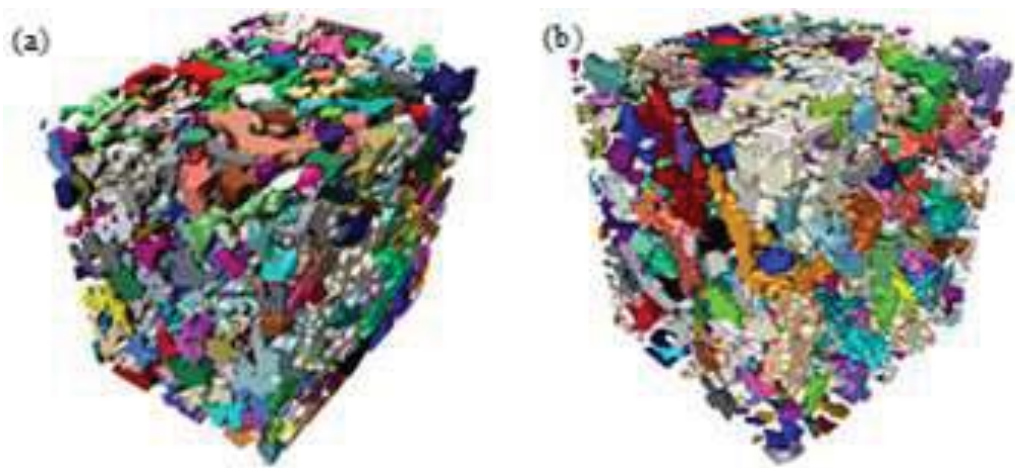

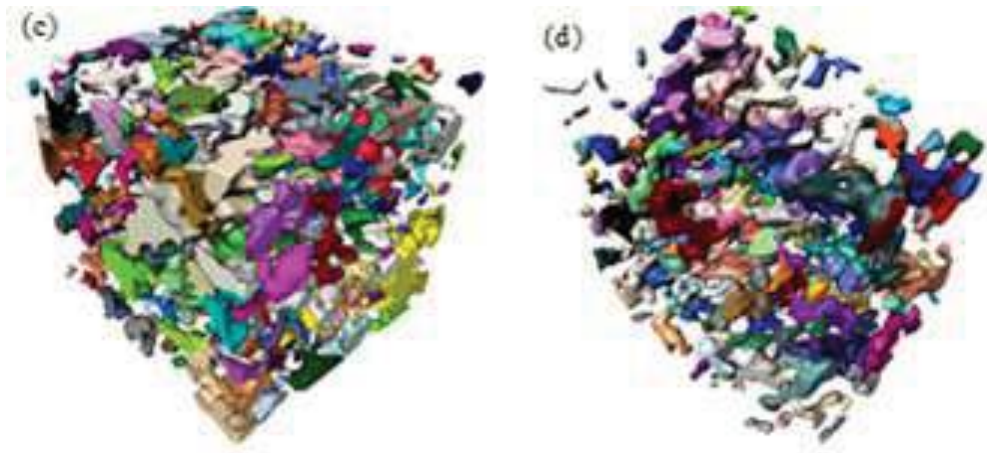

*Fig. 7-47 3D surface view of pores after object separation for (a) LSCF-9, (b) LSCF-10, (c) LSCF-11, (d) LSCF-12*

### 7.5.3.2 Spatial Relationship between Separated Particles

A detailed insight of the spatial relationship of the separated particles which were originally interconnected is demonstrated here by an example showing two selected particles in a LSCF-10 film. Their system numbering and spatial relationship are shown in Table 7-4 and microstructures and positioning shown in Fig. 7-48.

*Table 7-4 Example of relationships of particles and necks in the Avizo analysis*

| Patch No. | Material No. | Particle | Neck | Particle Pair for Neck |
|-----------|--------------|----------|------|------------------------|
| P972 | M365 | Yes | No | N/A |
| P1020 | M382 | Yes | No | N/A |
| P1023 | N/A | No | Yes | M362/M382 |

A series of numbers was given by the system to distinguish the separated physical components (noted as *patch*) required for analysis in the test microstructure. Two types of components are identified and investigated, namely *particles* and interfacial *necks* between adjacent particles. When a separated patch represents a solid material (i.e. particle), a corresponding material number is given to the patch. In the example shown in Table 7-4, there are two particles (M365 and M382) which formed a particle pair M365/M382 linked by an interfacial neck P1023. Note that a neck which is between two joining particles could have more than one interfacial facet. Note also that in Fig. 7-48, the solid particles are illustrated as hollow bodies.

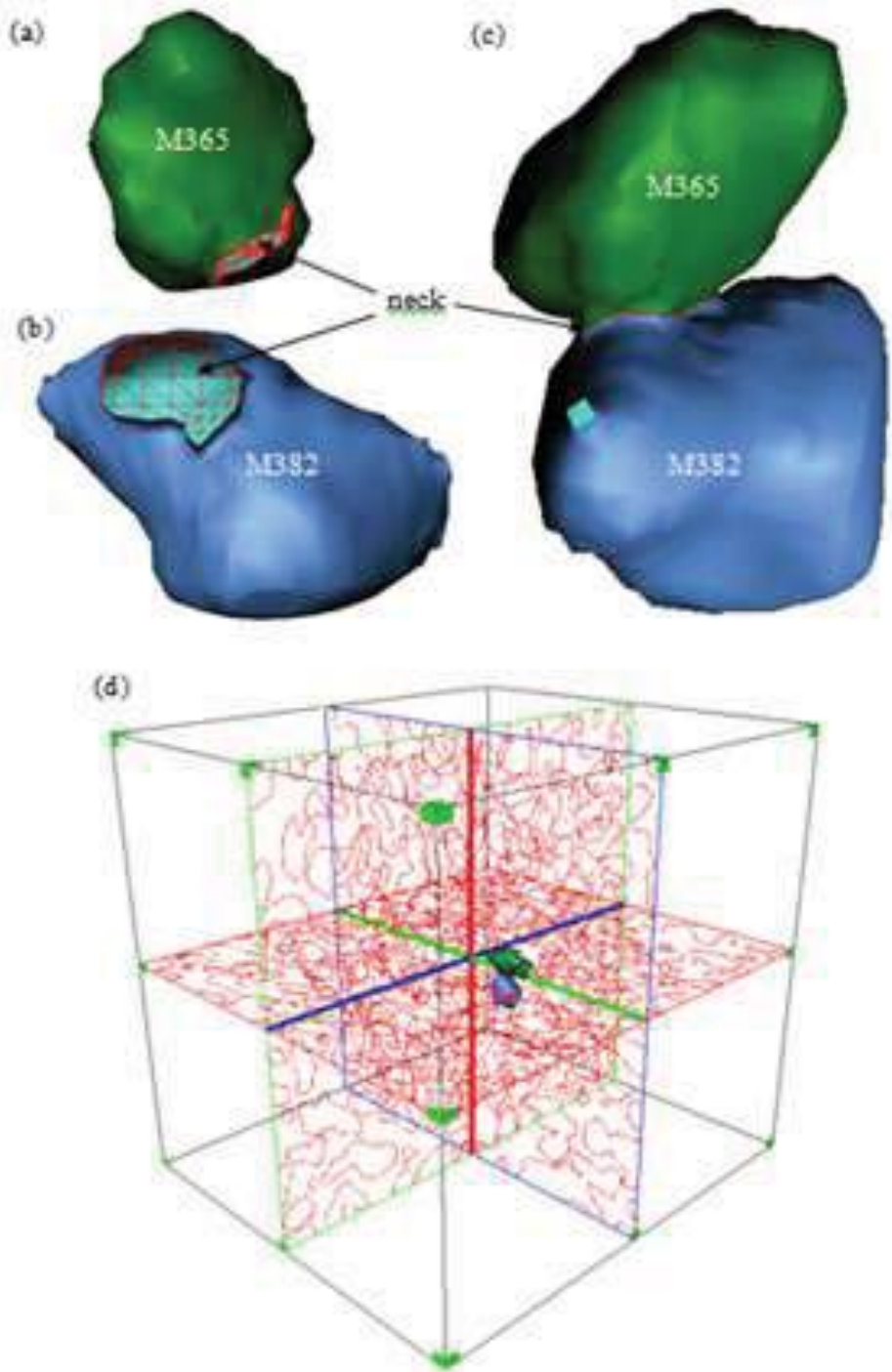

*Fig. 7-48 Example of particles and interfacial neck: (a) particle M365; (b) particle M382; (c) particle pair M365/M382; (d) positioning of the particles in the sample volume. Note that the red meshed areas in (a), (b) and (c) indicate the same interfacial neck*

### 7.5.3.3 Tortuosity Factor of Particle Phase and Pore Phase

The tortuosities of the solid particle and pore phases were measured individually for comparison of all datasets of films sintered over the temperature range, as shown in Fig. 7-49, where 3D total porosities are also plotted.

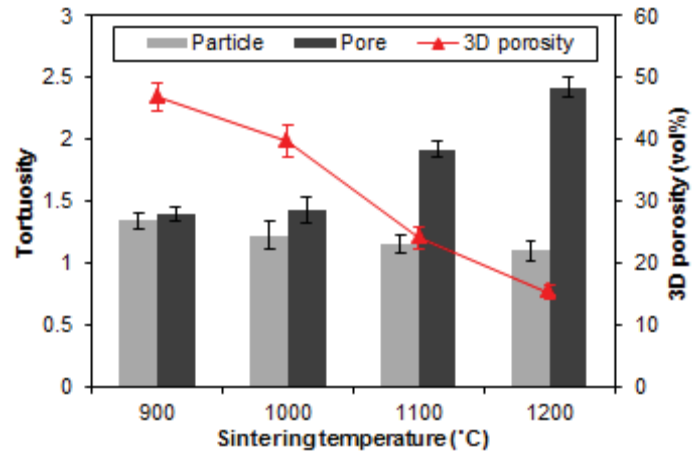

*Fig. 7-49 Comparison of tortuosities for particle and pore phases in all sintered at different temperatures*

Similar tortuosities were found for both particle and pore phases in LSCF-9 films, with a value of approximately 1.4. This similarity was expected as the porosity was close to 50 vol%. As the porosity decreased with the increasing sintering temperature, the tortuosity of the pore phase significantly increased, reflecting much more tortuous pore phase in the films after sintering at higher temperatures. On the other hand, the tortuosities of the solid particle phase experienced a slight decrease, which was attributed to the major part occupied by solid in these films.

### 7.5.3.4 Separated Particles: ESD, Sphericity and Orientation

The resulting Equivalent Spherical Diameters (ESDs) of the separated particles for each sample type were plotted as size distribution histograms, as shown in Fig. 7-50. Note that compared with the distribution of the other types of samples, the plot for LSCF-12 (Fig. 7-50(d)) exhibits a slightly broader ESD distribution, which might be attributed to the higher volume fraction of solid phase.

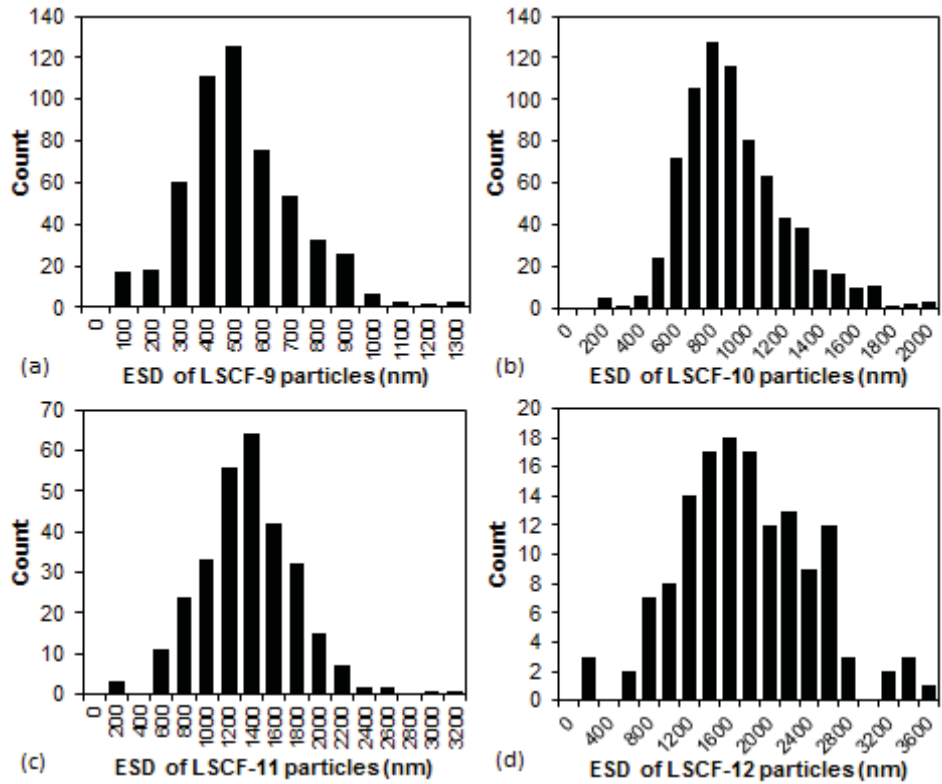

*Fig. 7-50 ESD of separated particles in the samples: (a) LSCF-9, (b) LSCF-10, (c) LSCF-11, (d) LSCF-12*

The average particle ESD determined based on Avizo's particle separation algorithm of each sample was and is compared in Table 7-5 with the average grain size estimated from 2D SEM images of the corresponding film top surface shown in Fig. 4.29. Both approaches show the significant increase of average size with increasing sintering temperature. However, the average ESD from 3D analysis are typically 2-3 times the average grain size for each sample. This suggests that at least two grains were incorporated in each "particle" which had been generated by partitioning using Avizo's *Separate Objects* algorithm. Although the average ESD values can be used in an indicative way for comparison, it is worth noticing that the large standard deviation shown in Table 7-5 suggests that it is more appropriate to represent the particle size as a probability distribution (Fig. 7-50) rather than looking at the mean particle size only.

*Table 7-5 Average particle ESD vs. average grain size*

| Sample type | Average ESD (nm) | Average grain size (nm) |
|---|---|---|
| LSCF-9 | 470±180 | 200 |
| LSCF-10 | 850±290 | 270 |
| LSCF-11 | 1280±-430 | 450 |
| LSCF-12 | 1700±630 | 690 |

The distribution histograms of sphericity of the separated particles for each sample type are also drawn in Fig. 7-51. The average sphericity was measured to be 0.61±0.11, 0.57±0.11, 0.66±0.08 and 0.69±0.06 for LSCF-9, LSCF-10, LSCF-11 and LSCF-12, respectively, which are close to the sphericity of a 4-faced tetrahedron (=0.671).

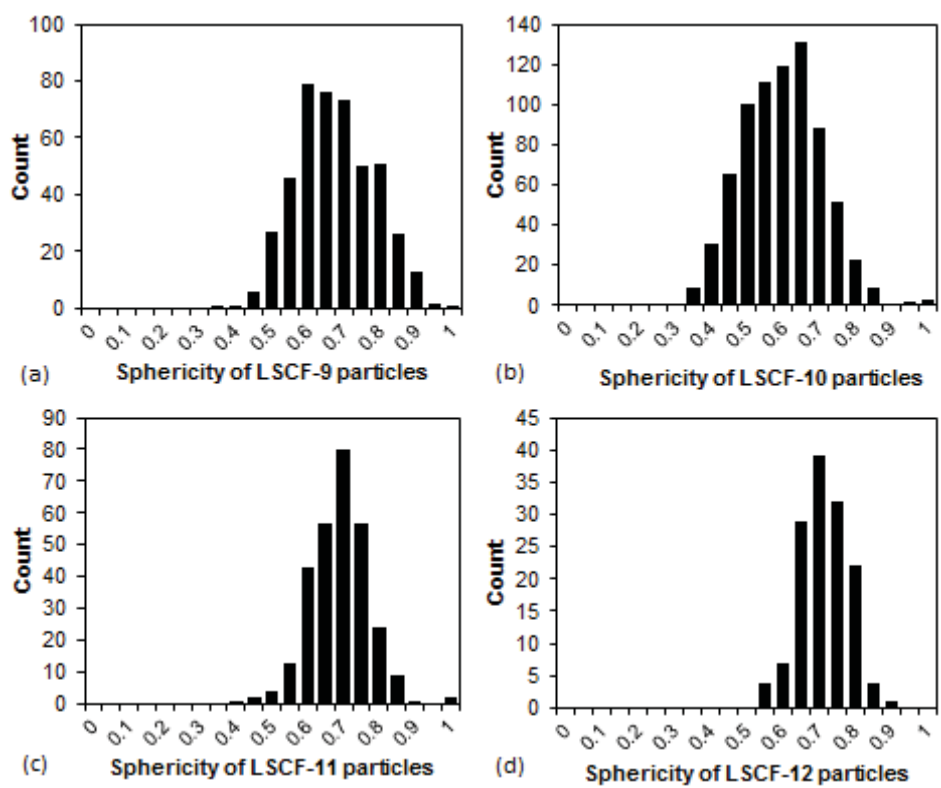

*Fig. 7-51 Sphericity of separated particles in the samples: (a) LSCF-9, (b) LSCF-10, (c) LSCF-11, (d) LSCF-12*

The orientation angles ($\phi$ and $\theta$) for the separated particles for each microstructure were calculated and their distribution histograms are shown in Fig. 7-52 and Fig. 7-53, respectively. The histograms in Fig. 7-52 all reveal that the majority of particles possessed orientation $\phi$ greater than 45˚, suggesting the orientation preference of particles is perpendicular to the Y axis (normal direction to the plane of the film). This would be consistent with the existence of larger neck size parallel to the X-Z plane and hence result in larger $E$ modulus in the Y direction. However, for the distributions of orientation $\theta$ shown in Fig. 7-53, a large number of 90˚-oriented (i.e. perpendicular to the X-Y plane, in the Z axis) particles are noticeable in LSCF-10 and LSCF-11. The development of preferred orientation in the plane of the films is not understood.

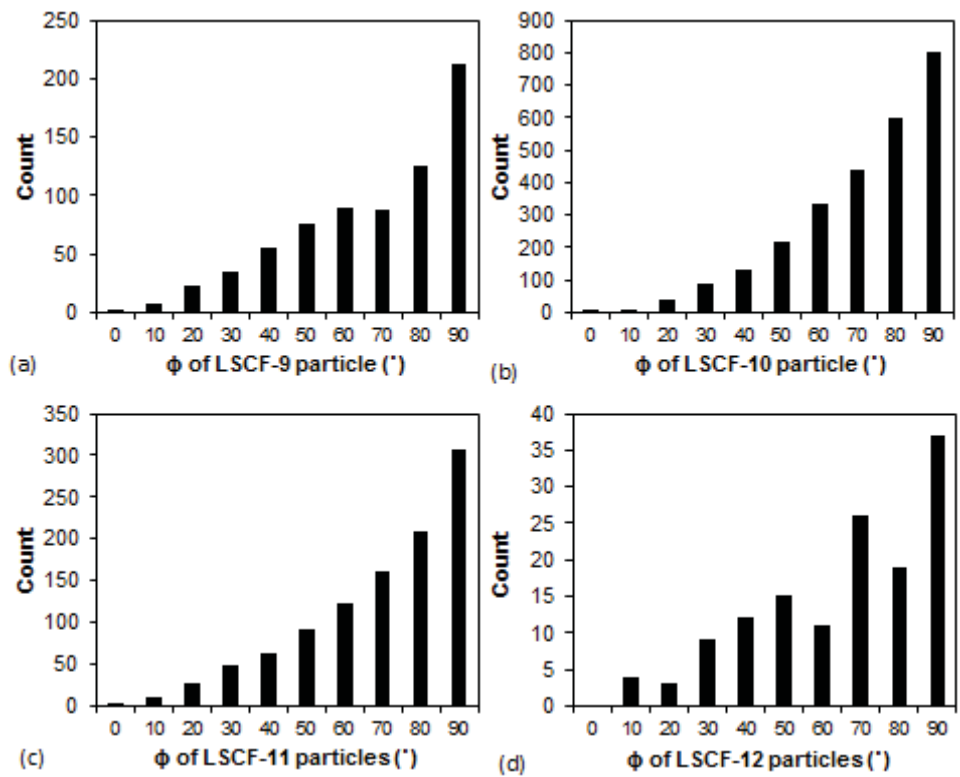

*Fig. 7-52 Particle orientation $\phi$ (out of the film plane) of separated particles in the samples: (a) LSCF-9, (b) LSCF-10, (c) LSCF-11, (d) LSCF-12*

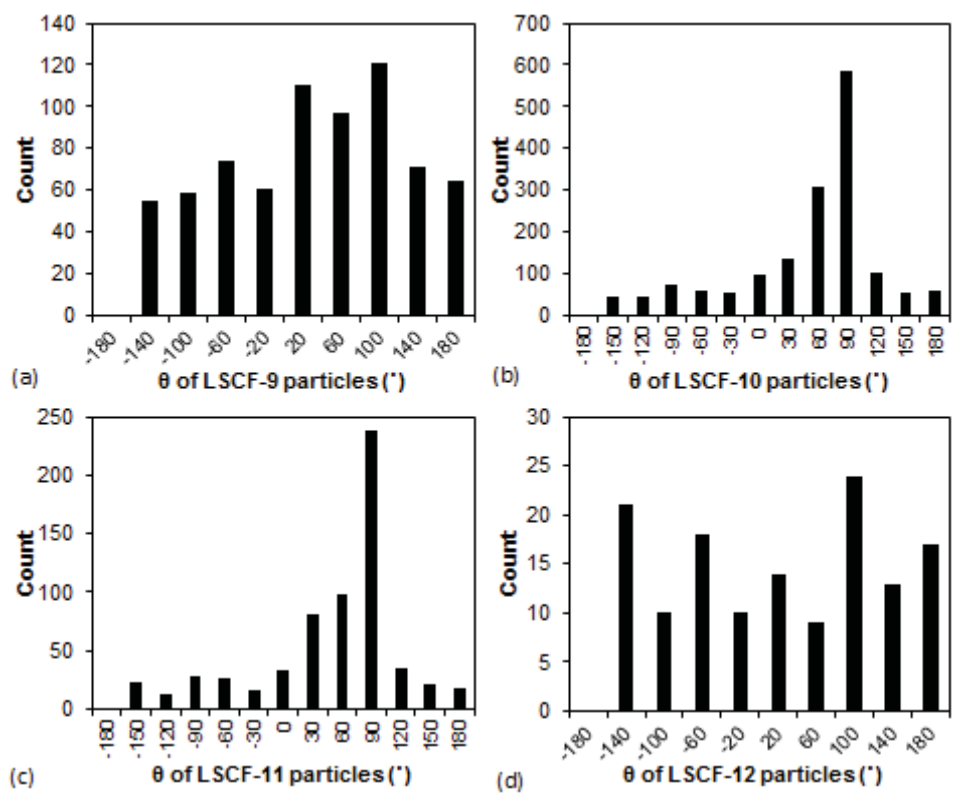

*Fig. 7-53 Particle orientation θ (within the film plane) of separated particles in the samples:*
*(a) LSCF-9, (b) LSCF-10, (c) LSCF-11, (d) LSCF-12*

### 7.5.3.5 ECD of Inter-particle Necks and Distribution

The distribution of Equivalent Circular Diameter (ECD) of the resulting necks between interconnected particles for each sample type is shown as histograms in Fig. 7-54.

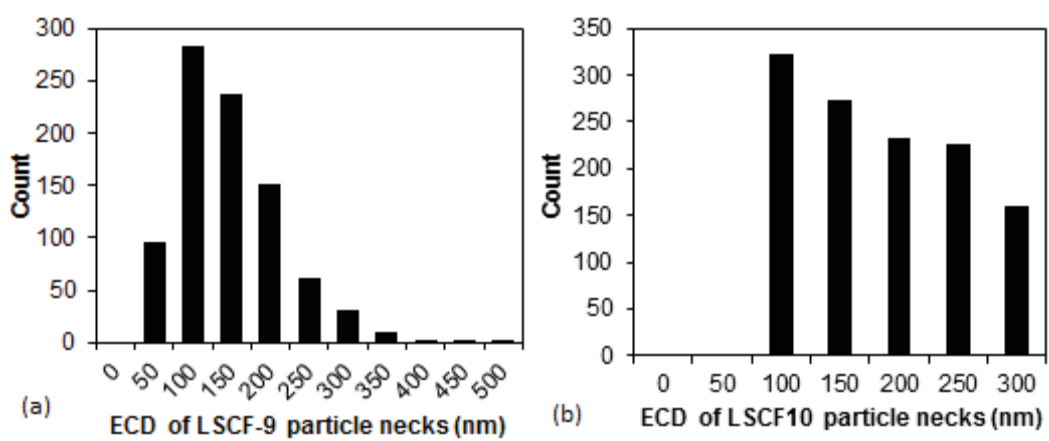

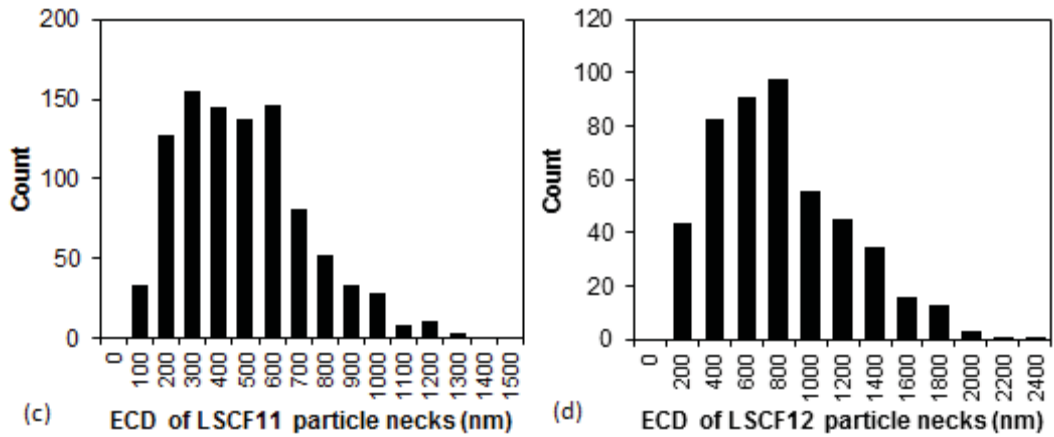

*Fig. 7-54 ECD of particle necks in the samples: (a) LSCF-9, (b) LSCF-10, (c) LSCF-11, (d) LSCF-12*

The average particle neck *ECD* measured was 123±66 nm, 191±101 nm, 440±245 nm and 706±418 nm, respectively for LSCF-9, LSCF-10, LSCF-11, and LSCF-12. Despite the large standard deviation (i.e. broad range of *ECD* distribution), the increasing trend of ECD with the increase of sintering temperature is consistent with the reduction of porosity in the samples.

### 7.5.3.6 Separated Pores: ESD, Sphericity and Orientation

The resulting ESD distribution and sphericity of the separated pores for each sample type were plotted in Fig. 7-55 and Fig. 7-56, respectively.

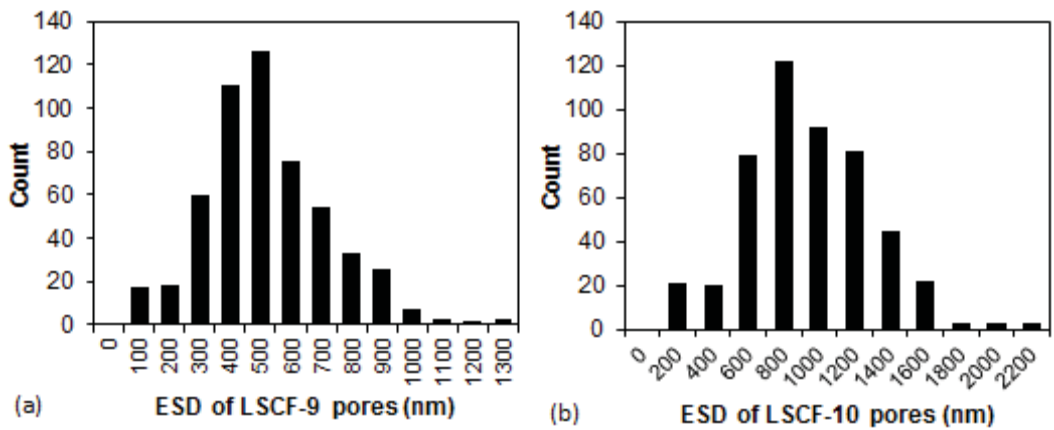

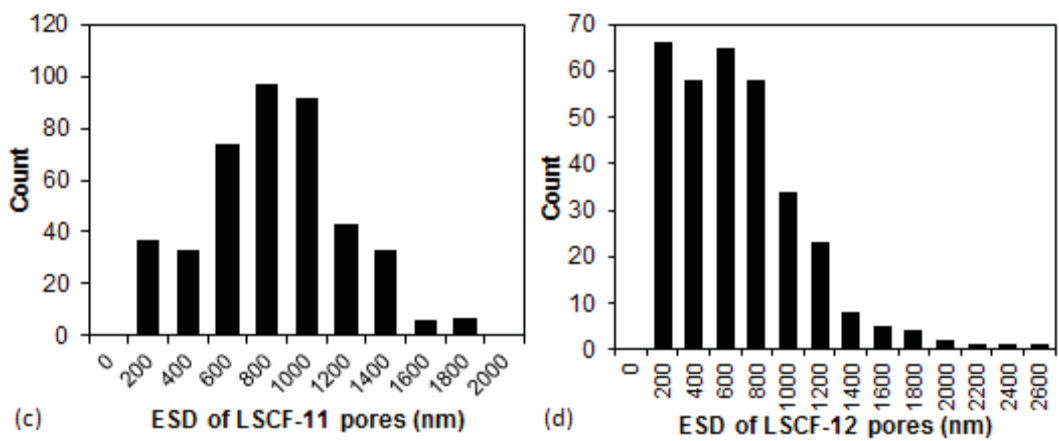

*Fig. 7-55 ESD of separated pores in the samples: (a) LSCF-9, (b) LSCF-10, (c) LSCF-11, (d) LSCF-12*

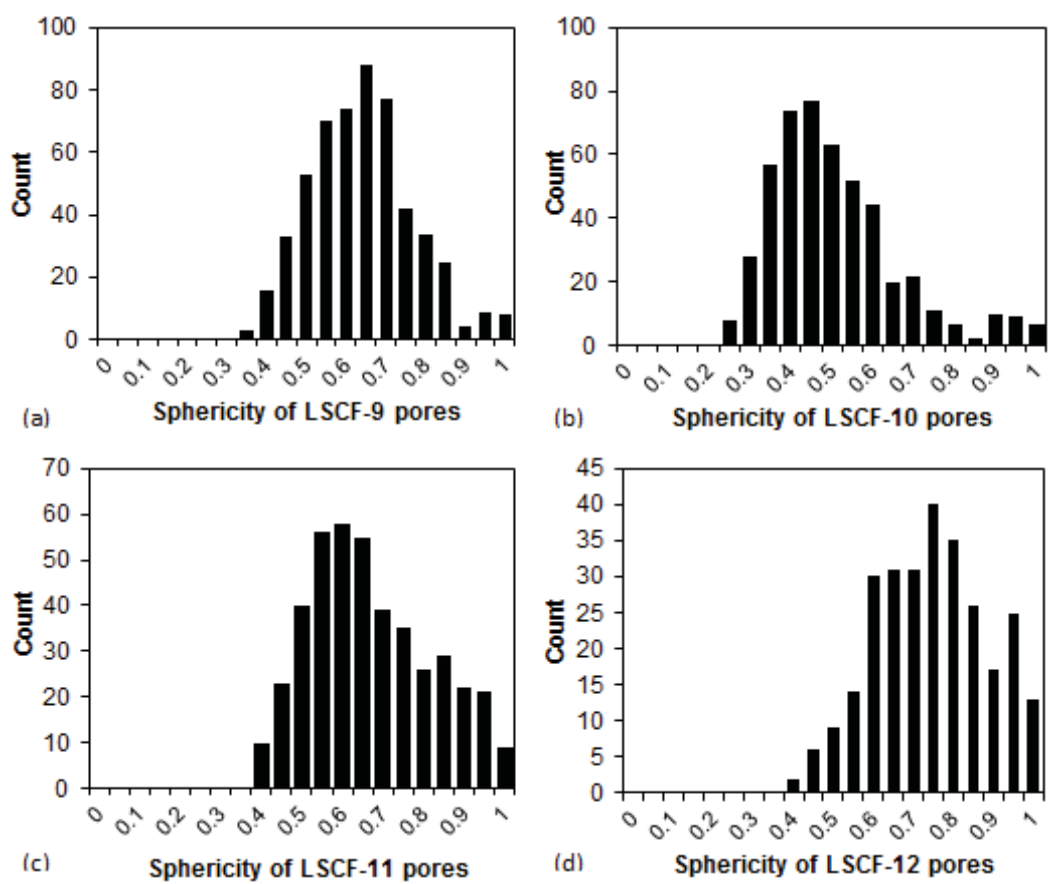

*Fig. 7-56 Sphericity of separated pores in the samples: (a) LSCF-9, (b) LSCF-10, (c) LSCF-11, (d) LSCF-12*

The average pore size for each type of microstructure was calculated to be 472±204 nm, 821±289 nm, 743±351 nm and 659±375 nm, respectively. The reduced porosity resulting from higher sintering temperatures was responsible for the decrease of pore size. Different from the other distribution plots in Fig. 7-55, it can be seen from Fig. 7-55 (d) that a large number of pores in the LSCF-12 microstructure had ESD size smaller than 800 nm. The average sphericity of the separated pores was measured to be 0.61±0.12, 0.51±0.13, 0.65±0.15 and 0.72±0.14 for LSCF-9, LSCF-10, LSCF-11 and LSCF-12, respectively. It is found that for lower sintering temperatures such as 900 and 1000 ˚C, the sphericity of pores was very close to that of particles due to the similar volume fraction of pore and solid phases. On the contrary for the specimen sintering temperature of 1200 ˚C, the pores had a higher sphericity than that of particles. This is attributed to the appearance of an increasing number of close pores in more spherical shape with the increase of densification degree at higher temperatures.

The orientation angles ($\phi$ and $\theta$) for the separated pores for each microstructure were calculated and their distribution histograms are shown in Fig. 7-52 and Fig. 7-58, respectively.

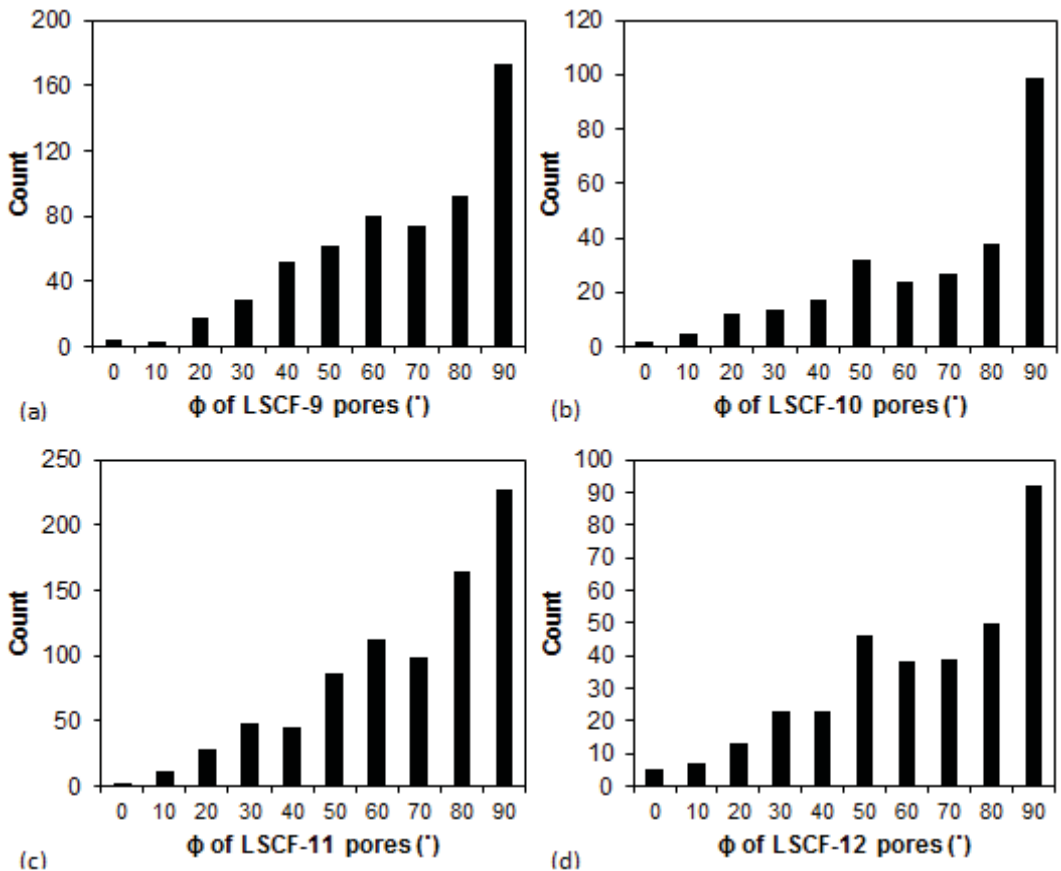

*Fig. 7-57 Orientation $\phi$ (out of film plane) of separated pores in the samples: (a) LSCF-9, (b) LSCF-10, (c) LSCF-11, (d) LSCF-12*

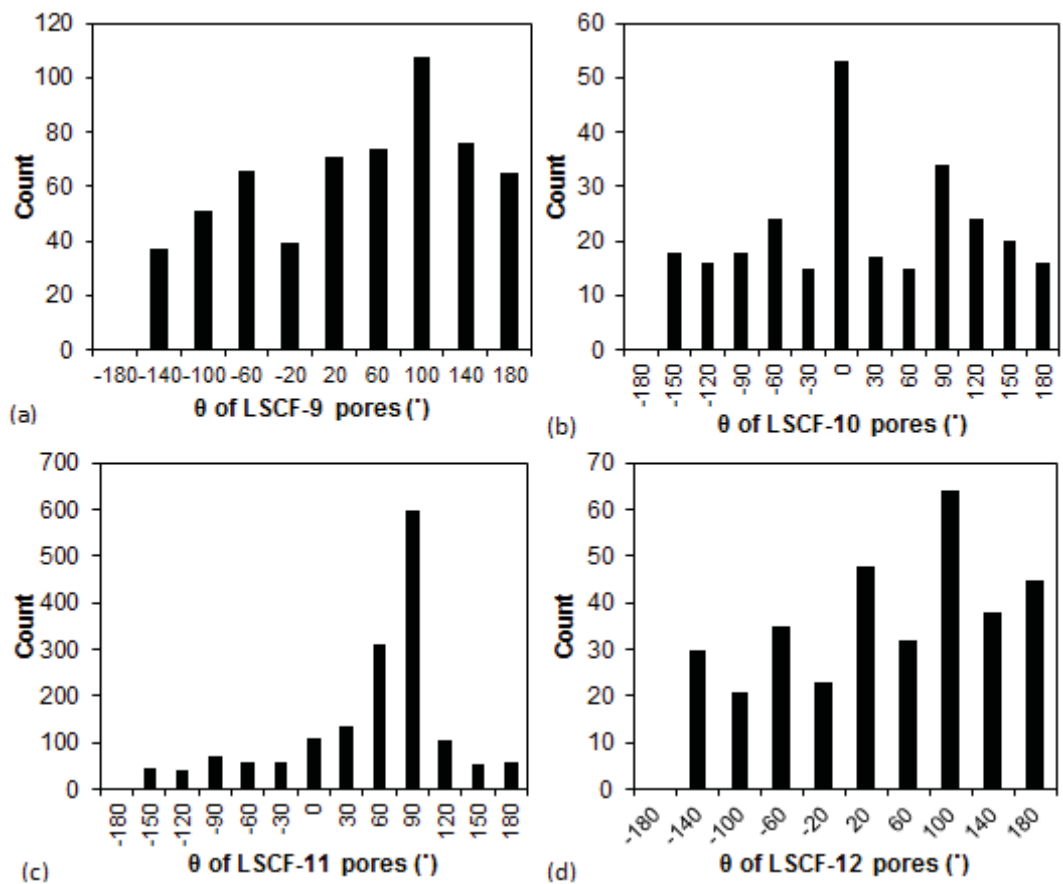

*Fig. 7-58 Orientation θ (in film plane) of separated pores in the samples: (a) LSCF-9, (b) LSCF-10, (c) LSCF-11, (d) LSCF-12*

Similar to the case of separated particles, the histograms in Fig. 7-52 also show that the majority of pores had orientation $\phi$ greater than 45˚, suggesting the pores were preferentially elongated in the X-Z plane of the film. However, the development of preferred orientation $\theta$ is not understood.

## *Summary*

In this Chapter, the artefacts (including the curtain effect, charging effect, low phase contrast, shadow effect, redeposition of sputtered material and image drift) arising during the FIB/SEM image acquisition were addressed. Efforts were made to minimise these disadvantageous effects, which could degrade the acquired image quality. The procedure and of the 3D microstructural reconstruction based on the acquired image stacks were described and the problems encountered were discussed. Mechanical simulations based on FEM were performed to calculate the effective elastic moduli and the Poisson's ratios of the as-reconstructed models. Despite the different degrees of anisotropy revealed for the models, a good agreement was found between the simulation-derived results and the previously measured data using nanoindentation. Important microstructural parameters, such as porosity, particle and pore size distribution and neck size of interconnected particles, were also determined for the 3D microstructures with the aid of Avizo software. Particular attention was given to describing the data processing, parameter quantification and analysis processes.

## *Chapter 7 References*

1.   Yao N: **Focused ion beam systems: basics and applications**: Cambridge University Press Cambridge, UK, and New York; 2007.
2.   Thydén K, Liu YL, Bilde-Sørensen JB: **Microstructural characterization of SOFC Ni–YSZ anode composites by low-voltage scanning electron microscopy**. *Solid State Ionics* 2008, **178**(39–40):1984-1989.
3.   Ko S-J, Lee YH: **Center weighted median filters and their applications to image enhancement**. *Circuits and Systems, IEEE Transactions on* 1991, **38**(9):984-993.
4.   Perona P, Malik J: **Scale-space and edge detection using anisotropic diffusion**. *Pattern Analysis and Machine Intelligence, IEEE Transactions on* 1990, **12**(7):629-639.
5.   **ScanIP software +FE +CAD reference guide**. In.: Simpleware Ltd.; 2011.
6.   Chou Y-S, Stevenson JW, TArmstrong TR, LPederson LR: **Mechanical Properties of La1-xSrxCo0.2Fe0.8O3-δ Mixed-Conducting Perovskites Made by the Combustion Synthesis Technique**. *Journal of American Ceramic Society* 2000, **83**(6):1457-1464.
7.   Wang X, Atkinson A: **Microstructure evolution in thin zirconia films: Experimental observation and modelling**. *Acta Materialia* 2011, **59**(6):2514-2525.
8.   **Avizo User Guide**. In.: Visualization Sciences Group, SAS; 2013.
9.   Mangan AP, Whitaker RT: **Partitioning 3D surface meshes using watershed segmentation**. *Visualization and Computer Graphics, IEEE Transactions on* 1999, **5**(4):308-321.
10.  Atwood R, Jones J, Lee P, Hench L: **Analysis of pore interconnectivity in bioactive glass foams using X-ray microtomography**. *Scripta Materialia* 2004, **51**(11):1029-1033.
11.  Rice RW: **Evaluation and extension of physical property-porosity models based on minimum solid area**. *Journal of Materials Science* 1996, **31**(1):102-118.

# 8  Numerical Modification of Microstructures and Influence on Elastic Modulus

## 8.1 Numerical Approach for Microstructural Modification

In order to investigate how parameters other than porosity would influence the elastic modulus of the porous microstructures, a cellular automaton (CA) algorithm based method developed by Wang *et al.* [1] was used in this study to artificially develop coarser microstructures with constant porosity based on the initial 3D microstructures reconstructed from the sequential 2D images acquired by FIB/SEM. The continuum solid phase (LSCF6428) was represented by discrete pixels and the chemical potential of a pixel located on a digital surface was calculated by the pixel-counting-based CA algorithm, which transformed the physical laws describing local surface energy into interaction between the material pixels. The numerical microstructure evolution of a material surface was controlled by local transition rules which simulate material transfer driven by the local excess energy. This corresponds to mass transport via the surface or vapour phase and does not result in densification. It therefore led to coarsening of the microstructure, but maintained constant porosity. During the 3D numerical microstructure evolution process, the change of state of a material surface voxel was manifest as discrete-time-dependent erosion/deposition of material from the original voxel to a vacant site at which the solid voxel would have lower energy. This was performed for all the solid surface voxels at conceptually the same time to give a modified microstructure. This was then repeated in the modified microstructure so that the microstructure evolved in a sequence of "time steps" (ts). An example of morphological changes of a LSCF-10 microstructure at 0, 2, 16 and 20 ts is given in 2D in Fig. 8-1.

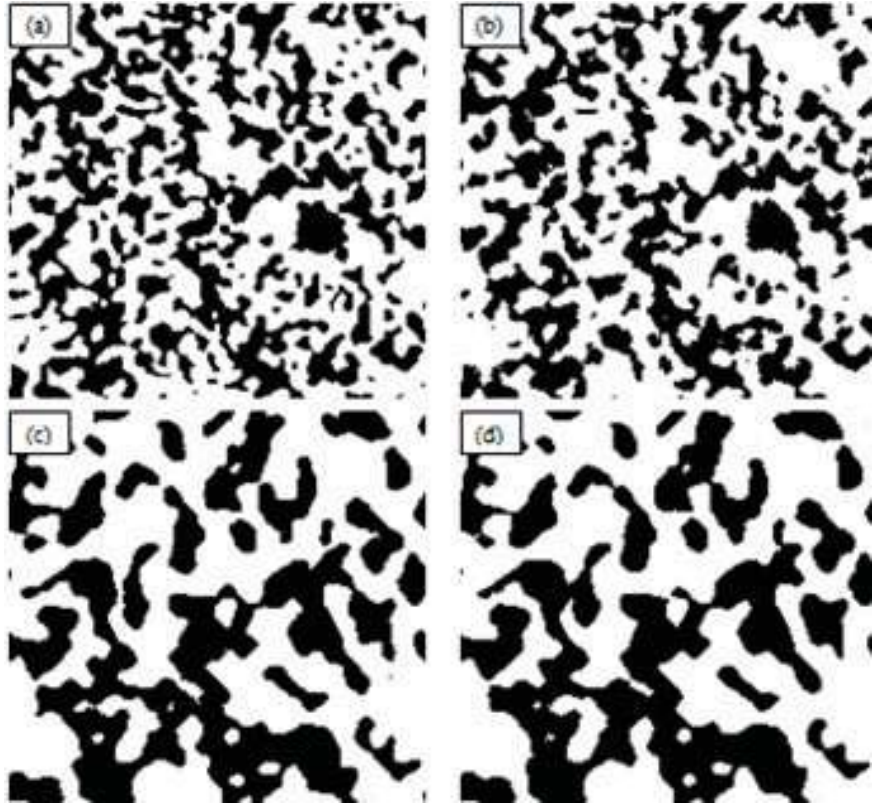

*Fig. 8-1 Microstructure evolution developed by the CA algorithm of the 3D LSCF-10 model microstructure (here shows slice #1) after (a) 0 ts, (b) 2 ts, (c) 16 ts and (d) 20 ts. Note here pixels in white represent solid and black is pore phase.*

A 2D illustration of microstructure evolution is shown in Fig. 8-2, where the erosion and deposition of material can be readily seen in the overlapping images which are separated by 20 ts. In this figure, the black pixels represent unchanged pore phase; the dark gray pixels denote where solid material has changed into pore phase; the white pixels indicate unchanged solid phase; and the light gray pixels show where pore space has become solid phase.

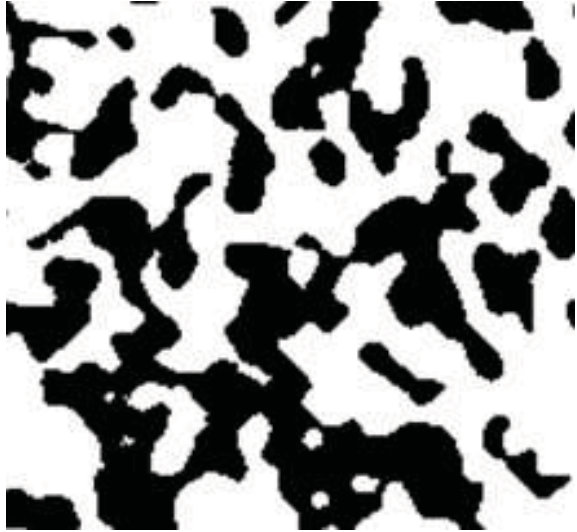

*Fig. 8-2 2D illustration of microstructure evolution by CA modification by overlap of images before and after 20 ts. The black pixels represent unchanged pore phase; the dark gray pixels denote where solid material has changed into pore phase; the white pixels indicate unchanged solid phase; and the light gray show where pore space has become solid phase*

It is worth noting that because the assumed mass transportation does not lead to densification, the neck size generally increases, as found in the above figures. On the other hand, as shown later, although the overall 3D porosity stayed constant, the 2D porosity (i.e. the surface area fraction of porous phase) of individual slices deviated from the initial ones after microstructure modification. The microstructural parameters and elastic moduli of the modified microstructures were then to determine how parameters other than porosity would influence elastic modulus.

### 8.2 Visualisation of the Modified Microstructures

Fig. 8-3, Fig. 8-4, Fig. 8-5 and Fig. 8-6 respectively demonstrate the 2D *Ortho View* and 3D *Surface View* of the microstructures obtained by CA evolution at 20 ts of the originally reconstructed microstructures (see Section 7.6.3.1) of LSCF6428 films sintered at 900-1200 ˚C. Compared to the original microstructure shown earlier, the coarsening of particles as well as the enlargement of pores can be readily seen in the following pictures. Note that the microstructures after modification at 20 ts are denoted as LSCF-9M, LSCF-10M, LSCF-11M and LSCF12M.

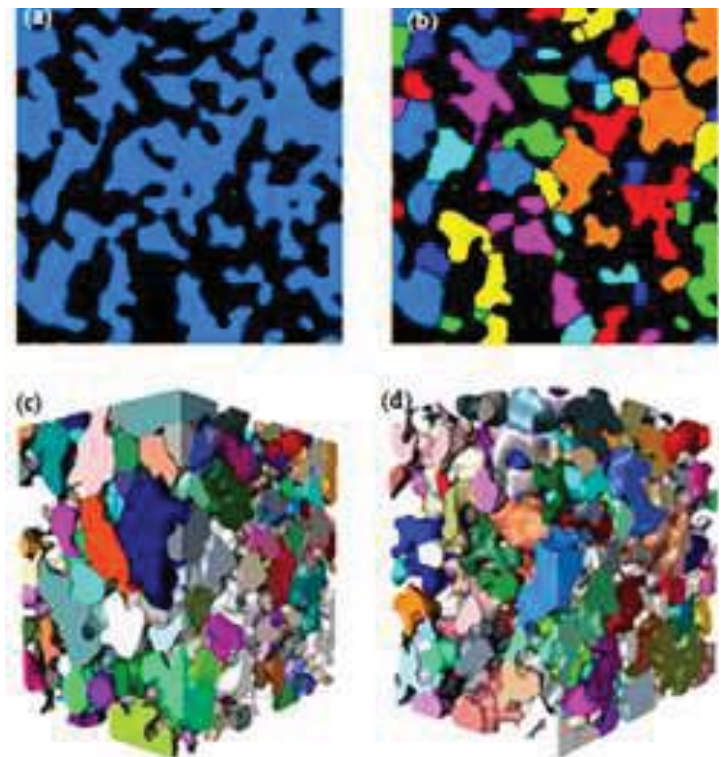

*Fig. 8-3 LSCF-9M dataset: Ortho view of an image (a) after "Auto Thresholding"; (b) after "Separate Objects" and colour labelling; 3D surface view of (c) particle microstructure after objects separation and (d) pore microstructure after objects separation*

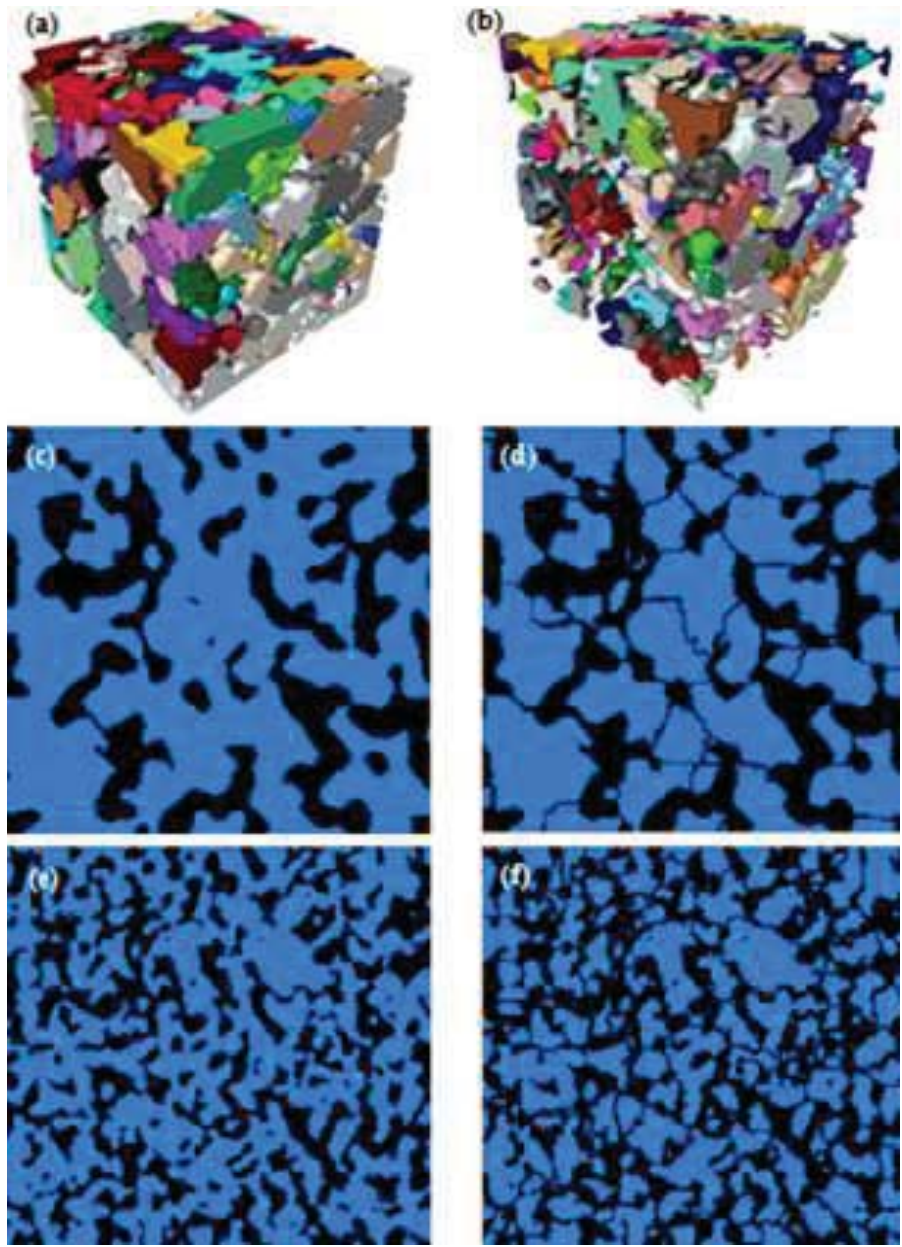

*Fig. 8-4 LSCF-10M dataset: 3D surface view after "Separate Objects" of (a) particle microstructure, (b) pore microstructure, 2D ortho view of a particle slice (c)before separation and (d) after separation. (e) and (f) show the original slice before and after separation for comparison*

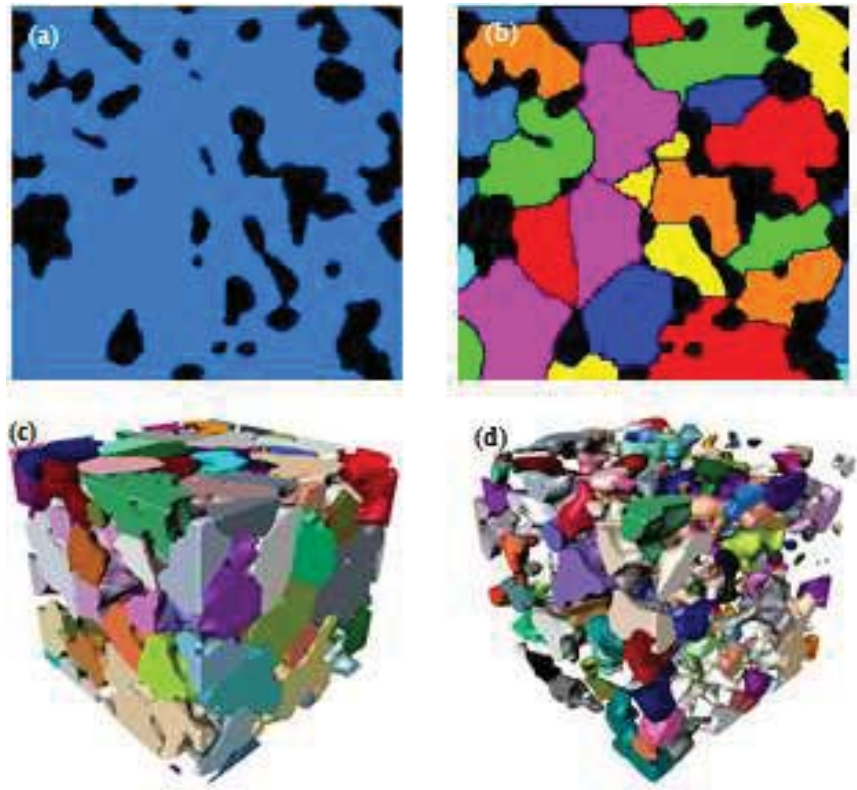

*Fig. 8-5 LSCF-11M dataset: 2D ortho view of a particle slice (a) before separation, (b) after separation and labelling, 3D surface view of (c) particle microstructure after separation and (d) pore microstructure after separation*

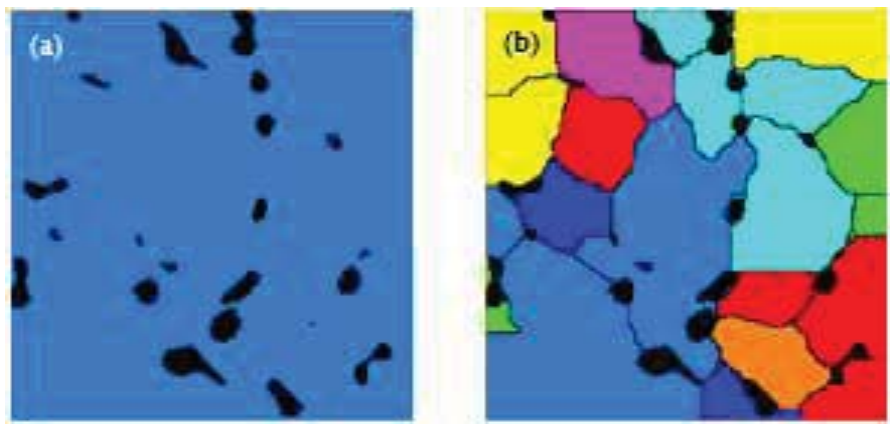

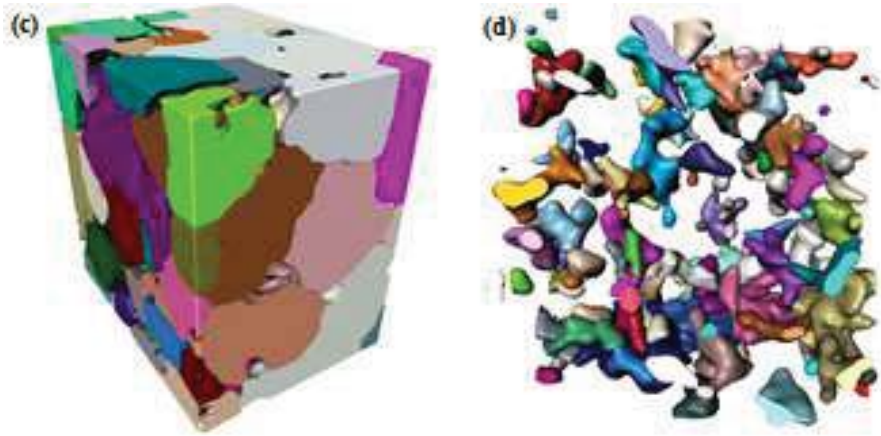

*Fig. 8-6 LSCF-12M dataset: 2D ortho view of a particle slice (a) before separation, (b) after separation and labelling, 3D surface view of (c) particle microstructure after separation and (d) pore microstructure after separation*

## 8.3 Quantification of Microstructural Parameters for the Modified Microstructures

### 8.3.1 Evaluation of 2D/3D Porosity and Specific Surface Area

The 2D porosities which represent the surface area fraction of pore phase in individual as-acquired slices of image along the Y axis of a dataset were also measured to help assess the uniformity of the volume fraction of pore phase in the analysed 3D volume. Fig. 8-7 shows the measured 2D porosities in each slice before and after numerical modification for different time steps (10 and 20 ts), as well as their mean values.

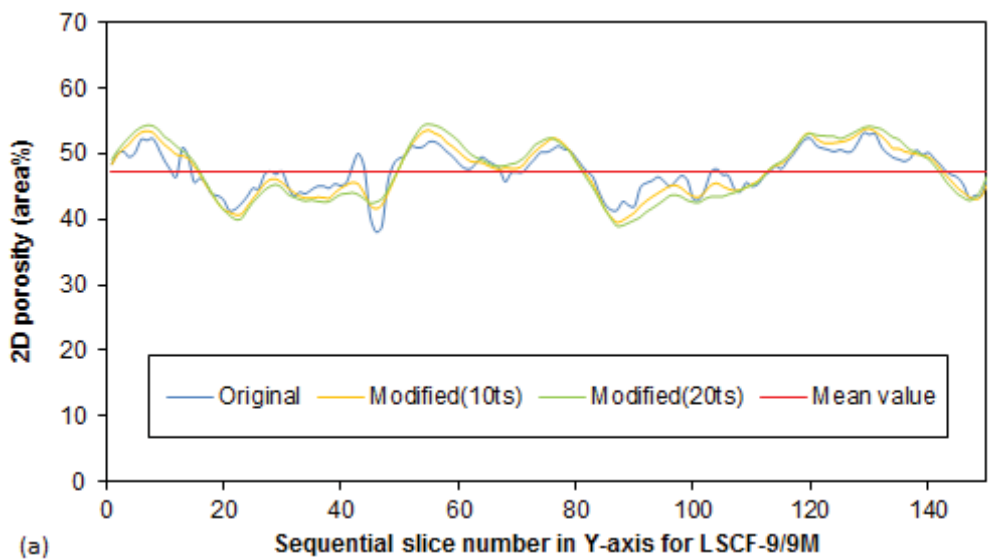

(a)

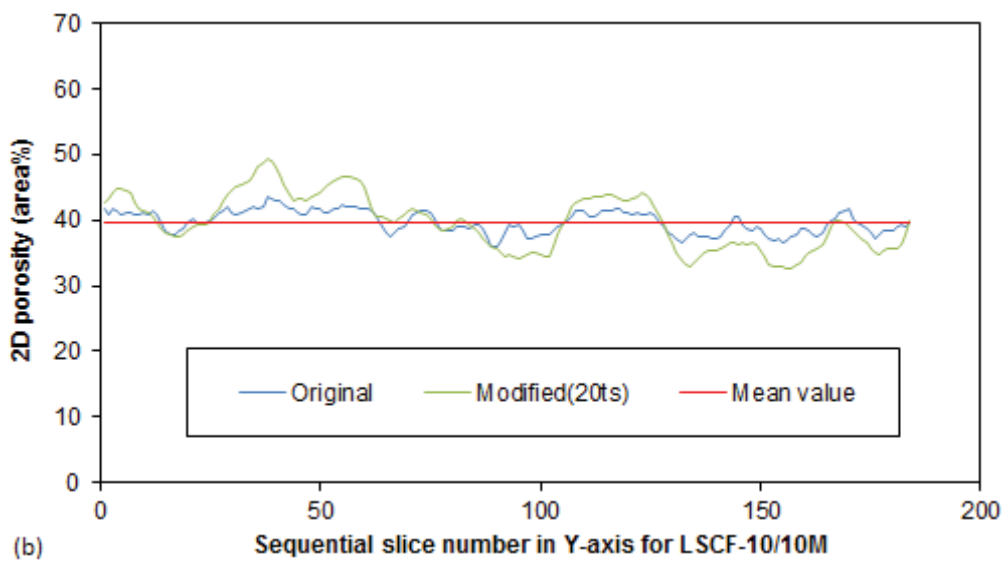

(b)

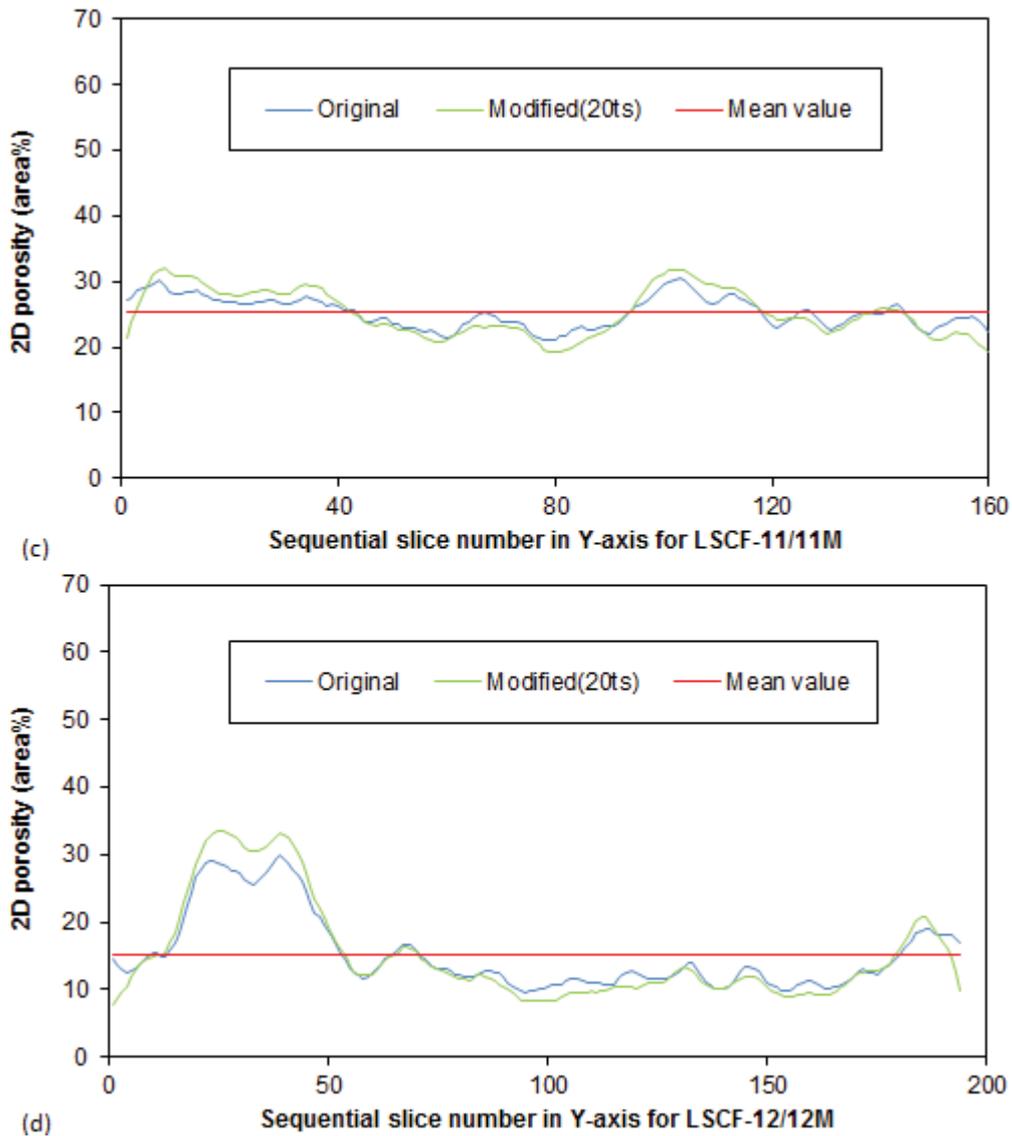

*Fig. 8-7 Comparison of 2D porosities for individual image slices in datasets before and after modification for different time steps (ts) for (a) LSCF-9/9M, (b) LSCF-10/10M, (c) LSCF-11/11M and (d) LSCF-12/12M*

A common feature demonstrated in Fig. 8-7 is that regardless of specimen sintering temperature both original and modified microstructures of each film had the same *mean value* of 2D porosity (as shown by the red straight curve in each figure) which was identical to the 3D porosity measured previously. This is a necessary consequence of the CA algorithm used for microstructural modification. Also in all the above figures, after modification, the slice-to-slice variability in the 2D porosity became larger. This observation may imply that regarding the spatial distribution of pores, original large pores became larger while small pores became even smaller. This might also be true for the solid phase as a result of the constant porosity in the microstructures.

Particularly in the case of LSCF-12/12M (Fig. 8-7 (d)), the 2D porosities of images with slice number below 50 significantly deviated from the mean value, which suggests a non-uniform distribution of pores (i.e. the existence of relatively large pores) in both microstructures. The above observations are also confirmed by the quantitative results shown in Table 8-1.

*Table 8-1 Comparison of 2D/3D porosities and 2D standard deviations before and after microstructure change at 20 ts*

| Sintering Temperature (°C) | Original 2D Porosity (%) | Modified 2D Porosity (%) | Mean Value of 3D Porosity (%) |
|---|---|---|---|
| 900 | 46.9±2.7 | 46.9±4.5 | 46.9 |
| 1000 | 39.7±2.2 | 39.7±4.2 | 39.7 |
| 1100 | 24.1±2.3 | 24.1±3.6 | 24.1 |
| 1200 | 15.2±5.6 | 15.2±7.3 | 15.2 |

It can be seen in the table for each specimen sintering temperature, the 2D porosities after microstructure modification show increasing standard deviations compared with that for the original ones, suggesting a less homogeneous pore distribution in the modified microstructures. As a result, the inter-particle neck sizes in the modified microstructures were expected to increase globally by the application of the CA algorithm. Hence, the FEM-derived elastic moduli for the modified microstructures would possibly increase.

In addition, the volume-normalised surface areas of both the original and modified microstructures (at 20ts) before the implementation of the *Separate Objects* algorithm were calculated and compared in Fig. 8-8.

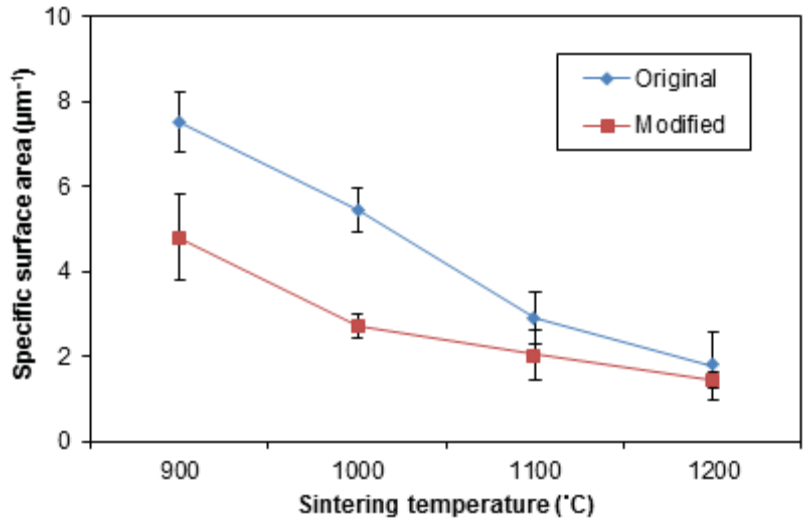

*Fig. 8-8 Comparison of specific surface areas of the microstructures before and after evolution for 20 ts*

The general trend found in Fig. 8-8 is that both specific surface areas experienced some decrease over the sintering temperature range, which is considered to be attributable to the increasing sintering degree, i.e. the increased coarsening of the solid phase and the densification of both original and modified microstructures. In addition, for each sintering temperature the original specific surface area always had higher specific surface area than the modified one. This is ascribed to another solid coarsening effect, which however resulted from the artificial evolution based on the CA algorithm rather than the actual sintering mechanism.

### 8.3.2 Tortuosity Factors

Tortuosity factors for all the original and modified microstructures including particles and pores are computed and compared in Fig. 8-9.

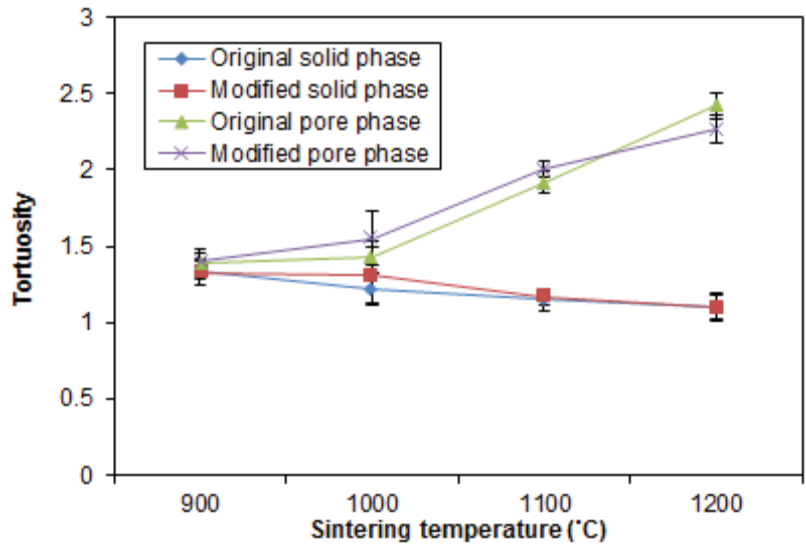

*Fig. 8-9 Comparison of tortuosity factors before and after microstructural modification*

The plot reveals that the microstructure evolution had little effect on tortuosity of either solid or pores. This would suggest that the evolution did not qualitatively change the microstructures, which were self-similar before and after the evolution. For both original and modified microstructures as the sintering temperature increased, tortuosities in pores quickly rose while in solid they experienced slow decrease. Thereby, there existed increasingly larger difference between tortuosities for pores and particles as sintering temperature rose. This is particularly true in films sintered at higher sintering temperatures, as less open but more closed pores were formed, which complicated the tortuous paths in the corresponding pore microstructures and hence the tortuosity factors increased. At the same time the solid phase occupied much more room, the tortuous paths got simpler with lower tortuosities reached towards 1. Particularly for the densest microstructures (i.e. lowest porosity) LSCF-12/M, the lowest tortuosity was reached (=1.1) in both original and modified particle microstructures of LSCF-12, while the highest was obtained (=2.4) in LSCF-12's original pore microstructure.

### *8.3.4 ESD of Separated Particles and Pores*

ESD distributions of the separated particles and pores (i.e. particle size distribution and pore size distribution) for each type of sample after microstructure modification at 20ts are plotted in Fig. 8-10 and Fig. 8-11, respectively.

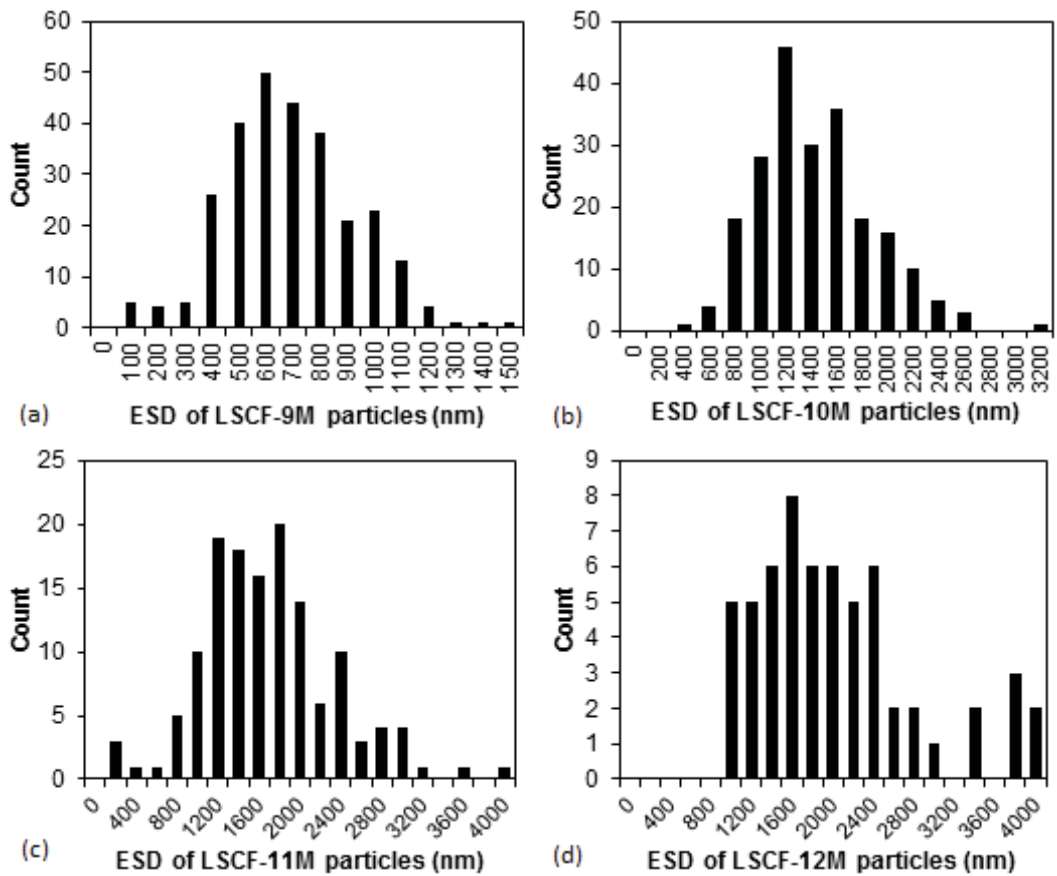

*Fig. 8-10 ESD distribution of the separated particles after microstructure modification at 20 ts: (a) LSCF-9M, (b) LSCF-10M, (c) LSCF-11M, (d) LSCF-12M*

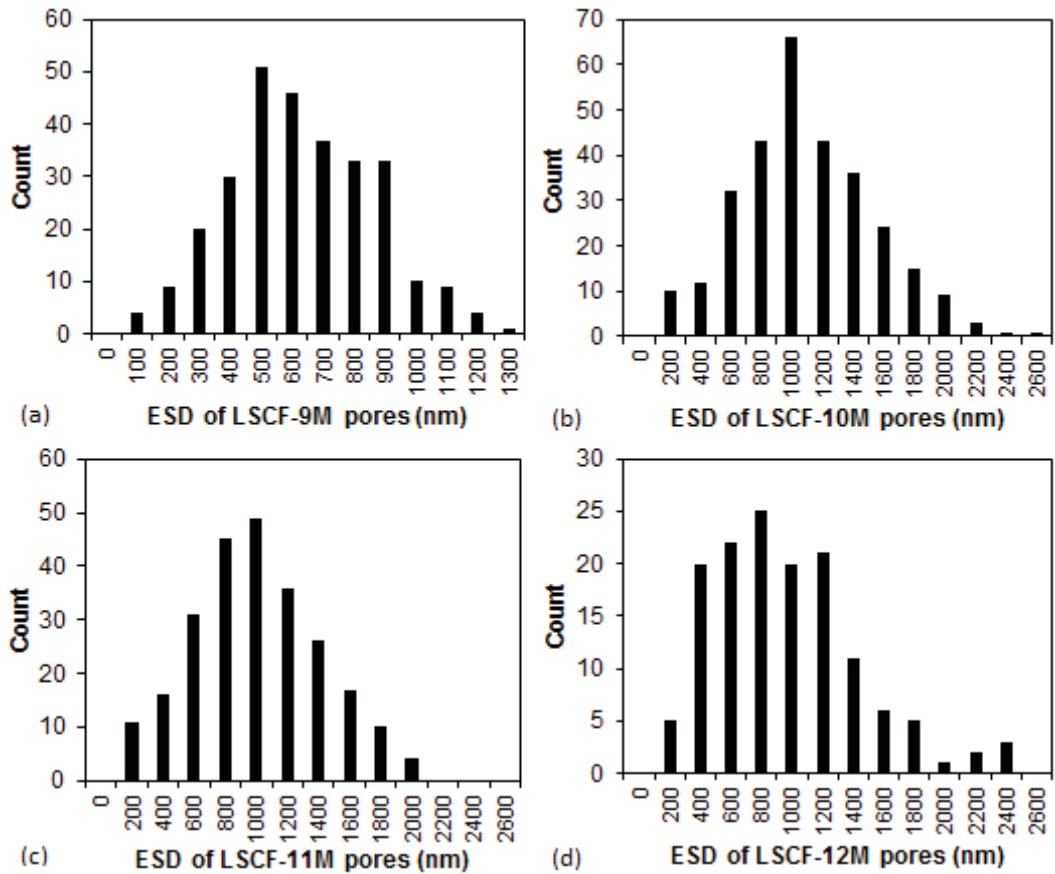

*Fig. 8-11 ESD distribution of the separated pores after microstructure modification at 20 ts: (a) LSCF-9M, (b) LSCF-10M, (c) LSCF-11M, (d) LSCF-12M*

It can be readily observed from Fig. 8-10 that the particle size increased while the total number of the separated particles decreased with the increasing sintering temperature. While for pore size distribution shown in Fig. 8-11, the trend cannot be readily seen. Nevertheless, LSCF-12M exhibited relatively more homogeneous size distribution in both cases (Fig. 8-10 (d) and Fig. 8-11 (d)). Note that this is not contradictory to the inhomogeneous *spatial* distribution of porosity in the microstructure as found earlier in Section 6.6.4.3.

The average ESDs of the computationally separated particles and pores in the modified microstructures were further compared with the original ones, as shown in Fig. 8-12.

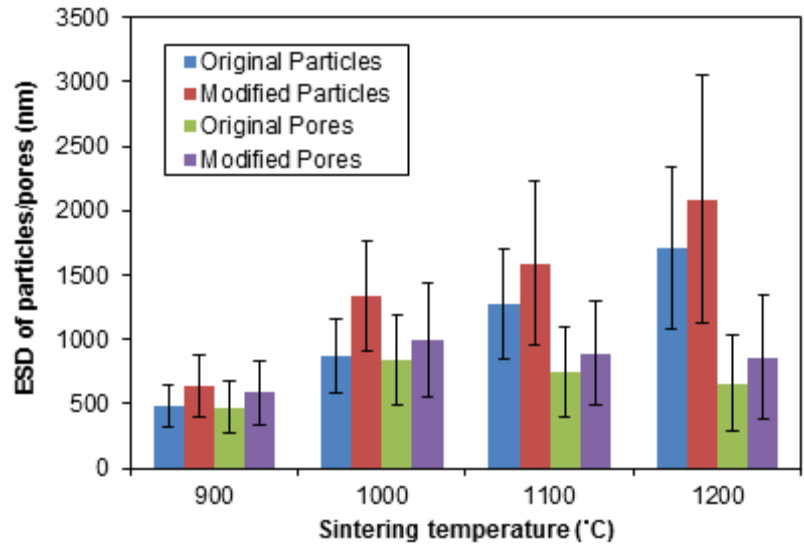

*Fig. 8-12 Overall comparison of ESD of the separated particles and pores for original and modified microstructures at 20ts*

Fig. 8-12 shows that for microstructures obtained at each sintering temperature, the modified ones always possessed both larger particles and pores, which is as expected on the basis of the CA principle and algorithm. Because the size of particles and pores were both larger and at the same time the 3D porosity had to remain constant, the numbers of particles and pores were reduced.

### 8.3.5 Sphericity of Separated Particles and Pores

The distribution of sphericity of separated particles and pores for the microstructure after modification at 20 ts is plotted in Fig. 8-13 and Fig. 8-14, respectively. Comparison with the original microstructures is made as shown in Fig. 8-15.

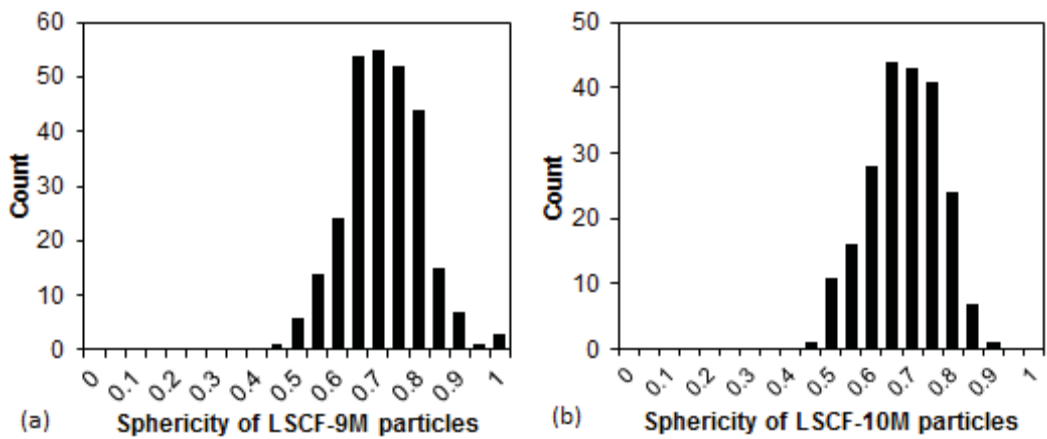

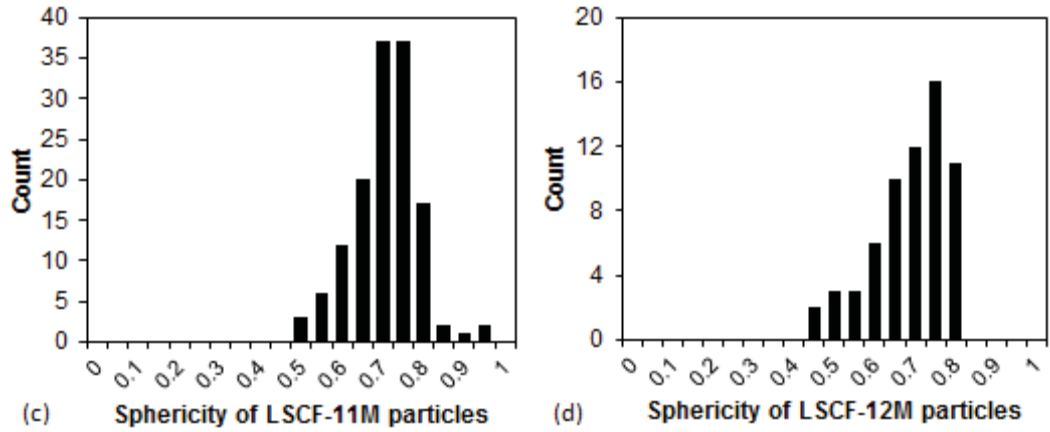

*Fig. 8-13 Sphericity of particles after microstructure modification at 20 ts: (a) LSCF-9M, (b) LSCF-10M, (c) LSCF-11M, (d) LSCF-12M*

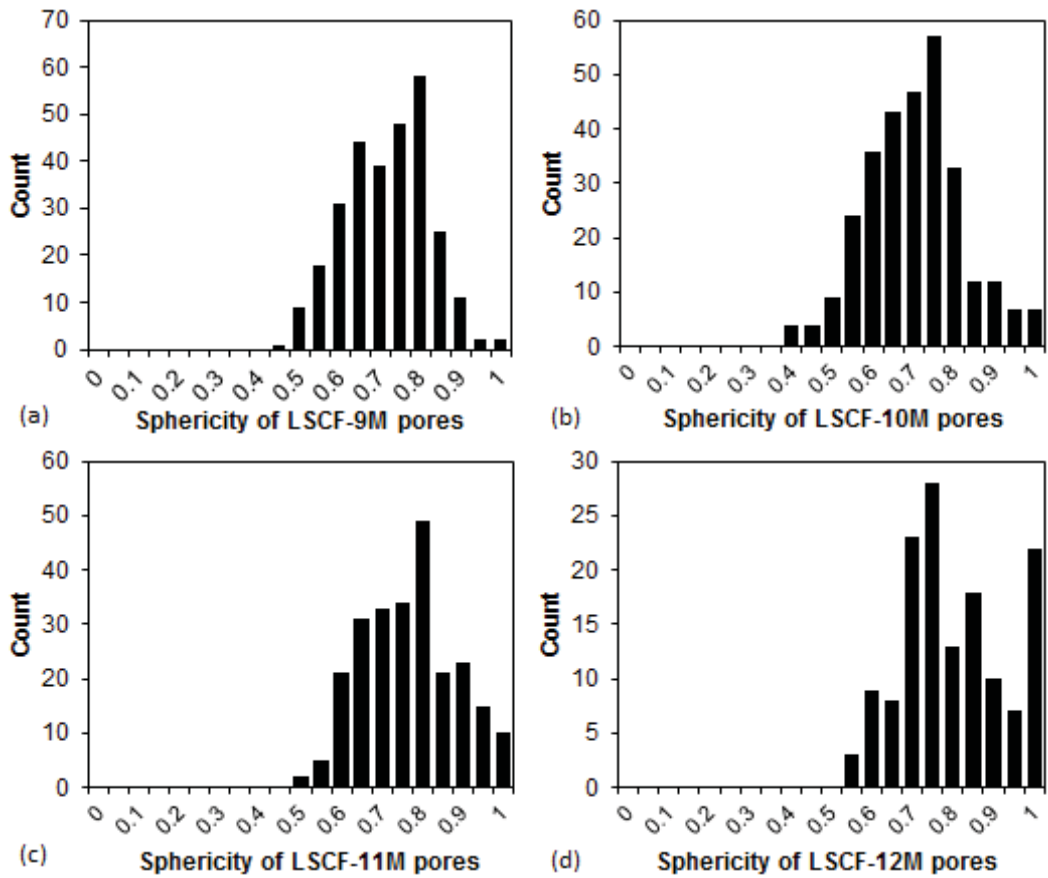

*Fig. 8-14 Sphericity of pores after microstructure modification at 20ts: (a) LSCF-9M, (b) LSCF-10M, (c) LSCF-11M, (d) LSCF-12M*

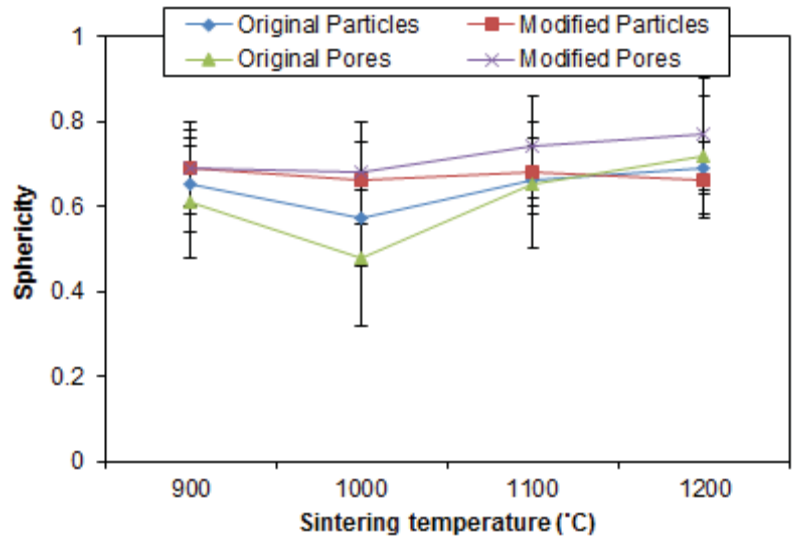

*Fig. 8-15 Overall comparison of sphericity for particles and pores before and after microstructure modification for 20 ts*

Fig. 8-15 shows that the modified particles and pores were more spherical than in the original microstructures. Furthermore, it is found in Fig. 8-13 and Fig. 8-15 that regardless of the original particle sphericity, the modified particles tended to have very similar sphericity, which was close to 0.7, across the range of all the film types. For modified pores, however, Fig. 8-14 shows that a large proportion had sphericity approaching 1 for microstructures generated by higher sintering temperature. This is due to an increasing number of isolated pores in regular shapes similar to spheres.

### 8.3.6 ECD of Inter-particle Necks

The distributions of ECD for inter-particle necks after microstructure modification at 20 ts are shown in Fig. 8-16. The comparison with ECD of original inter-particle necks is made in Fig. 8-17.

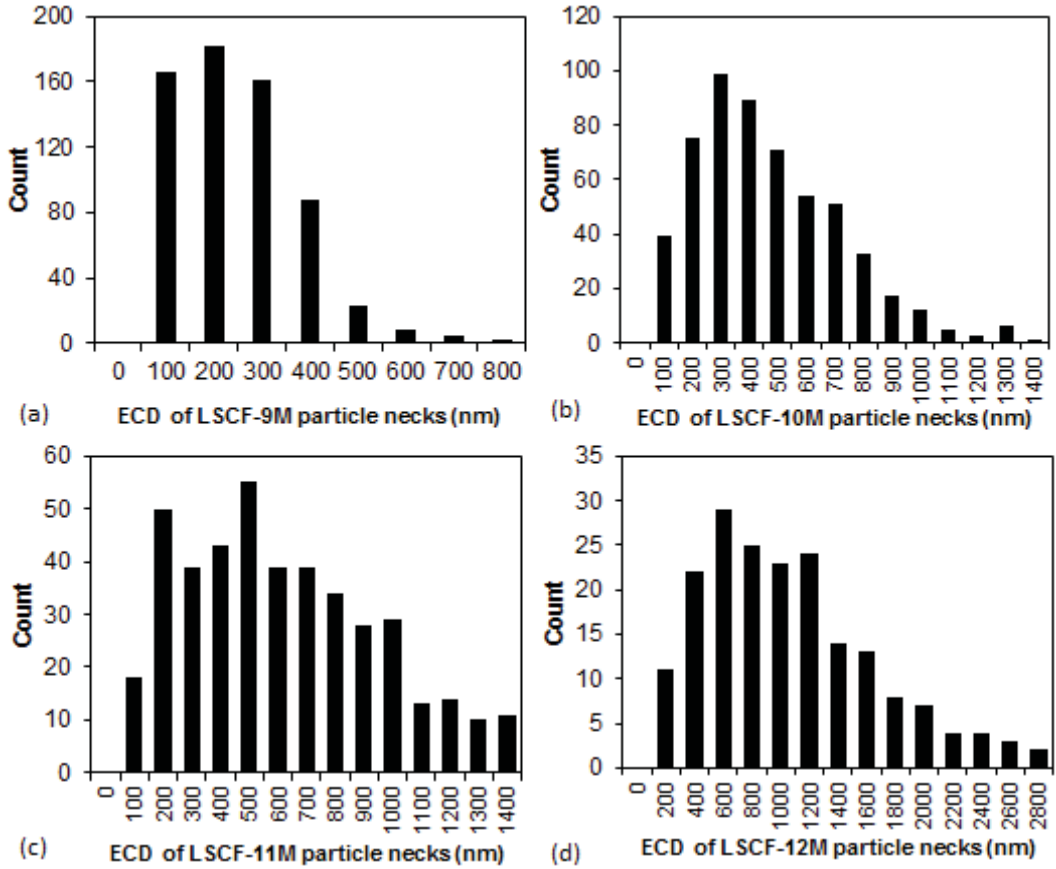

*Fig. 8-16 Distribution of ECD of inter-particle necks after microstructure evolution at 20 ts: (a) LSCF-9M, (b) LSCF-10M, (c) LSCF-11M, (d) LSCF-12M*

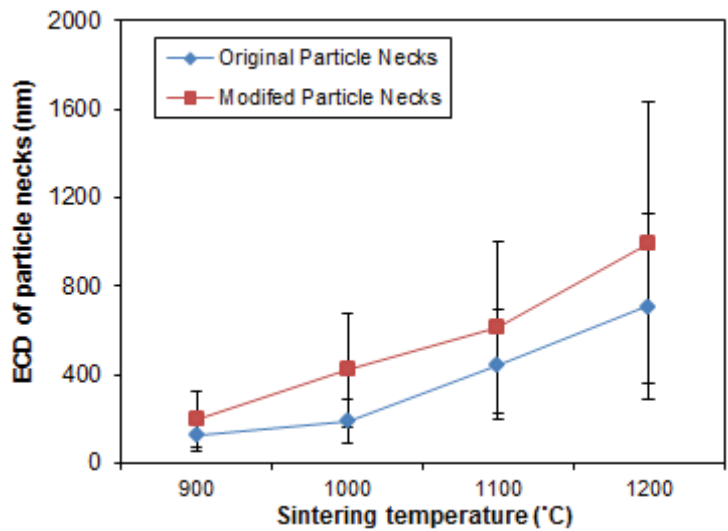

*Fig. 8-17 Comparison of ECD of inter-particle necks before and after microstructure evolution at 20 ts*

From the figure shown above, besides the evident increasing trend of neck size as a function of sintering temperature, it is also found that the ECD for the modified inter-particle necks was increased compared with that of the original ones.

Based on the above analyses of the changes in the microstructural parameters upon modification, the increase of particle and neck sizes in the modified microstructures were supposed to positively affect the elastic moduli by some growth, which might be similar to the work reported by Hardy and Green [2], whose experiments on partially sintered alumina showed significant increase in elastic modulus with even minimal densification and they ascribed this change to the neck growth by surface diffusion.

## *8.4 Elastic Modulus Affected by Microstructure Modification*

After microstructure change at different number of time steps (i.e. 2, 5, 10 and 20 ts) using CA based method, the as-modified microstructures were all subjected to FEM for elastic modulus and Poisson's ratio estimation. To facilitate the overall comparison, the estimated elastic moduli for each set of samples were averaged over the values of the three directions and then normalised relative to that of the original microstructures (i.e. at 0 ts), as plotted in Fig. 8-18.

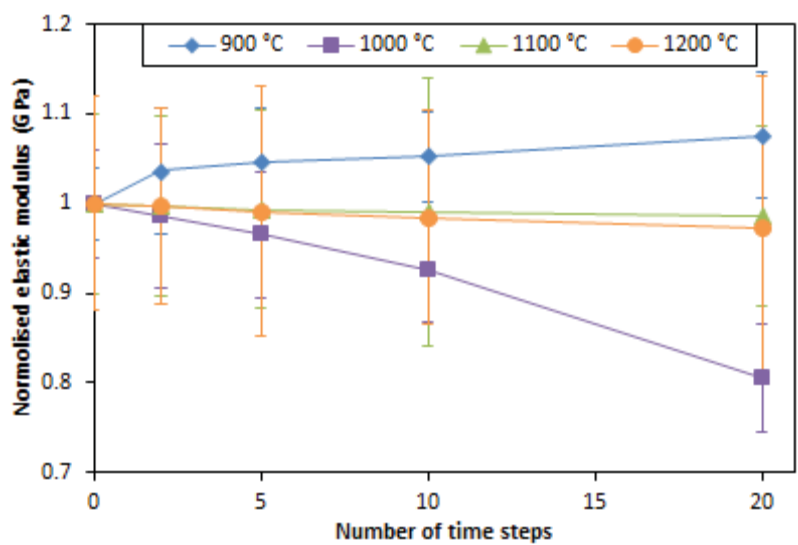

*Fig. 8-18 Comparison of the averaged normalised elastic moduli estimated by FEM for original and modified microstructures*

As mentioned earlier, it had been originally expected that all moduli would increase all the way with the time of modification because of the enlargement of particle and neck sizes. While surprisingly, the figure above readily shows that for different original sintering temperatures the corresponding set of microstructures experienced distinct variation trends as a function of the microstructure modification time steps. Particularly the elastic moduli for the 900 °C microstructures tended to increase (by less than 10% at 20 ts) while for the 1000 °C samples their moduli appeared to drop all the way over the time steps and retained 80% of the modulus of the original microstructure when 20 ts was reached. On the contrary, the modification for 1100 and 1200 °C samples seemed to have little effect on the modulus change. They did however experience very small amount of decrease (< 3%) over the entire range of time steps.

The results above suggest that a single metric such as the particle neck size might not be sufficient to account for the elastic modulus changes over the course of microstructure modification for these microstructure sets. There might be other parameters generated from the statistical microstructural geometry which allowed a more complete description of the elastic modulus. Therefore, an in-depth investigation was desired to access more comprehensive information of such a diversity of changes in elastic modulus of constant porosity microstructures.

## *8.5 Poisson's Ratio Affected by Microstructure Modification*

The calculated Poisson's ratios in individual directions of the as-modified microstructures are plotted in Fig. 8-19 for each specimen sintering temperature as a function of the number of time steps.

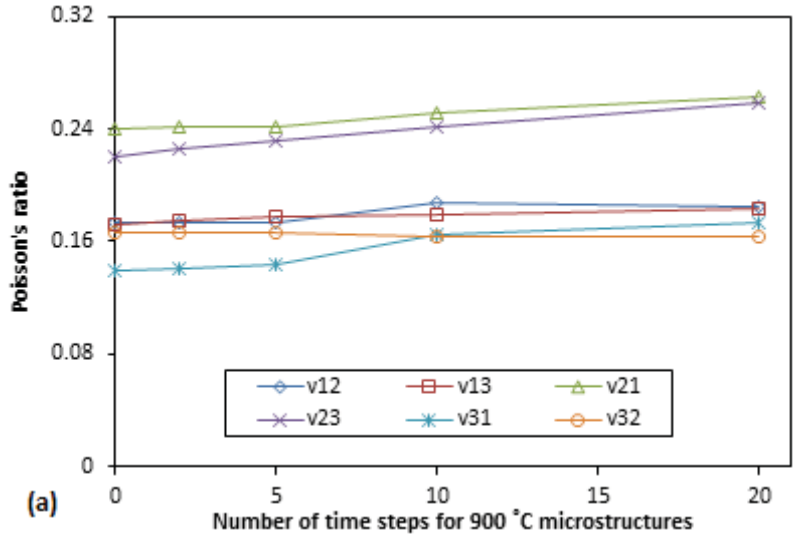

(a)

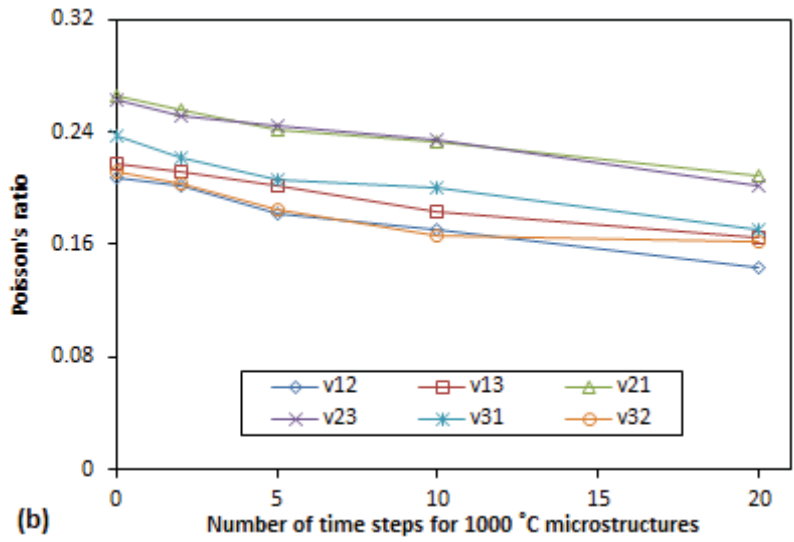

(b)

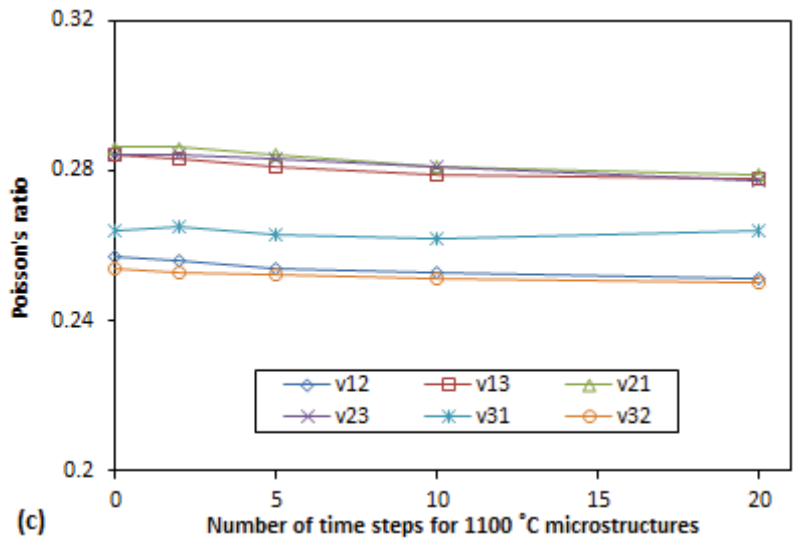

(c)

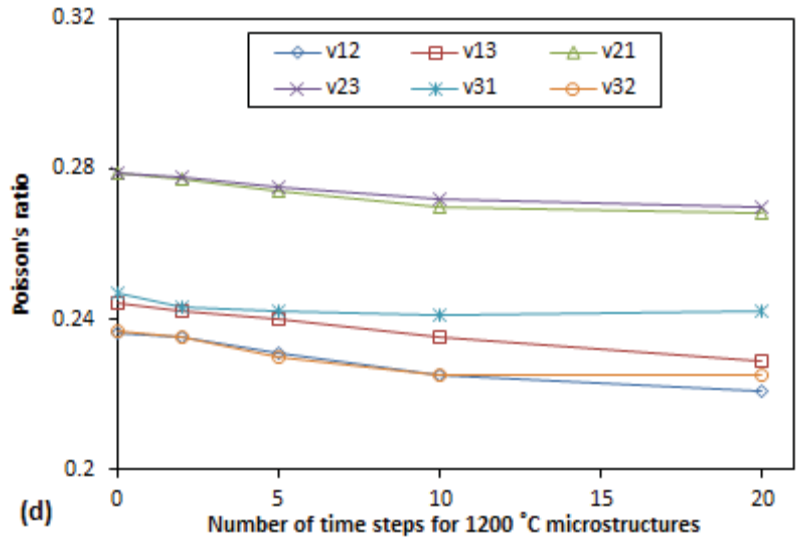

*Fig. 8-19 Evolution of the Poisson's ratio as a function of the number of time steps used for the microstructure modification. (a) 900 ˚C microstructures, (b) 1000 ˚C microstructures, (c) 1100 ˚C microstructures and (d) 1200 ˚C microstructures.*

It is found in the figures above that regardless of the specimen sintering temperature or the number of time steps used for the CA modification, the Poisson's ratios *v21* and *v23* (i.e. in the Y direction normal to the film plane) remained greater than the rest of the values of *v12, v13* in the X direction and *v31, v32* in the Z direction. The latter four, however, had very close values in all cases. Again, such observations in the results are attributable to the constrained sintering as also expected in the *E* estimation described earlier. On the other hand, except the 900 ˚C microstructures which had gradually increased Poisson's ratio as the number of time steps increased, the other three sets of plots all experienced some decrease in the Poisson's ratio values. This observation can be more readily seen when comparing the normalised Poisson's ratios for individual specimen sintering temperatures. That is the Poisson's ratios for each set of samples were averaged over the values of the three directions and then normalised relative to that of the original microstructures (i.e. at 0 ts), as plotted in Fig. 8-20.

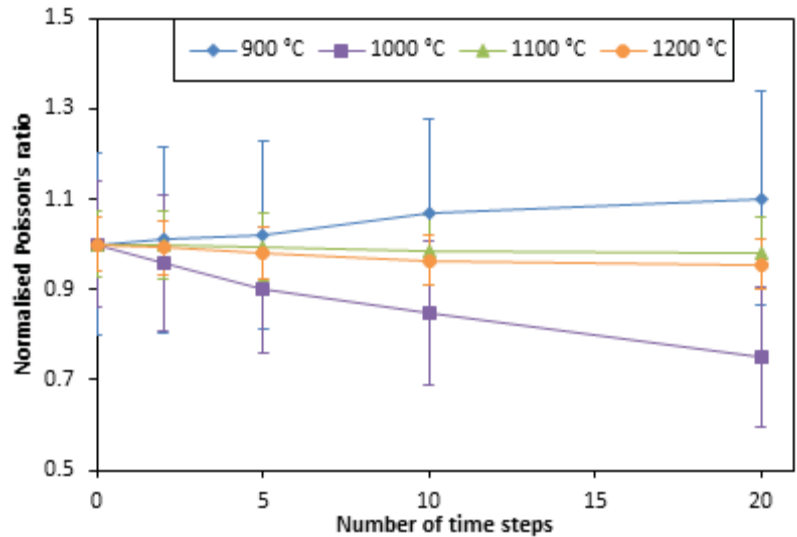

*Fig. 8-20 Comparison of the averaged normalised Poisson's ratios as a function of the number of time steps*

Compare Fig. 8-20 with Fig. 8-18 in which the normalised elastic moduli were plotted as a function of the number of time steps, we can find out that the evolution of the Poisson's ratios experienced almost the same trends as that of the elastic moduli for individual specimen sintering temperature upon the microstructure modification. Therefore, it can be concluded that microstructure modification had identical effect on both elastic modulus and the Poisson's ratio of the microstructures. The above results also indicate that a given elastic modulus or Poisson's ratio does not necessarily correspond to a particular microstructure.

## 8.6 Study of Numerical Modification on Artificial Microstructures

### 8.6.1 Creating Porous Microstructures Containing Randomly Distributed Mono-size Spherical Particles

In order to find out the cause of the unexpected variation of elastic modulus with the LSCF6428 microstructure changes and to gain better understanding of the controlling parameters over elastic modulus changes when the microstructure of a porous volume at constant porosity is modified, a simple artificial microstructure consisting of mono-size spheres representing solid phase was created and evolved using CA based method in the similar way to the LSCF6428 microstructures as carried out previously.

In the light of work done by Rhazaoui *et al.* [3], the starting artificial microstructure was generated based on Monte-Carlo process by randomly packing a cube of a volume size =

$200^3$ voxel (i.e. pixel$^3$) with two types of mono-size spheres (diameter = 20 pixel and 10 pixel, respectively) representing the solid and pore phases. Once the cube was densely packed with spheres in tangent contact, the spheres representing pores were removed, and together with the unfilled regions the total voids in the cube now could represent the real pore phase. This would result in a total porosity of ~ 44.6 vol% in the cubic microstructure. As mentioned above, the starting microstructure was subjected to evolve using CA algorithm based method while the porosity was kept constant. In this way, a series of modified microstructures was successively generated at different number of time steps: 2, 4, 20, 40, 70, 100 and 200. Fig. 8-21 shows the 2D cross-sectional images of the first slice in each microstructure (in which white represents solid and black is pore phase) of the same slice number (slice thickness = 1 pixel) and the corresponding 3D representations of the microstructures for the starting volume (i.e. at 0 ts) and the volume at 2, 4, 20, 40, 70, 100, and 200 ts of modification. The gradual microstructure changes and coarsening of particles and particle necks can be readily seen.

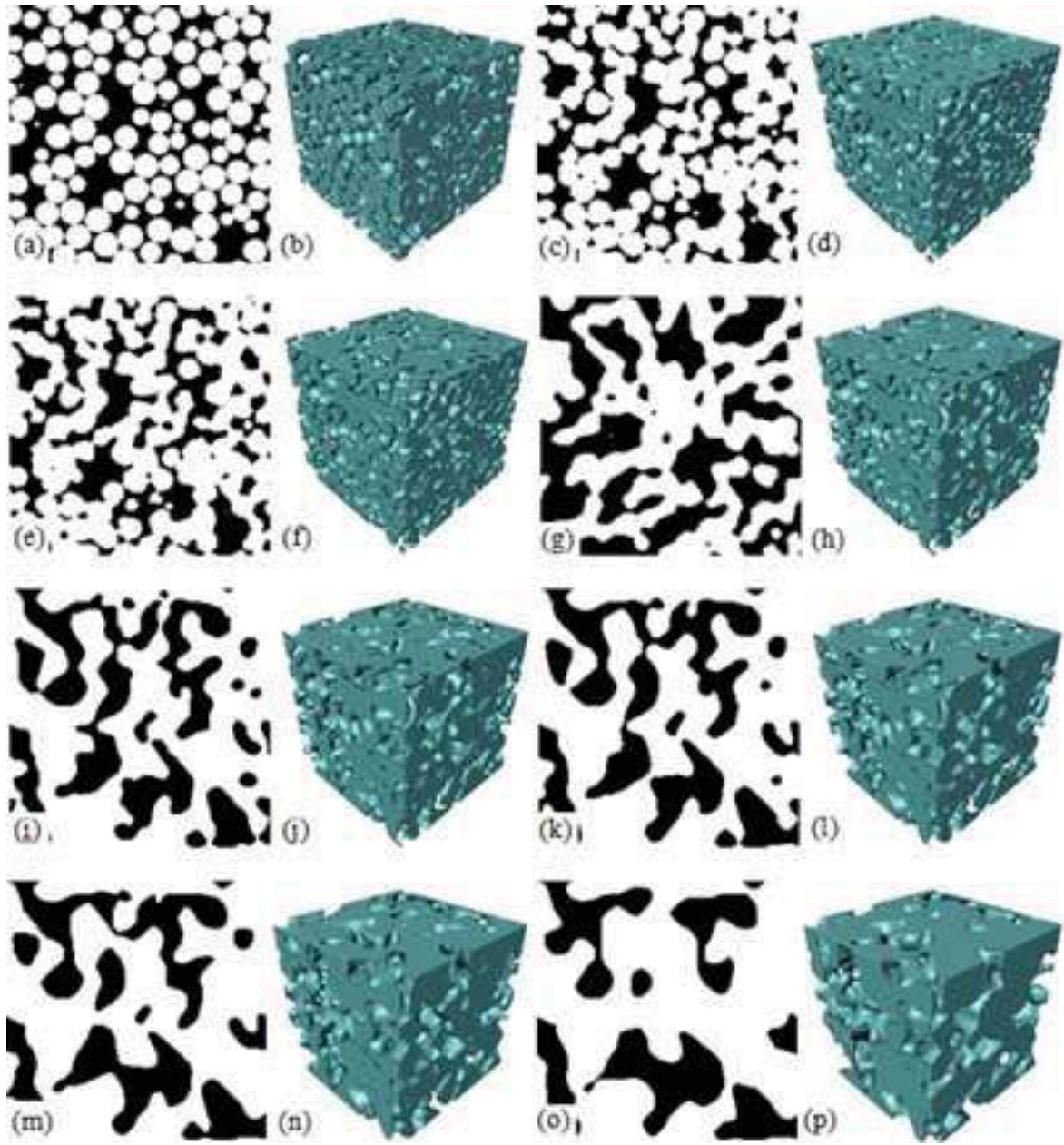

*Fig. 8-21 The 2D and 3D representations of the artificially created microstructures at 0 ts: (a, b) (i.e. starting microstructure) and evolved microstructures for 2 ts: (c, d), 4 ts (e, f), 20 ts (g, h), 40 ts: (i, j), 70 ts: (k, l), 100 ts: (m, n) and 200 ts: (o, p). Note that all microstructures have the same volume size: $200^3$ pixel$^3$*

## 8.6.2 Variation of Elastic Modulus and Poisson's Ratio due to Microstructure Changes

By using the same methods of 3D reconstruction and FEM as for LSCF6428 microstructures, the elastic moduli and the Poisson's ratios of each modified microstructure in all three directions were estimated, as shown in Fig. 8-22.

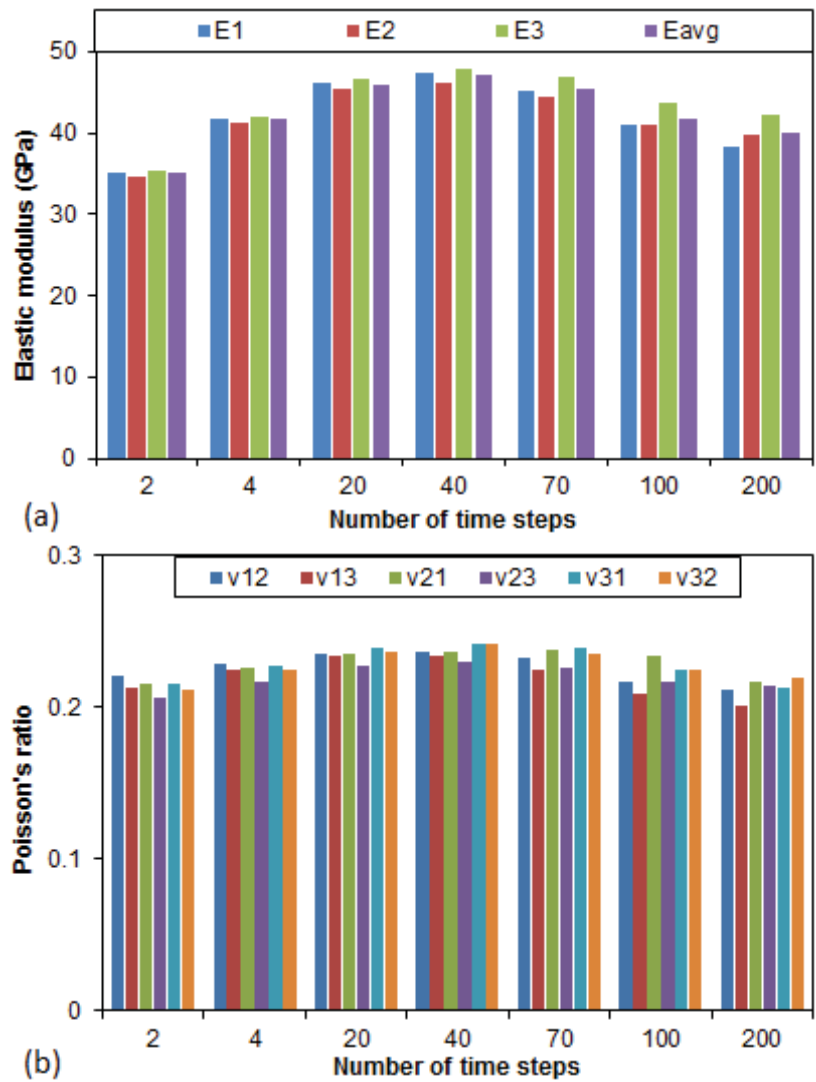

*Fig. 8-22 (a) Elastic moduli and (b) Poisson's ratios in all directions and the arithmetic means for the microstructures modified at different time steps*

In Fig. 8-22, some isotropy can be perceived for each microstructure as similar elastic modulus and Poisson's ratio were found in all directions of each microstructure, despite the fact that increasingly larger difference arose among elastic moduli in the three directions as time step increased. Nevertheless, this difference in each individual microstructure was well below 10 % of the average value. Therefore, for a more convenient way of comparison, the averaged elastic moduli and Poisson's ratios were used and normalised against the ones for the microstructure at 2 ts, as shown in Fig. 8-23.

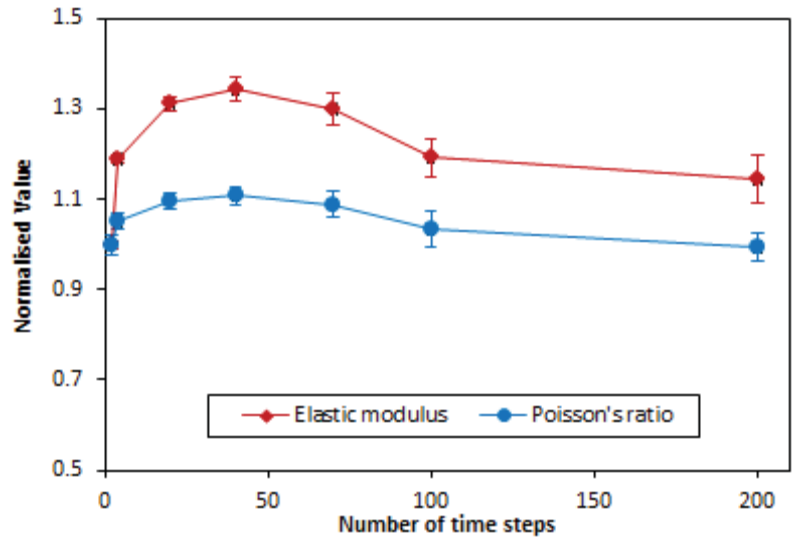

*Fig. 8-23 Normalised elastic modulus and Poisson's ratio as a function of the number of time steps*

Fig. 8-23 shows that the elastic modulus as a result of the microstructure changes experienced a rapid increase until the number of time steps reached 40, after which a gradual fall took place. Similar trend can be found for the evolution of the Poisson's ratio, while much less steep progress was experienced. This is indeed a very interesting feature of how the elastic modulus and Poisson's ratio varied, as the increase in the early stage of microstructure variation was much of our expectation (which was similar to that for the set of 900 °C LSCF6428 microstructures, see Fig. 8-18 and Fig. 8-20), whereas the later stage's decrease was more or less like the decline of elastic modulus and the Poisson's ratio in the case of 1000, 1100 and 1200 °C microstructures. Therefore, the next step was to investigate what could be the factors controlling the above variation in elastic modulus during the course of microstructure changes.

### 8.6.3 Effect of Inter-particle Neck Size

The purpose of the study of the evolution on the artificial microstructures was to gain more thorough understanding of the dependence of elastic modulus on factors other than porosity of porous microstructures. Since the porosity was kept constant, the most critical microstructure feature needed to be firstly taken into account would always be the inter-particle necks in the microstructures. Due to their previously-mentioned importance to the elastic modulus determination, it had been expected that in a porous single phase microstructure such as the actual LSCF6428 microstructures in our study, larger neck size would generally lead to a higher effective elastic modulus. It is therefore hard to believe that at constant porosity an increase in the particle neck size would result in the falling of the microstructure's elastic modulus. Nevertheless, Fig. 8-23 shows the other side of the picture: an increase of elastic modulus took place before its dropping.

It should be emphasised that despite the clear illustration of neck and its influence on elastic modulus in a 2D model as shown in Fig. 7-40, there is however no precise definition on the solid or pore size, shape or neck in complex 3D porous random microstructures [4], like in the current LSCF6428 microstructures. As a result, the measurement of the "neck size" might be sometimes ambiguous, particularly when the microstructure was extremely irregular. Therefore it was not simple to attribute the elastic modulus results to these features or parameters, although they were somehow "measurable" with the aid of current Avizo software in this study. Nevertheless, the measured results along with some general morphological observations could be used in an interpretive way to correlate with the mechanical behaviours of the microstructures.

In the light of the analysis above, we have estimated parameters related to particles and necks in individual modified microstructures as a function of the number of time steps, as shown in Fig. 8-24 and Fig. 8-25.

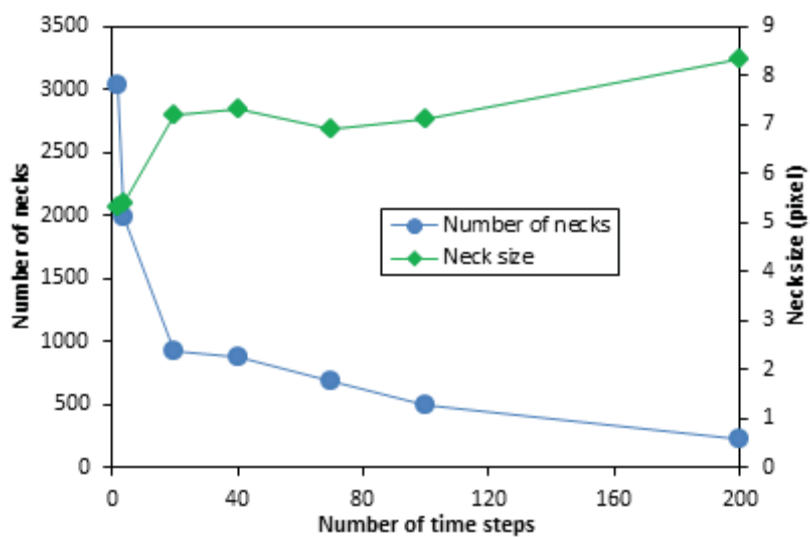

*Fig. 8-24 Number of necks and average neck size of individual modified microstructures as a function of the number of time steps*

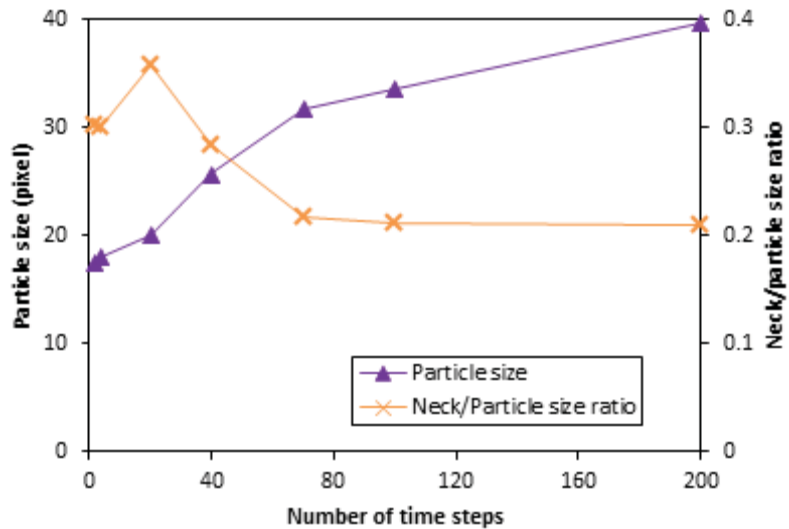

*Fig. 8-25 Average particle size and average neck to particle size ratio of individual modified microstructures as a function of the number of time steps*

Fig. 8-24 indicates that the total number of necks dropped drastically before a smooth derease took over at the time step of 20, while the average neck size experienced a significant increase before levelling off from 20 ts. It can be seen in Fig. 8-25 that the average particle size gradually increased with the augmented time steps. The average ratio of neck to particle size reached a maximum at 20 ts and continued to decrease as time steps went higher. When the trend of the elastic modulus change in Fig. 8-23 is reexamined by taking into account the above analysis, it is almost certain to confirm that the elastic modulus was largely influenced by the neck size at the early stage of microstructure evolution (time steps < 40). In other words, the rise of elastic modulus was mainly contributed by the increase of neck size. At some point its effectiveness was no longer sufficient to dominate the change of elastic modulus. As a result, the role was taken over by some other factors, which required further investigation and should also be applicable to the case of LSCF6428 film microstructures studied earlier.

### 8.6.4 Issues of "Heterogeneity" in the Porous Microstructures

As it is believed that larger particle neck size would always result in higher elastic modulus under the context of changing microstructure at constant porosity, the unexpected drop in the elastic modulus could possibly be related to factors that possibly were not directly related to the microscopic characteristics but rather on macro-scale features of the structure. One of the possible reasons we have considered to account for the drop of elastic modulus was the aggravated "heterogeneity", i.e. the degree of systematic or random variations of individual microstructural constituents [5], of the microstructure arising from the prolonged time steps of microstructure change. In other words, longer time of microstructure modification generates less homogeneous (i.e. more disordered) spatial distribution of components in the microstructure. This would result in a more localised property being measured, which can deviate significantly from the overall statistically effective property.

In order to minimise such a disadvantageous effect, a sufficiently large volume to be statistically representative of the porous material and to effectively include a sampling of all microstructural heterogeneities existed in the porous material [6]. This type of volume is defined as a representative volume element (denoted as RVE) of the material, which is an important notion for solving homogenisation problems in the mechanics and physics of periodic and random heterogeneous materials with a view to predicting their effective properties [7]. Any volume larger than the critical minimum RVE can be regarded as representative. However, a small enough RVE is preferable as it can save significant computation cost when dealing with very large datasets and the computation power is a concern.

The determination of the minimum RVE size of a non-uniform microstructure containing randomly distributed particles has been extensively studied in the past years, using various statistical-numerical analyses [6-12]. The general idea is to link the size of a RVE to the size of inclusions such as particles, based on the fact that the quantity of the macro-scale effective property under consideration is estimated to be consistent within a given precision (usually of relative error ~5%) over different volumes larger than the RVE. For example, Van Mier [12] in his study of concrete reported the RVE size to be approximately 7-8 times the largest inclusion or particle size. This would result in an *L/D* ratio of ~ 2 in a cubic sample.

As said, ideally, for a volume larger than the minimum RVE size, the overall property computed would be insensitive to the change of volume size, the sampling location and the number of samples. However, for real porous media the realisation of a minimum RVE would always depend on the precision wanted of the overall property. The precision includes the bias of the averaged values from the convergent value and the statistical fluctuation of the scattering values (i.e. standard deviation). The convergence of overall property can be achieved by increasing the volume size beyond RVE, provided that there is no constraint in computational power. Since generally there is no periodicity in real random media, the RVE can only be approached approximately on finite scales [10].

Ostoja-Starzewski [10] has demonstrated that the determination of RVE also depends on the investigated morphological or physical property, the contrast in the properties of the constituents, and their volume fractions. In our cases of both the LSCF6428 and the artificial microstructures, there are merely two phases (i.e. solid and pores). Therefore the volume fractions of particles/pores may be of importance.

On the other hand, there is currently not a direct quantitative indicator of the so-called "heterogeneity", which is often universally neglected when evaluating microstructure dependence of properties [5]. Nevertheless, for a non-uniform porous microstructure with a definite dimension, interpretations on the determination of volume fraction related parameters of solid and pores can be made to gain insight into it, instead of rough topological observation of the microstructure. The parameters considered here could be 2D porosity in the sequential image slices (or the standard deviation of it) and similarly the interception length distribution in the images.

Fig. 8-26 shows the 2D surface area ratio of pores over each individual image slice (i.e. 2D porosity of each image slice) in every microstructure. The red line in the figure represents the mean value of the 2D porosity for each microstructure which should be the same and equivalent to the total 3D porosity of each microstructure: 44.6 %. As found earlier for the evolution of the actual LSCF microstructures, the parts of curves swinging above or below the red line reflected a higher or lower portion of pores in the corresponding images, whereas less or more solid could be found. The amplitudes of the curves deviating from the red line also suggest an increasingly noticeable bias was produced with the increase of the number of time steps. This bias can be demonstrated as an evolution of standard deviation (SD) of the 2D porosity, or more precisely the ratio between it and average particle size, as shown in Fig. 8-27. It can be readily seen that both parameters increased with increasing time steps, which does suggest that microstructures with greater extent of heterogeneity were generated as the number of time steps increased.

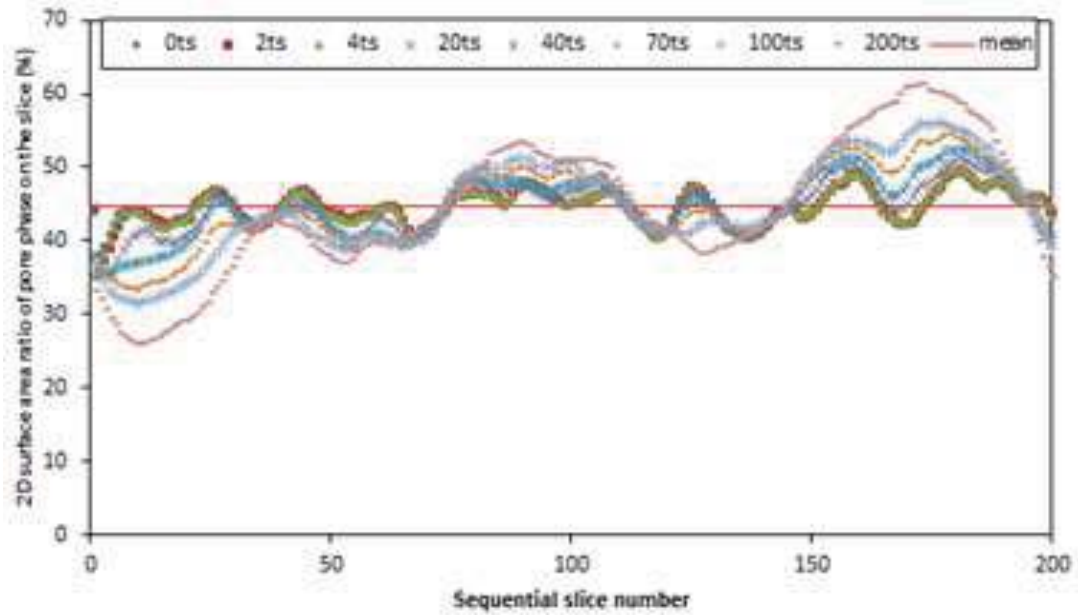

*Fig. 8-26 Evolution of 2D surface area ratio of pore phase (i.e. 2D porosity) on each slice for each microstructures*

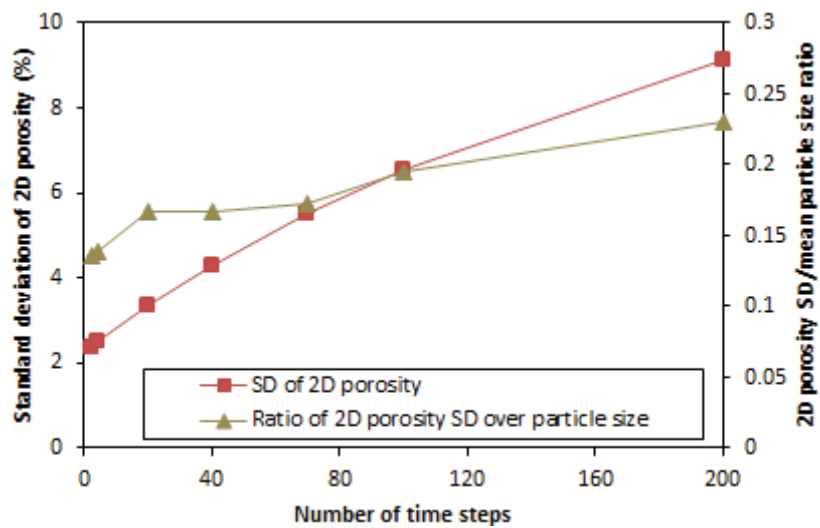

*Fig. 8-27 Standard deviation of 2D porosity in individual microstructures as a function of the number of time steps*

### 8.6.5 Determination of the Minimum RVE Size (i.e. the Minimum L/D Ratio)

As long as the results above have confirmed the existence of the aggravation in "heterogeneity" for the samples with prolonged time steps of microstructure change using the CA based method, the next step was to determine the size of the critical minimum RVE size, based on which the minimum $L/D$ ratio could also be estimated, where $L$ is the lateral length of the cubic volume and $D$ the mean particle diameter.

As previously explained, in order that the minimum RVE size could be determined, the desired property was estimated using different volume sizes. We started to look into the modified microstructure at 20 ts (volume size = $200^3$ pixel$^3$).

The original 20 ts microstructure was partitioned into different sets of equal size cubes: i.e. a set of $100^3$ pixel$^3$ cubes and the other set $50^3$ pixel$^3$. Examples of their 3D microstructures are shown in Fig. 8-28.

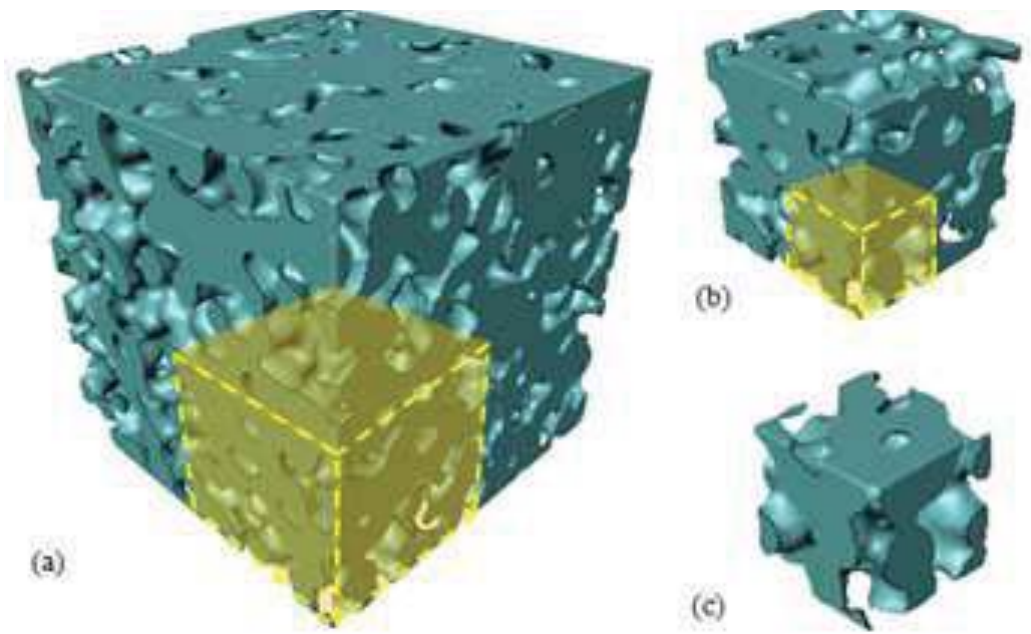

*Fig. 8-28 Examples of the two partitioning volumes: (a) original 20 ts microstructure (volume size = $200^3$ pixel$^3$), (b) $100^3$ pixel$^3$ volume, (c) $50^3$ pixel$^3$ volume. Yellow transparent boxes mark the partitioning volumes from larger volumes.*

Note that to save computational time, only 11 randomly chosen microstructures were analysed for $50^3$ pixel$^3$ microstructures. The partition was performed in a way that almost no overlapping region was allowed to ensure independent volume-based measurements of individual samples can be achieved.

The 2D surface area ratio of pore phase in the sequential slices, 3D porosity and the elastic modulus of individual microstructures within each set were measured and plotted in Fig. 8-29 and Fig. 8-30, respectively. Note that the properties of the original 20 ts microstructure were also compared (denoted as 20ts_orig in these figures).

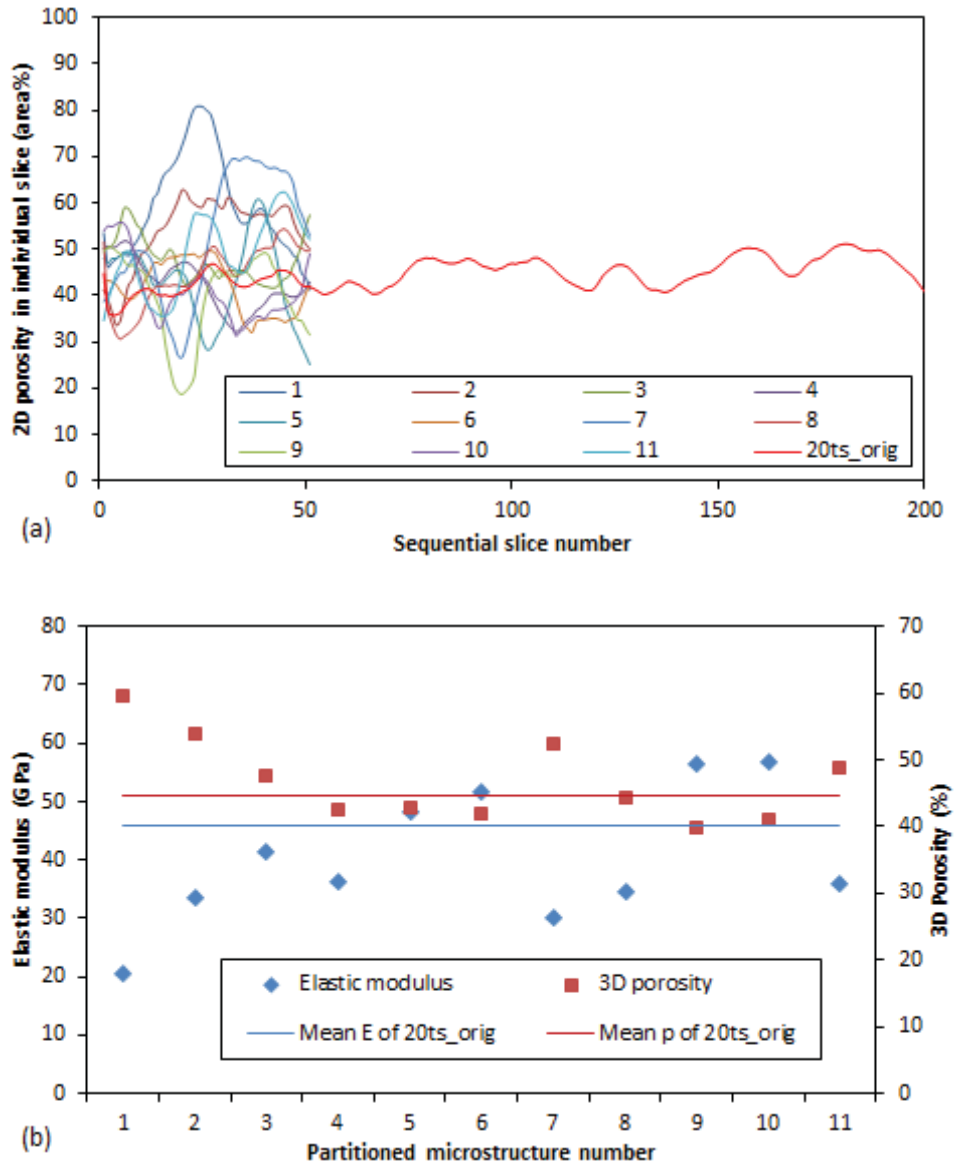

*Fig. 8-29 Results of (a) 2D porosity in sequential slices and (b) elastic modulus of individual microstructures of the $50^3$ pixel$^3$ volume set*

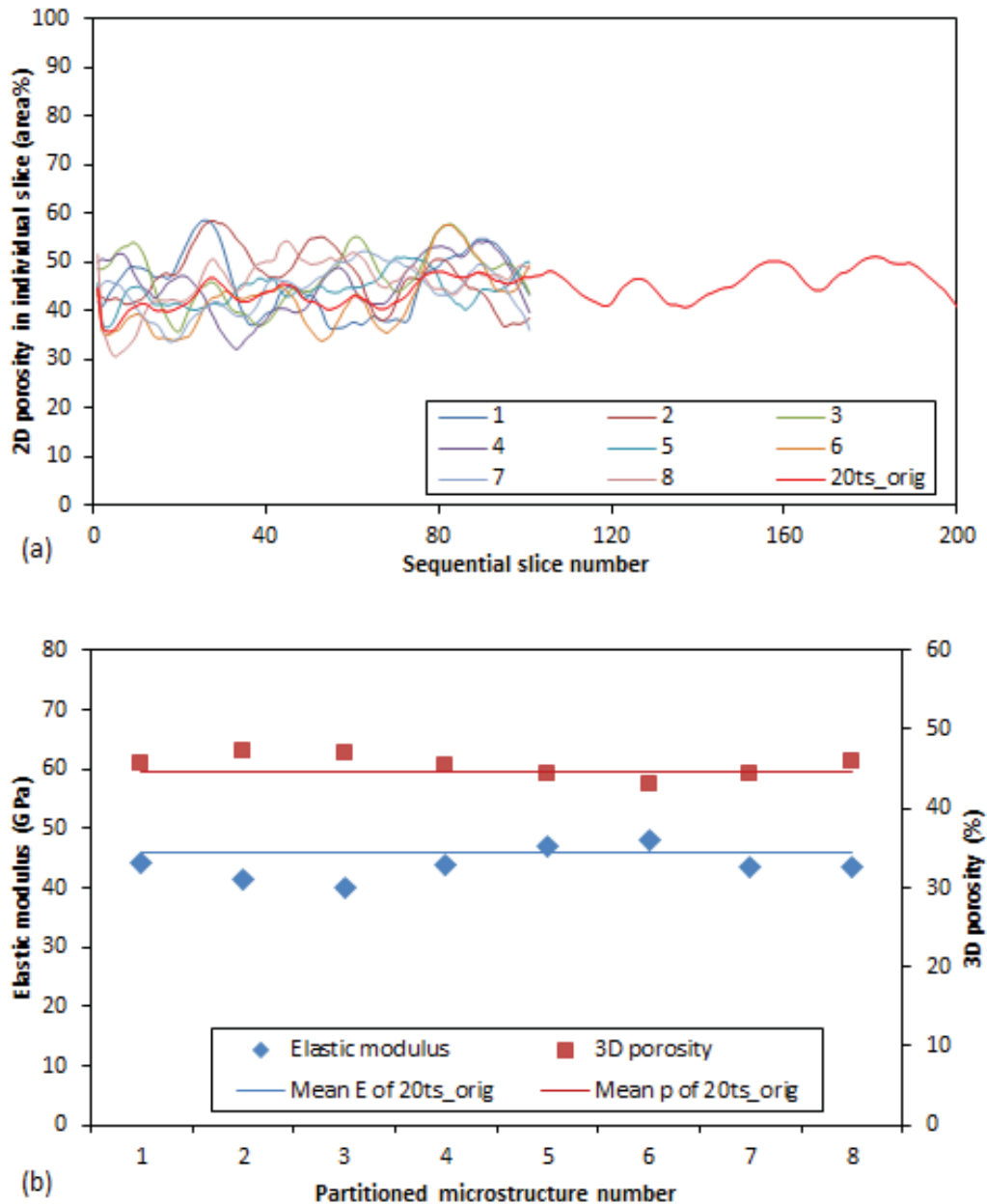

*Fig. 8-30 Results of (a) 2D porosity in sequential slices and (b) elastic modulus of individual microstructures of the $100^3$ pixel$^3$ volume set*

Significantly larger variation in the 2D porosity is found for $50^3$ pixel$^3$ microstructures in Fig. 8-29 (a) due to the smaller and thus less representative volume used, compared with that for $100^3$ pixel$^3$ samples as shown in Fig. 8-30 (a). Nevertheless, the original 20 ts microstructure tended to exhibit smoother fluctuation over the sequential slice range (i.e. curves shown in red). As a result, 3D porosities much larger than that of the original 20 ts microstructure were obtained for the $50^3$ pixel$^3$ volume set, as shown in Fig. 8-29 (b), which led to most of the estimated elastic moduli being much lower than that of the original 20 ts microstructure. On the contrary, in the case of $100^3$ pixel$^3$ volume set shown in Fig. 8-30 (b), merely subtle changes are observed in 3D porosities and elastic moduli, which are almost identical to that of the original 20 ts microstructure. From another perspective Fig. 8-29 (b) and Fig. 8-30 (b) both also confirm the importance of variable porosity on the determination of elastic modulus. The averaged values were further compared in Table 8-2.

*Table 8-2 Averaged estimation of 3D porosity and elastic modulus for different sets of volume*

| Volume size (pixel$^3$) | 3D porosity (vol %) | Elastic modulus (GPa) |
|---|---|---|
| $50^3$ | 46.8±6.3 | 40.5±11.6 |
| $100^3$ | 45.3±1.4 | 43.9±2.6 |
| $200^3$ | 44.6 | 46 |

As shown in the table above, despite fairly close 3D porosities obtained, a bias in the estimated mean elastic modulus values for smaller volumes can be observed, namely smaller volumes tended to yield smaller mean elastic moduli. In addition, smaller samples also induced much greater statistical fluctuations (i.e. standard deviations) in the results. Nevertheless, a relatively slight error (approximately 6% relative error) was generated by using microstructures consisting of a volume size of $100^3$ pixel$^3$, for which the particle diameter was 20 pixels, suggesting that the microstructure size was statistically representative of the material for an *L/D* ratio of greater than 5.

Finally, by taking into account the inflection point (i.e. at 40 ts) of the elastic modulus evolution (Fig. 8-23) and the average particle size estimated in Fig. 8-25 (= 25.7 pixel at 40 ts), it can be concluded that the minimum *L/D* ratio to be considered as a RVE of the entire material in terms of elastic modulus is *L/D* = 200/25.7 = 7.8. This value is fairly close to the literature data of studying SOFC electrode performance: Cai *et al.* [13] suggested an *L/D* of RVE = 7.5 for Ni-YSZ anode structures at 21.5% porosity when evaluating their TPB percolation and a minimum *L/D* = 7 was reported by Joos *et al.* [14] for LSCF cathode structures containing 30-50% porosity when calculating the their area specific resistance. This way, sufficiently accurate simulation with relative error falling below 5% might be expected.

It is speculated that there should be a bias and standard deviation (which were supposed to be much smaller than in the case of $100^3$ microstructure) for the modulus of the $200^3$ microstructure as well, although they had not been estimated due to lack of sample number (which is only one in the current study). However, this estimated modulus value is supposed to be close to the asymptotic limit of the overall effective modulus, as in this case the ratio of $L/D$ reached 10 (= 200/20), which was higher than the minimum one (= 7.8) and thus more accurate results were obtained. By plotting the minimum $L$ size required (i.e. minimum RVE lateral size = 7.8 times average particle size) as a function of the number of evolution time steps, as shown in one can estimate that approximately > $300^3$ pixel$^3$ volume size is needed of a microstructure at 200 time steps of evolution to be statistically representative for much accurate elastic modulus simulation.

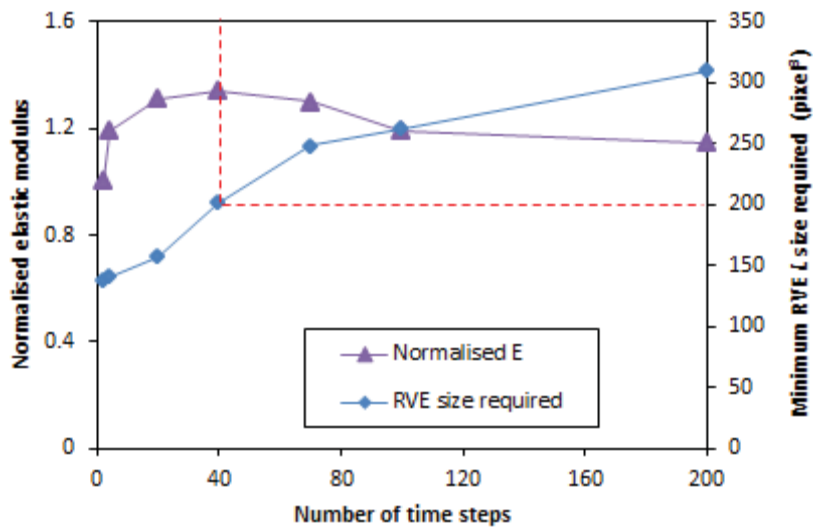

*Fig. 8-31 Minimum RVE lateral size required as a function of number of time steps, for a better understanding, the normalised modulus is also plotted. Red dashed lines indicate the threshold values at 40 ts.*

In addition, we also measured the total neck surface area density, which was calculated as the ratio of the total surface area of all necks in a microstructure relative to the volume of the microstructure. The result is shown in Fig. 8-32 and the normalised elastic modulus is shown again for correlating with the neck area density. The observation indicates a direct dependence is most likely, particularly for the early rising stage where the volume sizes were larger than the minimum RVE size required. The decline of the elastic modulus at the later stage might be attributable to the much dramatic drop of the neck area density. However the most fundamental reason to account for the decrease in modulus was the fact that the microstructure volumes were no longer representative as they became increasingly smaller than the minimum RVE size required. Nevertheless, the figure from another perspective further confirms the importance of the overall neck size to the elastic modulus during the early microstructure modification period before the insufficient volumes became a key to rule the modulus change. As a result, the moduli calculated by simulation for the volumes larger than the minimum RVE size were considered statistically equivalent to the microstructures' constitutive property.

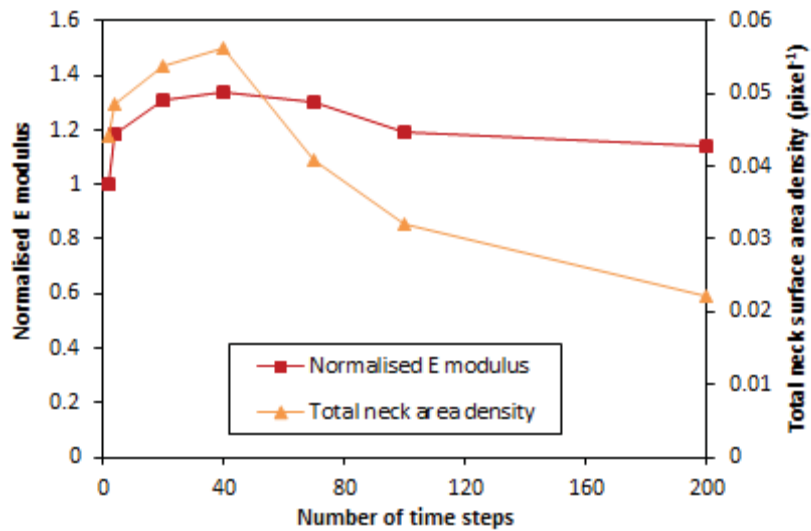

*Fig. 8-32 The total neck surface area density as a function of the number of time steps. The normalised elastic modulus is also plotted to show correlation.*

## 8.7 Applying the Minimum L/D Ratio to LSCF6428 Microstructures

As far as the effect of the modified microstructures of LSCF6428 was concerned, the actual ratio of *L/D* was calculated for each individual microstructure at different time steps. The results were plotted in Fig. 8-33 along with the minimum *L/D* ratio (shown as a red line) required for a volume to be representative.

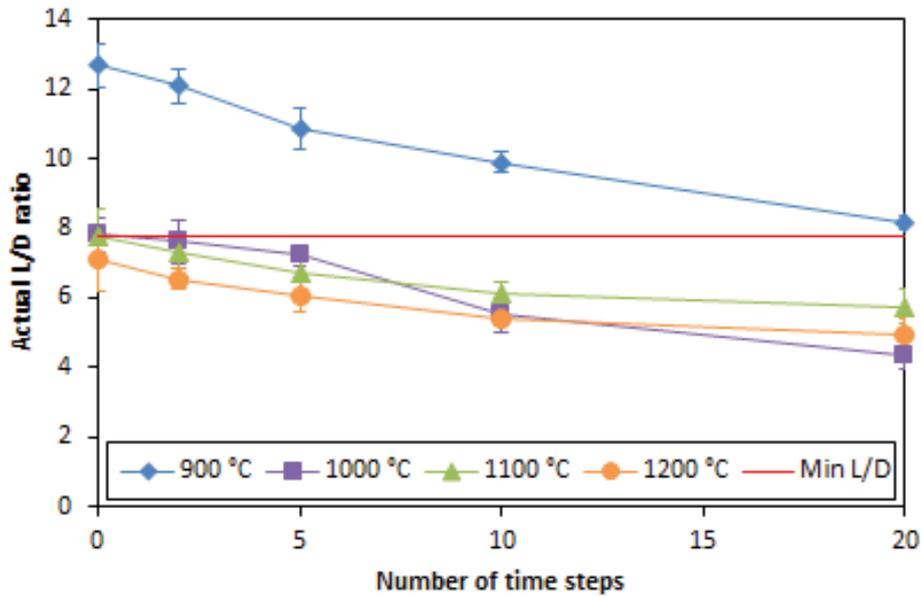

*Fig. 8-33 Comparison of the actual L/D ratios for all sets of microstructures and the previously derived Minimum ratio (Min L/D = 7.8, shown as a red line).*

The figure above shows clearly that the ratios for each individual microstructure set experienced gradual drop as the time steps increased, implying that generally coarser microstructures require larger RVE sizes. It is also noticeable that the ratios for the 900 °C microstructures are all above the critical minimum *L/D* ratio (= 7.8), whereas for the other three sample sets (1000, 1100 and 1200 °C) the values were all below 7.8 and became increasingly small over the period of modification time steps, despite the close values to 7.8 found for the initial microstructures at 0 ts. The above observation suggests that the volume sizes of the 900 °C microstructures were all larger than the minimum RVE required, so that the properties derived could be regarded statistically representative and reliable. In this case the increase of neck size (or the overall neck surface area to sample volume ratio) contributed to the raise of elastic modulus. On the contrary, the *L/D* ratios below 7.8 for the other three sets of microstructures after modification indicate insufficient volume sizes for simulation, namely smaller than the minimum RVE size required so that the "heterogeneity" problem with the microstructures appeared to be increasingly predominant, which as a result led to the increase in the elastic modulus estimation. Nevertheless, as the initial 1000, 1100 and 1200 °C microstructures possessed *L/D* ratios close enough to 7.8, their volumes remained fairly representative, so that the validity of the FEM-derived elastic moduli of these initial microstructures was little affected.

The analysis above once again confirms that under the condition of sample volume being larger than the minimum RVE size, the elastic modulus change of the 900 °C microstructures (see Fig. 8-18) in the course of CA based microstructure modification behaved as in the early stage (going up) of the curve shown in Fig. 8-31. On the other hand, for the other sets of microstructures, their elastic modulus changes behaved more like in the middle/later stage (going down) of that curve, due to the volume sizes being less than the minimum RVE size, particularly at longer time steps.

The ratios of the total surface area of necks relative to the corresponding sample volumes were also estimated, as shown in Fig. 8-34.

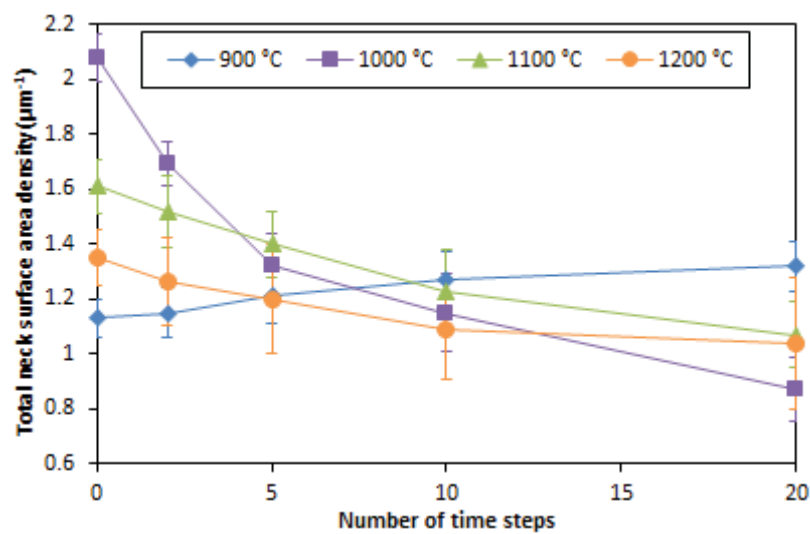

*Fig. 8-34 Comparison of the total neck surface area densities as a function of number of time steps*

The figure shows increased ratios with the increase of time steps for 900 °C microstructures, whereas reduction was observed for the other three sets of microstructures. Although the volume sizes of the 1000, 1100 and 1200 °C microstructures were insufficient to be representative at relatively higher time steps, the figure above suggests that the contribution of the variation of total neck surface area "density" (i.e. total neck surface area per unit volume) to the elastic modulus change was the most likely.

A number of conclusions can be drawn from this study,

(1) The elastic moduli of the porous LSCF films produced in this study were calculated using FEM based on the 3D microstructures reconstructed using FIB/SEM tomography, and good agreement was found with the previously measured nanoindentation results.

(2) Porosity determined the elastic modulus of the porous microstructures to a large extent. Generally larger porosity led to smaller elastic modulus. On the other hand, it was shown that porosity alone was not sufficient to account for the mechanical properties of the partially sintered material.

(3) CA based method was used in the study of both the actual porous LSCF microstructures and the artificially created microstructures, to evaluate how factors other than porosity affected the elastic modulus and the Poisson's ratio when microstructures were modified at constant porosity. Results revealed fairly identical dependence of these two elastic properties on the microstructure modification.

(4) Since FEM generally took a definite volume of material for consideration, the heterogeneity could always be an issue in determining the statistically effective properties. It was found that there existed a critical minimum RVE size (or $L/D$ ratio) in terms of determining a statistically representative elastic modulus. The heterogeneity of the porous microstructures required the volume sizes not to be smaller than the minimum RVE size in the numerical simulations for calculating statistically representative elastic moduli.

(5) The minimum $L/D$ ratio was found to be 7.8 based on the analysis of the artificially created microstructure evolution. The elastic modulus evolution on the artificially created mono-size microstructures followed the following two patterns continuously with the increase of microstructure modification time steps.

(a) When the volume stayed statistically representative (i.e. >= minimum RVE or $L/D$ >= 7.8 in the current study), the neck size (or more specifically the total neck surface area per unit volume) governed the elastic modulus of the microstructure, i.e. volumes with larger neck size tended to yield greater elastic moduli. This means that ideally, CA based modification at constant porosity of such volumes would result in increasing elastic modulus computed using FEM as time steps increased. The results could be regarded as statistically representative of each individual microstructure's constitutive properties.

(b) When the volume was not statistically representative (i.e. < minimum RVE or $L/D$ < 7.8 in the current study), heterogeneity took over the role and ruled the elastic modulus of the microstructure, i.e. poorer heterogeneity resulted in weak elastic modulus. This means that, CA based modification at constant porosity of such volumes would result in decreasing elastic modulus computed using FEM as time steps increased.

(6) For the actual porous LSCF microstructures generated by sintering at 900-1200 ˚C which were then modified at constant porosity, the minimum RVE size (or the minimum $L/D$ ratio) theory was proved to be applicable to the different sets of porous LSCF microstructures.

Their behaviours of the changes in elastic modulus and the Poisson's ratio experienced either the first increase (case (a)) or the later decrease (case (b)) due to the modified microstructure volumes being continuously (or no longer) representative at longer time steps. Because of the contribution of particle neck size and the RVE-related issues in numerical modelling apart from porosity, a given elastic modulus or Poisson's ratio does not necessarily correspond to a particular microstructure.

## *Summary*

The aim of this Chapter was to investigate the dependence of the elastic modulus of the porous 3D microstructure on factors other than porosity. To achieve this, the actual as-sintered 3D microstructures collected were numerically modified at constant porosity using a cellular automaton based method, which only involved coarsening without densification of the solid phase. Comparisons were made of the important microstructural parameters measured between the as-sintered microstructures and the modified ones. The negligible change in tortuosity was thought to be attributed to the self-similarity of the microstructures before and after modification. The quantification of other important parameters (such as particle size and neck size) suggested that an increase in elastic modulus would be expected after microstructure modification. However, FEM-derived results showed surprisingly noticeable variability (increase and/or decrease) in elastic moduli and the Poisson's ratios for different specimen sintering temperatures and over different microstructure evolution time steps.

An in-depth study was carried out to address the unexpected results. A simple artificial microstructure consisting of mono-size spheres representing solid phase was created and evolved using CA based method in the similar way to the as-sintered LSCF6428 microstructures as carried out earlier. Further analyses suggested that at constant porosity, the neck surface area density and the heterogeneity of microstructures (with a critical minimum ratio of specimen side length to average particle size determined) were responsible for such variations. Finally a number of key criteria were proposed to explain and predict the variation in elastic modulus for porous materials.

## *Chapter 8 References*

1.  Wang X, Atkinson A: **Cellular Automata Modelling of Microstructure Evolution of Ni Cermet Anode**. *ECS Transactions* 2013, **57**(1):2465-2473.
2.  Hardy D, Green DJ: **Mechanical properties of a partially sintered alumina**. *Journal of the European Ceramic Society* 1995, **15**(8):769-775.
3.  Rhazaoui K: **SOFC microstructure and performance modeling**. *PhD Thesis.* London, UK: Imperial College London; 2014.
4.  Roberts AP, Garboczi EJ: **Elastic Properties of Model Porous Ceramics**. *Journal of the American Ceramic Society* 2000, **83**(12):3041-3048.
5.  Rice RW: **Porosity of Ceramics: Properties and Applications**: CRC Press; 1998.
6.  Kanit T, Forest S, Galliet I, Mounoury V, Jeulin D: **Determination of the size of the representative volume element for random composites: statistical and numerical approach**. *International Journal of Solids and Structures* 2003, **40**(13):3647-3679.
7.  Pelissou C, Baccou J, Monerie Y, Perales F: **Determination of the size of the representative volume element for random quasi-brittle composites**. *International Journal of Solids and Structures* 2009, **46**(14):2842-2855.
8.  Kari S, Berger H, Rodriguez-Ramos R, Gabbert U: **Computational evaluation of effective material properties of composites reinforced by randomly distributed spherical particles**. *Composite structures* 2007, **77**(2):223-231.
9.  Gitman I, Askes H, Sluys L: **Representative volume: existence and size determination**. *Engineering Fracture Mechanics* 2007, **74**(16):2518-2534.
10. Ostoja-Starzewski M: **Material spatial randomness: From statistical to representative volume element**. *Probabilistic Engineering Mechanics* 2006, **21**(2):112-132.
11. Drugan W, Willis J: **A micromechanics-based nonlocal constitutive equation and estimates of representative volume element size for elastic composites**. *Journal of the Mechanics and Physics of Solids* 1996, **44**(4):497-524.
12. van Mier JG: **Microstructural effects on fracture scaling in concrete, rock and ice**. In: *IUTAM symposium on scaling laws in ice mechanics and ice dynamics: 2001*: Springer; 2001: 171-182.
13. Cai Q, Adjiman CS, Brandon NP: **Modelling the 3D microstructure and performance of solid oxide fuel cell electrodes: Computational parameters**. *Electrochimica Acta* 2011, **56**(16):5804-5814.
14. Joos J, Ender M, Carraro T, Weber A, Ivers-Tiffée E: **Representative volume element size for accurate solid oxide fuel cell cathode reconstructions from focused ion beam tomography data**. *Electrochimica Acta* 2012, **82**(0):268-276.

# 9 Conclusions and Future Work

## 9.1 Conclusions

Mixed ionic-electronic conductive perovskite LSCF6428 has been considered one of the most attractive cathode materials for IT-SOFC applications. The motivation of the present research was to investigate the mechanical properties (i.e. elastic modulus, hardness and fracture toughness) of LSCF6428 cathode films and bulk samples fabricated by varying high temperature sintering using nanoindentation experiments and analyses. The results were interpreted in terms of 3D microstructural parameters obtained from 3D reconstruction based on FIB/SEM tomography which were used to simulate the elastic properties (elastic modulus and Poisson's ratio) using finite element modelling.

Firstly LSCF6428 and CGO bulk samples were fabricated from raw powder following conventional ceramic processing routes. LSCF6428 films were prepared by deposition of the inks onto LSCF6428 and CGO dense substrates. Important physical properties such as density, porosity, and 2D microstructural morphologies of these as-fabricated samples were investigated. Abnormally large agglomerates were found in the commercially provided LSCF6428 powder which was therefore subjected to further mechanical milling to generate a powder with finer particle size suitable to be used in the study, particularly to make bulk samples rather than inks for film making.

A readymade LSCF6428 ink was chosen for film deposition due to its finer and homogeneous particle size distribution. The thickness of the sintered LSCF6428 films was found to reach a maximum as a function of the number of screen-printing passes. More importantly, we have shown that sintered porous thin LSCF6428 films fabricated using a typical screen-printing LSCF6428 ink were unacceptably rough and extensively cracked, which caused extreme inconsistency and errors in attempts to measure the film elastic modulus by nanoindentation. Various processing parameters were investigated in order to eliminate these defects. Different film deposition methods and thermal treatments showed little effect on the prevention of the defect formation. Furthermore the cracks were shown not to arise from thermal expansion mismatch between film and substrate and it was concluded that the defects initiated during drying of the ink and were then made more severe by the sintering process. Thus, sintered porous thin LSCF6428 cathode films were successfully fabricated without any crack or surface asperities by using a much less viscous ink. Results showed that increasing the amount of terpineol solvent added to the as-received ink improved the wetting and self-levelling of the ink upon deposition and thus a more homogeneous packing of LSCF6428 particles in the films was achieved. As a result there was more homogeneous shrinkage in the film during the drying and thermal treatments that followed and lower local stresses in the films. With the defect-free films, consistent and reliable elastic modulus values of the films could then be obtained using nanoindentation.

Consequently, the measurements of important mechanical properties including elastic modulus, hardness and fracture toughness of LSCF6428 in both bulk and film forms were investigated after they were sintered at 900-1200 °C.

For elastic modulus and hardness measurements, a spherical indenter was used since its contact area was much greater than the scale of the porous microstructure. The elastic modulus of the bulk samples was found to increase from 34-174 GPa and hardness from 0.64-5.3 GPa as the porosity decreased from 45-5% after sintering at 900-1200 °C. The results were found to be sensitive to surface roughness at shallow indentation depth, while stable values were obtained at larger depths. Densification under the indenter was found to have little influence on the measured elastic modulus. However, the elastic modulus measured by indentation of the nominally dense bulk specimens sintered at 1200 °C, 174 GPa, was significantly greater than that measured by impulse excitation, 147 GPa. This was shown to possibly be due to residual porosity of approximately 5% in the nominally dense specimens which influenced the long range elastic modulus measured by impulse excitation. Therefore the higher value is characteristic of fully dense material. No evidence of a ferroelastic contribution to the load-deflection indentation response was found.

Crack-free LSCF6428 films of acceptable surface roughness for indentation were also prepared by sintering at 900-1200 °C. The porosities of the films were in the range 15-47%. The influence of surface roughness at shallow depth was due to the granular nature of the porous films, while the influence of the substrate at greater depth was due to formation of a "plastic" zone of crushed, higher density, material under the indent which touched the substrate if the indentation was too deep. The experiments in this study showed that for this type of porous film and using a spherical indenter, relatively reliable measurements of the true properties of the films were obtained by data extrapolation provided that the ratio of indentation depth to film thickness was in the range 0.1 (which was approximately equivalent to 5 times the roughness) to 0.2, acknowledging the fact that the substrate effect persisted but at a lower level in this depth range. Thus it was found that the elastic modulus of the films increased from 32 to 121 GPa and hardness from 0.37 to 1.97 GPa as the sintering temperature increased from 900 to 1200 °C. Comparison with bulk specimens clearly showed that the porous films behaved very similarly to the porous bulk specimens in terms of the dependency of elastic modulus on porosity. Microstructures obtained by FIB/SEM slice and view of the film specimens, after indentation revealed the nature of the "plastic" deformation zone and how this affected the measurement of elastic modulus when it reached the substrate.

Based on the crack length measurements from micrographs obtained for Berkovich-indented samples, the fracture toughness of bulk porous LSCF6428 was determined to increase from 0.51 to 0.99 MPa·m$^{1/2}$ after sintering at increasing temperatures from 900 to 1200 °C. The surface and subsurface crack morphologies at the indents were investigated using SEM and FIB/SEM slice and view. This revealed the presence of radial and half-penny crack systems induced by Berkovich indentation and justified the applicability of the conventional bulk fracture toughness expression in the current porous material. Similar experiments on films suggested that the generation of distinguishable indentation-induced cracks could be very difficult for porous thin films sintered at temperatures below 1200 °C, and cracks were only detectable in the films sintered at 1200 °C, which gave a toughness of 0.16 MPa·m$^{1/2}$, much smaller than that of bulk material with even larger porosity.

In order to further correlate the elastic behaviour of these as-sintered porous LSCF6428 films with their microstructural characteristics, FIB/SEM tomography was used to reconstruct digitally their actual 3D microstructures. Some issues and artefacts originating from the image acquisition process, such as the curtain effect, ion charging, low phase contrast, the shadow effect, redeposition of sputtered material and image drift, were investigated in order to evaluate their influence on the quality of the images acquired. Measures were subsequently taken to reduce these arising side effects. High resolution and high quality images thus obtained were then used to implement the 3D microstructure reconstruction, followed by volumetric mesh generation to create FE models. Finite element modelling was carried out on the digitised and meshed microstructures to calculate the elastic modulus in three orthogonal directions. The FEM results showed excellent agreement with the data produced using nanoindentation, despite the existence of some anisotropy in modulus discerned in the models. The anisotropy was attributable to constrained sintering, which eventually resulted in larger stiffness normal to the film plane and smaller isotropic modulus in the film plane. Comprehensive quantification of the microstructural parameters (such as porosity, volume specific surface area, phase tortuosity, distribution of particle and pore size and orientation, and particle neck area) was carried out. The actual 3D microstructures, and artificial microstructures formed of contacting spherical particles, were numerically modified at constant porosity by applying a cellular automaton algorithm that simulated coarsening by vapour phase transport. This enabled the influence of factors other than porosity on elastic modulus to be evaluated. The results on the artificial microstructures showed that neck growth in the initial stage of coarsening increased with time. Further analyses suggested that at constant porosity, the neck size and the heterogeneity of microstructures (with a critical minimum ratio of volume side length to average particle size determined) were responsible for the different changes in elastic properties of all the coarsened actual LSCF6428 microstructures,

## *9.2  Future Work*

The experiments revealed that the films and bulk material had well-defined hardness controlled by crushing of the porous structure. However, there is currently no theory available to interpret hardness of porous materials quantitatively. Therefore, the hardness behaviours and data found in this study must be further explored and interpreted.

On the other hand, finite element modelling would also be a part of the future work to assist a better understanding of fracture behaviours in the LSCF6428 films and bulk samples.

The study involved a number of FIB/SEM tomographic 3D microstructural reconstructions, but relatively small volumes of interest were processed. Although results showed that the volumes were more or less representative, larger volumes were supposed to generate more accurate results. However, the compromise between accuracy and computation cost should always be considered. Further work is required to provide guidelines for optimising the trade-off between these conflicting goals.

In this study, numerical methods have been applied to vary the characteristics of the original 3D reconstructed microstructures by simulation of coarsening and evaluate the effect on elastic modulus. However, a more detailed and quantitative correlation should be established to account for the decrease of elastic modulus when the criterion of representative volume element size is not met. Additionally, a more realistic experimental approach will need to be implemented to validate the models. In other words, the modification of real microstructures should be investigated, for example, by varying the microstructures with powder characteristics or by adding pore-forming additives. Thereby, the development of more predictable and controllable microstructures can thus be achieved for further understanding and optimisation of mechanical properties.

Future work is also required to test the electrochemical performance of cells fabricated consisting of the LSCF6428 cathode with desired microstructural features. The balance of trade-off between the electrochemical performance and the ability to withstand mechanical constraints must be achieved.

# *List of Publications*

## *Journal Papers*

1. <u>Chen Z</u>, Wang X, Bhakhri V, Giuliani F, Atkinson A: **Nanoindentation of porous bulk and thin films of La0.6Sr0.4Co0.2Fe0.8O3−δ**. *Acta Materialia* 2013, **61**(15):5720-5734.

2. Wang X, <u>Chen Z</u>, Atkinson A: **Crack formation in ceramic films used in solid oxide fuel cells**. *Journal of the European Ceramic Society* 2013, **33**(13-14):2539-2547.

3. <u>Chen Z</u>, Wang X, Giuliani F, Atkinson A: **Surface quality improvement of porous thin films suitable for nanoindentation**. *Ceramics International* 2014, **40**(3):3913-3923

4. Wang X, He F, <u>Chen Z</u>, Atkinson A: **Porous LSCF/dense 3YSZ interface fracture toughness measured by single cantilever beam wedge test**. *Journal of the European Ceramic Society* 2014, **34**(10):2351-2361

5. Tariq F, Yufit V, Eastwood DS, Merla Y, Biton M, Wu Billy, <u>Chen Z</u>, Freedman K, Offer G, Peled E, Lee P D, Golodnitsky D, Brandon N: **In-Operando X-ray Tomography Study of Lithiation Induced Delamination of Si Based Anodes for Lithium-Ion Batteries**. *ECS Electrochemistry Letters* 2014, **3** (7): A76-A78

6. <u>Chen Z</u>, Wang X, Giuliani F, Atkinson A: **Analyses of microstructural and elastic properties of porous SOFC cathodes based on focused ion beam tomography**. *Journal of Power Sources* 2015, 273: 486-494

7. <u>Chen Z</u>, Wang X, Giuliani F, Atkinson A: **Determination of elastic moduli for porous SOFC cathode films using nanoindentation and FEM**. *Advances in Bioceramics and Porous Ceramics VII: Ceramic Engineering and Science Proceedings* 2015, 35 (5): 111-120

8. <u>Chen Z</u>, Wang X, Giuliani F, Atkinson A: **Fracture toughness of porous material of LSCF in bulk and film forms**. *Journal of American Ceramic Society*, In press.

9. <u>Chen Z</u>, Wang X, Giuliani F, Atkinson A: **Microstructural characteristics and elastic modulus of porous solids**. *Acta Materialia*, In press.

## *Conference Presentations*

1. <u>Chen Z</u>, Wang X, Bhakhri V, Giuliani F, Atkinson A: **Elastic modulus and 3D microstructures of porous LSCF films** *(Oral presentation)*, 10th European SOFC Forum, 26-29 June 2012, Lucerne, Switzerland.

2. <u>Chen Z</u>, Wang X, Giuliani F, Atkinson A: **Characterisation of microstructures and Young's modulus for LSCF cathodes** *(Poster presentation)*, Summer Research Symposium, 13 July 2012, Graduate School, Imperial College London, UK.

3. <u>Chen Z</u>, Wang X, Giuliani F, Atkinson A: **Nanoindentation and modelling on LSCF films** *(Oral presentation)*, UK 1 Day Research Meeting on Advanced Ceramics, 3 May 2013, Imperial College London, UK

4. <u>Chen Z</u>, Wang X, Giuliani F, Atkinson A: **Evaluation of porosity-dependent Young's modulus of La0.6Sr0.4Co0.2Fe0.8O3-δ films: experiments and modelling** *(Oral presentation)*, 13th International Conference of the European Ceramic Society, 23-26 June 2013, Limoges, France.

5. <u>Chen Z</u>, Wang X, Giuliani F, Atkinson A: **Determination of elastic moduli for porous ceramic films using nanoindentation and FEM** *(Oral presentation)*, 38th International Conference on Advanced Ceramics and Composites, 26-31 January 2014, Daytona Beach, USA.